\documentclass[english,aps,pre,reprint,superscriptaddress,a4paper]{quantumarticle}
\pdfoutput=1

\usepackage[T1]{fontenc}
\usepackage[latin9]{inputenc}
\usepackage{color}
\usepackage{babel}
\usepackage{amsmath}
\usepackage{amssymb}
\usepackage{graphicx}
\usepackage{float}
\usepackage[unicode=true,
 bookmarks=true,bookmarksnumbered=false,bookmarksopen=false,
 breaklinks=false,pdfborder={0 0 0},pdfborderstyle={},backref=false,colorlinks=true]
 {hyperref}
\hypersetup{
 allcolors=violet}

\usepackage{xkeyval}
\usepackage{etoolbox}
\usepackage{geometry}
\usepackage{xcolor}
\usepackage{fancyhdr}
\usepackage{tikz}
\usepackage{hyperref}
\usepackage{ltxgrid}
\usepackage{ltxcmds} 

\usepackage[numbers,sort&compress]{natbib}

\newcommand*\LyXThinSpace{\,\hspace{0pt}}

\usepackage{braket}
\usepackage{dsfont}

\begin{document}
\title{Catalytic transformations with finite-size environments: applications
to cooling and thermometry}
\author{Ivan Henao}
\email{ivan.henao@mail.huji.ac.il }

\author{Raam Uzdin}
\email{raam@mail.huji.ac.il}

\affiliation{Fritz Haber Research Center for Molecular Dynamics,Institute of Chemistry,
The Hebrew University of Jerusalem, Jerusalem 9190401, Israel}
\begin{abstract}
The laws of thermodynamics are usually formulated under the assumption
of infinitely large environments. While this idealization facilitates
theoretical treatments, real physical systems are always finite and
their interaction range is limited. These constraints have consequences
for important tasks such as cooling, not directly captured by the
second law of thermodynamics. Here, we study catalytic transformations
that cannot be achieved when a system exclusively interacts with a
finite environment. Our core result consists of constructive conditions
for these transformations, which include the corresponding global
unitary operation and the explicit states of all the systems involved.
From this result we present various findings regarding the use of
catalysts for cooling. First, we show that catalytic cooling is always
possible if the dimension of the catalyst is sufficiently large. In
particular, the cooling of a qubit using a hot qubit can be maximized
with a catalyst as small as a three-level system. We also identify
catalytic enhancements for tasks whose implementation is possible
without a catalyst. For example, we find that in a multiqubit setup
catalytic cooling based on a three-body interaction outperforms standard
(non-catalytic) cooling using higher order interactions. Another advantage
is illustrated in a thermometry scenario, where a qubit is employed
to probe the temperature of the environment. In this case, we show
that a catalyst allows to surpass the optimal temperature estimation
attained only with the probe. 
\end{abstract}
\maketitle

\section{Introduction}

In the field of Chemistry, catalysts are substances that can be used
to assist a chemical reaction without being consumed in the process.
This simple but powerful principle has also found applications in
areas of quantum information \citep{1catalytic-entanglement,key-2catal-majorization,key-3Mathematical-structure-of-entanglement-catalysis,key-4Catalytic-transformations-for-bipartite-pure-states,key-5Necessary-conditions-for-entanglement-catalysts,key-6Catalytic-Coherence,key-7Catalytic-coherence-transformations,key-8RTs-with-catalysts,key-8.1Muller-unitarity,key-8.2Rethinasamy,key-8.3Boes-catal-randomness,key-8.4Renner-catal-decoupling,key-9asymmetry-with-correlated-catalysts,key-9.1Wilming-entropy-and-rev-catalysis}
and quantum thermodynamics \citep{key-10second-laws,11Limits-to-catalysis-in-quantum-thermodynamics,12Oppenheim-passivity,13Third-Law-of-Thermodynamics-as-a-Single-Inequality,14Muller-catalysts,15universal-cataysts,15.1Lost-stochastic-independence,15.2Bypassing-FTs,15.3Karen,15.4Sagawa},
where catalysts are quantum systems that enable the implementation
of otherwise impossible transformations. For example, transformations
that are forbidden under local operations and classical communication
(LOCC) become possible once a suitable entangled state is employed
as catalyst \citep{1catalytic-entanglement}. Regarding the technical
aspect, catalytic transformations have often been addressed using
the concept of ``catalytic majorization'' \citep{key-2catal-majorization,key-3Mathematical-structure-of-entanglement-catalysis,key-4Catalytic-transformations-for-bipartite-pure-states}
and related extensions \citep{14Muller-catalysts,15.5Muller-and-Pastena-Major}.
A distinctive feature of this approach is that it provides general
conditions for the existence of a catalyst state that enables the
transformation. However, typically such a state and the corresponding
implementation are not explicitly given. 

In quantum thermodynamics, it has been shown that catalysts extend
the set of state transitions that a system can undergo in the presence
of a thermal environment \citep{key-10second-laws,14Muller-catalysts}.
These transformations are performed through global unitaries that
couple the environment with the rest of the system, and preserve the
energy of the total setup. Originally introduced without the inclusion
of catalysts, such conditions define maps on the system known as ``thermal
operations'' \citep{16RT-Horodecki-and-Oppenheim,17RT-Brandao,18Lostaglio-coherence-RT,20Korzekwa-work-and-coherence,21Lostaglio-element-TOs,22Lostaglio-review,23quantum-thermo-review,24Anders-Quant-Therm}.
Although a few studies have addressed thermal operations with finite
environments \citep{21Lostaglio-element-TOs,24.1Alhambra-HBAC,24.2Muller-finite-baths},
the most general transformations derived within this framework rely
on the possibility of interactions with arbitrarily large baths (determined
by arbitrary bath Hamiltonians)\citep{16RT-Horodecki-and-Oppenheim,22Lostaglio-review}
\textemdash{} an assumption that has also been adopted in the case
of catalytic thermal operations \citep{key-10second-laws,13Third-Law-of-Thermodynamics-as-a-Single-Inequality,14Muller-catalysts,15universal-cataysts,15.4Sagawa}.
Here, we consider catalytic transformations where the main system
interacts with a catalyst and a finite environment. The transformations
result from the application of non-energy preserving unitaries on
the total system, and the enviroment may start in a generic state.
Moreover, they are \textit{explicit},\textit{ }in the sense that explicit
unitaries and the corresponding catalyst state are obtained. From
a thermodynamic viewpoint, the main motivation is the characterization
of conditions to catalytically circumvent cooling limitations due
to the finite character of the environment. 

The interest in the formalization and quantification of the fundamental
limits for cooling has seen a resurgence in the last years \citep{34Paz-review-cooling,34.1Challeng-unat-principle}.
One approach is to cast these limits as bounds on the duration of
continuous \citep{35Masanes-Oppenh-third-law,35.1Kosloff-third-law}
or discrete cooling processes \citep{21Lostaglio-element-TOs,35.2polariz-and-HBAC-with-n-steps,35.3limit-HBAC},
in the spirit of the celebrated unattainability principle \citep{34Paz-review-cooling,36Paz-fund-limits-for-cooling}.
Alternatively, bounds for maximum cooling have been derived for different
kinds of refrigerators operating in the steady-state regime \citep{37Huber-cooling-bound1}
and heat bath algorithmic cooling protocols \citep{39.5satirability-of-asymp-limit-in-HBAC,39.6HBAC-and-correlated-environm}.
In principle, the saturability of these bounds requires an infinite
amount of time (or an infinite number of discrete steps), and therefore
time emerges as a fundamental resource in this context. 

Naturally, the maximum degree of cooling depends not only on the available
time but also on other physical resources \citep{34Paz-review-cooling,38Huber-cooling-bound2,38.1Gaussian-TOs,39.7cooling-improv-with-memory-effects}.
In addition to standard thermodynamic resources such as work and heat,
other factors that have been proven useful for \textit{assisting}
cooling are the access to non-equilibrium states \citep{13Third-Law-of-Thermodynamics-as-a-Single-Inequality},
in the framework of catalytic thermal operations, the (Hilbert space)
dimension of microscopic refrigerators \citep{38.2Refrigerator-dimension,39.2Smallest-refrigerators},
and quantum properties such as coherence \citep{39.3Mitchison-cooling-and-coherence}
and entanglement \citep{39.4entang-and-cooling}. In contrast, we
are interested in a \textit{limiting} factor that can dramatically
restrict the ability to cool. Namely, the limited access to the environment
where heat is dissipated. Although this issue has been addressed in
previous studies \citep{24.2Muller-finite-baths,36.1Karen-limits-cooling,39.8No-go-theorem-for-purity,39.9cooling-with-virtual-environment,40Wolf-Improved-Landauer},
we consider extreme cases where cooling is not even possible. For
example, if the environment is very small and very hot \citep{40.1Raam-PD}.
After introducing some notational conventions, this limitation is
illustrated with a simple and intuitive example in Sect. II. Using
the notion of passivity \citep{40.2KMS-passive-states,41work-and-passive-states,42Paul-Sk-passivity-and-virt-temp,43Raam},
we characterize the impossibility to cool using a small environment
in Sect. III. Next, we establish sufficient conditions to lift this
restriction in Sect. IV (Corollary 1), by means of a \textit{finite-dimensional}
\textit{and single-copy} catalyst. In the same section we also present
a graphical method that provides an intuitive picture of the studied
catalytic transformations. 

On the technical side, the results of Sect. IV are applicable to a
broader class of catalytic transformations, which we dub ``non-unital
transformations'' (Definition 3). These transformations are characterized
in Corollary 2 and include cooling as a particular case. Moreover,
we show that two established results on catalysts fall within this
general class: catalytic extraction of work from a passive state \citep{12Oppenheim-passivity},
and a catalytic violation of the Jarzynski fluctuation theorem \citep{15.2Bypassing-FTs}. 

In Sect. V, we present a theorem (Theorem 2) where the dimension of
the catalyst is shown to play a key role for both cooling and non-unital
transformations in general. This theorem states that such transformations
can be implemented for almost any initial state of the system and
the environment, if the catalyst dimension is sufficiently large.
These findings are illustrated in Sect. VI through several examples.
First, we consider the task of optimizing the cooling of a two-level
system using another two-level system as environment, and catalysts
of different dimensions. On the one hand, it is shown that the catalyst
dimension required to bypass the passivity constraint increases as
the system approaches its ground state, given a fixed state of the
environment. If the system state and the environment state are close
to each other, we instead see that the smallest catalyst (i.e. another
two-level system) suffices to cool. Next, we show that it is also
possible to catalytically increase the ground population of the system,
which is yet another example of non-unital transformation. Section
VI is concluded by showing that, in cases where cooling is possible
without a catalyst, a sufficiently large catalyst provides a cooling
\textit{enhancement}. We present a general statement of this advantage,
for the cooling of a two-level system using an environment of even
dimension (Theorem 3). In the case of odd dimension, we illustrate
a cooling enhancement through a two-level catalyst, after maximum
cooling of a qubit has been performed using a three-level environment. 

Sections VII and VIII are also framed within the context of catalytic
enhancements. In Sect. VII we address the problem of cooling a group
of qubits, using environments composed of different numbers of identical
qubits. We find that the introduction of a two-level catalyst can
provide a two-fold advantage. First, if the environment is not too
large compared to the number of cooled qubits, the catalyst allows
to extract more heat. Secondly, cooling without the catalyst requires
potentially much more complex interactions. In Sect. VIII we illustrate
an application to thermometry \citep{44Thermometry-review}. Thermometry
makes part of the broader field of metrology \citep{45Giovanetti-metrology,46Advances-metrology,47Paris-metrology,48quantum-sensing}
and aims at estimating the temperature of some environment at thermal
equilibrium. In particular, temperature information can be transferred
to a probe that undergoes a suitable coupling with the environment
\citep{48.1Brunelli-Qubit-therm-for-microm-resonators,48.2Brunelli-Qubit-therm-of-a-harm-oscillator,48.3Single-qubit-therm,48.4temp-estim-with-sequential-meas,48.5Non-eq-themometry-with-qubit-probes,49Correa-thermometry,54fermometer,54.1therm-via-strong-coupling,55ancilla-assisted-thermometry,56collisional-thermom,57Karen-CoarseGr-thermometry}.
We show that this information can be increased via a two-level catalyst,
thereby reducing the error in the temperature estimation. Finally,
we present the conclusions and outlook in Sect. IX. 

\section{Preliminaries}

Henceforth we will often call the system to be cooled and the environment
``cold object'' and ``hot object'', respectively. Moreover, the
ground state of these systems and the catalyst will be denoted using
the label ``1'' instead of ''0''. This choice is convenient to
simplify the notation of other physical quantities that will be defined
later. States that describe the total system (formed by the catalyst,
the cold object and the hot object) are written without labels, as
well as the corresponding unitary operations. This also simplifies
notation and does not generate ambiguity, since this is the only three-partite
setup considered. 

The eigendecompositions of local states read 
\begin{equation}
\rho_{x}=\sum_{i=1}^{d_{x}}p_{i}^{x}|i_{x}\rangle\langle i_{x}|,\label{eq:0 eigendecomposition of local states}
\end{equation}
where the label $x$ can refer to the cold object ($x=c$), the hot
object ($x=h$), or the catalyst ($x=v$). Moreover, $d_{x}$ stands
for the dimension of the corresponding Hilbert space $\mathcal{H}_{x}$.
We also adopt a convention of non-increasing eigenvalues, $p_{i}^{x}\geq p_{i+1}^{x}$
for all $1\leq i\leq d_{x}$, which is useful for the description
of passive states. 

\begin{figure}
	\centering{}\includegraphics[scale=0.45]{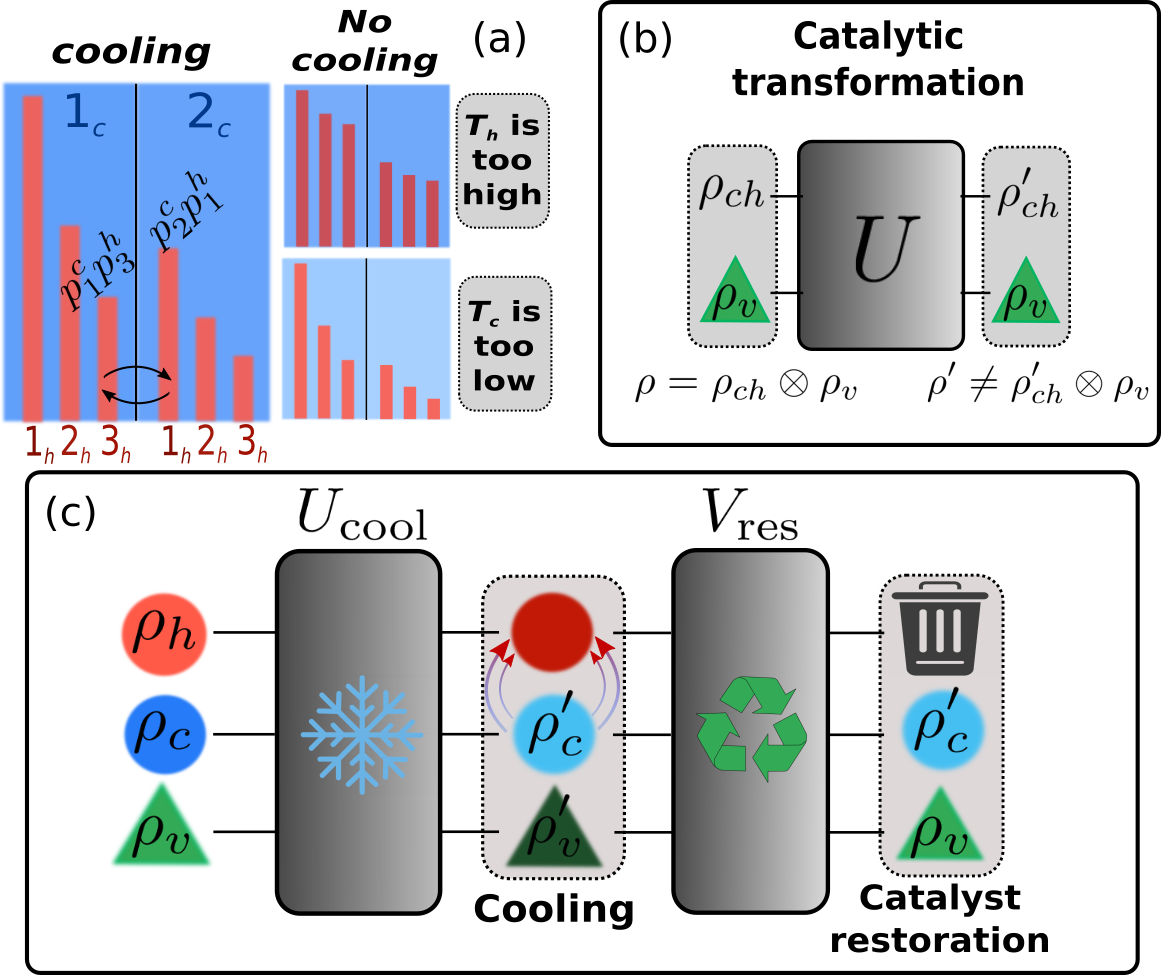}\caption{Framework for catalytic and cooling transformations. (a) Illustrative
		example: a three-level system in the initial state $\rho_{h}$ is
		used as hot object to cool a (cold) qubit in the initial state $\rho_{c}$.
		The eigenvalues of the joint state $\rho_{ch}=\rho_{c}\otimes\rho_{h}$
		are obtained by rescaling the eigenvalues of $\rho_{h}$ (orange bars)
		with the eigenvalues of $\rho_{c}$. If $p_{2}^{c}p_{1}^{h}>p_{1}^{c}p_{3}^{h}$
		cooling is possible by swapping (black arrows) the eigenstates $|1_{c}3_{h}\rangle$
		and $|2_{c}1_{h}\rangle$. If $\rho_{c}$ and $\rho_{h}$ are thermal
		states with respective temperatures $T_{c}$ and $T_{h}$, and $T_{h}$
		is too high or $T_{c}$ is too low, no unitary $U_{ch}$ applied on
		$\rho_{ch}$ can perform cooling. (b) This restriction can be bypassed
		through a global unitary $U$ applied on $\rho=\rho_{ch}\otimes\rho_{v}$,
		where $\rho_{v}$ is an appropriate catalyst state, and correlations
		are allowed in the final state $\rho'$. (c) In our scheme cooling
		is first implemented through a \textquotedblleft cooling unitary\textquotedblright{}
		$U_{\textrm{cool}}$, which perturbs the catalyst, and followed by
		a unitary $V_{\textrm{res}}$ that brings it back to $\rho_{v}$.
		Hence, we assume the form $U=V_{\textrm{res}}U_{\textrm{cool}}$. }
\end{figure}

Although most of the time we will be dealing with an initial uncorrelated
state $\rho=\rho_{c}\otimes\rho_{h}\otimes\rho_{v}$, some times we
will refer to the joint state of the cold and hot objects as $\rho_{ch}$.
In such a case, it must be understood that $\rho_{ch}$ is a \textit{general}
state, unless we explicitly write $\rho_{ch}=\rho_{c}\otimes\rho_{h}$.
The general eigendecomposition for $\rho_{ch}$ (including the special
case $\rho_{c}\otimes\rho_{h}$) is written as $\rho_{ch}=\sum_{\psi=1}^{d_{c}d_{h}}p_{\psi}^{ch}|\psi_{ch}\rangle\langle\psi_{ch}|$,
with non-increasing eigenvalues $p_{\psi}^{ch}\geq p_{\psi+1}^{ch}$.
The greek letters $\psi$ and $\varphi$ will also be used as indices
for the eigendecomposition of $\rho$. For example, $\rho=\sum_{\psi}p_{\psi}|\psi\rangle\langle\psi|$,
with the absence of the label ``$ch$'' distinguishing its eigenstates
from those of $\rho_{ch}$. For product states $\rho_{c}\otimes\rho_{h}\otimes\rho_{v}$
the global eigenstates are also denoted as $|i_{c}j_{h}k_{v}\rangle=|i_{c}\rangle\otimes|j_{h}\rangle\otimes|k_{v}\rangle$. 

Figure 1(a) illustrates a situation where cooling with a very small
hot object is forbidden. Here, the cold object is a qubit in the initial
state $\rho_{c}=\sum_{i=1}^{2}p_{i}^{c}|i_{c}\rangle\langle i_{c}|$,
and the hot object is a three-level system in the state $\rho_{h}=\sum_{j=1}^{3}p_{j}^{c}|j_{h}\rangle\langle j_{h}|$.
With the prescription of non-increasing eigenvalues, cooling is possible
if and only if $p_{1}^{c}p_{3}^{h}<p_{2}^{c}p_{1}^{h}$. The case
$p_{1}^{c}p_{3}^{h}\geq p_{2}^{c}p_{1}^{h}$ constitutes an example
of passivity in the context of cooling, characterized for generic
systems (with discrete Hamiltonians) in the next section (cf. Eq.
(\ref{eq:4 passiv cond with populations})). As explained in Sect.
IV, such a cooling limitation can be circumvented by adding a catalyst
in a proper initial state $\rho_{v}$, such that it allows cooling
and also remains unaltered by the transformation. The action of the
studied catalytic transformations on the compound of hot and cold
objects is illustrated in Fig. 1(b). In particular, the generation
of correlations between the catalyst and the rest of the total system
is allowed. This condition is characteristic of recent works on catalysts
\citep{12Oppenheim-passivity,14Muller-catalysts,15.4Sagawa} and is
also natural in our framework, where all the systems involved can
be arbitrarily small. A more detailed form of the global unitaries
that implement the transformations is shown in Fig. 1(c).

\section{Passivity and cooling}

The fundamental limits for cooling can be understood using the notion
of passivity. Passivity is essentially a condition whereby applying
unitary transformations to a system cannot decrease the mean value
of certain observables. While traditionally it has been associated
with the Hamiltonian and the impossibility of work extraction \citep{41work-and-passive-states,60.1Aicki-batteries,60.2Ext-work-from-correlations,60.3Marti-most-energetic-passive-states,60.4Gaussian-work-extr},
passivity can be extended to any hermitian operator that represents
an observable \citep{40.1Raam-PD,43Raam}. 

Since we will also be dealing with passive states of the cold object
in the traditional sense, it is important to characterize them before
introducing our extended definition, directly related to the task
of cooling. A state $\rho_{c}$ is called passive if and only if its
average energy cannot be decreased using local unitary evolutions
$U_{c}$ \citep{41work-and-passive-states}. That is, if $\Delta\bigl\langle H_{c}\bigr\rangle=\textrm{Tr}\left[H_{c}\left(U_{c}\rho_{c}U_{c}^{\dagger}-\rho_{c}\right)\right]\geq0$
for any $U_{c}$. In quantum thermodynamics, this means that it is
impossible to extract work from $\rho_{c}$ via any classical driving.
Importantly, any thermal state is passive but there are many passive
states that are non-thermal. Specifically, $\rho_{c}$ is passive
if and only if $[\rho_{c},H_{c}]=0$ and the eigenvalues of $\rho_{c}$
are non-increasing \textit{with respect to the eigenenergies of} $H_{c}$.
Assuming the sorting $p_{i}^{c}\geq p_{i+1}^{c}$, we have that $\rho_{c}$
is passive iff $H_{c}=\sum_{i}\varepsilon_{i}^{c}|i_{c}\rangle\langle i_{c}|$,
and $\varepsilon_{i+1}^{c}\geq\varepsilon_{i}^{c}$. 

\subsection{Passive states for cooling }

Consider now a bipartite system in the initial state $\rho_{c}\otimes\rho_{h}$,
where $\rho_{c}$ is a passive state. Since local unitaries on $\rho_{c}$
cannot decrease its average energy, we wonder if global unitaries
$U_{ch}$ on the compound $\rho_{c}\otimes\rho_{h}$ can do so. We
say that $\rho_{c}\otimes\rho_{h}$ is passive with respect to $H_{c}$,
iff
\begin{equation}
\Delta\bigl\langle H_{c}\bigr\rangle\geq0\textrm{ for any }U_{ch},\label{eq:passivity}
\end{equation}
where $\Delta\bigl\langle H_{c}\bigr\rangle=\textrm{Tr}\left[H_{c}\left(U_{ch}\rho_{c}\otimes\rho_{h}U_{ch}^{\dagger}-\rho_{c}\otimes\rho_{h}\right)\right]$.
Conversely, if $\Delta\left\langle H_{c}\right\rangle <0$ for some
$U_{ch}$, then $\rho_{c}\otimes\rho_{h}$ is non-passive with respect
to $H_{c}$. 

In Appendix A we characterize the conditions for passivity of any
state $\rho_{c}\otimes\rho_{h}$ in terms of its eigenvalues. We find
that $\rho_{c}\otimes\rho_{h}$ is also passive with respect to $H_{c}$
iff the following inequalites hold: 
\begin{equation}
\frac{p_{i}^{c}}{p_{i+1}^{c}}\geq\frac{p_{1}^{h}}{p_{d_{h}}^{h}}\textrm{ for all }i.\label{eq:4 passiv cond with populations}
\end{equation}

We remark that Eq. (\ref{eq:4 passiv cond with populations}) is obtained
without assuming anything on the Hamiltonian of the hot object or
the eigenstates of $\rho_{h}$. Interestingly, this expression tells
us that the only relevant parameter behind passivity is the ratio
between the highest and smallest eigenvalues of the hot object \citep{60.5Comment_on_PD}.
In particular, it allows us to understand why a thermal state $\rho_{h}$
at a very high temperature $T_{h}$ can prevent cooling. If $\{\varepsilon_{j}^{h}\}_{j=1}^{d_{h}}$
denotes the energy spectrum of the hot object, the ratio $p_{1}^{h}/p_{d_{h}}^{h}$
reads $p_{1}^{h}/p_{d_{h}}^{h}=e^{(\varepsilon_{d_{h}}^{h}-\varepsilon_{1}^{h})/T_{h}}$.
In the limit $\varepsilon_{d_{h}}^{h}-\varepsilon_{1}^{h}\ll T_{h}$,
this ratio tends to one and consequently all the inequalities (\ref{eq:4 passiv cond with populations})
are satisfied (keeping in mind the convention $p_{i}^{c}\geq p_{i+1}^{c}$).
Similarly, for a thermal state $\rho_{c}$ the ratios $p_{i}^{c}/p_{i+1}^{c}$
become larger the lower the corresponding temperature, making more
feasible the scenario (\ref{eq:4 passiv cond with populations}). 

\subsection{Remarks on cooling }

Traditionally, the task of cooling is defined as a heat exchange between
two systems at thermal equilibrium, where heat is extracted from the
system at \textit{lower} temperature. In this paper we will adopt
a more general approach, whose main requirement is the passivity of
the state $\rho_{c}$. In this view the only Hamiltonian that plays
a relevant role is that of the cold object, and the only energetic
transformation that we care about is the reduction of the average
energy $\bigl\langle H_{c}\bigr\rangle$. Due to the passive character
of $\rho_{c}$, cooling is only possible by attaching an ancillary
system to the cold object and performing a global unitary evolution
on the resulting compound. This ancillary system can be the hot object
alone (if $\rho_{c}\otimes\rho_{h}$ is non-passive), or the hot object
in combination with a catalyst.

It is important to note that cooling is also typically associated
with a \textit{work} investment, according to the second law of thermodynamics.
Another manifestation of this energetic cost is the heat dissipated
into the environment, which should be larger than the extracted heat
if the environment is equilibrated at a higher temperature. Since
we do not need to specify the Hamiltonian of the hot object (recall
that the hot object and the environment represent the same system),
neither to assume that it starts in a thermal state, predictions on
its energetic behavior are irrelevant for our analysis. Essentially,
its role is restricted to mediate the reduction of the average energy
$\bigl\langle H_{c}\bigr\rangle$. That being said, we also stress
this scenario coincides with the traditional characterization of cooling,
if $\rho_{c}$ and $\rho_{h}$ are both thermal states and their temperatures
satisfy $T_{c}\leq T_{h}$.

\section{Catalytic transformations}

\subsection{Catalytic transformations and cooling}

Given the passivity condition (\ref{eq:4 passiv cond with populations}),
our goal is to introduce a third system that enables cooling and works
as a catalyst. This means that if the catalyst is initially in a state
$\rho_{v}=\sum_{k=1}^{d_{v}}p_{k}^{v}|k_{v}\rangle\langle k_{v}|$,
at the end of the transformation it must be returned to the same state.
In addition, we assume that the catalyst starts uncorrelated from
the cold and objects, i.e. the initial total state is $\rho=\rho_{c}\otimes\rho_{h}\otimes\rho_{v}$.
The transformation on the cold object is implemented through a global
unitary map $U$ that acts on the total system. Denoting the final
total state as $\rho'$, a generic catalytic transformation satisfies
\begin{align}
\rho\rightarrow\rho' & =U(\rho_{c}\otimes\rho_{h}\otimes\rho_{v})U^{\dagger},\label{eq:3 catal transf}\\
\textrm{Tr}_{ch}(\rho') & =\rho_{v}.\label{eq:4 catalysis}
\end{align}

Note that Eq. (\ref{eq:4 catalysis}) guarantees ``catalysis'' (i.e.
the restoration of the catalyst to its initial state) but does not
say anything about the final correlations between the catalyst and
the rest of the total system. Allowing correlations has proven to
be useful in extending the transformations that a system can undergo
in the presence of a catalyst \citep{12Oppenheim-passivity,14Muller-catalysts,15.2Bypassing-FTs,15.3Karen,15.4Sagawa}.
In addition to this possibility, the repeated implementation of a
given transformation is naturally associated with the notion of catalyst.
For example, under the condition (\ref{eq:4 catalysis}) the transformation
(\ref{eq:3 catal transf}) can be performed as many times as desired,
using the \textit{same} catalyst, and provided that each transformation
involves a \textit{new copy} of the state $\rho_{c}\otimes\rho_{h}$.
This property will be of special importance for the results presented
in Sect. VII. 

\subsubsection{Cooling and catalyst restoration }

Given a passive state $\rho_{c}\otimes\rho_{h}$, the catalyst allows
to reduce the mean energy $\left\langle H_{c}\right\rangle $ as long
as the total state $\rho_{c}\otimes\rho_{h}\otimes\rho_{v}$ is non-passive
with respect to $H_{c}$. The verification of this condition is simple
but a bit subtle. Specifically, we can straightforwardly apply the
characterization of passivity (\ref{eq:4 passiv cond with populations})
to $\rho_{c}\otimes\rho_{h}\otimes\rho_{v}$, by considering $\rho_{h}\otimes\rho_{v}$
as the state of a new ``effective'' hot object. This amounts to
replace $p_{1}^{h}$ and $p_{d_{h}}^{h}$ by the largest and minimum
eigenvalues of $\rho_{h}\otimes\rho_{v}$, which are respectively
given by $p_{1}^{h}p_{1}^{v}$ and $p_{d_{h}}^{h}p_{d_{v}}^{v}$.
In this way, the catalyst allows to break down the passivity constraint
if and only if there exists $i'$ such that
\begin{equation}
\frac{p_{i'}^{c}}{p_{i'+1}^{c}}<\frac{p_{1}^{h}p_{1}^{v}}{p_{d_{h}}^{h}p_{d_{v}}^{v}}.\label{eq:5 cooling with the catalyst}
\end{equation}

Since the ratio at the r.h.s. of Eq. (\ref{eq:5 cooling with the catalyst})
is always larger than $p_{1}^{h}/p_{d_{h}}^{h}$, by a factor of $p_{1}^{v}/p_{d_{v}}^{v}$,
passivity with respect to $\rho_{c}\otimes\rho_{h}\otimes\rho_{v}$
can be violated, even if all the inequalities (\ref{eq:4 passiv cond with populations})
are satisfied. In particular, a divergent ratio results if $\rho_{v}=|1_{v}\rangle\langle1_{v}|$.
However, we will see later that the catalysis condition (\ref{eq:4 catalysis})
requires the use of catalysts in initial \textit{mixed} states. 

Once Eq. (\ref{eq:5 cooling with the catalyst}) is satisfied, $\rho$
is a non-passive state and cooling is possible by applying a global
unitary. For example, consider the unitary $\mathcal{U}_{|i'_{c}d_{h}d_{v}\rangle\leftrightarrow|(i'+1)_{c}1_{h}1_{v}\rangle}$,
which swaps the eigenstates $|i'_{c}d_{h}d_{v}\rangle$ and $|(i'+1)_{c}1_{h}1_{v}\rangle$
and acts as the identity on any other eigenstate of $\rho$. That
is, 
\begin{align}
\mathcal{U}_{|i'_{c}d_{h}d_{v}\rangle\leftrightarrow|(i'+1)_{c}1_{h}1_{v}\rangle}|i'_{c}d_{h}d_{v}\rangle & =|(i'+1)_{c}1_{h}1_{v}\rangle,\nonumber \\
\mathcal{U}_{|i'_{c}d_{h}d_{v}\rangle\leftrightarrow|(i'+1)_{c}1_{h}1_{v}\rangle}|(i'+1)_{c}1_{h}1_{v}\rangle & =|i'_{c}d_{h}d_{v}\rangle.\label{eq:5.1 cooling swap}
\end{align}
When applied on $\rho$, Eq. (\ref{eq:5.1 cooling swap}) yields a
state where the only effect is to exchange the eigenvalues of $|i'_{c}d_{h}d_{v}\rangle$
and $|(i'+1)_{c}1_{h}1_{v}\rangle$. In this way, (\ref{eq:5 cooling with the catalyst})
implies that

\begin{equation}
\Delta\bigl\langle H_{c}\bigr\rangle=\left(p_{i'}^{c}p_{d_{h}}^{h}p_{d_{v}}^{v}-p_{i'+1}^{c}p_{1}^{h}p_{1}^{v}\right)(\varepsilon_{i'+1}^{c}-\varepsilon_{i'}^{c})<0.\label{eq:5.2 delta<Hc> for cooling swap}
\end{equation}

However, the swap in Eqs. (\ref{eq:5.1 cooling swap}) also modifies
the initial state of the catalyst. Denoting as $\bigl\langle|k_{v}\rangle\langle k_{v}|\bigr\rangle=\textrm{Tr}\left[|k_{v}\rangle\langle k_{v}|\rho\right]$
the catalyst population corresponding to the eigenstate $|k_{v}\rangle$,
it increases $\bigl\langle|d_{v}\rangle\langle d_{v}|\bigr\rangle$
by $p_{i'}^{c}p_{d_{h}}^{h}p_{d_{v}}^{v}-p_{i'+1}^{c}p_{1}^{h}p_{1}^{v}$
and reduces $\bigl\langle|1_{v}\rangle\langle1_{v}|\bigr\rangle$
by the same amount. This example motivates the introduction of the
following definitions, which are crucial for our analysis of catalytic
and cooling transformations. In these definitions an initial state
of the standard form $\rho=\rho_{c}\otimes\rho_{h}\otimes\rho_{v}$
is assumed. 

\textbf{Definition 1 (Cooling unitary)}.\textit{ }A cooling unitary,
denoted as $U_{\textrm{cool}}$, is a unitary that satisfies $\textrm{Tr}(H_{c}U_{\textrm{cool}}\rho U_{\textrm{cool}}^{\dagger})<\textrm{Tr}(H_{c}\rho)$\textit{.
}An example of cooling unitary is the swap described in Eqs. (\ref{eq:5.1 cooling swap}). 

\textbf{Definition 2 (Restoring unitary)}. Given an initial transformation
$\rho\rightarrow\sigma$ that modifies the catalyst state (i.e. $\textrm{Tr}_{ch}(\sigma)\neq\rho_{v}$),
a restoring unitary $V_{\textrm{res}}$ is a unitary that satisfies\textit{
$\textrm{Tr}_{ch}(V_{\textrm{res}}\sigma V_{\textrm{res}}^{\dagger})=\rho_{v}$. }

\subsubsection{Unitary operations involved in catalytic transformations}

Henceforth, any unitary that restores the catalyst or contributes
to its restoration will be denoted using the symbol $V$, instead
of $U$. However, the following remarks and any subsequent comment
regarding a general unitary $U$ are also valid for $V$. 

Given a unitary $U$ and a Hilbert subspace $\mathcal{H}_{\textrm{sub}}\subseteq\mathcal{H}$,
being $\mathcal{H}_{\textrm{sub}}$ spanned by a subset of eigenstates
of $\rho$, we say that $U$ maps $\mathcal{H}_{\textrm{sub}}$ into
itself if:
\begin{enumerate}
\item $U|\psi\rangle=|\psi\rangle$ for $|\psi\rangle\notin\mathcal{H}_{\textrm{sub}}$. 
\item $U|\psi\rangle\in\mathcal{H}_{\textrm{sub}}$ for $|\psi\rangle\in\mathcal{H}_{\textrm{sub}}$. 
\end{enumerate}
Equivalently, if $U$ satisfies conditions 1 and 2 we also say that
``$U$ acts on $\mathcal{H}_{\textrm{sub}}$'' , which is symbolically
written as $U:\mathcal{H}_{\textrm{sub}}\rightarrow\mathcal{H}_{\textrm{sub}}$. 

Our interest will be on transformations of the form $\rho\rightarrow\rho'=U\rho U^{\dagger}$,
where 
\begin{equation}
U=V_{\textrm{res}}U_{\textrm{cool}}.\label{eq:22.0 VresUcool}
\end{equation}
Following Definition 2, these transformations are catalytic by construction,
since $V_{\textrm{res}}$ restores the catalyst state initially modified
by the transformation $\rho\rightarrow\sigma=U_{\textrm{cool}}\rho U_{\textrm{cool}}^{\dagger}$. 

While there may exist many cooling unitaries if $\rho$ is non-passive,
we will focus on the most basic unitary that allows cooling and admits
a ``simple'' solution to the problem of catalyst restoration. Namely,
a cooling unitary that acts on a two-dimensional subspace $\mathcal{H}_{\textrm{sub}}$,
which we term ``two-level unitary''. The swap (\ref{eq:5.1 cooling swap})
is an example of such an operation. Under the transformation $\rho\rightarrow\sigma$
implemented by this swap, it would be necessary to find a restoring
unitary that recovers the populations $\bigl\langle|1_{v}\rangle\langle1_{v}|\bigr\rangle$
and $\bigl\langle|d_{v}\rangle\langle d_{v}|\bigr\rangle$. In some
situations this may not be possible, and therefore it is convenient
to explore other two-level unitaries whose effect can be reverted
via a suitable $V_{\textrm{res}}$. This leads us to present an alternative
formulation of Eq. (\ref{eq:5 cooling with the catalyst}). Using
such a formulation, we will be able to broaden the possibilities for
catalyst restoration, by finding a variety of two-level cooling unitaries
that have different effects on the catalyst. 

Specifically, the inequality (\ref{eq:5 cooling with the catalyst})
can be rewritten in the equivalent way: 

\begin{equation}
\frac{p_{i'}^{c}}{p_{i'+1}^{c}}<\frac{p_{1}^{h}p_{l}^{v}}{p_{d_{h}}^{h}p_{l'+1}^{v}},\label{eq:22 recast of passivity violation with l and l'}
\end{equation}
for some set of indices $\{i^{\prime},l,l^{\prime}\}$ such that $i^{\prime}\in\{1,2,...,d_{c}-1\}$
and $l,l^{\prime}\in\{1,2,...,d_{v}-1\}$. Equation (\ref{eq:5 cooling with the catalyst})
implies Eq. (\ref{eq:22 recast of passivity violation with l and l'}),
because $\{i^{\prime},l,l^{\prime}\}=\{i^{\prime},1,d_{v}-1\}$ is
an example of such a set of indices. Conversely, since $\frac{p_{1}^{h}p_{l}^{v}}{p_{d_{h}}^{h}p_{l'+1}^{v}}\leq\frac{p_{1}^{h}p_{1}^{v}}{p_{d_{h}}^{h}p_{d_{v}}^{v}}$,
the inequality (\ref{eq:22 recast of passivity violation with l and l'})
implies Eq. (\ref{eq:5 cooling with the catalyst}). 

In Sect. IV-B2 we will apply Eq. (\ref{eq:22 recast of passivity violation with l and l'})
to identify a familiy of two-level cooling unitaries and the corresponding
restoring unitaries. These restoring unitaries act on a subspace 
\begin{equation}
\mathcal{H}_{n,N}\equiv\textrm{span}\{|\psi_{ch}\rangle\}_{\psi=n}^{N}\otimes\mathcal{H}_{v},\label{eq:21 subspace Hn,N}
\end{equation}
where $n$ and $N$ are two numbers such that $n\leq N-1$ and $n,N\in\{1,2,...,d_{c}d_{h}\}$.
We remark that the eigenstates in the set $\{|\psi_{ch}\rangle\}_{\psi=n}^{N}$
are arranged according to the sorting $p_{\psi}^{ch}\geq p_{\psi+1}^{ch}$.
For example, $|1_{ch}\rangle$ possess the largest eigenvalue $p_{1}^{ch}=p_{1}^{h}p_{1}^{c}$,
and therefore it is the ``first'' eigenstate of $\rho_{c}\otimes\rho_{h}$.
This observation is also important to understand that the eigenstates
$|n_{ch}\rangle$ and $|N_{ch}\rangle$ in Eqs. (\ref{eq:26.1 n and N for CC transformation})
and (\ref{eq:26.4 n and N for non-unital transf}) are consistent
with the condition $p_{n}^{ch}\geq p_{N}^{ch}$.

\subsection{Catalytic transformations and population currents }

The effect of the swap $\mathcal{U}_{|i'_{c}d_{h}d_{v}\rangle\leftrightarrow|(i'+1)_{c}1_{h}1_{v}\rangle}$
can be more conveniently described as a \textit{population transfer}
from the eigenstate $|1_{v}\rangle$ towards the eigenstate $|d_{v}\rangle$,
since the increment in $\left\langle |d_{v}\rangle\langle d_{v}|\right\rangle $
equals exactly the reduction in $\left\langle |1_{v}\rangle\langle1_{v}|\right\rangle $.
That is, $\Delta\left\langle |d_{v}\rangle\langle d_{v}|\right\rangle =-\Delta\left\langle |1_{v}\rangle\langle1_{v}|\right\rangle $.
In this way, a possible restoring unitary would be a two-level unitary
that transfers the same population in the opposite direction, i.e.
from $|d_{v}\rangle$ to $|1_{v}\rangle$. If the final state of the
catalyst is also diagonal in the eigenbasis $\{|k_{v}\rangle\}_{k}$,
such an operation is sufficient for its full recovery. Note also that
this could be trivially achieved with $V_{\textrm{res}}=\mathcal{U}_{|i'_{c}d_{h}d_{v}\rangle\leftrightarrow|(i'+1)_{c}1_{h}1_{v}\rangle}^{\dagger}$.
However, we are obviously interested in a recovery operation that
does not cancel the effect of the cooling unitary on the cold object. 

In general, it is not possible to perform a complete catalyst restoration
via a single two-level unitary. Notwithstanding, a composition of
many two-level unitaries can operate jointly towards the realization
of this goal. As an extension of the compensation between population
transfers mentioned before, the restoration mechanism in this case
can be understood as the result of a series of population transfers
that cancel each other. These population transfers are an example
of what we call ``local currents'', since they involve eigenstates
of the catalyst. Other forms of local currents refer to population
transfers between eigenstates of the cold object, or between eigenstates
of the hot object. It is important to stress that this notion of current
describes a \textit{net population transfer}, rather than some rate
of population exchanged per unit of time. With this observation in
mind, in what follows we introduce the basic tools for the characterization
of local currents, by connecting them with population transfers between
eigenstates of $\rho$, also termed ``global currents''. 

\subsubsection{Characterization of local and global currents }

Let us denote a general two-level as $U^{(2)}$, with the superscript
``$(2)$'' indicating that it maps a two-dimensional subspace into
itself. If $|\psi\rangle$ and $|\varphi\rangle$ are eigenstates
of $\rho$ that expand this subspace, $U^{(2)}$ performs a population
transfer between $|\psi\rangle$ and $|\varphi\rangle$ that we can
fully characterize through the equations 

\begin{align}
U^{(2)}|\psi\rangle & =\sqrt{1-r}|\psi\rangle+\sqrt{r}|\varphi\rangle,\label{eq:11 U|i>}\\
U^{(2)}|\varphi\rangle & =\sqrt{r}|\psi\rangle-\sqrt{1-r}|\varphi\rangle,\label{eq:12 U|j>}
\end{align}
where $0\leq r\leq1$ is the probability that $|\psi\rangle$ and
$|\varphi\rangle$ are swapped, or ``swap intensity''. For $0\leq r<1$,
we say that $U^{(2)}$ is a ``partial swap'', and also denote it
as $U_{|\psi\rangle\leftrightarrow|\varphi\rangle}$. For $r=1$,
the corresponding total swap (or simply ``swap'') is denoted as
$\mathcal{U}^{(2)}$ or $\mathcal{U}_{|\psi\rangle\leftrightarrow|\varphi\rangle}$.
Importantly, the swap intensity is not affected by the addition of
local or global phases in Eqs. (\ref{eq:11 U|i>}) and (\ref{eq:12 U|j>}).
Since $r$ is the only parameter we care about, we restrict ourselves
(and without loss of generality) to the description of two-level unitaries
given by these equations. Note also that the swap intensity is invariant
if we exchange $\psi$ and $\varphi$. Hence, both $U_{|\psi\rangle\leftrightarrow|\varphi\rangle}$
and $U_{|\varphi\rangle\leftrightarrow|\psi\rangle}$ are two-level
unitaries that swap $|\psi\rangle$ and $|\varphi\rangle$ with probability
$r$. 

Although $U^{(2)}$ in Eqs. (\ref{eq:11 U|i>}) and (\ref{eq:12 U|j>})
is defined on (a subspace of) the total Hilbert space $\mathcal{H}$,
we remark that the action of two-level unitaries acting on local Hilbert
spaces can be analogously formulated. In particular, we will also
consider two-level unitaries that act on the Hilbert spaces $\mathcal{H}_{c}\otimes\mathcal{H}_{h}$
and $\mathcal{H}_{c}\otimes\mathcal{H}_{v}$. These maps can be described
by simply replacing $|\psi\rangle$ and $|\varphi\rangle$ in (\ref{eq:11 U|i>})
and (\ref{eq:12 U|j>}) by eigenstates of $\rho_{c}\otimes\rho_{h}$,
in the first case, or by eigenstates of $\rho_{c}\otimes\rho_{v}$,
in the second case. For example, a two-level unitary $U_{|\psi_{ch}\rangle\leftrightarrow|\varphi_{ch}\rangle}$
swaps the eigenstates $|\psi_{ch}\rangle$ and $|\varphi_{ch}\rangle$
with probability $r$. 

For initial populations $p_{\psi}=\textrm{Tr}(|\psi\rangle\langle\psi|\rho)$
and $p_{\varphi}=\textrm{Tr}(|\varphi\rangle\langle\varphi|\rho)$,
Eqs. (\ref{eq:11 U|i>}) and (\ref{eq:12 U|j>}) yield the final population
\begin{align}
p_{\psi}^{\prime} & =\textrm{Tr}[|\psi\rangle\langle\psi|U^{(2)}\rho U^{(2)\dagger}]\nonumber \\
 & =p_{\psi}+r(p_{\varphi}-p_{\psi}).\label{eq:13 p'j}
\end{align}
We identify the population transfer $\Delta p_{\psi}=p_{\psi}^{\prime}-p_{\psi}$
with a ``global current'' $J_{|\varphi\rangle\rightarrow|\psi\rangle}=\Delta p_{\psi}$,
from the eigenstate $|\varphi\rangle$ towards the eigenstate $|\psi\rangle$.
If $J_{|\varphi\rangle\rightarrow|\psi\rangle}$ is negative, it indicates
that the transfer takes place in the opposite direction. From Eq.
(\ref{eq:13 p'j}) we obtain 

\begin{equation}
J_{|\varphi\rangle\rightarrow|\psi\rangle}=r(p_{\varphi}-p_{\psi}).\label{eq:14 current}
\end{equation}
The maximum population transfer corresponds to a swap $\mathcal{U}_{|\psi\rangle\leftrightarrow|\varphi\rangle}$,
and the associated current is denoted as $\mathcal{J}_{|\varphi\rangle\rightarrow|\psi\rangle}$.
That is, 
\begin{equation}
\mathcal{J}_{|\varphi\rangle\rightarrow|\psi\rangle}=p_{\varphi}-p_{\psi}.\label{eq:18 maxJ(i-->j)}
\end{equation}

To establish the connection between the global current (\ref{eq:14 current})
and local currents let us write the eigenstates $|\psi\rangle$ and
$|\varphi\rangle$ as $|\psi\rangle=|i_{c}^{\prime}j_{h}^{\prime}k_{v}^{\prime}\rangle$
and $|\varphi\rangle=|i_{c}^{\prime\prime}j_{h}^{\prime\prime}k_{v}^{\prime\prime}\rangle$.
Moreover, let $J_{|k_{v}^{\prime\prime}\rangle\rightarrow|k_{v}^{\prime}\rangle}$
denote the local current from $|k_{v}^{\prime\prime}\rangle$ to $|k_{v}^{\prime}\rangle$,
and let us denote the currents from $|i_{c}^{\prime\prime}\rangle$
to $|i_{c}^{\prime}\rangle$ and from $|j_{h}^{\prime\prime}\rangle$
to $|j_{h}^{\prime}\rangle$ as $J_{|i_{c}^{\prime\prime}\rangle\rightarrow|i_{c}^{\prime}\rangle}$
and $J_{|j_{h}^{\prime\prime}\rangle\rightarrow|j_{h}^{\prime}\rangle}$,
respectively. In Appendix C2, we apply Eqs. (\ref{eq:11 U|i>}) and
(\ref{eq:12 U|j>}) to show that 
\begin{align}
J_{|i_{c}^{\prime\prime}\rangle\rightarrow|i_{c}^{\prime}\rangle} & =r\mathcal{J}_{|\varphi\rangle\rightarrow|\psi\rangle}\leq\left|\mathcal{J}_{|i_{c}^{\prime\prime}\rangle\rightarrow|i_{c}^{\prime}\rangle}\right|,\label{eq:19.1 local current 1}\\
J_{|j_{h}^{\prime\prime}\rangle\rightarrow|j_{h}^{\prime}\rangle} & =r\mathcal{J}_{|\varphi\rangle\rightarrow|\psi\rangle}\leq\left|\mathcal{J}_{|j_{h}^{\prime\prime}\rangle\rightarrow|j_{h}^{\prime}\rangle}\right|,\label{eq:19.2 local current 2}\\
J_{|k_{v}^{\prime\prime}\rangle\rightarrow|k_{v}^{\prime}\rangle} & =r\mathcal{J}_{|\varphi\rangle\rightarrow|\psi\rangle}\leq\left|\mathcal{J}_{|k_{v}^{\prime\prime}\rangle\rightarrow|k_{v}^{\prime}\rangle}\right|,\label{eq:19.3 local current 3}
\end{align}
where the equalities imply that the local currents coincide with $J_{|\varphi\rangle\rightarrow|\psi\rangle}$
(cf. Eqs. (\ref{eq:14 current}) and (\ref{eq:18 maxJ(i-->j)})).
In addition, each bound is the absolute value of the local current
obtained by applying the swap $\mathcal{U}_{|\psi\rangle\leftrightarrow|\varphi\rangle}$. 

\subsubsection{Two important classes of catalytic transformations }

In combination with Eqs. (\ref{eq:14 current}), (\ref{eq:18 maxJ(i-->j)})
and (\ref{eq:19.1 local current 1}), we can now apply Eq. (\ref{eq:22 recast of passivity violation with l and l'})
to show that the non-passivity of $\rho$ is tantamount to the existence
of a two-level cooling unitary characterized below. Specifically,
consider the unitary 
\begin{equation}
U_{i^{\prime},l,l'}\equiv U_{|i_{c}^{\prime}d_{h}(l'+1)_{v}\rangle\leftrightarrow|(i^{\prime}+1)_{c}1_{h}l_{v}\rangle},\label{eq:23 Ui',l,l'}
\end{equation}
where $i^{\prime}\in\{1,2,...,d_{c}-1\}$ and $l,l^{\prime}\in\{1,2,...,d_{v}-1\}$.
From Eqs. (\ref{eq:14 current}), (\ref{eq:18 maxJ(i-->j)}) and (\ref{eq:19.1 local current 1}),
it follows that $U_{i^{\prime},l,l'}$ generates a positive current
$J_{|(i^{\prime}+1)_{c}\rangle\rightarrow|i_{c}^{\prime}\rangle}$
(i.e. $J_{|(i^{\prime}+1)_{c}\rangle\rightarrow|i_{c}^{\prime}\rangle}>0$)
if and only if $p_{i'+1}^{c}p_{1}^{h}p_{l}^{v}>p_{i'}^{c}p_{d_{h}}^{h}p_{l'+1}^{v}$,
which is a rearrangement of the inequality (\ref{eq:22 recast of passivity violation with l and l'}).
Since $J_{|(i^{\prime}+1)_{c}\rangle\rightarrow|i_{c}^{\prime}\rangle}$
represents a population transfer from a high-energy level $\varepsilon_{i'+1}^{c}$
towards a low-energy level $\varepsilon_{i'}^{c}$, it cools down
the cold object by the amount
\begin{equation}
\Delta\bigl\langle H_{c}\bigr\rangle=J_{|(i^{\prime}+1)_{c}\rangle\rightarrow|i_{c}^{\prime}\rangle}\left(\varepsilon_{i'}^{c}-\varepsilon_{i'+1}^{c}\right).\label{eq:23.1 Delta<Hc> with cooling current}
\end{equation}
Hence, we conclude that (\ref{eq:22 recast of passivity violation with l and l'})
is equivalent to the existence of a cooling unitary $U_{i^{\prime},l,l'}$. 

Using the tools introduced in appendices B-D, in Appendix E we prove
a theorem (Theorem 1) that characterizes the existence of catalytic
transformations $\rho\rightarrow U\rho U^{\dagger}$, where $\rho{\color{red}{\normalcolor =\rho_{c}\otimes\rho_{h}\otimes\rho_{v}}}$,
\begin{equation}
U=U_{i^{\prime},l,l'}\oplus V_{\textrm{res},n,N},\label{eq:25 explicit U for catal transform}
\end{equation}
and $V_{\textrm{res},n,N}:\mathcal{H}_{n,N}\rightarrow\mathcal{H}_{n,N}$
is a restoring unitary that maps $\mathcal{H}_{n,N}$ into itself.
Notice that the direct sum structure in Eq. (\ref{eq:25 explicit U for catal transform})
implies that $U_{i^{\prime},l,l'}$ and $V_{\textrm{res},n,N}$ act
on orthogonal subspaces. Hence, $[U_{i^{\prime},l,l'},V_{\textrm{res},n,N}]=0$,
and consequently Eq. (\ref{eq:25 explicit U for catal transform})
can also be written as $V_{\textrm{res},n,N}U_{i^{\prime},l,l'}$,
in agreement with (\ref{eq:22.0 VresUcool}). The fact that $U_{i^{\prime},l,l'}$
and $V_{\textrm{res},n,N}$ act on orthogonal subspaces also excludes
the trivial restoration given by $V_{\textrm{res},n,N}=U_{i^{\prime},l,l'}^{\dagger}$. 

The restoring unitary $V_{\textrm{res},n,N}$ is \textit{explicitly}
given by a direct sum of partial swaps
\begin{equation}
V_{\textrm{res},n,N}=\oplus_{k=l}^{l'}V_{|n_{ch}(k+1)_{v}\rangle\leftrightarrow|N_{ch}k_{v}\rangle},\label{eq:24 explicit Vres}
\end{equation}
with swap intensities properly tuned to guarantee that $\rho_{v}=\textrm{Tr}_{ch}\left(U\rho U^{\dagger}\right)$.
Noting the dependence of $V_{\textrm{res},n,N}$ on $l$ and $l'$,
we can see that the restoring unitary is adapted to the cooling unitary
$U_{i^{\prime},l,l'}$. Theorem 1 (cf. Appendix E) indicates that
the direct sum $\oplus_{k=l}^{l'}V_{|n_{ch}(k+1)_{v}\rangle\leftrightarrow|N_{ch}k_{v}\rangle}$
is a genuine restoring unitary if and only if \textit{all} its partial
swaps generate positive currents $J_{|(k+1)_{v}\rangle\rightarrow|k_{v}\rangle}$.
On top of that, it also states that no direct sum of two-level unitaries
acting on $\mathcal{H}_{n,N}$ can recover the catalyst if this condition
is not satisfied. 

It is also worth stressing that $U$ in (\ref{eq:25 explicit U for catal transform})
is not necessarily itself a cooling unitary, since the total change
in $\Delta\bigl\langle H_{c}\bigr\rangle$ includes the contribution
(\ref{eq:23.1 Delta<Hc> with cooling current}) and the contribution
due to $V_{\textrm{res},n,N}$. Taking this into account, in what
follows we characterize two kinds of interesting catalytic transformations
that can be obtained from Eq. (\ref{eq:25 explicit U for catal transform}).
Apart from transformations where $U$ is also cooling, we consider
the broader class of \textit{non-unital} transformations. Both classes
of transformations are possible by imposing certain restrictions on
the subspace $\mathcal{H}_{n,N}$, which in turn restrict the states
connected by the partial swaps in Eq. (\ref{eq:24 explicit Vres}). 

\textbf{Definition 3 (non-unital transformation)}. Let $\rho_{c}$
be a general (not necessarily passive) state of the cold object, and
let $\{U_{c}^{(i)}\}_{i}$ denote a set of local unitaries acting
on $\mathcal{H}_{c}$. A non-unital transformation is a transformation
$\rho_{c}\rightarrow\rho'_{c}$ that cannot be expressed by applying
a random unitary map \citep{43.1random-unitaries} on $\rho_{c}$.
That is, $\rho'_{c}\neq\sum_{i}\lambda_{i}U_{c}^{(i)}\rho_{c}U_{c}^{(i)\dagger}$,
for any $\{U_{c}^{(i)}\}_{i}$ and discrete probability distribution
$\{\lambda_{i}\}_{i}$ ($0\leq\lambda_{i}\leq1$ and $\sum_{i}\lambda_{i}=1$). 

Definition 3 is independent of the relation between $\rho_{c}$ and
$H_{c}$. However, if we further assume that $\rho_{c}$ is passive,
it implies that any transformation that cools down the cold object
must be non-unital. Otherwise, $\rho'_{c}$ could be written as $\sum_{i}\lambda_{i}U_{c}^{(i)}\rho_{c}U_{c}^{(i)\dagger}$
, and since $\Delta\bigl\langle H_{c}\bigr\rangle=\sum_{i}\lambda_{i}\textrm{Tr}\left(U_{c}^{(i)}\rho_{c}U_{c}^{(i)}-\rho_{c}\right)$
then $\Delta\bigl\langle H_{c}\bigr\rangle\geq0$. The term ``non-unital''
is employed because $\rho'_{c}\neq\sum_{i}\lambda_{i}U_{c}^{(i)}\rho_{c}U_{c}^{(i)\dagger}$
if and only if $\rho'_{c}\neq\mathcal{E}_{\textrm{un}}(\rho_{c})$
\citep{60Nielsen-majorization,61Unital-maps} for any unital map $\mathcal{E}_{\textrm{un}}$,
which is defined as a map that satisfies $\mathcal{E}_{\textrm{un}}(\mathbb{I}_{c})=\mathbb{I}_{c}$.
Hence, a non-unital transformation cannot be expressed by applying
a unital map on $\rho_{c}$ either. We also stress that the aforementioned
equivalence does not mean that $\mathcal{E}_{\textrm{un}}(\cdot)$
and $\sum_{i}\lambda_{i}U_{c}^{(i)}(\cdot)U_{c}^{(i)\dagger}$ are
equivalent maps (in fact, the set of unital maps is strictly larger
and only coincides with the set of random unitary maps in dimension
2 \citep{61Unital-maps}), but only that for the \textit{specific
state} $\rho_{c}$ they yield the same output $\rho'_{c}$. 

While our main focus is on catalytic and cooling transformations,
(more general) non-unital transformations deserve separate attention.
Apart from the aforementioned relation with cooling, they have been
illustrated in a thermodynamic context by showing catalytic work extraction
from a passive state \citep{12Oppenheim-passivity}, and a catalytic
violation of the Jarzynski equality \citep{15.2Bypassing-FTs}. In
the first case non-unitality can be deduced using the same argument
that we applied for cooling. Namely, any unital transformation can
only increase the average of an initially passive state, and therefore
it is also useless for work extraction. On the other hand, it is explicitly
shown in Ref. \citep{15.2Bypassing-FTs} that the Jarzynski equality
is valid for any evolution that can be expressed in terms of a unital
map. 

Intuitively, non-unital transformations capture physical evolutions
that cannot be performed through an external driving, even if various
external fields are applied with different probabilities $\lambda_{i}$.
Hence, they may be meaningful in any scenario where the interaction
with another quantum system is necessary to achieve a desired effect.
With the aim of setting a framework for a more general study of these
transformations, in Corollary 2 we present sufficient conditions for
their implementation. The proof of this corollary is provided in Appendix
F. Corollaries 1 and 2 are consequences of Theorem 1. 

\textbf{Corollary 1 (catalytic and cooling transformations)}.\textit{
Let $\rho_{c}\otimes\rho_{h}$ be a passive state and let $\rho=\rho_{c}\otimes\rho_{h}\otimes\rho_{v}$
be a non-passive state. If $|n_{ch}\rangle$ and $|N_{ch}\rangle$
in }(\ref{eq:24 explicit Vres})\textit{ satisfy 
\begin{equation}
{\color{red}{\normalcolor |n_{ch}\rangle=|i_{c}^{\prime\prime}1_{h}\rangle\textrm{ and }|N_{ch}\rangle=|i_{c}^{\prime\prime}d_{h}\rangle,}}\label{eq:26.1 n and N for CC transformation}
\end{equation}
where $i^{\prime\prime}\in\{1,2,...,d_{c}\}$, the catalytic transformation
$\rho\rightarrow\rho'=U\rho U^{\dagger}$ implemented by $U$ (cf.
Eq. }(\ref{eq:25 explicit U for catal transform})\textit{) is also
a cooling transformation. This kind of transformation will be denoted
as} $\rho\overset{\textrm{CC}}{\longrightarrow}\rho'$.

\textit{Proof}. For $|n_{ch}\rangle$ and $|N_{ch}\rangle$ in (\ref{eq:26.1 n and N for CC transformation}),
the restoring unitary (\ref{eq:24 explicit Vres}) is composed of
partial swaps $V_{|i_{c}^{\prime\prime}1_{h}(k+1)_{v}\rangle\leftrightarrow|i_{c}^{\prime\prime}d_{h}k_{v}\rangle}$.
Using Eqs. (\ref{eq:11 U|i>}) and (\ref{eq:12 U|j>}), we can write
\begin{align}
V_{|i_{c}^{\prime\prime}1_{h}(k+1)_{v}\rangle\leftrightarrow|i_{c}^{\prime\prime}d_{h}k_{v}\rangle} & =|i_{c}^{\prime\prime}\rangle\langle i_{c}^{\prime\prime}|\otimes V_{|1_{h}(k+1)_{v}\rangle\leftrightarrow|d_{h}k_{v}\rangle}\nonumber \\
 & \quad+\left(\mathbb{I}_{c}-|i_{c}^{\prime\prime}\rangle\langle i_{c}^{\prime\prime}|\right)\otimes\mathbb{I}_{hv},\label{eq:26.2 partial swap V for corollary 1}
\end{align}
where $V_{|1_{h}(k+1)_{v}\rangle\leftrightarrow|d_{h}k_{v}\rangle}$
is a partial swap between $|1_{h}(k+1)_{v}\rangle$ and $|d_{h}k_{v}\rangle$,
and $\mathbb{I}_{c}$$(\mathbb{I}_{hv})$ is the identity operator
on $\mathcal{H}_{c}$($\mathcal{H}_{c}\otimes\mathcal{H}_{v}$). 

Therefore, Eq. (\ref{eq:24 explicit Vres}) yields the restoring unitary
\begin{align}
V_{\textrm{res},|i_{c}^{\prime\prime}\rangle} & \equiv|i_{c}^{\prime\prime}\rangle\langle i_{c}^{\prime\prime}|\otimes\left(\oplus_{k=l}^{l'}V_{|1_{h}(k+1)_{v}\rangle\leftrightarrow|d_{h}k_{v}\rangle}\right)\nonumber \\
 & \quad+\left(\mathbb{I}_{c}-|i_{c}^{\prime\prime}\rangle\langle i_{c}^{\prime\prime}|\right)\otimes\mathbb{I}_{hv},\label{eq:26.3 Vres for corollary 1}
\end{align}
which is a controlled unitary with the cold object operating as control.
Since $[V_{\textrm{res},|i_{c}^{\prime\prime}\rangle},H_{c}]=0$,
it follows that $\Delta\bigl\langle H_{c}\bigr\rangle=\textrm{Tr}\left[H_{c}\left(U\rho U^{\dagger}-\rho\right)\right]=\textrm{Tr}\left[H_{c}\left(U_{i^{\prime},l,l'}\rho U_{i^{\prime},l,l'}^{\dagger}-\rho\right)\right]<0$. 

Importantly, the positivity of the current $J_{|(k+1)_{v}\rangle\rightarrow|k_{v}\rangle}$
generated by $V_{|i_{c}^{\prime\prime}1_{h}(k+1)_{v}\rangle\leftrightarrow|i_{c}^{\prime\prime}d_{h}k_{v}\rangle}$
does \textit{not} depend on $i^{\prime\prime}$. This is a simple
consequence of applying Eqs. (\ref{eq:18 maxJ(i-->j)}) and (\ref{eq:19.3 local current 3}),
which imply that $J_{|(k+1)_{v}\rangle\rightarrow|k_{v}\rangle}>0$
iff $p_{i^{\prime\prime}}^{c}(p_{1}^{h}p_{k+1}^{v}-p_{d_{h}}^{h}p_{k}^{v})>0$
iff $p_{1}^{h}p_{k+1}^{v}-p_{d_{h}}^{h}p_{k}^{v}>0$. Accordingly,
the unitaries $\{V_{\textrm{res},|i_{c}^{\prime\prime}\rangle}\}_{i^{\prime\prime}=1}^{d_{c}}$
are \textit{all} valid restoring unitaries if the condition of Corollary
1 is satisfied. 

\textbf{Corollary 2 (catalytic and non-unital transformations)}.\textit{
Let $\rho=\rho_{c}\otimes\rho_{h}\otimes\rho_{v}$ be a state that
satisfies Eq. }(\ref{eq:22 recast of passivity violation with l and l'})\textit{.
If $|n_{ch}\rangle$ and $|N_{ch}\rangle$ in }(\ref{eq:24 explicit Vres})\textit{
satisfy 
\begin{align}
|n_{ch}\rangle & =|1_{c}1_{h}\rangle\textrm{ and }|N_{ch}\rangle=|i'_{c}d_{h}\rangle,\nonumber \\
\textrm{or }|n_{ch}\rangle & =|(i'+1)_{c}1_{h}\rangle\textrm{ and }|N_{ch}\rangle=|d_{c}d_{h}\rangle,\label{eq:26.4 n and N for non-unital transf}
\end{align}
the catalytic transformation $\rho\rightarrow\rho'=U\rho U^{\dagger}$
implemented by $U$ (cf. Eq. }(\ref{eq:25 explicit U for catal transform})\textit{)
is also a non-unital transformation with respect to $\rho_{c}$ (cf.
Definition 3). This kind of transformation will be denoted as} $\rho\overset{\textrm{CNU}}{\longrightarrow}\rho'$. 

Note that no reference to passivity is given in Corollary 2. This
is not surprising, as we have already emphasized that the definition
of non-unital transformation is independent of this property. Although
the reference to Eq. (\ref{eq:22 recast of passivity violation with l and l'})
seemingly indicates that $\rho=\rho_{c}\otimes\rho_{h}\otimes\rho_{v}$
is non-passive, such an assertion only makes sense once the form of
the Hamiltonian $H_{c}$ is specified. In contrast, Corollary 2 tells
us that the sole relation between eigenvalues manifested by (\ref{eq:22 recast of passivity violation with l and l'})
suffices for the implementation a non-unital transformation.

\subsubsection{Graphical representation of currents }

To gain physical insight on the catalytic transformations generated
by the unitary (\ref{eq:25 explicit U for catal transform}), we introduce
a method for the graphical representation of global currents and the
corresponding local currents. The fundamental tool for the application
of this method is the $\textrm{ln}(p^{ch})\times\textrm{ln}(p^{v})$\textbf{
}Diagram, which we describe below and illustrate in Fig. 2.

\begin{figure}
	\centering{}\includegraphics[scale=0.48]{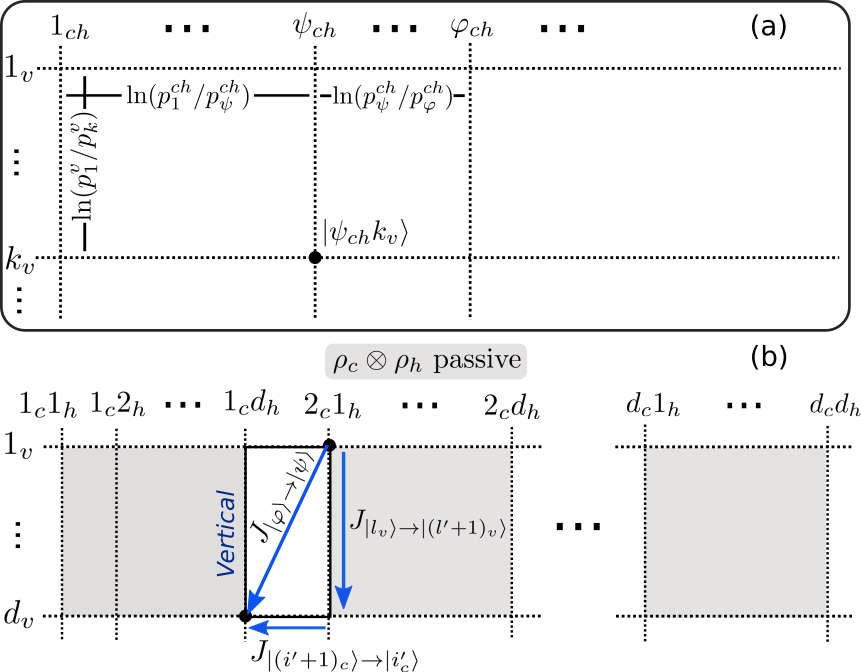}\caption{(a) $\textrm{ln}(p^{ch})\times\textrm{ln}(p^{v})$ diagram for the
		state $\rho_{ch}\otimes\rho_{v}$, where $\rho_{ch}$ and $\rho_{v}$
		have eigendecompositions $\rho_{ch}=\sum_{\psi}p_{\psi}^{ch}|\psi_{ch}\rangle\langle\psi_{ch}|$
		and $\rho_{v}=\sum_{k}p_{k}^{v}|k_{v}\rangle\langle k_{v}|$. Vertical
		(horizontal) lines depict eigenstates of $\rho_{ch}$ ($\rho_{v}$),
		and their intersection illustrates a global eigenstate $|\psi_{ch}k_{v}\rangle$.
		Assuming $p_{\psi}^{ch}\protect\geq p_{\psi+1}^{ch}$ and $p_{k}^{v}\protect\geq p_{k+1}^{v}$,
		the distance between $1_{ch}$ and any \textquotedblleft column\textquotedblright{}
		$\psi_{ch}$ reads $\textrm{ln}(p_{\psi}^{ch}/p_{\varphi}^{ch})$,
		and the distance between $1_{v}$ and any \textquotedblleft row\textquotedblright{}
		$k_{v}$ reads $\textrm{ln}(p_{1}^{v}/p_{k}^{v})$. (b) $\textrm{ln}(p^{ch})\times\textrm{ln}(p^{v})$
		diagram for $\rho_{ch}\otimes\rho_{v}$, assuming $\rho_{ch}=\rho_{c}\otimes\rho_{h}$
		and $\rho_{ch}$ passive with respect to $H_{c}$. In this case passivity
		and the distance convention implies that each group of columns $\{i_{c}j_{h}\}_{j=1}^{d_{h}}$
		is located at the left of $\{(i+1)_{c}j_{h}\}_{j=1}^{d_{h}}$. The
		blue arrows depict the global current $J_{|\varphi\rangle\rightarrow|\psi\rangle}$
		and the corresponding local currents (Eqs. (\ref{eq:19.1 local current 1})-(\ref{eq:19.3 local current 3}))
		generated by $U_{|\varphi\rangle\rightarrow|\psi\rangle}=U_{i',l,l'}$,
		with $\{i',l,l'\}=\{1,1,d_{v}-1\}$.\textcolor{red}{{} }}
\end{figure}

Using this diagram we obtain the depiction of currents shown in Fig. 3.
In particular, the structure termed ``loop'' is formally defined
in Appendix D. In Fig. 3 this structure is composed of the local current
$J_{|l_{v}\rangle\rightarrow|(l'+1)_{v}\rangle}$ (blue arrow), generated
by $U_{i',l,l'}$, and the currents $\{J_{|(k+1)_{v}\rangle\rightarrow|k_{v}\rangle}\}_{k=l}^{l'}$
(green arrows), generated by the restoring unitary $V_{\textrm{res},n,N}$. If all these currents have the same magnitude, they give rise to a
mutual cancellation of population transfers that keep the state of
the catalyst unchanged (cf. Appendix D).

$\textrm{ln}(p^{ch})\times\textrm{ln}(p^{v})$ \textit{Diagram}. This
diagram is a structure that contains all the information about the
eigenvalues of an initial product state $\rho_{ch}\otimes\rho_{v}$.
Here, the eigenstates of the state $\rho_{ch}$ are depicted as vertical
lines, dubbed ``columns'', and the eigenstates of $\rho_{v}$ are
depicted as horizontal lines, dubbed ``rows''. Moreover, the intersection
between the column $|\psi_{ch}\rangle$ and the row $|k_{v}\rangle$
depicts the global eigenstate $|\psi_{ch}k_{v}\rangle=|\psi_{ch}\rangle\otimes|k_{v}\rangle$. 

The distance between two columns is determined as follows. First of
all, a column is at the right hand side of another column if and only
if its eigenvalue (i.e. the eigenvalue of the associated eigenstate)
is smaller or equal than the eigenvalue of the column located at the
left. Taking into account the non-increasing order of the eigenvalues
$\{p_{\psi}^{ch}\}_{\psi=1}^{d_{c}d_{h}}$, this implies that the
left-most column corresponds to $|1_{ch}\rangle$, the next column
corresponds to $|2_{ch}\rangle$, and so forth. If two columns have
eigenvalues $p_{\psi}^{ch}$ and $p_{\varphi}^{ch}\leq p_{\psi}^{ch}$,
the distance between them is $\textrm{ln}(p_{\psi}^{ch}/p_{\varphi}^{ch})$.
Hence, the $\psi$th column ($|\psi_{ch}\rangle$) is located at a
distance $\textrm{ln}(p_{1}^{ch}/p_{\psi}^{ch})$ from the ``reference
column'' $|1_{ch}\rangle$. Similarly, rows with larger eigenvalues
are located above rows with smaller eigenvalues, and the distance
of the $k$th row ($|k_{v}\rangle$) from the ``reference row''
$|1_{v}\rangle$ is given by $\textrm{ln}(p_{1}^{v}/p_{k}^{v})$.

Given a partial swap $U_{|\psi\rangle\leftrightarrow|\varphi\rangle}$,
the current $J_{|\varphi\rangle\rightarrow|\psi\rangle}$ is depicted
in the $\textrm{ln}(p^{ch})\times\textrm{ln}(p^{v})$ diagram as an
arrow that connects the eigenstates $|\psi\rangle$ and $|\varphi\rangle$,
oriented according to the direction of the population transfer. In
Fig 2(b) we illustrate the global current generated by a two-level
unitary $U_{i',l,l'}$, assuming that Eq. (\ref{eq:22 recast of passivity violation with l and l'})
is valid for $\{i',l,l'\}=\{1,1,d_{v}-1\}$. Notable features are
the direction of the arrow and the fact that it is enclosed by a \textit{vertical}
rectangle, i.e. a rectangle whose height is larger that its width.
The shape of this rectangle is a graphical characterization of Eq.
(\ref{eq:22 recast of passivity violation with l and l'}). This follows
by rearranging (\ref{eq:22 recast of passivity violation with l and l'})
as $\frac{p_{i'}^{c}p_{d_{h}}^{h}}{p_{i'+1}^{c}p_{1}^{h}}<\frac{p_{l}^{v}}{p_{l'+1}^{v}}$
and applying the natural logarithm at both sides, which yields the
inequality: 

\begin{equation}
\textrm{ln}\left(\frac{p_{i'}^{c}p_{d_{h}}^{h}}{p_{i'+1}^{c}p_{1}^{h}}\right)<\textrm{ln}\left(\frac{p_{l}^{v}}{p_{l'+1}^{v}}\right).\label{eq:20 log version of passivity violation}
\end{equation}
When translated into the \textbf{$\textrm{ln}(p^{ch})\times\textrm{ln}(p^{v})$}
diagram, such inequality means that the distance between $|i'_{c}d_{h}\rangle$
and $|(i'+1)_{c}1_{h}\rangle$ is smaller than the distance between
$|l_{v}\rangle$ and $|(l'+1)_{v}\rangle$. 

Since $U_{1,1,d_{v}-1}$ is by assumption a cooling unitary, population
is transferred from $|2_{c}1_{h}1_{v}\rangle$ towards $|1_{c}d_{h}d_{v}\rangle$,
and from $|2_{c}\rangle$ towards $|1_{c}\rangle$, as indicated by
the arrows corresponding to $J_{|\varphi\rangle\rightarrow|\psi\rangle}$
and $J_{|(i'+1)_{c}\rangle\rightarrow|i'_{c}\rangle}$, respectively.
We remark that the length of these arrows must not be interpreted
as the magnitude of the associated currents. In fact, all the currents
illustrated in Fig. 2(b) must have the same magnitude, according to
Eqs. (\ref{eq:14 current}) and (\ref{eq:19.1 local current 1})-(\ref{eq:19.3 local current 3}). 

\begin{figure}[h]
	\centering{}
\includegraphics[scale=0.68]{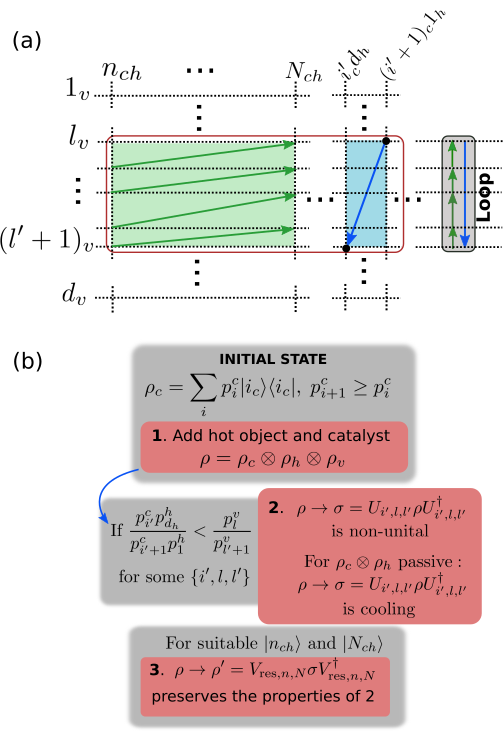}
	\centering{}\caption{(a) $\textrm{ln}(p^{ch})\times\textrm{ln}(p^{v})$ diagram for an
		example of catalytic transformation $\rho\rightarrow\rho'=U\rho U^{\dagger}$,
		where $\rho{\color{red}{\normalcolor =\rho_{c}\otimes\rho_{h}\otimes\rho_{v}}}$
		and $U=U_{i',l,l'}\oplus V_{\textrm{res},n,N}$ (cf. (\ref{eq:25 explicit U for catal transform})).
		Vertical (horizontal) ellipsis indicate that there can be rows (columns)
		in between. The unitary $U_{i^{\prime},l,l'}$ induces a global current
		(blue inclined arrow) and an associated local current $J_{|l_{v}\rangle\rightarrow|(l'+1)_{v}\rangle}$
		(blue vertical arrow) that modifies the catalyst. This effect can
		be reverted by $V_{\textrm{res},n,N}$ (cf. (\ref{eq:24 explicit Vres})),
		if there is a sequence of horizontal rectangles \textquotedblleft sandwiched\textquotedblright{}
		between the rows $l_{v}$ and $(l'+1)_{v}$ (green area). In this
		case the partial swaps $V_{|n_{ch}(k+1)_{v}\rangle\leftrightarrow|N_{ch}k_{v}\rangle}$
		can generate the global currents enclosed by such rectangles (green
		inclined arrows), whose corresponding local currents $J_{|(k+1)_{v}\rangle\rightarrow|k_{v}\rangle}$
		(green vertical arrows) cancel the effect of $J_{|l_{v}\rangle\rightarrow|(l'+1)_{v}\rangle}$.
		(b) Elemental structure of catalytic transformations studied here.
		1, 2 and 3 represent logical steps to achieve a desired transformation.
		The gray backgrounds contain underlying conditions for each step. }
\end{figure}

In Fig. 3(a) we apply the $\textrm{ln}(p^{ch})\times\textrm{ln}(p^{v})$
diagram to illustrate the currents generated by a catalytic unitary
$U_{i^{\prime},l,l'}\oplus V_{\textrm{res},n,N}$. Since each $V_{|n_{ch}(k+1)_{v}\rangle\leftrightarrow|N_{ch}k_{v}\rangle}$
produces a positive current $J_{|(k+1)_{v}\rangle\rightarrow|k_{v}\rangle}$
iff $p_{n}^{ch}p_{k+1}^{v}>p_{N}^{ch}p_{k}^{v}$, as per Eqs. (\ref{eq:14 current})
and (\ref{eq:19.3 local current 3}), it follows that: 
\begin{equation}
\textrm{ln}\left(\frac{p_{n}^{ch}}{p_{N}^{ch}}\right)>\textrm{ln}\left(\frac{p_{k}^{v}}{p_{k+1}^{v}}\right).\label{eq:26 ln depiction of currents J|k+1>-->|k>}
\end{equation}
This inequality implies that the distance between the columns $|n_{ch}\rangle$
and $|N_{ch}\rangle$ is larger than the distance between the rows
$|k_{v}\rangle$ and $|(k+1)_{v}\rangle$. Therefore, for $l\leq k\leq l'$
all the global currents $J_{|n_{ch}(k+1)_{v}\rangle\rightarrow|N_{ch}k_{v}\rangle}$
are enclosed by \textit{horizontal} rectangles (with height larger
than its width), as shown in Fig. 3(a). Moreover, the associated local
currents $J_{|(k+1)_{v}\rangle\rightarrow|k_{v}\rangle}$ are \textit{upward-oriented},
thus generating the loop with the current $J_{|l_{v}\rangle\rightarrow|(l'+1)_{v}\rangle}$.

Before moving to the next section, let us make some important comments: 
\begin{itemize}
\item The depictions in the $\textrm{ln}(p^{ch})\times\textrm{ln}(p^{v})$
diagram are not intended to be quantitatively precise, but to provide
sufficient information at the qualitative level. This means that separations
between rows and columns can be imprecise as long as the figure informs
correctly which distances are larger and which are smaller. 
\item Equation (\ref{eq:26 ln depiction of currents J|k+1>-->|k>}) is equivalent
to the positivity of $J_{|(k+1)_{v}\rangle\rightarrow|k_{v}\rangle}$,
given the monotonic character of the natural logarithm. It also implies
that if the ratio $\frac{p_{k}^{v}}{p_{k+1}^{v}}$ is \textit{too}
\textit{large}, for some $k\in\{l,l+1,...,l'\}$, it may impossible
to satisfy (\ref{eq:26 ln depiction of currents J|k+1>-->|k>}) for
any $n,N\in\{1,2,...,d_{c}d_{h}\}$. In particular, this is always
true if $p_{k}^{v}=1$ and $p_{k+1}^{v}=0$, which describes a \textit{pure}
state $\rho_{v}=|k_{v}\rangle\langle k_{v}|$. We can thus conclude
that a catalytic transformation $\rho\rightarrow U\rho U^{\dagger}$,
where $U$ satisfies Eq. (\ref{eq:25 explicit U for catal transform}),
is possible only if $\rho_{v}$ is a \textit{mixed} state. 
\item Figure 3(b) summarizes the essential features of the catalytic transformations
addressed in this article. Importantly, Corollaries 1 and 2 provide
conditions for $V_{\textrm{res},n,N}$ to preserve the cooling or
non-unital character of the initial transformation $\rho\rightarrow U_{i^{\prime},l,l'}\rho U_{i^{\prime},l,l'}^{\dagger}$.
We also stress that while $V_{\textrm{res},n,N}$ possess the specific
form (\ref{eq:24 explicit Vres}), its existence is also necessary
for the existence of other restoring unitaries characterized in Theorem
1. 
\end{itemize}

\section{Catalyst dimension as a resource for cooling and non-unital transformations}

Theorem 1 and the derived corollaries (1 and 2) provide sufficient
conditions for catalytic transformations, given a \textit{fixed} state
$\rho=\rho_{c}\otimes\rho_{h}\otimes\rho_{v}$. Now, we ask ourselves
the following question: given a fixed state $\rho_{c}\otimes\rho_{h}$,
is there a catalyst state $\rho_{v}$ such that $\rho\overset{\textrm{CC}}{\longrightarrow}U\rho U^{\dagger}$
or $\rho\overset{\textrm{CNU}}{\longrightarrow}U\rho U^{\dagger}$
are possible? This question is intimately related to the dimension
of the catalyst, as seen in the following theorem. The proof can be
consulted in Appendix G. 

\textbf{Theorem 2 (catalyst size and catalytic transformations)}.\textit{
Given a catalyst dimension $d_{v}=d_{v}^{\ast}$ (where $d_{v}^{\ast}$
is a sufficiently large and explicit dimension derived in Appendix
G) and a suitable catalyst state $\rho_{v}=\sum_{k=1}^{d_{v}^{\ast}}p_{k}^{v}|k_{v}\rangle\langle k_{v}|$,
the following transformations are possible:}
\begin{enumerate}
\item \textbf{\textit{Catalytic and cooling transformations}}\textit{: If
$\rho_{c}\otimes\rho_{h}$ is a passive state, where $\rho_{h}$ is
not fully mixed (i.e. ${\normalcolor p_{1}^{h}>p_{d_{h}}^{h}}$),
there exists a explicit state $\rho_{v}$ such that for $\rho=\rho_{c}\otimes\rho_{h}\otimes\rho_{v}$
a transformation} $\rho\overset{\textrm{CC}}{\longrightarrow}U\rho U^{\dagger}$
\textit{can be implemented. }
\item \textbf{\textit{Catalytic and non-unital transformations}}\textit{:
If $\rho_{c}$ satisfies $p_{1}^{c}>p_{i'}^{c}$ or $p_{i'+1}^{c}>p_{d_{c}}^{c}$
for some $i'\in\{1,2,...,d_{c}-1\}$, and $d_{c}\geq3$, there exists
a explicit state $\rho_{v}$ such that for $\rho=\rho_{c}\otimes\rho_{v}$
a transformation} $\rho\overset{\textrm{CNU}}{\longrightarrow}U\rho U^{\dagger}$
\textit{can be implemented}.\textit{ }
\end{enumerate}
According to Theorem 2, a sufficiently large catalyst allows catalytic
transformations for almost any initial state $\rho_{c}\otimes\rho_{h}$.
In particular, Statement 1 implies that any hot object with non-degenerate
energy spectrum and finite temperature suffices to perform catalytic
cooling. Furthermore, Statement 2 tells us that, in the case of non-unital
transformations, the hot object can be ignored if $\rho_{c}$ satisfies
the mentioned properties. In other words, there exists a unitary $U$
that performs the transformation and acts on $\mathcal{H}_{c}\otimes\mathcal{H}_{v}$
, as shown in Appendix G2. 

It is also worth pointing out that a harmonic oscillator constitutes
an example of universal catalyst, in the sense that it can be prepared
in any required state $\rho_{v}$. To that end, we only need to populate
$d_{v}^{\ast}$ of its levels with the eigenvalues of $\rho_{v}$,
irrespective of how large is $d_{v}^{\ast}$. 

\section{Examples of catalytic cooling}

In the first two parts of this section we illustrate different catalytic
and cooling transformations that stem from Theorem 2, some of which
rely on restoring unitaries that generalize those based on Eq. (\ref{eq:24 explicit Vres}).
The last section (VI-C) explores the new scenario of cooling \textit{enhancement}
via a catalyst, not considered until now. Before proceeding with the
examples, we shall briefly explain how the aforementioned generalization
takes place. 

The essential idea is that, depending on the eigenvalues of $\rho$,
there could be various restoring unitaries of the form (\ref{eq:24 explicit Vres}).
This would occur if for several pairs of states $(|n_{ch}\rangle,|N_{ch}\rangle)$
\textit{all} the partial swaps in the set $\{V_{|n_{ch}(k+1)_{v}\rangle\leftrightarrow|N_{ch}k_{v}\rangle}\}_{k=l}^{l'}$
can generate positive currents $J_{|(k+1)_{v}\rangle\rightarrow|k_{v}\rangle}$.
Under this condition, we show in Appendix H that another restoring
unitary can be obtained as 
\begin{align}
V_{\textrm{res}} & =\oplus_{n,N}V_{\textrm{res,}n,N}\nonumber \\
 & =\oplus_{k=l}^{l'}\left(\oplus_{n,N}V_{|n_{ch}(k+1)_{v}\rangle\leftrightarrow|N_{ch}k_{v}\rangle}\right).\label{eq:26.5 sum of Vres}
\end{align}
In Eq. (\ref{eq:26.5 sum of Vres}), as well as in Eq. (\ref{eq:24 explicit Vres}),
the swap intensities are implicit and can take any value $0<r\leq1$.
However, as long as $p_{n}^{ch}p_{k+1}^{v}-p_{N}^{ch}p_{k}^{v}>0$
for any pair $(|n_{ch}\rangle,|N_{ch}\rangle)$ in the sum, it is
always possible to tune these intensities in such a ways that $U=U_{i',l,l'}\oplus V_{\textrm{res}}$
gives rise to a catalytic transformation (cf. Appendix H). 

To illustrate the usefulness of Eq. (\ref{eq:26.5 sum of Vres}) we
now generalize the controlled unitaries (\ref{eq:26.3 Vres for corollary 1})
(assuming that they are valid restoring unitaries, as per Corollary
1) to a restoring unitary $V_{\textrm{res}}$ that acts on $\mathcal{H}_{h}\otimes\mathcal{H}_{v}$.
Such a property is important because it means that in this case the
cold object is not involved in the restoration of the catalyst, and
therefore a two-body interaction with the hot object is sufficient.
Since the set $\{V_{\textrm{res},|i_{c}^{\prime\prime}\rangle}\}_{i^{\prime\prime}=1}^{d_{c}}$
constitutes a family of restoring unitaries, $\oplus_{i^{\prime\prime}}V_{\textrm{res},|i_{c}^{\prime\prime}\rangle}$
is also a restoring unitary of the form (\ref{eq:26.5 sum of Vres}),
where $|n_{ch}\rangle$ and $|N_{ch}\rangle$ are related to $i^{\prime\prime}$
through (\ref{eq:26.1 n and N for CC transformation}). In addition,
from (\ref{eq:26.3 Vres for corollary 1}) it readily follows that
\begin{align}
\oplus_{i^{\prime\prime}}V_{\textrm{res},|i_{c}^{\prime\prime}\rangle} & =\mathbb{I}_{c}\otimes\left(\oplus_{k=l}^{l'}V_{|1_{h}(k+1)_{v}\rangle\leftrightarrow|d_{h}k_{v}\rangle}\right)\nonumber \\
 & =V_{\textrm{res},hv},\label{eq:26.6 Vres,hv}
\end{align}
where the subindex $hv$ indicates that $V_{\textrm{res},hv}$ acts
on $\mathcal{H}_{h}\otimes\mathcal{H}_{v}$. 

\subsection{Optimal catalytic cooling of a qubit using another qubit as hot object}

Based on Statement 1 of Theorem 2, our goal now is to explore how
catalysts of different dimensions perform to cool a qubit using as
hot object another qubit, with respective initial states $\rho_{c}=\sum_{i=1}^{2}p_{i}^{c}|i_{c}\rangle\langle i_{c}|$
and $\rho_{h}=\sum_{j=1}^{2}p_{i}^{h}|j_{h}\rangle\langle j_{h}|$.
In this case, the passivity constraint for $\rho_{c}\otimes\rho_{h}$
yields the simple inequality $p_{2}^{c}\leq p_{2}^{h}$. Moreover,
the cooling effect is due to the unitary $U_{i',1,d_{v}-1}$ (cf.
Eq. (\ref{eq:23 Ui',l,l'})), which only admits the value $i'=1$
for dimension $d_{c}=2$. Assuming $\varepsilon_{1}^{c}=0$, Eq. (\ref{eq:23.1 Delta<Hc> with cooling current})
yields 
\begin{align*}
\Delta\bigl\langle H_{c}\bigr\rangle & =\textrm{Tr}\left[H_{c}\left(U\rho U^{\dagger}-\rho\right)\right]\\
 & =-J_{\textrm{cool}}\varepsilon_{2}^{c},
\end{align*}
where $U=U_{1,1,d_{v}-1}\oplus V_{\textrm{res},hv}$ and $J_{\textrm{cool}}=J_{|2_{c}\rangle\rightarrow|1_{c}\rangle}$
is the cooling current induced by $U_{1,1,d_{v}-1}$. 

While in Theorem 2 we refer to a certain catalyst state that lifts
the passivity constraint by enabling cooling (Statement 1), here we
are interested in an \textit{optimal} state. The associated optimization
means that, if cooling is possible for a certain dimension $d_{v}^{\ast}$,
we maximize it over the eigenvalues $\{p_{k}^{v}\}_{k}$ of \textit{full-rank}
states $\rho_{v}=\sum_{k=1}^{d_{v}^{\ast}}p_{k}^{v}|k_{v}\rangle\langle k_{v}|$,
which amounts to maximize $J_{\textrm{cool}}$ over $\{p_{k}^{v}\}_{k}$.
Full-rank states are chosen because the explicit state $\rho_{v}$
that allows us to prove Theorem 2 is of this form. 

\begin{figure}[H]
	\centering{}
	\includegraphics[scale=0.55]
	{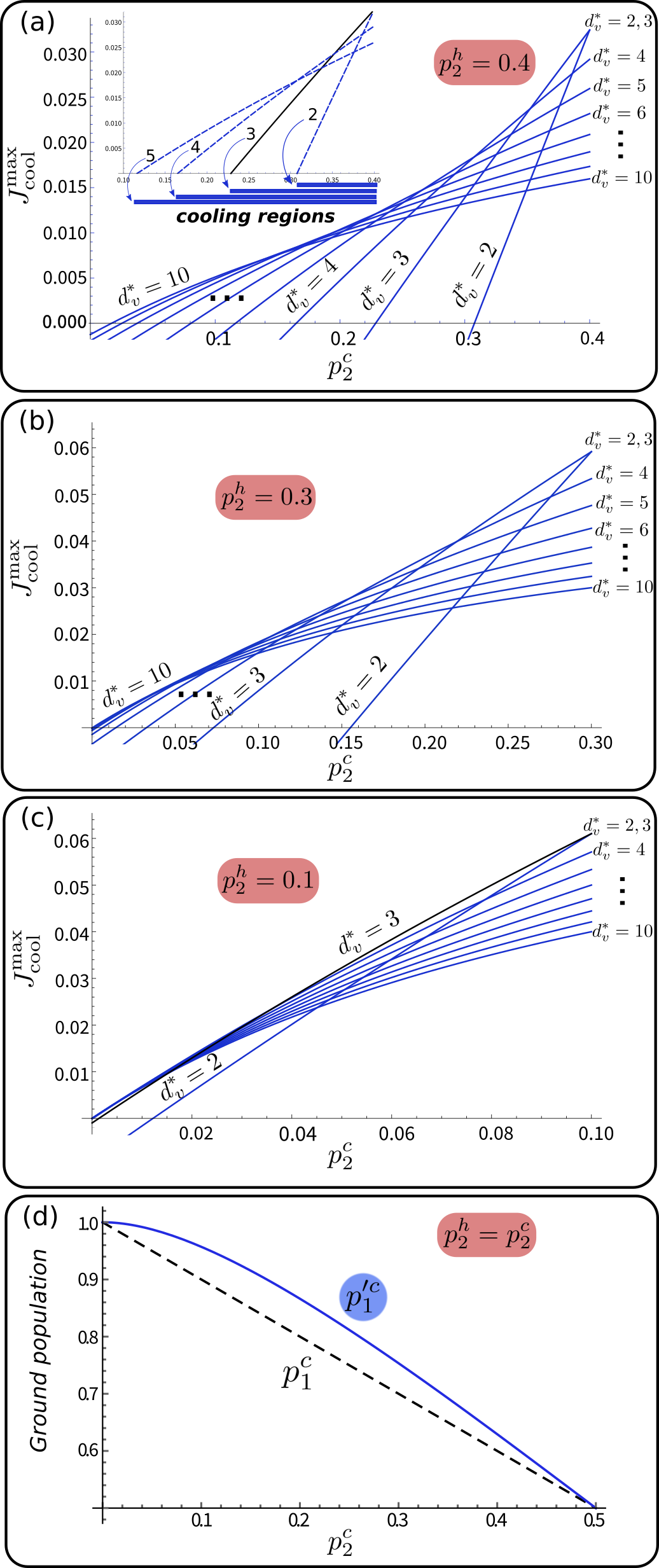}\caption{Optimal catalytic cooling of a qubit using another qubit as hot object.
		(a), (b) and (c) show the cooling currents obtained with catalyst
		states $\rho_{v}$ of dimensions $2\protect\leq d_{v}^{\ast}\protect\leq10$,
		and hot qubits with initial (excited) populations $p_{2}^{h}=0.4$,
		$p_{2}^{h}=0.3$, and $p_{2}^{h}=0.1$, respectively. The inset in
		(a) shows the cooling regions corresponding to $2\protect\leq d_{v}^{\ast}\protect\leq5$,
		where $d_{v}^{\ast}$ is sufficiently large to allow cooling. Since
		the maxima of $J_{\textrm{cool}}^{\textrm{max}}$ in (a), (b) and
		(c) are achieved for $p_{2}^{c}=p_{2}^{h}$ and $d_{v}^{\ast}=2,3$,
		in (d) we consider the cooling using a two-level catalyst ($d_{v}^{\ast}=2$)
		and a hot qubit such that $p_{2}^{c}=p_{2}^{h}$. The black dashed
		line depicts the initial ground population of the cold qubit and the
		blue curve is the corresponding final population.}
\end{figure}

In Appendix I we obtain the maximum current $J_{\textrm{cool}}$:
\begin{equation}
J_{\textrm{cool}}^{\textrm{max}}=\left[\frac{(\zeta-1)p_{2}^{h}\left(p_{2}^{c}\zeta^{d_{v}^{\ast}}-p_{1}^{c}\right)}{\zeta^{d_{v}^{\ast}-1}(\zeta-1)+p_{1}^{c}(\zeta^{d_{v}^{\ast}-1}-1)}\right]\bar{p}_{1}^{v},\label{eq:36 max Jcool}
\end{equation}
where $\zeta\equiv\frac{p_{1}^{h}}{p_{2}^{h}}$ and the superscripts
in $\zeta$ indicate powers.The optimal eigenvalues are denoted as
$\{\bar{p}_{k}^{(v)}\}_{k}$ and are also derived in the same appendix.
For these eigenvalues, the unitary that generates the current (\ref{eq:36 max Jcool})
is given by 

\begin{align}
U & =\mathcal{U}_{1,1,d_{v}^{\ast}-1}\oplus V_{\textrm{res},hv},\label{eq:34 optimal CC unitary}\\
V_{\textrm{res},hv} & =\oplus_{k=1}^{d_{v}^{\ast}-1}\mathcal{V}_{|1_{h}(k+1)_{v}\rangle\leftrightarrow|2_{h}k_{v}\rangle}.\label{eq:34.1 optimal rest unitary}
\end{align}

This implies that optimal cooling is achieved by setting $r=1$ for
all the two-level unitaries that compose $U$. This results in a direct
sum of swaps, and consequently $U$ in Eq. (\ref{eq:34 optimal CC unitary})
is a \textit{permutation} of the eigenstates of $\rho$. 

Figure 4 shows plots of $J_{\textrm{cool}}^{\textrm{max}}$ and the
final ground population $p_{1}^{\prime c}$, for different values
of $p_{2}^{h}$ and catalyst of dimensions $2\leq d_{v}^{\ast}\leq10$.
Each solid curve in Figs. 4(a)-(c) depicts the maximum cooling current
corresponding to a different value of $d_{v}^{\ast}$. Moreover, $J_{\textrm{cool}}^{\textrm{max}}$
in Eq. (\ref{eq:36 max Jcool}) is plotted as a function of $0\leq p_{2}^{c}\leq p_{2}^{h}$,
which constitutes the interval where $\rho_{c}\otimes\rho_{h}$ is
passive. In Fig. 4(a) we can see that as $d_{v}^{\ast}$ increases
the interval of $p_{2}^{c}$ where $J_{\textrm{cool}}^{\textrm{max}}$
is positive also increases. Since $J_{\textrm{cool}}^{\textrm{max}}<0$
means that population would be transferred from the ground state to
the excited state of the cold qubit, thereby heating it up, the ``cooling
region'' is described by the condition $J_{\textrm{cool}}^{\textrm{max}}\geq0$.
The inset in Fig. 4(a) shows more clearly the cooling regions (blue
bars) corresponding to states of dimensions $2\leq d_{v}^{\ast}\leq5$.
The enlargement of these regions as $d_{v}^{\ast}$ increases indicates
that \textit{larger catalysts may allow cooling in regimes not accessible
to small catalysts}, characterized by $p_{2}^{c}\ll p_{2}^{h}$. On
the other hand, for $p_{2}^{c}=p_{2}^{h}$ it is remarkable that $J_{\textrm{cool}}^{\textrm{max}}$
is maximized by $d_{v}^{\ast}=2$ and $d_{v}^{\ast}=3$, and decreases
for larger values of $d_{v}^{\ast}$. This implies that in such a
case \textit{the smallest possible catalyst, corresponding to a two-level
system, is sufficient to achieve maximum cooling}. Moreover, it is
also worth noting that \textit{the cooling current corresponding to
$d_{v}^{\ast}=3$ always surpasses the current corresponding to $d_{v}^{\ast}=2$}
(except for $p_{2}^{c}=p_{2}^{h}$). 

Figures 4(b) and 4(c) display the same pattern that characterizes
Fig. 4(a). In particular, notice that in both cases a catalyst of
dimension $d_{v}^{\ast}=10$ allows to cool for almost any value of
$p_{2}^{c}$. In Fig. 4(c) we also see that a catalyst with $d_{v}^{\ast}=3$
(black curve) is essentially as effective as any catalyst of dimension
$4\leq d_{v}^{\ast}\leq10$. Accordingly, in this case a\textit{ three-level
catalyst is optimal for almost any value of $p_{2}^{c}$}. Figure
4(d) shows the initial and final ground populations as a function
of $p_{2}^{c}$, if the populations of the hot and cold qubits always
coincide. The final population is computed as $p_{1}^{\prime c}=p_{1}^{c}+J_{\textrm{cool}}^{\textrm{max}}$,
where $J_{\textrm{cool}}^{\textrm{max}}$ is the cooling current attained
for $d_{v}^{\ast}=2$ (or $d_{v}^{\ast}=3$).

\subsection{Catalytic increment of the ground population of the cold object}

Reducing the average energy of the cold object is not the only approach
for cooling. Alternatively, increasing the ground population of a
quantum system has also been considered as a way to cool it \citep{37Huber-cooling-bound1}.
As the following proposition shows, such an increment constitutes
yet another example of useful non-unital transformation. 

\textbf{Proposition 1}. \textit{Any transformation $\rho_{c}\rightarrow\rho'_{c}$
such that} $\textrm{Tr}\left(|1_{c}\rangle\langle1_{c}|\rho'_{c}\right)>\textrm{Tr}\left(|1_{c}\rangle\langle1_{c}|\rho_{c}\right)$
\textit{is a non-unital transformation.}

\textit{Proof}. To prove this proposition we use the fact that the
aforementioned transformation can always be cast as a cooling transformation,
given a suitable energy spectrum $\{\varepsilon_{i}^{c}\}_{i}$. Specifically,
we can consider a Hamiltonian $H_{c}=\sum_{i=1}^{d_{c}}\varepsilon_{i}^{c}|i_{c}\rangle\langle i_{c}|$
with eigenvalues that satisfy $\varepsilon_{1}^{c}<\varepsilon_{2}^{c}$,
and $\varepsilon_{2}^{c}=\varepsilon_{i}^{c}$ for $2\leq i\leq d_{c}$.
These eigenvalues ensure that the state $\rho_{c}=\sum_{i}p_{i}^{c}|i_{c}\rangle\langle i_{c}|$
(with $p_{i}^{c}\geq p_{i+1}^{c}$) is passive. Denoting the population
variation corresponding to $|i_{c}\rangle$ as $\Delta p_{i}^{c}$,
and applying probability conservation $\Delta p_{1}^{c}=-\sum_{i=2}^{d_{c}}\Delta p_{i}^{c}$,
we have that $\sum_{i=2}^{d_{c}}\Delta p_{i}^{c}\varepsilon_{i}^{c}=-\Delta p_{1}^{c}\varepsilon_{2}^{c}$.
Accordingly, 
\begin{align}
\Delta\bigl\langle H_{c}\bigr\rangle & =\Delta p_{1}^{c}\varepsilon_{1}^{c}+\sum_{i=2}^{d_{c}}\Delta p_{i}^{c}\varepsilon_{i}^{c}\nonumber \\
 & =\Delta p_{1}^{c}(\varepsilon_{1}^{c}-\varepsilon_{2}^{c}),\label{eq:35 Delta<Hc> for transformation that increases p1c}
\end{align}
which is negative for any increment $\Delta p_{1}^{c}=\textrm{Tr}\left(|1_{c}\rangle\langle1_{c}|(\rho'_{c}-\rho_{c})\right)>0$.
Since any cooling transformation is non-unital (cf. Definition 3 and
subsequent comments), any transformation that increases the ground
population is non-unital. 

In Appendix G2 (Corollary 3) we show how the existence of a catalytic
transformation that satisfies $\Delta p_{1}^{c}>0$ follows from the
constructive proof for Statement 2 of Theorem 2. Accordingly, a catalytic
increment of $\bigl\langle|1_{c}\rangle\langle1_{c}|\bigr\rangle$
can be performed via a transformation $\rho_{c}\otimes\rho_{v}\rightarrow U\rho_{c}\otimes\rho_{v}U^{\dagger}$,
using a sufficiently large catalyst, and without requiring the hot
object. Corollary 3 indicates that for this to be possible it suffices
to consider a cold object whose eigenvalues satisfy $p_{2}^{c}>p_{d_{c}}^{c}$.

\subsection{Catalyst-aided \textit{enhancement} of cooling }

The usefulness of catalysts is not restricted to the implementation
of transformations that are forbidden without the utilization of these
systems. Here we show that cooling can be catalytically enhanced,
even if the hot object is sufficient to achieve a certain level of
cooling. This is formally stated in the following theorem, whose proof
consists of two parts and is given in Appendix J. First, we derive
a global unitary that provides optimal cooling without using the catalyst,
and then construct a catalytic transformation that yields the enhancement.
We also remark that optimal cooling unitaries for a qubit interacting
with a finite environment have been shown in Ref. \citep{36.1Karen-limits-cooling}.
However, we present a derivation based on passivity, in order to maintain
a self-contained structure.

\textbf{Theorem 3 (cooling enhancement with a catalyst)}. \textit{Let
$\rho_{c}$ be a passive state of a qubit, and let $\rho_{c}\otimes\rho_{h}$
be a }\textbf{\textit{non-passive}}\textit{ state, where $\rho_{h}=\sum_{j=1}^{d_{h}}p_{j}^{h}|j_{h}\rangle\langle j_{h}|$
is the state of a hot object of even dimension $d_{h}\geq4$. If $d_{v}$
is sufficiently large and $p_{1}^{h}>p_{d_{h}/2}^{h}$ or $p_{d_{h}/2}^{h}>p_{d_{h}}^{h}$,
there exist a explicit catalyst state that }\textbf{\textit{increases}}\textit{
the optimal cooling achieved with the hot object alone. That is, there
exists a catalytic transformation} $\rho=\rho_{c}\otimes\rho_{h}\otimes\rho_{v}\overset{\textrm{CC}}{\longrightarrow}\rho'=U\rho U^{\dagger}$,
\textit{where $\rho_{v}=\sum_{k=1}^{d_{v}}|k_{v}\rangle\langle k_{v}|$
and} $\textrm{Tr}\left(H_{c}\rho'\right)<\textrm{min}_{U_{ch}}\textrm{Tr}\left(H_{c}U_{ch}\rho_{c}\otimes\rho_{h}U_{ch}^{\dagger}\right)$,
\textit{for arbitrary unitaries $U_{ch}$ acting on $\mathcal{H}_{c}\otimes\mathcal{H}_{h}$. }

\begin{figure}[H]
\centering{}
\includegraphics{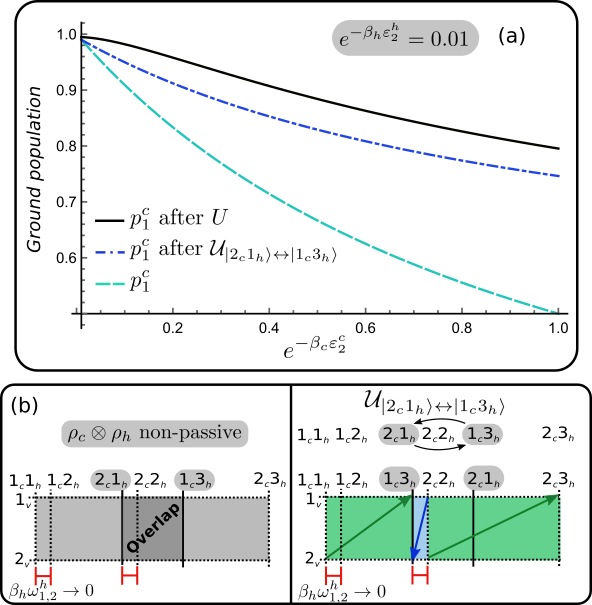}\caption{Catalytic enhancement of cooling using a qubit as catalyst and a three-level
system as hot object. (a) Initial ground population of the cold qubit
(cyan dashed), ground population after optimal cooling without the
catalyst (blue dashed-dotted), and ground population after a subsequent
catalytic and cooling transformation (black solid). The parameter
$e^{-\beta_{h}\varepsilon_{2}^{h}}=0.01$ is fixed. (b) Schematic
representation of the catalytic transformation. (Left) $\textrm{ln}(p^{ch})\times\textrm{ln}(p^{v})$
diagram for the \textit{non-passive} state $\rho_{c}\otimes\rho_{h}$
(note that $1_{c}3_{h}$ is at the right of $2_{c}1_{h}$). The distance
$\beta_{h}\omega_{1,2}^{h}\equiv\beta_{h}(\varepsilon_{2}^{h}-\varepsilon_{1}^{h})$
tends to zero to comply with the degeneracy $\varepsilon_{2}^{h}=\varepsilon_{1}^{h}$.
(Right) After the optimal cooling using $\mathcal{U}_{|2_{c}1_{h}\rangle\leftrightarrow|1_{c}3_{h}\rangle}$
the unitary $U$ in (\ref{eq:37  U catalytic}) is applied on the
resulting state. The resulting currents are depicted in the diagram
below. }
\end{figure}

While Theorem 3 concerns catalytic enhancement of cooling using hot
objects of even dimension, we also show that this is possible by means
of a three-level hot object. The following example is based on a three-level
system with Hamiltonian $H_{h}=\sum_{j=1}^{3}\varepsilon_{j}^{h}|j_{h}\rangle\langle j_{h}|$,
with degeneracy $\varepsilon_{1}^{h}=\varepsilon_{2}^{h}=0$ and a
non-null energy gap $\omega_{2,3}^{h}=\varepsilon_{3}^{h}-\varepsilon_{2}^{h}>0$.
We characterize cooling in terms of the final ground population of
the cold qubit, keeping in mind that in this case the minimization
of the average energy amounts to maximize the ground population. 

In Fig. 5(a) we show the maximum cooling attainable via a hot object
prepared in the thermal state $\rho_{h}=\frac{e^{-\beta_{h}H_{h}}}{\textrm{Tr}(e^{-\beta_{h}H_{h}})}$,
as well as an additional cooling through a transformation that employs
a qubit as catalyst. The total transformation is thus a composition
$\rho\rightarrow\sigma_{ch}\otimes\rho_{v}\rightarrow\rho'$, where
$\rho=\rho_{c}\otimes\rho_{h}\otimes\rho_{v}$, $\sigma_{ch}=U_{ch}\rho_{c}\otimes\rho_{h}U_{ch}^{\dagger}$,
being $U_{ch}$ a unitary that optimally cools $\rho_{c}$ using $\rho_{h}$,
and $\rho'=U\sigma_{ch}\otimes\rho_{v}U^{\dagger}$, being $U$ a
catalytic and cooling unitary. Following Eq. (\ref{eq:4 passiv cond with populations}),
$\rho_{c}\otimes\rho_{h}$ is non-passive with respect to $H_{c}$
iff $\frac{p_{1}^{c}}{p_{2}^{c}}<\frac{p_{1}^{h}}{p_{3}^{h}}$, which
implies also that the swap $\mathcal{U}_{|2_{c}1_{h}\rangle\leftrightarrow|1_{c}3_{h}\rangle}$
cools down the cold qubit by the amount $\Delta\bigl\langle H_{c}\bigr\rangle=(p_{2}^{c}p_{1}^{h}-p_{1}^{c}p_{3}^{h})(\varepsilon_{1}^{c}-\varepsilon_{2}^{c})$.
In fact, it is not difficult to corroborate that this swap corresponds
to the optimal cooling unitary $U_{ch}^{\textrm{opt}}$. To that end
we show that the application of $\mathcal{U}_{|2_{c}1_{h}\rangle\leftrightarrow|1_{c}3_{h}\rangle}$
yields a passive state with respect to $H_{c}$. Since the only effect
of $\mathcal{U}_{|2_{c}1_{h}\rangle\leftrightarrow|1_{c}3_{h}\rangle}$
is to exchange the eigenvalues of $|2_{c}1_{h}\rangle$ and $|1_{c}3_{h}\rangle$,
the resulting state $\sigma_{ch}$ reads

\begin{align}
\sigma_{ch} & =|1_{c}\rangle\langle1_{c}|\otimes\left(\sum_{j=1}^{2}p_{1}^{c}p_{j}^{h}|j_{h}\rangle\langle j_{h}|+p_{2}^{c}p_{1}^{h}|3_{h}\rangle\langle3_{h}|\right)\nonumber \\
 & \quad+|2_{c}\rangle\langle2_{c}|\otimes\left(p_{1}^{c}p_{3}^{h}|1_{h}\rangle\langle1_{h}|+\sum_{j=2}^{3}p_{2}^{c}p_{j}^{h}|j_{h}\rangle\langle j_{h}|\right).\label{eq:36 passive state Sigma(ch)}
\end{align}

This state is such that all the eigenvalues in the first line of Eq.
(\ref{eq:36 passive state Sigma(ch)}) are larger or equal than the
eigenvalues in the second line: Clearly, $p_{1}^{c}p_{2}^{h}\geq p_{1}^{c}p_{3}^{h}$
and $p_{1}^{c}p_{2}^{h}\geq p_{2}^{c}p_{2}^{h}$, which guarantees
that the aforementioned property holds when comparing the eigenvalues
in the sum $\sum_{j=1}^{2}p_{1}^{c}p_{j}^{h}|j_{h}\rangle\langle j_{h}|$
with all the eigenvalues of the second line. Furthermore, $p_{2}^{c}p_{1}^{h}\geq p_{2}^{c}p_{2}^{h}$,
which guarantees that $p_{2}^{c}p_{1}^{h}$ is larger or equal than
the eigenvalues in $\sum_{j=2}^{3}p_{2}^{c}p_{j}^{h}|j_{h}\rangle\langle j_{h}|$,
and $p_{2}^{c}p_{1}^{h}>p_{1}^{c}p_{3}^{h}$ is equivalent to the
non-passivity of $\rho_{c}\otimes\rho_{h}$. In this way, the passivity
of $\sigma_{ch}$ can be concluded by noting that $[H_{c}\otimes\mathbb{I}_{h},\sigma_{ch}]=0$
and that the eigenvalues of $H_{c}\otimes\mathbb{I}_{h}$ regarding
eigenstates in the first (second) line of (\ref{eq:36 passive state Sigma(ch)})
are all equal to $\varepsilon_{1}^{c}$ ($\varepsilon_{2}^{c}$).
Hence, the eigenvalues of $\sigma_{ch}$ are non-increasing with respect
to those of $H_{c}\otimes\mathbb{I}_{h}$. 

In Fig. 5(a) we set $e^{-\beta_{h}\varepsilon_{3}^{h}}=0.01$, thereby
fixing the eigenvalues of $\rho_{h}$ (taking into account the degeneracy
$\varepsilon_{1}^{h}=\varepsilon_{2}^{h}$). The blue dash-dotted
curve depicts the ground population after the initial transformation
$\rho\rightarrow\sigma_{ch}\otimes\rho_{v}$, and the black solid
curve stands for the final population achieved with the subsequent
transformation $\sigma_{ch}\otimes\rho_{v}\rightarrow\rho'$. This
transformation is executed through a permutation
\begin{align}
U & =\mathcal{U}_{|2_{c}2_{h}1_{v}\rangle\leftrightarrow|1_{c}3_{h}2_{v}\rangle}\oplus V_{\textrm{res}},\label{eq:37  U catalytic}\\
V_{\textrm{res}} & =\mathcal{V}_{|1_{c}1_{h}2_{v}\rangle\leftrightarrow|1_{c}3_{h}1_{v}\rangle}\oplus\mathcal{V}_{|2_{c}2_{h}2_{v}\rangle\leftrightarrow|2_{c}3_{h}1_{v}\rangle}\nonumber \\
 & =\left[|1_{c}\rangle\langle1_{c}|\otimes\mathcal{V}_{|1_{h}2_{v}\rangle\leftrightarrow|3_{h}1_{v}\rangle}+|2_{c}\rangle\langle2_{c}|\otimes\mathbb{I}_{hv}\right]\nonumber \\
 & \quad\oplus\left[|2_{c}\rangle\langle2_{c}|\otimes\mathcal{V}_{|2_{h}2_{v}\rangle\leftrightarrow|3_{h}1_{v}\rangle}+|1_{c}\rangle\langle1_{c}|\otimes\mathbb{I}_{hv}\right],\label{eq:38 Vres for U catalytic}
\end{align}
which is derived in Appendix J2 . Noting that $[V_{\textrm{res}},H_{c}]=0$,
we have that the only contribution to $\Delta_{U}\bigl\langle H_{c}\bigr\rangle\equiv\textrm{Tr}\left[H_{c}\left(U\sigma_{ch}\otimes\rho_{v}U^{\dagger}-\sigma_{ch}\otimes\rho_{v}\right)\right]$
comes from the swap $\mathcal{U}_{|2_{c}2_{h}1_{v}\rangle\leftrightarrow|1_{c}3_{h}2_{v}\rangle}$.
Since $\mathcal{U}_{|2_{c}2_{h}1_{v}\rangle\leftrightarrow|1_{c}3_{h}2_{v}\rangle}$
reduces $\bigl\langle H_{c}\bigr\rangle$, as shown in J2, it follows
that $\Delta_{U}\bigl\langle H_{c}\bigr\rangle<0$. Specifically,
this variation is given by 
\begin{equation}
\Delta_{U}\bigl\langle H_{c}\bigr\rangle=\left(\frac{p_{2}^{h}p_{1}^{c}-p_{3}^{h}p_{2}^{c}}{(1+p_{2}^{c})p_{2}^{h}+p_{2}^{c}}\right)p_{2}^{c}p_{2}^{h}(\varepsilon_{1}^{c}-\varepsilon_{2}^{c}).\label{eq:40 Jcool with total restoring current}
\end{equation}

\begin{figure}[t]
	\centering{}
	\includegraphics[scale=0.6]{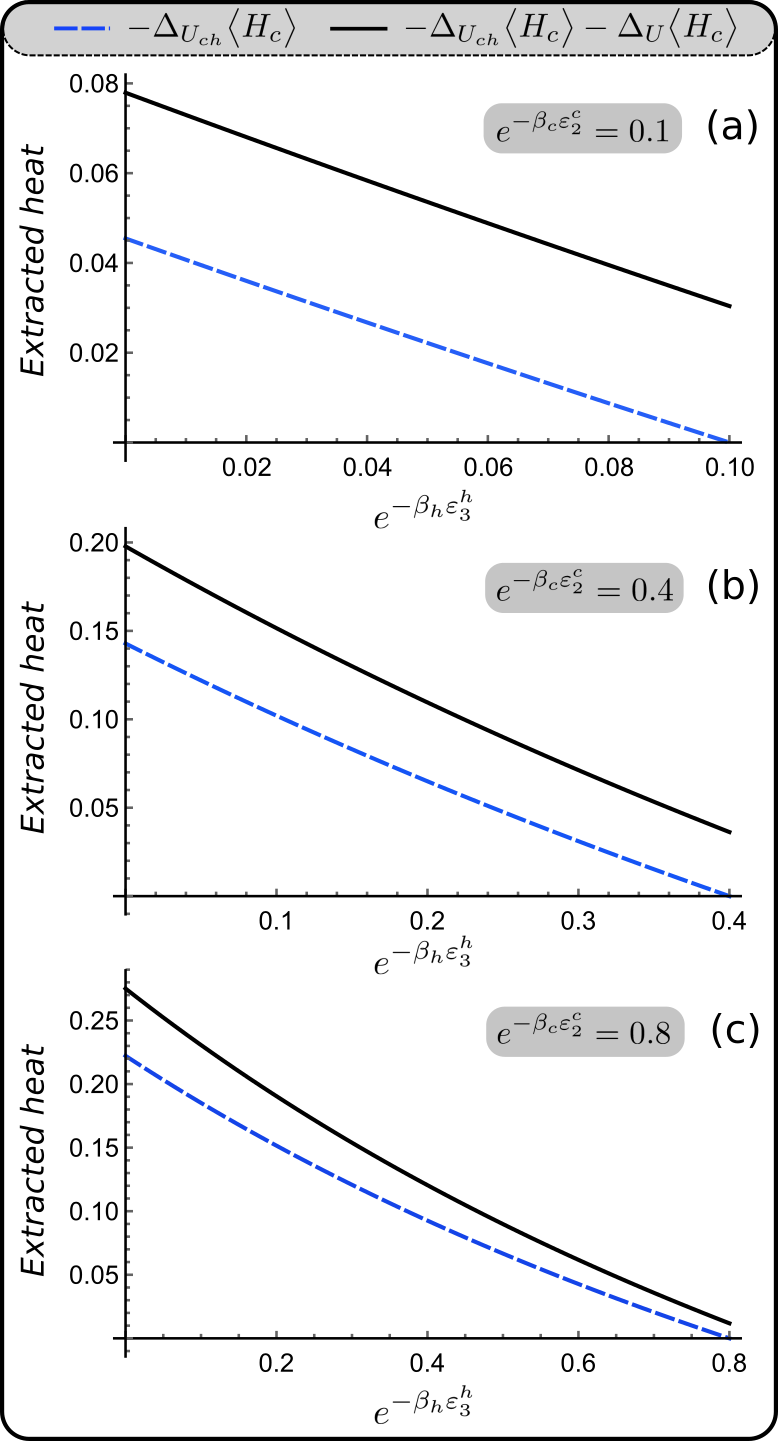}\caption{Initially extracted heat (dashed blue curves) and total extracted
		heat (solid black curves), including the contribution from the catalytic
		transformation. The parameter $e^{-\beta_{c}\varepsilon_{2}^{c}}$
		has the fixed values 0.1 (a), 0.4 (b), and 0.8 (c). }
\end{figure}

Remarkably, we see from Fig. 5(a) that for low temperatures ($\beta_{c}$
large) the increment of $p_{1}^{c}$ due to the catalytic transformation
is comparable to that achieved via optimal cooling without the catalyst.
Moreover, the cooling enhancement provided by the catalyst is significant
in all the temperature range. 

Figure 5(b) illustrates the global currents generated by all the swaps
in $U$, using a $\textrm{ln}(p^{ch})\times\textrm{ln}(p^{v})$ diagram
of the state $\sigma_{ch}\otimes\rho_{v}$ (diagram at the right hand
side). The columns in this diagram are ordered taking into account
the sorting corresponding to the initial state $\rho_{c}\otimes\rho_{h}$
(top arrangement). In this way, the sorting associated with $\sigma_{ch}$
is obtained by simply exchanging the columns $2_{c}1_{h}$ and $1_{c}3_{h}$,
which describes the effect of the swap $U_{ch}=\mathcal{U}_{|2_{c}1_{h}\rangle\leftrightarrow|1_{c}3_{h}\rangle}$.
The swap $\mathcal{U}_{|2_{c}2_{h}1_{v}\rangle\leftrightarrow|1_{c}3_{h}2_{v}\rangle}$
generates the cooling current (blue arrow), and $\mathcal{V}_{|1_{c}1_{h}2_{v}\rangle\leftrightarrow|1_{c}3_{h}1_{v}\rangle}$
($\mathcal{V}_{|2_{c}2_{h}2_{v}\rangle\leftrightarrow|2_{c}3_{h}1_{v}\rangle}$)
generates the current depicted by the left (right) green arrow.

In Fig. 6 we plot the initially extracted heat $-\Delta_{U_{ch}}\bigl\langle H_{c}\bigr\rangle\equiv\textrm{Tr}[H_{c}(\rho_{c}\otimes\rho_{h}-U_{ch}\rho_{c}\otimes\rho_{h}U_{ch}^{\dagger})]$,
and the total extracted heat $-\Delta_{U_{ch}}\bigl\langle H_{c}\bigr\rangle-\Delta_{U}\bigl\langle H_{c}\bigr\rangle$,
obtained after the application of $U$. In these plots $e^{-\beta_{c}\varepsilon_{2}^{c}}$
is fixed, and we instead vary the parameter $e^{-\beta_{h}\varepsilon_{3}^{h}}$.
The maximum of $e^{-\beta_{h}\varepsilon_{3}^{h}}$ corresponds to
$e^{-\beta_{c}\varepsilon_{2}^{c}}$, where $\Delta_{U_{ch}}\bigl\langle H_{c}\bigr\rangle=0$
and it is impossible to cool without the catalyst (i.e. where $\rho_{c}\otimes\rho_{h}$
becomes passive). Although the catalytic contribution is again more
significant at low temperatures, evidently the relative contribution
with respect to $-\Delta_{U_{ch}}\bigl\langle H_{c}\bigr\rangle$
is larger at higher temperatures, where the state $\rho_{c}\otimes\rho_{h}$
approaches the passive configuration. 

\section{Cooling of many qubits and catalytic advantage}

In this section we present another example of catalytic enhancement
for cooling. This example is special in the sense that it illustrates
how the reusable character of the catalyst can be fully exploited
in a scenario that involves the cooling of a large number of qubits.
Similarly to the problem considered in Sect. VI-C, the cooling of
these systems can be performed without a catalyst. However, under
certain circumstances the catalyst allows to extract as much heat
as twice what is possible if it is not used. In addition, we will
see that such a catalytic advantage takes place through cooling cycles
that require at most three-body interactions, while arbitrary many-body
interactions are assumed in the cooling scenario that does not involve
the catalyst. 

\subsection{Catalytic cooling vs. cooling using many-body interactions}

Consider the scenario schematically depicted in Fig. 7. The goal is
to cool as much as possible a group of $N_{c}$ qubits, using a group
of $N_{h}$ qubits that play the role of a hot environment. All the
qubits start at the same inverse temperature $\beta$ and have identical
energy spectrum. Assuming zero ground eigenenergy and energy gap equal
to one, the Hamiltonians of the $i$th cold and hot qubits are respectively
$H_{c}^{(i)}=|2_{c}\rangle_{i}\langle2_{c}|$ and $H_{h}^{(i)}=|2_{h}\rangle_{i}\langle2_{h}|$.
The total Hamiltonian for the $X=C,H$ group is $H_{X}=\sum_{i=1}^{N_{x}}|2_{x=c,h}\rangle_{i}\langle2_{x=c,h}|$,
and the global initial state is a product of thermal states $\rho_{CH}=\rho_{C}\otimes\rho_{H}$,
where $\rho_{X}=\frac{e^{-\beta H_{X}}}{\textrm{Tr}(e^{-\beta H_{X}})}=\rho_{x}^{\otimes N_{x}}$
and $\rho_{x=c,h}=\sum_{i=1}^{2}p_{i}^{x}|i_{x}\rangle\langle i_{x}|$.
Given a fixed number qubits $N=N_{c}+N_{h}$, we now describe the
two cooling strategies illustrated in Fig. 7.

1. Many-body cooling (MBC) strategy: subsets of $2\leq k\leq N_{h}$
qubits from the hot group are used to \textit{optimally} cool \textit{individual}
qubits in the cold group, through \textit{optimal unitary transformations}.
Each qubit is cooled down only one time and the hot qubits pertaining
to different subsets are all different (this implies that hot qubits
are also used only once). Note also that $k\geq2$, since all the
qubits have identical states and therefore cooling is forbidden for
$k=1$ (since $\rho_{c}^{(i)}\otimes\rho_{h}^{(i)}$ is passive). 

2. Catalytic cooling (CC) strategy: a catalyst is employed to cool
down \textit{single} qubits from the cold group, using only \textit{one}
hot qubit per cold qubit. As with the MBC strategy, there is no re-usage
of hot qubits and each cold qubit is cooled down only one time. 
\begin{figure}[t]
	\centering{}
	\includegraphics[scale=0.7]{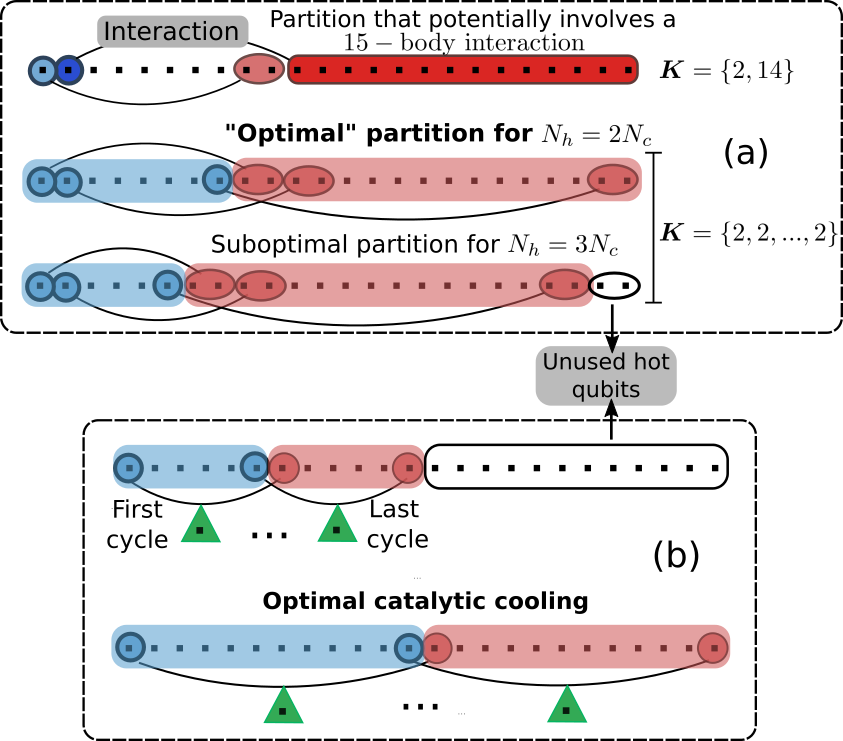}\caption{Many-body cooling (a) and catalytic cooling (b) strategies, to cool
		a group of $N_{c}$ qubits using $N_{h}$ qubits as hot environment.
		Each sequence of dots represents the total number of qubits $N=N_{c}+N_{h}$.
		From top to bottom: $N_{c}=8$ (first two sequences), $N_{c}=6$ (third
		and fourth sequences), and $N_{c}=12$. (a) Different partitions of
		the environment into cooling subsets. In the top sequence two qubits
		are cooled using two subsets of $k_{0}=2$ and $k_{1}=14$ hot qubits.
		The next sequence shows the optimal partition $\boldsymbol{K}=\{2,2,...,2\}$
		for $N_{c}=N_{h}/2=8$, if the conjecture (\ref{eq:44.1 conjecture})
		is true. The same partition is suboptimal for the third sequence,
		since two hot qubits are unused. (b) In the CC strategy a catalyst
		(green triangle) allows to cool each qubit using a \textit{single}
		hot qubit at a time. In this case the extracted heat is maximized
		for $N_{c}=N_{h}$.}
\end{figure}

In the MBC strategy \textit{the optimal cooling using a subset of
$k$ hot qubits involves ($k+1$)-body interactions} between these
qubits and the corresponding cold qubit. More specifically, such couplings
are described by an interaction Hamiltonian that contains products
of the form $\otimes_{i=1}^{k+1}B_{i}$, where $B_{i}$ is a non-trivial
(i.e. different from the identity) operator on the Hilbert space of
the $i$th qubit. On the other hand, the CC strategy is based on the
repeated application of the unitary $U$ in Eq. (\ref{eq:34 optimal CC unitary}),
for the case $d_{v}^{\ast}=2$. This means that each cycle implements
the optimal cooling of a single qubit using a two-level catalyst and
one hot qubit. Importantly, the corresponding restoring unitary involves
only a two-body interaction between the catalyst and the hot qubit,
while $\mathcal{U}_{1,1,d_{v}^{\ast}-1}$ requires a three-body interaction. 

The purpose of any of the described strategies is to reduce as much
as possible the total average energy $\bigl\langle H_{C}\bigr\rangle$
of the cold qubits. Depending on the value of $N_{c}$, the number
of qubits that can be cooled may be smaller than $N_{c}$. This limitation
is determined by two factors. Namely, the amount of hot qubits available
to cool, and the division of these qubits into cooling subsets. For
example, if $N_{h}=2$ only one qubit can be cooled using the MBC
strategy, while the introduction of the catalyst increases this number
to two. That being said, it is important to remark that the following
analysis covers all the possible values $1\leq N_{c}\leq N-1$. Therefore,
it provides a full picture of the task at hand, including also the
situations where all the$N_{c}$ qubits can be cooled. With this observation
in mind, the total heat extracted is given by

\begin{equation}
Q_{C}\equiv-\Delta\left\langle H_{C}\right\rangle =\sum_{i=1}^{n_{c}}\Delta p_{1}^{(i)},\label{eq:41 total heat Q_C}
\end{equation}
where $n_{c}\leq N_{c}$. 

\subsection{Characterization of MBC }

In the case of MBC, the maximum extractable heat $Q_{C}$ can be conveniently
addressed by introducing a coefficient that characterizes how efficient
is the cooling of a single qubit, with respect to the number of hot
qubits employed. This is a natural figure of merit in our scenario,
taking into account that the hot qubits constitute a limited resource.
Specifically, we define the ``$k$-cooling coefficient'' $\xi_{\textrm{cool}}^{(k)}$
as 

\begin{equation}
\xi_{\textrm{cool}}^{(k)}\equiv\frac{Q_{C}^{(k)}}{k},\label{eq:42 cooling efficiency}
\end{equation}
where $Q_{C}^{(k)}$ is the heat extracted by using a subset of $k\leq N_{h}$
hot qubits. 

In the MBC strategy there are many ways in which the $N_{h}$ hot
qubits can be divided into cooling subsets. Two of such possibilities
are illustrated by the two upper sequences in Fig. 7(a), assuming
$N_{c}=8$ and $N_{h}=16$. For the top sequence, one qubit is cooled
down using two hot qubits and the cooling of a second qubit resorts
to fourteen hot qubits. Intuitively, the second qubit should end up
in a colder state because more qubits are invested in its cooling.
This also leads us to wonder if it is more profitable to cool less
qubits using larger cooling subsets, or more qubits using smaller
cooling subsets. Since we are interested in the total heat $Q_{C}$,
and not on maximizing the cooling of single qubits, the answer to
this puzzle is convoluted. However, as anticipated by the second sequence
in Fig. 7(a), at least for $N_{h}=2N_{c}$ using the \textit{smallest}
cooling subsets seems to be the optimal choice. This depends on the
validity of a conjecture that we will shortly present (Eq. (\ref{eq:44.1 conjecture})). 

By resorting to the cooling coefficient (\ref{eq:42 cooling efficiency}),
we can express the total extracted heat as 

\begin{equation}
Q_{C}=\sum_{k\in\boldsymbol{K}}Q_{C}^{(k)}=\sum_{k\in\boldsymbol{K}}\xi_{\textrm{cool}}^{(k)}k,\label{eq:43 Q_C in terms of cooling efficiency}
\end{equation}
where $\boldsymbol{K}=\{k_{0},k_{1},...\}$ describes a certain partition
of the hot group into cooling subsets. In particular, we note that
$\sum_{k\in\boldsymbol{K}}k=N_{h}$, and that it is perfectly legitimate
to have subsets of different sizes $k_{i}\neq k_{j}$, see Fig 7(a).
Given a \textit{fixed} partition, we also have the bound 
\begin{equation}
Q_{C}\leq\left(\textrm{max}_{k\in\boldsymbol{K}}\xi_{\textrm{cool}}^{(k)}\right)\sum_{k\in\boldsymbol{K}}k=\left(\textrm{max}_{k\in\boldsymbol{K}}\xi_{\textrm{cool}}^{(k)}\right)N_{h}.\label{eq:44 upper bound on Q_C}
\end{equation}

While the heat $Q_{C}^{(k)}$ is by construction a non-decreasing
function of $k$, Fig. 8 provides numerical evidence that $\xi_{\textrm{cool}}^{(k)}$
is maximum for $k=2$. For very large values of $k$ it is also naturally
expected that $\xi_{\textrm{cool}}^{(k)}$ tends to zero, since otherwise
$Q_{C}^{(k)}$ would be an unbounded quantity (cf. Eq. (\ref{eq:42 cooling efficiency})).
Therefore, we conjecture that 
\begin{equation}
\xi_{\textrm{cool}}^{(k)}\leq\xi_{\textrm{cool}}^{(2)}=\left(\frac{1-2p_{2}^{c}}{2}\right)p_{1}^{c}p_{2}^{c},\label{eq:44.1 conjecture}
\end{equation}
for all $k\geq2$ and for any $\beta$, which is satisfied for $2\leq k\leq14$
in Fig. 8. The explicit expression for $\xi_{\textrm{cool}}^{(2)}$
is derived in Appendix G. 
\begin{figure}[t]
\centering{}\includegraphics[scale=0.56]{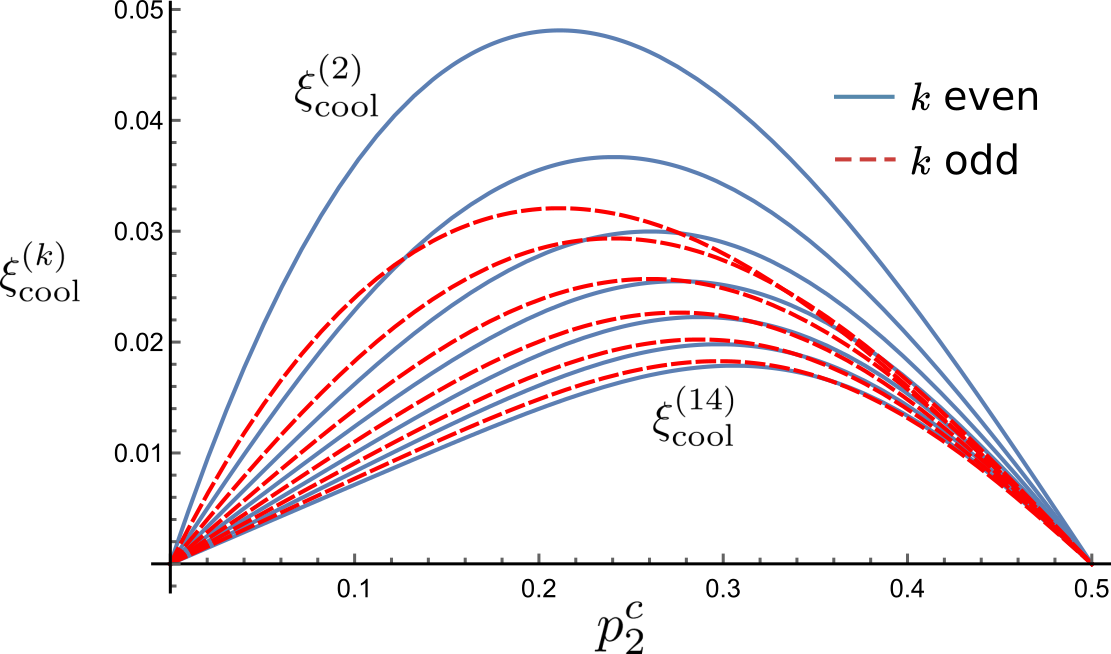}\caption{Cooling coefficient (\ref{eq:42 cooling efficiency}) curves for the
cooling of a cold qubit using $2\protect\leq k\protect\leq14$ hot
qubits. Blue solid (red dashed) curves stand for $k$ even (odd).
The highest and lowest curves correspond respectively to $\xi_{\textrm{cool}}^{(2)}$
and $\xi_{\textrm{cool}}^{(14)}$. Since $\xi_{\textrm{cool}}^{(k)}<\xi_{\textrm{cool}}^{(2)}$
for any value of $p_{2}^{c}$, this plot shows that the conjecture
(\ref{eq:44.1 conjecture}) is true for $2\protect\leq k\protect\leq14$. }
\end{figure}

Assuming the validity of the conjecture (\ref{eq:44.1 conjecture}),
the bound (\ref{eq:44 upper bound on Q_C}) is saturated if two conditions
are met. Namely, if $N_{h}$ is an even number, such that it can be
divided into cooling subsets of two qubits, and if \textit{all} these
cooling subsets can be put to use (second sequence in Fig. 7(a)).
In such a case a partition $\boldsymbol{K}=\{2,2,...,2\}$ maximizes
the extracted heat. The second condition requires that $N_{h}\leq2N_{c}$,
since otherwise $N-2N_{c}$ hot qubits would be left unused. For example,
the two unused qubits in Fig. 7(a) (third sequence) could be combined
with another pair of qubits to extract more heat from one of the qubits
in the cold group.

The third sequence in Fig. 7(a) also illustrates that for $N_{h}\geq2N_{c}+1$
the partition $\boldsymbol{K}=\{2,2,...,2\}$ allows to cool \textit{all}
the $N_{c}$ qubits. From Eq. (\ref{eq:42 cooling efficiency}), the
heat extracted in this way would be $2\xi_{\textrm{cool}}^{(2)}N_{c}$.
Since we already mentioned that such a partition is suboptimal, we
also have the bound $Q_{C}=\sum_{k\in\boldsymbol{K}}\xi_{\textrm{cool}}^{(k)}k\geq2\xi_{\textrm{cool}}^{(2)}N_{c}$.
Summarizing, 

\begin{align}
\textrm{max}_{\boldsymbol{K}}Q_{C} & =\xi_{\textrm{cool}}^{(2)}N_{h},\textrm{ if }N_{h}\leq2N_{c},\label{eq:45 maxQ_C for Nc>=00003DN/3}\\
2\xi_{\textrm{cool}}^{(2)}N_{c}\leq\textrm{max}_{\boldsymbol{K}}Q_{C} & \leq\xi_{\textrm{cool}}^{(2)}N_{h},\textrm{ if }N_{h}\geq2N_{c}+1.\label{eq:46 bound on maxQ_C over all partitions}
\end{align}
We emphasize that Eq. (\ref{eq:45 maxQ_C for Nc>=00003DN/3}) and
the upper bound in (\ref{eq:46 bound on maxQ_C over all partitions})
depend on the validity of (\ref{eq:44.1 conjecture}). Hence, our
following analysis is subject to this condition. 

\subsection{Advantage of the CC strategy }

\subsubsection{Characterization}

Let us denote as $Q_{C}^{(\textrm{CC})}$ the total extracted heat
in this case, to distinguish it from the heat $Q_{C}$ extracted via
the MBC strategy. Thanks to the reusable character of the catalyst,
the CC strategy operates through cooling cycles where each cycle involves
a different pair of qubits, yet the cooling is mediated by the \textit{same}
catalyst. This procedure is depicted in Fig. 7(b).

We consider a two-level catalyst (green triangle in Fig. 7(b)), which
allows us to apply Eq. (\ref{eq:36 max Jcool}) to obtain a simple
expression for the heat extracted per cycle. Since $d_{v}^{\ast}=2$
in this case, after setting $p_{2}^{c}=p_{2}^{h}$ (hot qubits identical
to the cold qubit) the cooling current (\ref{eq:36 max Jcool}) takes
the simple form $J_{\textrm{cool}}^{\textrm{max}}=\frac{p_{2}^{c}\left(1-2p_{2}^{c}\right)}{1+p_{2}^{c}}\bar{p}_{1}^{v}$.
The catalyst population $\bar{p}_{1}^{v}=1-\bar{p}_{2}^{v}$ can be
computed from the formula (\ref{eq:32.19 pdv in terms of p1}) in
Appendix I, which is valid for any $d_{v}^{\ast}\geq2$. This formula
yields $\bar{p}_{1}^{v}=\frac{p_{1}^{c}(1+p_{2}^{c})}{1+2p_{2}^{c}p_{1}^{c}}$.
In this way, after $n$ cycles the total extracted heat reads 
\begin{equation}
Q_{C}^{(\textrm{CC})}=nJ_{\textrm{cool}}^{\textrm{max}}=n\left(\frac{1-2p_{2}^{c}}{1+2p_{1}^{c}p_{2}^{c}}\right)p_{1}^{c}p_{2}^{c},\label{eq:47 Q_C in CC}
\end{equation}
being $J_{\textrm{cool}}^{\textrm{max}}$ the heat extracted per cycle. 

As with the MBC strategy, we are now going to derive expressions that
characterize the extracted heat given different relations between
$N_{c}$ and $N_{h}$. If $N_{h}\geq N_{c}$, \textit{all} the $N_{c}$
cold qubits can be catalytically cooled using $N_{c}$ hot qubits
(see Fig. 7(b)). In contrast, for $N_{h}\leq N_{c}-1$ we can only
cool $N_{h}$ cold qubits but \textit{all} the $N_{h}$ hot qubits
are consumed. Keeping in mind that both scenarios correspond to $n=N_{c}$
and $n=N_{h}$, respectively, Eq. (\ref{eq:47 Q_C in CC}) yields
\begin{equation}
Q_{C}^{(\textrm{CC})}=\begin{cases}
\begin{array}{c}
N_{c}\left(\frac{1-2p_{2}^{c}}{1+2p_{1}^{c}p_{2}^{c}}\right)p_{1}^{c}p_{2}^{c},\textrm{ if }N_{h}\geq N_{c},\\
N_{h}\left(\frac{1-2p_{2}^{c}}{1+2p_{1}^{c}p_{2}^{c}}\right)p_{1}^{c}p_{2}^{c},\textrm{ if }{\color{green}{\normalcolor N_{h}\leq N_{c}-1}},
\end{array}\end{cases}\label{eq:48 Q_CC}
\end{equation}

\subsubsection{Comparison between CC and MBC}

To perform the comparison between CC and MBC we introduce the relative
performance ratio 
\begin{equation}
\gamma\equiv\frac{Q_{C}^{(\textrm{CC})}}{\textrm{max}_{\boldsymbol{K}}Q_{C}},\label{eq:49 performance ratio}
\end{equation}
where the numerator and the denominator must be evaluated for the
same pair $(N_{c},N_{h})$ and the same population $p_{2}^{c}$ (which
fully characterizes the individual state of all the qubits). 
\begin{itemize}
\item For $N_{h}\geq2N_{c}+1$, $Q_{C}^{(\textrm{CC})}$ is given by the
first line of Eq. (\ref{eq:49 performance ratio}). Moreover, $\textrm{max}_{\boldsymbol{K}}Q_{C}$
is bounded from below by $2\xi_{\textrm{cool}}^{(2)}N_{c}$, according
to Eq. (\ref{eq:46 bound on maxQ_C over all partitions}). This leads
to the upper bound 
\begin{equation}
\gamma\leq\frac{Q_{C}^{(\textrm{CC})}}{2\xi_{\textrm{cool}}^{(2)}N_{c}}=\frac{1}{1+2p_{1}^{c}p_{2}^{c}},\textrm{ if }N_{h}\geq2N_{c}+1,\label{eq:50 performance ratio Nc<=00003DN/3-1}
\end{equation}
where we have also expressed $\xi_{\textrm{cool}}^{(2)}$ as in (\ref{eq:44.1 conjecture}).
Clearly, Eq. (\ref{eq:50 performance ratio Nc<=00003DN/3-1}) implies
that $\gamma\leq1$ and therefore MBC outperforms CC for $N_{h}\geq2N_{c}+1$. 
\item For $N_{h}\leq2N_{c}$,$\textrm{max}_{\boldsymbol{K}}Q_{C}$ is given
by Eq. (\ref{eq:45 maxQ_C for Nc>=00003DN/3}). If it also holds that
$N_{h}\geq N_{c}$, $Q_{C}^{(\textrm{CC})}$ obeys the first line
of Eq. (\ref{eq:48 Q_CC}). Otherwise, $N_{h}\leq N_{c}-1$ and $Q_{C}^{(\textrm{CC})}$
obeys the second line of this equation. In this way, Eqs. (\ref{eq:49 performance ratio})
and (\ref{eq:44.1 conjecture}) yield 
\begin{align}
\gamma & =\frac{Q_{C}^{(\textrm{CC})}}{\xi_{\textrm{cool}}^{(2)}N_{h}}=\frac{N_{c}}{N_{h}}\left(\frac{2}{1+2p_{1}^{c}p_{2}^{c}}\right),\textrm{ if }N_{c}\leq N_{h}\leq2N_{c},\label{eq:51 gamma for N/3=00003D<Nc=00003D<N/2-1}\\
\gamma & =\frac{Q_{C}^{(\textrm{CC})}}{\xi_{\textrm{cool}}^{(2)}N_{h}}=\frac{2}{1+2p_{1}^{c}p_{2}^{c}}\nonumber \\
 & \quad\geq\frac{4}{3},\textrm{ if }N_{h}\leq N_{c}-1.\label{eq:52 gamma N/2+1=00003D<Nc-1}
\end{align}
where the lower bound in the second line of (\ref{eq:52 gamma N/2+1=00003D<Nc-1})
is obtained from the maximum $\textrm{max}_{\beta}p_{1}^{c}p_{2}^{c}$.
Accordingly, for $N_{h}\leq N_{c}-1$ we have that $\gamma>1$ and
thus the CC strategy outperforms the MBC strategy. 
\end{itemize}
Let us now address how the relative performance ratio behaves in the
remaining interval $N_{c}\leq N_{h}\leq2N_{c}$. Following Eq. (\ref{eq:51 gamma for N/3=00003D<Nc=00003D<N/2-1}),
in this case $\gamma>1$ if and only if 
\begin{equation}
N_{h}<\frac{2N_{c}}{1+2p_{1}^{c}p_{2}^{c}}.\label{eq:53 lower bound on Nc/N to have gamma>1-1}
\end{equation}

In the limit $p_{2}^{c}\rightarrow0$, corresponding to $\beta\rightarrow\infty$,
this inequality is equivalent to $N_{h}<2N_{c}$. In the opposite
limit $p_{2}^{c}\rightarrow1/2$, corresponding to $\beta\rightarrow0$,
the previous inequality reads $N_{h}<4N_{c}/3$. Hence, we can conclude
that \textit{for extremely low temperatures the CC strategy outperforms
the MBC strategy in a wider regime}. This regime is characterized
by the total interval $N_{h}\in[1,N_{c}-1]\cup[N_{c},2N_{c}-1]$,
where the first interval is associated with Eq. (\ref{eq:52 gamma N/2+1=00003D<Nc-1})
and the second one is associated with (\ref{eq:51 gamma for N/3=00003D<Nc=00003D<N/2-1}).
On the other hand, the interval $N_{h}\in[1,N_{c}-1]\cup[N_{c},4N_{c}/3-1]$
characterizes the catalytic advantage in the limit of very large temperatures. 

\begin{figure}
	
	\begin{centering}
		\includegraphics[scale=0.7]{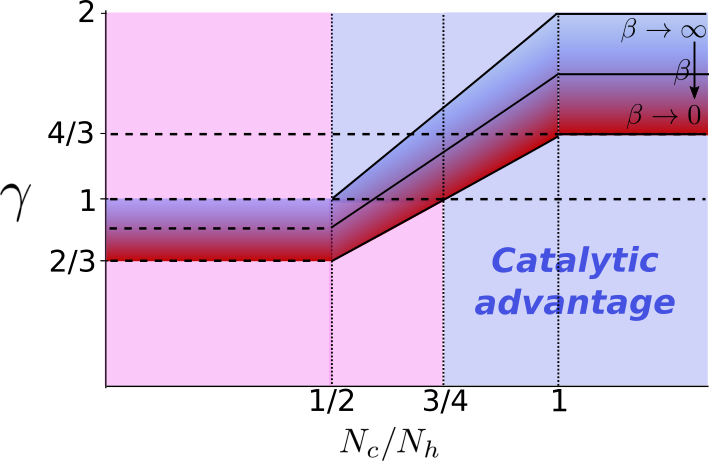}\caption{Performance ratio $\gamma$ (\ref{eq:49 performance ratio}) as a
			function of $N_{c}/N_{h}$. The top and bottom boundaries of the darker
			region correspond respectively to the limits $\beta\rightarrow\infty$
			and $\beta\rightarrow0$. The curve inside this region is representative
			of a fixed $\beta\in(0,\infty)$ and is characterized as follows:
			the continuous segments ($N_{c}/N_{h}\protect\geq1/2$) give the exact
			value of $\gamma$, and the dashed segment ($N_{c}/N_{h}<1/2$) constitutes
			an upper bound. The CC strategy outperforms the MBC strategy where
			$\gamma>1$ (light blue area). In particular, $\gamma>1$ for $N_{c}/N_{h}>3/4$,
			\textit{irrespective} of the value of $\beta$. For any $N_{c}/N_{h}>1$,
			$4/3\protect\leq\gamma\protect\leq2$. }
		\par\end{centering}
\end{figure}

Figure 9 depicts the ratio $\gamma$ and the associated bounds given
by Eqs. (\ref{eq:50 performance ratio Nc<=00003DN/3-1})-(\ref{eq:52 gamma N/2+1=00003D<Nc-1}).
The intervals in these equations are re-expressed in terms of $N_{c}/N_{h}$,
which is utilized as plotting variable. The light blue area shows
the regime where the catalyst provides an advantage with respect to
the MBC strategy. The interval $N_{c}/N_{h}\geq1$ corresponds to
Eq. (\ref{eq:52 gamma N/2+1=00003D<Nc-1}), where the advantage is
maximum. The ratio in (\ref{eq:52 gamma N/2+1=00003D<Nc-1}) is monotonically
increasing with respect to $\beta$, reaching its maximum $\gamma=2$
for $\beta\rightarrow\infty$ ($p_{2}^{c}=0$). For $1/2\leq N_{c}/N_{h}\leq1$,
$\gamma$ increases linearly with $N_{c}/N_{h}$ (cf. (\ref{eq:51 gamma for N/3=00003D<Nc=00003D<N/2-1})),
with a slope that varies between 4/3 for $p_{2}^{c}=1/2$ and 2 for
$p_{2}^{c}=0$. In particular, this implies that in the interval $N_{c}/N_{h}\in(1/2,3/4)$
the catalytic advantage takes place only if $\beta$ is sufficiently
large. For $N_{c}/N_{h}\leq1/2$, the upper bound (\ref{eq:50 performance ratio Nc<=00003DN/3-1})
varies between 2/3 and 1, corresponding to the high-temperature and
low-temperature limits, respectively.

\subsubsection{Maximum heat $Q_{C}^{(\textrm{CC})}$ and additional catalytic advantage}

We note that, in contrast with the MCB strategy, we have not considered
environment partitions in the case of catalytic cooling. Intentionally,
we have restricted ourselves to cooling cycles that involve a \textit{single}
hot qubit. The reason is that for these cycles only three qubits (including
the catalyst) must interact between each other, and therefore at most
a three-body interaction is required for their implementation. This
also implies that the CC strategy is not only capable of surpassing
the MBC strategy, but also that\textit{ it can do it with a lower
degree of control over the environment}. 

On the other hand, what we can do is to search for the optimal partition
$\{N_{c},N_{h}\}$ that maximizes the extracted heat in the case of
CC. This is equivalent to maximize the heat $Q_{C}^{(\textrm{CC})}$
with respect to $N_{c}=N-N_{h}$, given a \textit{fixed} value of
$N$. Equation (\ref{eq:48 Q_CC}) provides all the necessary ingredients
to accomplish this task. For $N_{h}\geq N_{c}$, the expression in
the first line is maximized if $N_{c}=N_{h}$, and for $N_{h}\leq N_{c}-1$,
the expression in the second line is maximized if $N_{h}=N_{c}-1$.
For $N$ fixed, we obtain $N_{c}=N/2$ in the first case and $N_{h}=(N-1)/2$
in the second case. Accordingly, 
\begin{equation}
\textrm{max}_{N_{c}}Q_{C}^{(\textrm{CC})}=\bigl\lfloor N/2\bigr\rfloor\left(\frac{1-2p_{2}^{c}}{1+2p_{1}^{c}p_{2}^{c}}\right)p_{1}^{c}p_{2}^{c},\label{eq:54 max of Q(CC) over Nc}
\end{equation}
where $\bigl\lfloor N/2\bigr\rfloor$ is the floor function applied
on $N/2$. 

Next, we want to derive bounds on the performance of the MBC strategy
by varying $N_{c}$, and to check if these bounds do not preclude
that $\textrm{max}_{\boldsymbol{K}}Q_{C}<\textrm{max}_{N_{c}}Q_{C}^{(\textrm{CC})}$.
Such a situation would show that, even if the the extracted heat is
maximized with respect to $N_{c}$, the introduction of the catalyst
can still be beneficial. In what follows we show that this is indeed
the case if $N$ is even and $\textrm{max}_{\boldsymbol{K}}Q_{C}$
is maximized in the restricted interval $N_{c}\geq N/3$. 

Using Eq. (\ref{eq:45 maxQ_C for Nc>=00003DN/3}), it follows that
in the regime $N_{h}\leq2N_{c}$ the quantity $\textrm{max}_{\boldsymbol{K}}Q_{C}$
takes its maximum value if $N_{h}=2N_{c}$, which corresponds to $N_{c}=N/3$
and reads $\textrm{max}_{\boldsymbol{K}}Q_{C}=\xi_{\textrm{cool}}^{(2)}(N/3)$.
By substituting $\xi_{\textrm{cool}}^{(2)}$ by the expression in
Eq. (\ref{eq:44.1 conjecture}), we thus have that 
\begin{equation}
\textrm{max}_{\boldsymbol{K}}Q_{C}\leq\frac{N\left(1-2p_{2}^{c}\right)p_{1}^{c}p_{2}^{c}}{3},\textrm{ if }N_{c}\geq N/3.\label{eq:55 upper bound for MBC if Nc>=00003DN/3}
\end{equation}
For $N_{h}\geq2N_{c}+1$ (equivalently $N_{c}\leq(N_{h}-1)/2$), the
lower bound $2\xi_{\textrm{cool}}^{(2)}N_{c}\leq\textrm{max}_{\boldsymbol{K}}Q_{C}$
in (\ref{eq:46 bound on maxQ_C over all partitions}) is maximized
if $N_{c}=(N_{h}-1)/2$, which corresponds to $N_{c}=(N-1)/3$. In
this way, 
\begin{equation}
\textrm{max}_{\boldsymbol{K}}Q_{C}\geq\frac{(N-1)\left(1-2p_{2}^{c}\right)p_{1}^{c}p_{2}^{c}}{3},\textrm{ if }N_{c}\leq(N-1)/3.\label{eq:56 Lower bound if Nc<=00003D(N-1)/3}
\end{equation}

Importantly, by construction the bound (\ref{eq:55 upper bound for MBC if Nc>=00003DN/3})
is saturated for $N_{c}=N/3$ (assuming that $N$ is divisible by
3). For $N$ even, the inequality is strict and we can apply it to
determine another catalytic advantage in the regime $N_{c}\geq N/3$.
Specifically, under these conditions the ratio between Eqs. (\ref{eq:54 max of Q(CC) over Nc})
and (\ref{eq:55 upper bound for MBC if Nc>=00003DN/3}) yields 

\begin{equation}
\frac{\textrm{max}_{N_{c}}Q_{C}^{(\textrm{CC})}}{\textrm{max}_{\boldsymbol{K}}Q_{C}}>\frac{3}{2(1+2p_{1}^{c}p_{2}^{c})}\geq1,\label{eq:56.1 catalytic advantage for Nc>=00003DN/3}
\end{equation}
where the lower bound at the r.h.s. follows by considering the maximum
$\textrm{max}_{\beta}p_{1}^{c}p_{2}^{c}=1/4$. This allows us to conclude
that: \textit{for $N_{c}\geq N/3$ and $N$ even, MBC extracts less
heat than the maximum heat extracted via CC}. 

\section{Catalytic thermometry}

In this section we study an example where a catalyst is applied for
precision enhancement in thermometry \citep{44Thermometry-review},
where the goal is to estimate the temperature of a certain environment
at thermal equilibrium. Let $\rho_{e}=\frac{e^{-\beta H_{e}}}{\textrm{Tr}\left(e^{-\beta H_{e}}\right)}$
denote the state of an environment with Hamiltonian $H_{e}=\sum_{j}\varepsilon_{j}^{e}|j_{e}\rangle\langle j_{e}|$,
equilibrated at inverse temperature $\beta$. Essentially, a temperature
estimation consists of assigning temperature values $\hat{T}_{i}$
to the different outcomes of a properly chosen observable $O$. In
this way, the set $\{\hat{T}_{i}\}$ defines a temperature estimator
$\hat{T}$, and the precision is assessed through the estimation error
\begin{equation}
\sqrt{\bigl\langle(\hat{T}-T)^{2}\bigr\rangle}\equiv\sqrt{\sum_{i}p_{i}(\hat{T}_{i}-T)^{2}},\label{eq:57 estimation error}
\end{equation}
where $T=\beta^{-1}$ is the actual temperature and $p_{i}$ is the
probability of measuring the outcome $i$. 

The traditional approach to characterize the thermometric precision
and also the precision in the estimation of more general physical
parameters is based on the Fisher information \citep{47Paris-metrology}.
This quantity determines a lower bound on the estimation error, known
as the Cramer-Rao bound. In the case of thermometry, it is known that
the Cramer-Rao bound is always saturated if $O=H_{e}$ \citep{49Correa-thermometry}.
That is, if the temperature estimation is carried out by directly
performing energy measurements on the environment. Here we consider
a different scenario, where an auxiliary system or ``probe'' is
used to extract temperature information via an interaction with the
environment. Such a technique may be useful for example if the environment
is very large and direct energy measurements are hard to implement.
However, our main motivation is to show that the estimation error
can be reduced below the minimum value attained only with the probe,
by including an additional interaction with a catalyst. We consider
a three-level environment with degeneracy $\varepsilon_{1}^{e}=\varepsilon_{2}^{e}=0$,
which is probed by a two-level system in the initial state ${\color{red}{\normalcolor \rho_{P}=p_{1}^{P}|1_{P}\rangle\langle1_{P}|+p_{2}^{P}|2_{P}\rangle\langle2_{P}|}}$,
with $p_{1}^{P}>p_{2}^{P}$. Moreover, the catalyst is also a two-level
system in the initial state $\rho_{v}$. This setup is illustrated
in Fig. 10, and is related to the physical configuration studied in
Sect. VI-C, with the probe and the environment taking respectively
the roles of the cold qubit and the hot object. As we will see, under
suitable conditions the same catalytic transformation that allowed
cooling enhancement also enables precision enhancement in the temperature
estimation. 
\begin{figure}
\centering{}\includegraphics[scale=0.8]{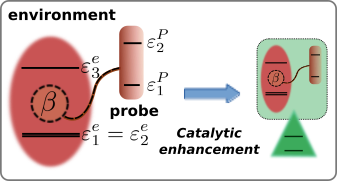}\caption{Thermometric setup. Initially a two-level system optimally probes
the temperature of a three-level environment with degeneracy $\varepsilon_{1}^{e}=\varepsilon_{2}^{e}$
(left). Subsequently, a joint interaction with a two-level catalyst
(green triangle) reduces the minimum estimation error previously achieved
(right).}
\end{figure}

We assume that $\hat{T}$ is an unbiased estimator, which means that
its expectation value coincides with the actual temperature: $\bigl\langle\hat{T}\bigr\rangle=T$.
It is important to mention that the assumption of unbiased estimators
is common not only in thermometry but also for metrology in general.
In particular, the Cramer-Rao bound limits the precision attained
with this kind of estimators. If $\bigl\langle\hat{T}\bigr\rangle=T$
it follows that $\bigl\langle(\hat{T}-T)^{2}\bigr\rangle=\textrm{Var}(\hat{T})$,
where $\textrm{Var}(\hat{T})=\bigl\langle\hat{T}^{2}\bigr\rangle-\bigl\langle\hat{T}\bigr\rangle^{2}$
is the variance of $\hat{T}$. Moreover, it can be shown that if the
temperature to be estimated belongs to a small interval $(T-\delta T,T+\delta T)$
(this is the so called ``local estimation regime'', most often studied
in thermometry and other areas of metrology \citep{61.1.1a}), the
estimation error using the observable $O$ reads \citep{44Thermometry-review}
\begin{equation}
\Delta T=\sqrt{\textrm{Var}(\hat{T})}=\frac{\sqrt{\textrm{Var}(O)}}{\bigl|\partial_{T}\left\langle O\right\rangle \bigr|},\label{eq:57.1 DeltaT}
\end{equation}
where $\textrm{Var}(O)=\bigl\langle O^{2}\bigr\rangle-\bigl\langle O\bigr\rangle^{2}$
and $\partial_{T}\left\langle O\right\rangle =\frac{\partial}{\partial T}\left\langle O\right\rangle $.
For the sake of convenience, we shall consider an ``inverse temperature
estimator'' $\hat{\beta}$ instead of $\hat{T}$. The errors $\Delta T$
and $\Delta\beta=\sqrt{\textrm{Var}(\hat{\beta})}$ are connected
through the simple relation $\Delta T=T^{2}\Delta\beta$, which follows
from the chain rule $\partial_{\beta}\left\langle O\right\rangle =-T^{2}\partial_{T}\left\langle O\right\rangle $.

\subsection{Optimal precision and catalytic enhancement}

In our example the observable $O_{P}=o_{1}^{P}|1_{P}\rangle\langle1_{P}|+o_{2}^{P}|2_{P}\rangle\langle2_{P}|$
describes a projective measurement on the probe, with eigenvalues
$o_{1}^{P}$ and $o_{2}^{P}$. Information about $\beta$ is encoded
in the probe state $\sigma_{P}=\textrm{Tr}_{e}\left[U_{Pe}(\rho_{P}\otimes\rho_{e})U_{Pe}^{\dagger}\right]$,
obtained after a unitary evolution $U_{Pe}$ that couples the probe
to the environment. It is straightforward to check that in this case
the estimation error reads 
\begin{equation}
\Delta\beta(\sigma_{P})=\frac{\sqrt{q_{1}^{P}q_{2}^{P}}}{\bigl|\partial_{\beta}q_{1}^{P}\bigr|},\label{eq:58 est error for beta}
\end{equation}
where $q_{1}^{P}=\textrm{Tr}(|1_{P}\rangle\langle1_{P}|\sigma_{P})$
is the ground population after the application of $U_{Pe}$. 

The ratio at the r.h.s. of Eq. (\ref{eq:58 est error for beta}) constitutes
the figure of merit in our analysis. On the one hand, under certain
conditions one can find a unitary $U_{Pe}$ that minimizes the product
$q_{1}^{P}q_{2}^{P}$, and at the same time maximizes the quantity
$\partial_{\beta}q_{1}^{P}$ \citep{61.2a}. In such a case the inequality
\begin{equation}
\textrm{min}_{U_{Pe}}\Delta\beta(\sigma_{P})\geq\frac{\textrm{min}_{U_{Pe}}\sqrt{q_{1}^{P}q_{2}^{P}}}{\textrm{max}_{U_{Pe}}\bigl|\partial_{\beta}q_{1}^{P}\bigr|}\label{eq:59 minimum est error}
\end{equation}
is saturated and the r.h.s. of (\ref{eq:59 minimum est error}) constitutes
the minimum error for unitary evolutions on the probe and the environment,
and measurements on the probe.

On the other hand, we will see that when the bound (\ref{eq:59 minimum est error})
is saturable it is possible to perform a catalytic transformation
such that 
\begin{equation}
\frac{\sqrt{p_{1}^{\prime P}p_{2}^{\prime P}}}{\bigl|\partial_{\beta}p_{1}^{\prime P}\bigr|}<\textrm{min}_{U_{Pe}}\left(\frac{\sqrt{q_{1}^{P}q_{2}^{P}}}{\bigl|\partial_{\beta}q_{1}^{P}\bigr|}\right)=\frac{\textrm{min}_{U_{Pe}}\sqrt{q_{1}^{P}q_{2}^{P}}}{\textrm{max}_{U_{Pe}}\bigl|\partial_{\beta}q_{1}^{P}\bigr|},\label{eq:59.1 catalytic enhancement}
\end{equation}
where $p_{1}^{\prime P}=\textrm{Tr}(|1_{P}\rangle\langle1_{P}|\rho'_{P})$
is the final ground population. Let us denote the unitary that minimizes
the upper bound (\ref{eq:59.1 catalytic enhancement}) as $U_{Pe}^{\textrm{opt}}$.
The total evolution is thus generated by a composition $UU_{Pe}^{\textrm{opt}}$,
where $U$ is a unitary that involves the interaction with a two-level
catalyst in the state $\rho_{v}$. This also gives rise to a composition
of transformations $\rho\rightarrow\sigma_{Pe}\otimes\rho_{v}\rightarrow\rho'$,
where $\rho=\rho_{P}\otimes\rho_{e}\otimes\rho_{v}$, $\sigma_{Pe}=U_{Pe}^{\textrm{opt}}\rho_{P}\otimes\rho_{e}U_{Pe}^{\textrm{opt}\dagger}$,
and $\rho'=U\sigma_{Pe}\otimes\rho_{v}U^{\dagger}$. 

\subsubsection*{Connection between optimal cooling and optimal precision}

In order to demonstrate a catalytic enhancement, we need first to
identify a scenario where the inequality (\ref{eq:59 minimum est error})
can be saturated. Interestingly, we will see that this occurs if the
state $\rho_{P}\otimes\rho_{e}$ is \textit{non-passive} with respect
to $H_{P}$, where $H_{P}=\varepsilon_{1}^{P}|1_{P}\rangle\langle1_{P}|+\varepsilon_{2}^{P}|2_{P}\rangle\langle2_{P}|$
is the Hamiltonian of the probe. Not only that, but it turns out that
\textit{the minimum} $\textrm{min}_{U_{Pe}}\Delta\beta$ \textit{is
attained by a unitary} $U_{Pe}$ \textit{that optimally cools down
the probe using only the environment}. In fact, the numerator at the
r.s.h. of (\ref{eq:59 minimum est error}) is minimized via a maximization
of $q_{1}^{P}$, which corresponds to optimal cooling. The saturation
of the bound (\ref{eq:59 minimum est error}) follows because the
\textit{same} unitary $U_{Pe}$ also maximizes the corresponding denominator,
as shown in the next section. 

Optimal cooling of the probe implies that the state $\sigma_{Pe}=U_{Pe}^{\textrm{opt}}\rho_{P}\otimes\rho_{e}U_{Pe}^{\textrm{opt}\dagger}$
must be passive with respect $H_{P}$. Keeping in mind that $\rho_{e}$
describes a three-level system with degeneracy $\varepsilon_{1}^{e}=\varepsilon_{2}^{e}=0$,
we can directly apply the results obtained in Sect. VI-C to characterize
such a state. Specifically, we have that 

\begin{align}
\sigma_{Pe} & =|1_{P}\rangle\langle1_{P}|\otimes\left(\sum_{j=1}^{2}p_{1}^{P}p_{j}^{e}|j_{e}\rangle\langle j_{e}|+p_{2}^{P}p_{1}^{e}|3_{e}\rangle\langle3_{e}|\right)\nonumber \\
 & \quad+|2_{P}\rangle\langle2_{P}|\otimes\left(p_{1}^{P}p_{3}^{e}|1_{e}\rangle\langle1_{e}|+\sum_{j=2}^{3}p_{2}^{P}p_{j}^{e}|j_{e}\rangle\langle j_{e}|\right),\label{eq:60 Sigma(Pe)}
\end{align}
which is obtained by replacing the labels $c$ and $h$ in Eq. (\ref{eq:36 passive state Sigma(ch)})
by $P$ and $e$, respectively. The corresponding unitary was also
derived in Sect. VI-C and (in this context) corresponds to the swap
\begin{equation}
{\color{green}{\normalcolor U_{Pe}^{\textrm{opt}}=\mathcal{U}_{|2_{P}1_{e}\rangle\leftrightarrow|1_{P}3_{e}\rangle}.}}\label{eq:60.1 Optimal swap U(Pe)}
\end{equation}
Later on, we will also show that the same substitution of labels in
Eqs. (\ref{eq:37  U catalytic}) and (\ref{eq:38 Vres for U catalytic})
yields a unitary $U$ that provides a catalytic enhancement of precision. 

\subsection{Maximization of the population sensitivity in terms of passivity }

In what follows we will refer to $\partial_{\beta}q_{1}^{P}$ as the
``population sensitivity'', as it quantifies how the population
$q_{1}^{P}$ varies with respect to temperature changes. Since $U_{Pe}$
and $\rho_{P}$ are both independent of $\beta$, we can write 
\begin{align}
\partial_{\beta}q_{1}^{P} & =\partial_{\beta}\textrm{Tr}\left[|1_{P}\rangle\langle1_{P}|U_{Pe}\rho_{P}\otimes\rho_{e}U_{Pe}^{\dagger}\right]\nonumber \\
 & =\textrm{Tr}\left[|1_{P}\rangle\langle1_{P}|U_{Pe}\left(\rho_{P}\otimes\partial_{\beta}\rho_{e}\right)U_{Pe}^{\dagger}\right].\label{eq:60.2 Initial population sensitivity}
\end{align}
The operator $\rho_{P}\otimes\partial_{\beta}\rho_{e}$ has real eigenvalues
$p_{i}^{P}\lambda_{j}^{e}$, where 
\begin{equation}
\lambda_{j}^{e}\equiv\partial_{\beta}p_{j}^{e}=p_{j}^{e}\left(\bigl\langle H_{e}\bigr\rangle-\varepsilon_{j}^{e}\right),\label{eq:61 lambdas}
\end{equation}
and $\bigl\langle H_{e}\bigr\rangle=\textrm{Tr}(H_{e}\rho_{e})$.
This property allows us to define a positive semidefinite operator
$A\equiv\rho_{P}\otimes\partial_{\beta}\rho_{e}-\textrm{min}_{i,j}\left(p_{i}^{P}\lambda_{j}^{e}\right)\mathbb{I}_{Pe}$,
where $\mathbb{I}_{Pe}$ is the identity operator on $\mathcal{H}_{P}\otimes\mathcal{H}_{e}$. 

The introduction of the operator $A$ is useful to cast the optimization
of $\partial_{\beta}q_{1}^{P}$ as the problem of finding a passive
state that minimizes the mean value of $|2_{P}\rangle\langle2_{P}|$.
To show this, note first that Eq. (\ref{eq:60.2 Initial population sensitivity})
can be re-expressed as 
\begin{align}
\partial_{\beta}q_{1}^{P} & =\textrm{Tr}(A)\textrm{Tr}\left[|1_{P}\rangle\langle1_{P}|U_{Pe}\frac{A}{\textrm{Tr}\left(A\right)}U_{Pe}^{\dagger}\right]\nonumber \\
 & \quad+\textrm{min}_{i,j}\left(p_{i}^{P}\lambda_{j}^{e}\right).\label{eq:62 Initial population sensitivity explicit}
\end{align}

Since $\frac{A}{\textrm{Tr}\left(A\right)}$ constitutes an effective
density matrix (i.e. its eigenvalues describe a probability distribution),
maximizing $\partial_{\beta}q_{1}^{P}$ is tantamount to maximize
the population $\bigl\langle|1_{P}\rangle\langle1_{P}|\bigr\rangle$
over unitaries $U_{Pe}$ applied on this state. Furthermore, probability
conservation $\bigl\langle|1_{P}\rangle\langle1_{P}|\bigr\rangle+\bigl\langle|2_{P}\rangle\langle2_{P}|\bigr\rangle=1$
implies that this operation is also equivalent to minimize the mean
value of $|2_{P}\rangle\langle2_{P}|$. In summary, we have that:
\begin{itemize}
\item The population sensitivity is maximized by minimizing the mean value
of $|2_{P}\rangle\langle2_{P}|$, which describes an effective Hamiltonian
$H_{P}^{\textrm{eff}}=|2_{P}\rangle\langle2_{P}|$ with eigenenergies
$\varepsilon_{1}^{P}=0$ and $\varepsilon_{2}^{P}=1$. 
\item However, the initial state in this optimization is \textit{not} $\rho_{P}\otimes\rho_{e}$,
but the effective state $\frac{A}{\textrm{Tr}\left(A\right)}$. This
implies that for minimizing $\bigl\langle|2_{P}\rangle\langle2_{P}|\bigr\rangle$
we must find a unitary that yields a passive state when applied on
$\frac{A}{\textrm{Tr}\left(A\right)}$. 
\end{itemize}
Let us see now that the swap $U_{Pe}^{\textrm{opt}}$ (cf. Eq. (\ref{eq:60.1 Optimal swap U(Pe)}))
satisfies this condition. To check that it leads to a passive state
(with respect to $|2_{P}\rangle\langle2_{P}|$) we only need to characterize
the effect of $U_{Pe}^{\textrm{opt}}$ on $\rho_{P}\otimes\partial_{\beta}\rho_{e}$,
since this characterization fully determines how the eigenvalues of
$U_{Pe}^{\textrm{opt}}\frac{A}{\textrm{Tr}\left(A\right)}U_{Pe}^{\textrm{opt}\dagger}$
are ordered (according to the definition of $A$). Using the identity
$\sigma_{Pe}=U_{Pe}^{\textrm{opt}}\rho_{P}\otimes\rho_{e}U_{Pe}^{\textrm{opt}\dagger}$
we have that $U_{Pe}^{\textrm{opt}}\rho_{P}\otimes\partial_{\beta}\rho_{e}U_{Pe}^{\textrm{opt}\dagger}={\normalcolor \partial_{\beta}\sigma_{Pe}}$,
and, from Eq. (\ref{eq:60 Sigma(Pe)}), 

\begin{align}
{\normalcolor \partial_{\beta}\sigma_{Pe}} & {\normalcolor =|1_{P}\rangle\langle1_{P}|\otimes\left(\sum_{j=1}^{2}p_{1}^{P}\lambda_{j}^{e}|j_{e}\rangle\langle j_{e}|+p_{2}^{P}\lambda_{1}^{e}|3_{e}\rangle\langle3_{e}|\right)}\nonumber \\
{\normalcolor } & {\normalcolor \quad+|2_{P}\rangle\langle2_{P}|\otimes\left(p_{1}^{P}\lambda_{3}^{e}|1_{e}\rangle\langle1_{e}|+\sum_{j=2}^{3}p_{2}^{P}\lambda_{j}^{e}|j_{e}\rangle\langle j_{e}|\right).}\label{eq:63 Derivative of Sigma(Pe) w.r.t. beta}
\end{align}

In this way, the eigenvalues of $\frac{A}{\textrm{Tr}\left(A\right)}$
are non-increasing with respect to those of $|2_{P}\rangle\langle2_{P}|$
if and only if the eigenvalues of ${\normalcolor \partial_{\beta}\sigma_{Pe}}$
satisfy this property. That is, if all the eigenvalues in the first
line of Eq. (\ref{eq:63 Derivative of Sigma(Pe) w.r.t. beta}) are
larger or equal than all the eigenvalues in the second line. We can
easily verify this condition by noting that (cf. Eq. (\ref{eq:61 lambdas}))
\begin{align}
\lambda_{1}^{e} & =\lambda_{2}^{e}=p_{1}^{e}\bigl\langle H_{e}\bigr\rangle>\lambda_{3}^{e}=p_{3}^{e}\left(\bigl\langle H_{e}\bigr\rangle-\varepsilon_{3}^{e}\right),\nonumber \\
p_{2}^{P}\lambda_{1}^{e} & =p_{2}^{P}p_{1}^{e}\bigl\langle H_{e}\bigr\rangle>p_{1}^{P}\lambda_{3}^{e}=p_{1}^{P}p_{3}^{e}\left(\bigl\langle H_{e}\bigr\rangle-\varepsilon_{3}^{e}\right),\label{eq:64 inequalities between lambdas}
\end{align}
where the second inequality is a consequence of the non-passivity
of $\rho_{P}\otimes\rho_{e}$ (i.e. $p_{2}^{P}p_{1}^{e}>p_{1}^{P}p_{3}^{e}$).
By applying the inequalities (\ref{eq:64 inequalities between lambdas}),
we can check the aforementioned ordering and thereby conclude that
$\frac{A}{\textrm{Tr}\left(A\right)}$ is passive with respect to
$|2_{P}\rangle\langle2_{P}|$. This also means that $U_{Pe}^{\textrm{opt}}$
saturates the bound (\ref{eq:59 minimum est error}). Using (\ref{eq:63 Derivative of Sigma(Pe) w.r.t. beta}),
the optimal population sensitivity reads: 

\begin{align}
\partial_{\beta}q_{1}^{P} & =\textrm{Tr}\left[|1_{P}\rangle\langle1_{P}|\partial_{\beta}\left(U_{Pe}^{\textrm{opt}}\rho_{P}\otimes\rho_{e}U_{Pe}^{\textrm{opt\ensuremath{\dagger}}}\right)\right]\nonumber \\
 & =\textrm{Tr}\left[|1_{P}\rangle\langle1_{P}|\partial_{\beta}\sigma_{Pe}\right]\nonumber \\
 & =(1+p_{1}^{P})\lambda_{1}^{e}.\label{eq:65 Initial population sensitivity explicit 1}
\end{align}

\subsection{Catalytic enhancement }

\begin{figure}
\centering{}\includegraphics[scale=0.43]{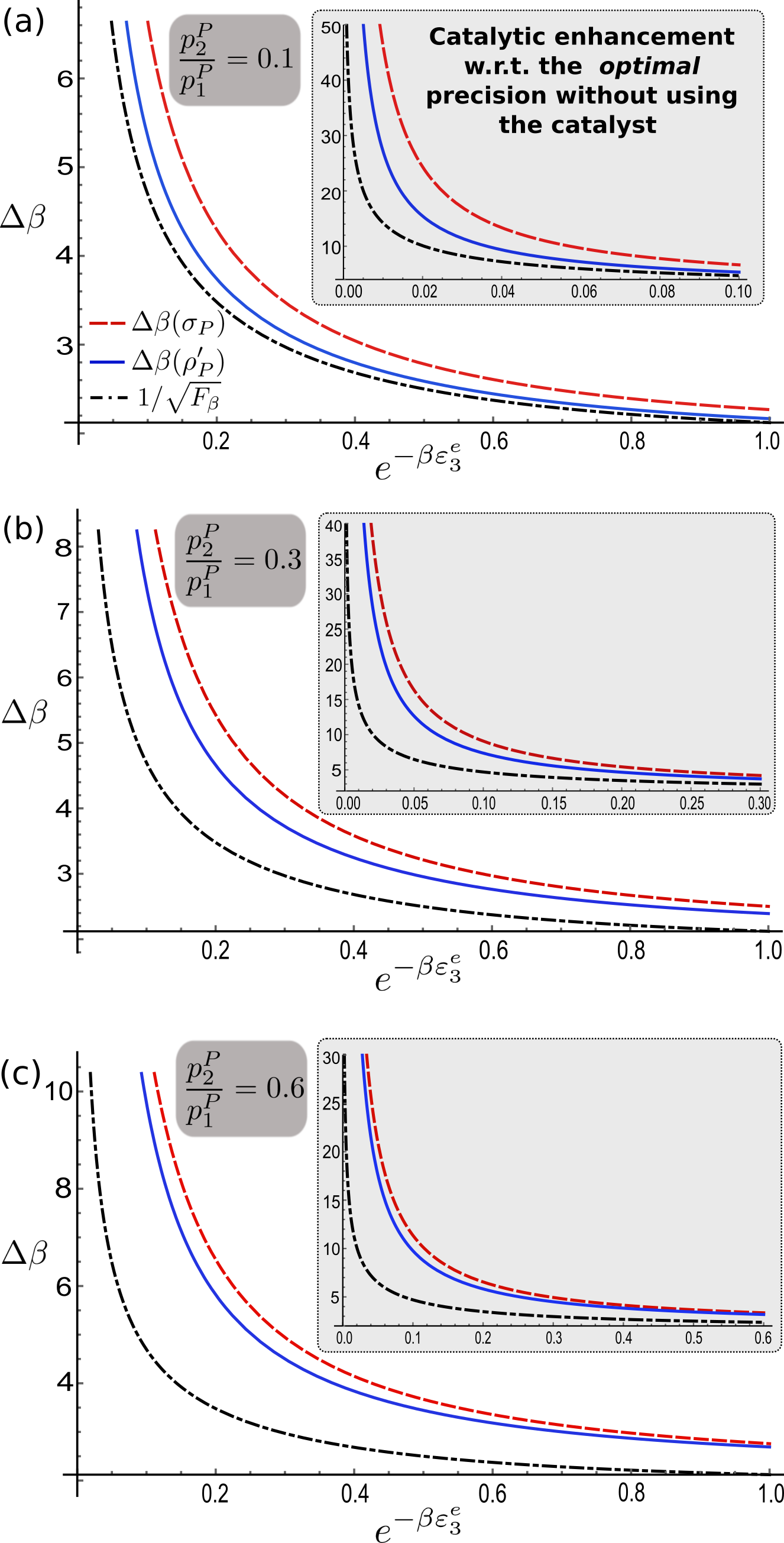}\caption{Estimation error $\Delta\beta$ for three initial probe states, characterized
by the population ratios $p_{2}^{P}/p_{1}^{P}=0.1$ (a), $p_{2}^{P}/p_{1}^{P}=0.3$
(b), and $p_{2}^{P}/p_{1}^{P}=0.6$ (c). Red dashed and blue solid
curves stand respectively for the initial error $\Delta\beta(\sigma_{P})$
and the final error $\Delta\beta(\rho'_{P})$, obtained after the
interaction with the catalyst. The black (dash-dotted) curves show
the thermal Cramer-Rao bound of the environment $1/\sqrt{F_{\beta}}$,
where $F_{\beta}=\bigl\langle H_{e}^{2}\bigr\rangle-\bigl\langle H_{e}\bigr\rangle^{2}$
is the associated Fisher information. The insets correspond to the
interval $e^{-\beta\varepsilon_{3}^{e}}\protect\leq p_{2}^{P}/p_{1}^{P}$,
where $\Delta\beta(\sigma_{P})$ is the \textit{minimum estimation
error} under general unitary evolutions that act jointly on the probe
and the environment. }
\end{figure}

We start this section by stressing that in the saturation of \ref{eq:59 minimum est error}
the non-passive character of $\rho_{P}\otimes\rho_{e}$ is crucial.
Otherwise, the numerator $\sqrt{q_{1}^{P}q_{2}^{P}}$ can only increase
under unitaries $U_{Pe}$, and its minimum value would correspond
to the trivial operation $U_{Pe}=\mathbb{I}_{Pe}$. Since in this
case $q_{1}^{P}=p_{1}^{P}$, it follows that $\partial_{\beta}q_{1}^{P}=0$
and it is clear that $\mathbb{I}_{Pe}$ cannot maximize the population
sensitivity. That being said, let us now focus on the precision enhancement
produced by a suitable catalytic transformation.

Consider a two-level catalyst $\rho_{v}=\sum_{k=1}^{2}p_{k}^{v}|k_{v}\rangle\langle k_{v}|$
and a transformation $\sigma_{Pe}\otimes\rho_{v}\rightarrow U\sigma_{Pe}\otimes\rho_{v}U^{\dagger}$,
where 

\begin{align}
U & =\mathcal{U}_{|2_{P}2_{e}1_{v}\rangle\leftrightarrow|1_{P}3_{e}2_{v}\rangle}\oplus V_{\textrm{res}},\nonumber \\
V_{\textrm{res}} & =\mathcal{V}_{|1_{P}1_{e}2_{v}\rangle\leftrightarrow|1_{P}3_{e}1_{v}\rangle}\oplus\mathcal{V}_{|2_{P}2_{e}2_{v}\rangle\leftrightarrow|2_{P}3_{e}1_{v}\rangle}.\label{eq:66 catalytic unitary}
\end{align}
The unitary $U$ is identical to $U$ in (\ref{eq:37  U catalytic}),
after the substitution of labels $c\rightarrow P$ and $h\rightarrow e$.
Now, suppose that the initial state $\rho_{P}\otimes\rho_{e}\otimes\rho_{v}$
is also identical to the state $\rho_{c}\otimes\rho_{h}\otimes\rho_{v}$,
considered in Sect. VI-C, which means that $p_{i}^{P}=p_{i}^{c}$
and $p_{j}^{e}=p_{j}^{h}$. In such a case, we can conclude that $\sigma_{Pe}$
is identical to $\sigma_{ch}$ (cf. (\ref{eq:36 passive state Sigma(ch)})),
and that $\sigma_{Pe}\otimes\rho_{v}\rightarrow U\sigma_{Pe}\otimes\rho_{v}U^{\dagger}$
is a catalytic and cooling (with respect to the probe) transformation. 

The results presented in Fig. 11 are derived under the assumption
of identical states stated above. The enhancement of precision stems
from two facts: 
\begin{enumerate}
\item Since $U$ is cooling by construction, it further reduces the product
$\sqrt{q_{1}^{P}q_{2}^{P}}$, i.e. $\sqrt{p_{1}^{\prime P}p_{2}^{\prime P}}<\sqrt{q_{1}^{P}q_{2}^{P}}$. 
\item $U$ also increases the population sensitivity, as we show below.
Therefore, $\bigl|\partial_{\beta}p_{1}^{\prime P}\bigr|>|\partial_{\beta}q_{1}^{P}|.$ 
\end{enumerate}
The combination of conditions 1 and 2 leads to the catalytic enhancement
manifested by the inequality (\ref{eq:59.1 catalytic enhancement}).
To show the increment in the population sensitivity, we write $\partial_{\beta}\left(U\sigma_{Pe}\otimes\rho_{v}U^{\dagger}\right)$
as $U\left(\partial_{\beta}\sigma_{Pe}\otimes\rho_{v}\right)U^{\dagger}$,
and apply Eqs. (\ref{eq:63 Derivative of Sigma(Pe) w.r.t. beta})
and (\ref{eq:66 catalytic unitary}) to obtain:
\begin{align*}
\partial_{\beta}p_{1}^{\prime P} & =\textrm{Tr}\left[|1_{P}\rangle\langle1_{P}|\partial_{\beta}\left(U\sigma_{Pe}\otimes\rho_{v}U^{\dagger}\right)\right]\\
 & =\textrm{Tr}\left[|1_{P}\rangle\langle1_{P}|U\left(\partial_{\beta}\sigma_{Pe}\otimes\rho_{v}\right)U^{\dagger}\right]\\
 & =\partial_{\beta}q_{1}^{P}+p_{2}^{P}\left(p_{1}^{v}\lambda_{2}^{e}-p_{2}^{v}\lambda_{1}^{e}\right)\\
 & =\partial_{\beta}q_{1}^{P}+p_{2}^{P}\left(p_{1}^{v}-p_{2}^{v}\right)\lambda_{1}^{e}.
\end{align*}
Due to the positivity of $\lambda_{1}^{e}$, we have that $\partial_{\beta}p_{1}^{\prime P}>\partial_{\beta}q_{1}^{P}$
(cf. Eq. (\ref{eq:65 Initial population sensitivity explicit 1}))
for any state $\rho_{v}$ such that $p_{1}^{v}>p_{2}^{v}$. 

Figure 11 shows the thermometric precision with and without the catalyst,
as quantified by the error $\Delta\beta$. The main plots depict this
error in the \textit{full temperature range} $0\leq e^{-\beta\varepsilon_{3}^{e}}\leq1$,
and the insets correspond to the restricted range $e^{-\beta\varepsilon_{3}^{e}}\leq p_{2}^{P}/p_{1}^{P}$,
where the state $\rho_{P}\otimes\rho_{e}$ is passive. In all the
cases the total transformation is implemented by $UU_{Pe}^{\textrm{opt}}$.
The red dashed curves give the error $\Delta\beta(\sigma_{P})$ in
Eq. (\ref{eq:58 est error for beta}), obtained after the application
of the swap (\ref{eq:60.1 Optimal swap U(Pe)}), and the blue solid
curves stand for the final error $\Delta\beta(\rho'_{P})$, with $\rho'_{P}=\textrm{Tr}_{ev}\left(UU_{Pe}^{\textrm{opt}}\rho_{P}\otimes\rho_{e}U_{Pe}^{\textrm{opt}\dagger}U^{\dagger}\right)$.
Interestingly, even for $e^{-\beta\varepsilon_{3}^{e}}>p_{2}^{P}/p_{1}^{P}$
the main plots show that the thermometric precision is increased after
the catalytic transformation. However, we can be certain that this
enhancement surpasses the optimal transformation without the catalyst,
\textit{only in the interval where} $\rho_{P}\otimes\rho_{e}$ \textit{is
passive}. In such a case the initial and final errors obey Eq. (\ref{eq:59.1 catalytic enhancement}). 

For comparison, the black dash-dotted curves show the thermal Cramer-Rao
bound for energy measurements on the environment, which constitutes
the minimum (attainable) error using any POVM (positive operator valued
measurement). In particular, no probe-based thermometric scheme can
surpass this fundamental limit, as illustrated in Fig. 11.

\section{Conclusions and outlook }

In this paper, we introduced tools for the systematic construction
of \textit{explicit} catalytic transformations on quantum systems
of finite size. Size limitations constrain tasks such as cooling using
a finite environment or thermometry with a very small probe. In the
case of cooling, we showed that the introduction of a catalyst lifts
cooling restrictions in two complementary ways: catalysts enable cooling
when it is impossible using only the environment, and enhance it when
the environment suffices to cool. These results were illustrated with
several examples regarding the cooling of a two-level system. In particular,
we found that small catalysts such as three-level systems allow maximum
cooling in wide temperature ranges. We also demonstrated that to cool
a system of any dimension a large enough catalyst and any environment
that starts in a non-fully mixed state are sufficient. In addition
to the reduction of the system mean energy, its ground population
can also be catalytically increased without the need of the environment.
Another catalytic advantage was shown in a setup consisting of many
qubits prepared in identical states, where a subset of qubits is employed
as environment to cool another subset. In this system, we found that
it is possible to outperform the cooling achieved through many-body
interactions with the environment, by including a two-level catalyst
that cools using at most three-body interactions. 

For thermometry we studied a simple example where a two-level system
extracts temperature information by probing a three-level environment.
We established a connection between the minimum estimation error and
the notion of passivity for cooling, and then showed that this error
can be further reduced if the probe-environment compound undergoes
a subsequent interaction with a two-level catalyst. As a matter of
fact, this is the smallest physical configuration where a catalyst
may improve the thermometric precision achieved through optimal probe-environment
interactions. For example, the state of any two-level environment
can be fully transferred to a two-level probe, thereby transferring
all its Fisher information. In the case of higher-dimensional environments,
bounds on the thermometric precision obtained via smaller probes have
been recently derived \citep{57Karen-CoarseGr-thermometry}. A natural extension of the results
presented here would be to examine if these bounds can be catalytically
surpassed. Unveiling further connections between thermometry and passivity
is another topic of interest. 

Throughout our presentation we have also emphasized the non-unital
character of many of the studied transformations, which admit a description
in terms of the impossibility to reproduce them through purely classical
drivings (see Definition 3 for a formal characterization). Apart from
practically motivated non-unital transformations such as cooling,
we also showed how catalysts can be used to perform more general transformations
of this kind. The identification of other non-unital transformations
with interesting applications and their implementation via catalysts
could be a subject of future research. The tools provided in the present
article could prove very useful in such endeavor. 

On the technical side, we list below various questions that remain
open: 
\begin{itemize}
\item \textbf{Maximum catalytic cooling}. The maximum catalytic cooling
of a qubit using another qubit as heat sink was studied in Ref. \citep{key-8.1Muller-unitarity}.
There, the authors showed that the purity can be increased to a value
arbitrarily close to one, if both qubits start in a common state with
(von Neumann) entropy smaller than 1/2. In this setting both the catalyst
state and the corresponding evolution are unknown. Therefore, it is
natural to assume that the dimension of the catalyst and the number
of subsystems that compose it enter as (unbounded) variables in the
optimization. \\
In Sect. VI-A we addressed the same problem, by optimizing over the
eigenvalues of a \textit{full-rank} catalyst state (Appendix I-2)
of \textit{fixed} dimension. Interestingly, we found that for $p_{2}^{c}\lesssim0.1$
a two-level catalyst suffices to bring the cooled qubit close to its
ground state (see Fig. 4(d)). In Appendix L we develop a protocol
that extends the scope of this type of optimization. This protocol
allows to explore a wider set of global unitaries for catalytic transformations,
thereby adding another variable to the optimization task. In particular,
we apply it to an example where a two-level system is cooled down
through a two-level catalyst and a four-level hot object. After a
first maximization with respect to the catalyst eigenvalues, the system
energy is reduced by the amount $\Delta\bigl\langle H_{c}\bigr\rangle=0.033(\varepsilon_{1}^{c}-\varepsilon_{2}^{c})$.
Subsequently, the application of the protocol leads to an improved
catalytic transformation that yields $\Delta\bigl\langle H_{c}\bigr\rangle=0.045(\varepsilon_{1}^{c}-\varepsilon_{2}^{c})$.
\\
Despite the usefulness of the protocol, it is still limited in several
aspects. For example, it is restricted to full-rank catalyst states
and evolutions given by direct sums of two-level unitaries. Hence,
the question of how \textit{finite-dimensional} catalysts can optimize
cooling is completely pertinent. The same question is of course valid
for the catalytic optimization of other interesting physical transformations. 
\item \textbf{The role of correlations in catalytic transformations}. Initial
studies on catalysts considered transformations where the catalyst
not only remains in its initial state but also does not develop correlations
with the rest of the system \citep{key-2catal-majorization,1catalytic-entanglement,key-10second-laws}.
Later on, it was shown that these correlations constitute a thermodynamic
resource that can be exploited to increase the set of accessible transformations
\citep{14Muller-catalysts} in microscopic thermodynamics. Thanks
to this possibility, the so called ``second laws'' \citep{key-10second-laws}
are replaced by the free energy reduction that characterizes the macroscopic
second law. \\
On the other hand, our results do not assume infinitely large environments,
as in Refs. \citep{key-10second-laws} and \citep{14Muller-catalysts},
and focus instead on environments as small as a two-level system.
Since the main system is also small, the generation of correlations
with the catalyst seems to be a natural prerequisite for the implementation
of non-trivial catalytic transformations. Nevertheless, it would be
interesting to corroborate this hypothesis, and to study the role
played by correlations in our framework and other scenarios where
catalysts can yield substantial advantages. 
\item \textbf{The role of the catalyst dimension in exact and inexact catalysis}.
Recently, it was shown that a large class of catalytic transformations
are possible without requiring finely tuned catalysts \citep{15universal-cataysts}.
That is, by employing instead sufficient copies of any catalyst state
and allowing a small disturbance in the final state. This relaxation
concerning the final state of the catalyst is known as inexact catalysis
\citep{11Limits-to-catalysis-in-quantum-thermodynamics}, as opposed
to exact catalysis (where the initial and final states must be identical).
\\
A question that remained open in Ref. \citep{15universal-cataysts}
was if multi-copy catalysts are essential or if a single-copy catalyst
of sufficiently large dimension is equally effective, and numerical
evidence in the second direction was given. In this paper we also
saw that sufficiently large and \textit{single-copy} catalysts enable
the implementation of cooling and (more general) non-unital transformations,
as expressed by Theorem 2. In addition, these results are valid under
mild assumptions regarding the initial states of the system and the
environment, and for the more stringent regime of \textit{exact} catalysis.
Nonetheless, we also stress that such findings cannot be directly
compared with those of Ref. \citep{15universal-cataysts}, since our
approach is based on very different physical constraints such as passivity.
With this observation in mind, our results constitute strong evidence
of the resourceful character of the catalyst dimension, which could
be further explored in other physical scenarios. 
\end{itemize}
\begin{acknowledgments}
We acknowledge fruitful discussions with Patryk Lipka-Bartosik. RU
is grateful for support from Israel Science Foundation (Grant No.
2556/20). 
\end{acknowledgments}

\appendix

\section*{Appendices }

The following appendices contain all the technical material referred
to in the main text. In Appendix A, we derive the stronger form of
passivity that leads to Eq. (\ref{eq:4 passiv cond with populations})
and is applied in the context of cooling. In Appendix B we introduce
some notation and describe the class of unitary operations that will
be considered subsequently, as well as their relation with the studied
transformations. The action of these unitary maps is characterized
in Appendix C. In particular, we introduce the key notion of population
transfers or ``currents'', which forms the basis for the analysis
of catalytic transformations. Building upon such a concept, we develop
in Appendix D a theoretical framework that contains all the ingredients
for the proof of Theorem 1 (Appendix E) and for later results regarding
this kind of transformations. 

The proofs of Corollary 2 and Theorem 2 are given in Appendix F and
Appendix G, respectively. In Appendix H we generalize the catalytic
transformations previously studied, by extending the set of unitary
operations that allows to implement them. Using the concept of ``loop
current'', presented in Appendix H, we optimize in Appendix I the
catalytic cooling of a two-level system using another two-level system
as hot object (cf. Sect. VI-A). The corresponding optimality proof
relies on the maximization of the loop current to derive the corresponding
maximum cooling. However, such a proof is not restricted to the aforementioned
cooling example, and is again applied in the last appendix (Appendix
L). 

The proof of Theorem 3 and the example of catalytic cooling enhancement
presented in Sect. VI-C are developed in Appendix J. In Appendix K
we derive a formula for the maximum cooling of a qubit using $k$
identical qubits, which is of great importance for the results of
Sect. VII. Finally, we introduce in Appendix L a protocol for the
systematic construction of catalytic unitaries that are aimed at increasing
the loop current. This protocol is illustrated with an example of
catalytic cooling, where it leads to a new transformation with enhanced
heat extraction. 

\section{EXTENDED NOTION OF PASSIVITY }

Here we perform the derivation of Eq. (\ref{eq:4 passiv cond with populations}).
To that end, we resort to the fact that the local Hamiltonian $H_{c}$
is also an operator acting on the joint Hilbert space $\mathcal{H}_{c}\otimes\mathcal{H}_{h}$,
if written as $H_{c}=H_{c}\otimes\mathbb{I}_{h}$. In this way, we
can apply known results \citep{43Raam} to determine when a state
that acts on $\mathcal{H}_{c}\otimes\mathcal{H}_{h}$ is passive with
respect to $H_{c}$. These results rely on a direct generalization
of the standard notion of passivity to arbitrary observables. In our
context, this means that, given a state $\rho_{ch}$ and some observable
(hermitian operation) $O_{ch}$ that act both on $\mathcal{H}_{c}\otimes\mathcal{H}_{h}$,
$\rho_{ch}$ is passive with respect to $O_{ch}$ if and only if $[\rho_{ch},O_{ch}]=0$
and the eigenvalues of $\rho_{ch}$ are non-increasing with respect
to those of $O_{ch}$. If $O_{ch}=H_{ch}$ is the Hamiltonian of the
joint system described by $\rho_{ch}$ standard passivity is recovered. 

Let us see how the aforementioned characterization manifests if $O_{ch}=H_{c}\otimes\mathbb{I}_{h}$
and $\rho_{ch}=\rho_{c}\otimes\rho_{h}$. Since $\mathbb{I}_{h}=\sum_{j=1}^{d_{h}}|j_{h}\rangle\langle j_{h}|$,
the eigendecomposition of $H_{c}\otimes\mathbb{I}_{h}$ reads 
\begin{equation}
H_{c}\otimes\mathbb{I}_{h}=\sum_{i=1}^{d_{c}}\sum_{j=1}^{d_{h}}\varepsilon_{i}^{c}|i_{c}j_{h}\rangle\langle i_{c}j_{h}|.\label{eq:Hc x Ih}
\end{equation}
First of all, note that $\rho_{c}\otimes\rho_{h}=\sum_{i=1}^{d_{c}}\sum_{j=1}^{d_{h}}p_{i}^{c}p_{j}^{c}|i_{c}j_{h}\rangle\langle i_{c}j_{h}|$
clearly commutes with $H_{c}\otimes\mathbb{I}_{h}$. According to
Eq. (\ref{eq:Hc x Ih}), the eigenvalues of $\rho_{ch}$ are non-increasing
with respect to those of $H_{c}\otimes\mathbb{I}_{h}$ if and only
if 
\begin{equation}
p_{i}^{c}p_{j}^{h}\geq p_{i'}^{c}p_{j'}^{h}\Leftrightarrow\varepsilon_{i'}^{c}\geq\varepsilon_{i}^{c},\label{eq:64.1 first passivity inequality}
\end{equation}
for any set of indices $\{i,i',j,j'\}$. 

Assuming (without loss of generality) that the eigenenergies are sorted
in non-decreasing order, then $\varepsilon_{i'}^{c}\geq\varepsilon_{i}^{c}$
only if $i'\geq i+1$. Under this sorting, the equivalence (\ref{eq:64.1 first passivity inequality})
is meaningful only for indices such that $i'\geq i+1$. In particular,
for $\{i,i',j,j'\}=\{i,i+1,d_{h},1\}$ it reads: $p_{i}^{c}p_{d_{h}}^{h}\geq p_{i+1}^{c}p_{1}^{h}\Leftrightarrow\varepsilon_{i+1}^{c}\geq\varepsilon_{i}^{c}$.
Such statement can be recast as 

\begin{equation}
\frac{p_{i}^{c}}{p_{i+1}^{c}}\geq\frac{p_{1}^{h}}{p_{d_{h}}^{h}}\Leftrightarrow\varepsilon_{i+1}^{c}\geq\varepsilon_{i}^{c},\label{eq:65 passivity of rhocxrhoh implies passiv of rhoc}
\end{equation}
for all $i\in\{1,2,...,d_{c}-1\}$. Since $\frac{p_{1}^{h}}{p_{d_{h}}^{h}}\geq1$,
Eq. (\ref{eq:65 passivity of rhocxrhoh implies passiv of rhoc}) tells
us that $p_{i}^{c}\geq p_{i+1}^{c}\Leftrightarrow\varepsilon_{i+1}^{c}\geq\varepsilon_{i}^{c}$,
i.e. that the eigenvalues of $\rho_{c}$ are non-increasing with respect
to those of $H_{c}$. In this way, the passivity of $\rho_{c}\otimes\rho_{h}$
leads to the passivity of $\rho_{c}$, as expected. 

On the other hand, Eq. (\ref{eq:65 passivity of rhocxrhoh implies passiv of rhoc})
actually coincides with our characterization of passivity (for $\rho_{c}\otimes\rho_{h}$)
(\ref{eq:4 passiv cond with populations}) in the main text. We will
show that it is equivalent to the original definition (\ref{eq:64.1 first passivity inequality}),
by proving that each side of (\ref{eq:65 passivity of rhocxrhoh implies passiv of rhoc})
is equivalent to the corresponding side in (\ref{eq:64.1 first passivity inequality}).

First, the inequalities $\varepsilon_{i+1}^{c}\geq\varepsilon_{i}^{c}$
(for all $i$) are clearly equivalent to the inequalities $\varepsilon_{i'}^{c}\geq\varepsilon_{i}^{c}$
(for any $i'$ such that $i'\geq i+1$). This establishes the equivalence
between the r.h.s. of (\ref{eq:65 passivity of rhocxrhoh implies passiv of rhoc})
and the r.h.s. of (\ref{eq:64.1 first passivity inequality}). Secondly,
$p_{i'}^{c}\leq p_{i+1}^{c}$ for all $i'\geq i+1$, because $\varepsilon_{i'}^{c}\geq\varepsilon_{i+1}^{c}$
for all $i'\geq i+1$ and we have seen that $\rho_{c}$ is passive.
This allows us to write 

\begin{equation}
\frac{p_{i}^{c}}{p_{i'}^{c}}\geq\frac{p_{i}^{c}}{p_{i+1}^{c}}\geq\frac{p_{1}^{h}}{p_{d_{h}}^{h}}\geq\frac{p_{j'}^{h}}{p_{j}^{h}},\label{eq:second passivity inequality}
\end{equation}
for all $\{i,i',j,j'\}$ such that $i'\geq i+1$. Noting that the
inequality connecting the outmost fractions in (\ref{eq:second passivity inequality})
can be rearranged as $p_{i}^{c}p_{j}^{h}\geq p_{i'}^{c}p_{j'}^{h}$,
we also have the equivalence between the l.h.s. of Eqs. (\ref{eq:65 passivity of rhocxrhoh implies passiv of rhoc})
and (\ref{eq:64.1 first passivity inequality}). This concludes the
proof. 

\section{NOTATION AND PRELIMINARIES }

Apart from the notation for initial states and their eigendecompositions,
described in Sect. II of the main text, we adopt the following conventions. 

\subsection{Notation for indices }

Let $\{1,2,..,n\}$ be a set of indices. Then, 
\begin{itemize}
\item Any operation over elements labeled by \textit{all} the indices $i\in\{1,2,...,n\}$
can be written in an explicit or implicit form, being both equivalent.
For example, a sum $\sum_{i=1}^{n}a_{i}$ (explicit form), where $a_{i}$
are elements of some set, has the equivalent expression $\sum_{i}a_{i}$
(implicit form). If the operation involves a subset of indices $I\subseteq\{1,2,...,n\}$,
this is explicitly written. For example, the sum $\sum_{i\in I}a_{i}$
specifies that only elements $a_{i}$ restricted by the subset $I$
are included. 
\item The previous convention is also applied to denote sets. For example,
the set $\{a_{i}\}_{i}$ is equivalent to the set $\{a_{i}\}_{i=1}^{n}$,
and the set $\{a_{i}\}_{i\in I}$ is contained into $\{a_{i}\}_{i}$.
In contrast, $\{a_{i}\}$ is the set that contains the \textit{single}
element $a_{i}$. 
\item Sometimes we will consider operations that ''run'' over a \textit{non-specific}
subset of indices. In such a case we add the term ``free'' to indicate
this property. For example, a sum $\sum_{i\textrm{ free}}a_{i}$ is
equivalent to a sum $\sum_{i\in I}a_{i}$, where $I$ could be \textit{any}
subset $I\subseteq\{1,2,...,n\}$. The possible subsets $I$ for which
the corresponding operation is meaningful will be clear from the context,
or explicitly indicated. 
\item The previous conventions are equally applicable to greek indices.
For example, the sum in the eigendecomposition $\rho=\sum_{\psi}p_{\psi}|\psi\rangle\langle\psi|$
runs over \textit{all} the indices $\psi\in\{1,2,...,d_{c}d_{h}d_{v}\}$
(note that if $\rho$ is not full-rank then $p_{\psi}=0$ for some
values of $\psi$). 
\end{itemize}

\subsection{Tranformations and unitary operations}

All the transformations studied possess an explicit form $\rho\rightarrow\rho'=U\rho U^{\dagger}$,
where $U$ is a unitary matrix. In this way, the subsystems $x=c,h,v$
undergo associated transformations $\rho_{x}\rightarrow\rho'_{x}$,
where $\rho_{x}=\textrm{Tr}_{\sim x}\left(\rho\right)$, $\rho'_{x}=\textrm{Tr}_{\sim x}\left(\rho'\right)$,
and $\sim x$ indicates the degrees of freedom different from $x$. 

\textbf{Remark 1}. In contrast with primed states (e.g. $\rho'$ and
$\rho'_{x}$), which denote final states, primed labels are used to
indicate the specific value taken by an index. For example, the label
$i'$ indicates that $i$ takes the value $i=i'$.

Given a unitary $U$ and a Hilbert subspace $\mathcal{H}_{\textrm{sub}}\subseteq\mathcal{H}$
(where $\mathcal{H}=\mathcal{H}_{c}\otimes\mathcal{H}_{h}\otimes\mathcal{H}_{v}$),
we say that $U$ maps $\mathcal{H}_{\textrm{sub}}$ into itself if:
\begin{enumerate}
\item $U|\psi\rangle=|\psi\rangle$ for $|\psi\rangle\notin\mathcal{H}_{\textrm{sub}}$. 
\item $U|\psi\rangle\in\mathcal{H}_{\textrm{sub}}$ for $|\psi\rangle\in\mathcal{H}_{\textrm{sub}}$. 
\end{enumerate}
If $U$ satisfies conditions 1 and 2 we also say that ``$U$ acts
on $\mathcal{H}_{\textrm{sub}}$'', and write it as $U:\mathcal{H}_{\textrm{sub}}\rightarrow\mathcal{H}_{\textrm{sub}}$.
In addition, we term $\mathcal{H}_{\textrm{sub}}$ the ``domain subspace''
of $U$, as per condition 1. 

\textbf{Definition 4 (two-level unitary)}. A two-level unitary $U^{(2)}$
is a unitary that acts on a two-dimensional subspace $\textrm{span}\{|\psi\rangle,|\varphi\rangle\}$,
where $|\psi\rangle,|\varphi\rangle\in\mathcal{H}$. 

\textbf{Remark 2}. If $U$ acts on $\mathcal{H}_{\textrm{sub}}$,
conditions 1 and 2 imply that $U$ also acts on any subspace $\mathcal{H}'_{\textrm{sub}}$
such that $\mathcal{H}_{\textrm{sub}}\subseteq\mathcal{H}'_{\textrm{sub}}$.
Moreover, these conditions can be straightforwardly extended to local
subspaces. For example, a local unitary $U_{ch}$ acts on a subspace
$\mathcal{H}_{\textrm{sub}}\subseteq\mathcal{H}_{c}\otimes\mathcal{H}_{h}$
if 1 and 2 hold, with $|\psi\rangle$ being replaced by $|\psi_{ch}\rangle$.

The unitaries addressed in this work have the form 

\begin{equation}
U=\oplus_{\alpha}U_{\alpha},\label{eq: 100 Direct sum of unitaries SM}
\end{equation}
where $\oplus$ is the direct sum operation. Hence, $U$ is a block-diagonal
matrix with blocks $\{U_{\alpha}\}_{\alpha}$. The most basic decomposition
of the form (\ref{eq: 100 Direct sum of unitaries SM}) is $U=U_{\textrm{cool}}\oplus V_{\textrm{res}}$,
where $U_{\textrm{cool}}$ is a cooling unitary and $V_{\textrm{res}}$
is a restoring unitary (see Definitions 1 and 2 in the main text).
In this case, $\{U_{\alpha}\}_{\alpha}=\{U_{\textrm{cool}},V_{\textrm{res}}\}$.
Later on we will consider other useful decompositions, where $U_{\textrm{cool}}$
and $V_{\textrm{res}}$ are themselves given by block-diagonal matrices. 

\section{MAIN QUANTITIES FOR THE CHARACTERIZATION OF CATALYTIC TRANSFORMATIONS }

\subsection{Average variations of observables}

Here we derive the general expression for the average variation of
an observable (hermitian operator) $O$, given a transformation $\rho\rightarrow\rho'=U\rho U^{\dagger}$,
where $U$ satisfies Eq. (\ref{eq: 100 Direct sum of unitaries SM}).
In combination with the notion of population currents, this will allow
us to present a description of population variations in the local
eigenbases $\{|i_{x=c,v}\rangle\}_{i}$, especially suited for the
analysis of catalytic transformations. 

To begin with, we express the unitary $U$ in the eigenbasis $\{|\psi\rangle\}_{\psi}$,
characteristic of the initial state $\rho$. Specifically, consider
the decomposition 
\begin{equation}
\rho=\sum_{\alpha}\rho^{(\alpha)},\label{eq:101 decomposition of rho}
\end{equation}
where the index $\alpha$ labels (disjoint) subsets of eigenstates
of $\rho$. Denoting these subsets as $E_{\alpha}$, $\rho^{(\alpha)}$
results from the projection of $\rho$ onto the corresponding subset.
Namely, $\rho^{(\alpha)}=\sum_{|\psi\rangle\in E_{\alpha}}|\psi\rangle\langle\psi|\rho\sum_{|\psi\rangle\in E_{\alpha}}|\psi\rangle\langle\psi|$.
Each subset $E_{\alpha}$ gives rise to a subspace $\mathcal{H}_{\alpha}=\textrm{span}\{|\psi\rangle\in E_{\alpha}\}$,
and $U_{\alpha}$ is a unitary that maps $\mathcal{H}_{\alpha}$ into
itself. 

The average variation of an observable $O$ is thus given by
\begin{equation}
\Delta\left\langle O\right\rangle =\textrm{Tr}\left[O(U\rho U^{\dagger}-\rho)\right]=\sum_{\alpha}\Delta_{\alpha}\left\langle O\right\rangle ,\label{eq:102 delta<O>}
\end{equation}
where 
\begin{align}
\Delta_{\alpha}\left\langle O\right\rangle  & \equiv\textrm{Tr}\left[O(U_{\alpha}\rho U_{\alpha}^{\dagger}-\rho)\right]\nonumber \\
 & =\textrm{Tr}\left[O(U_{\alpha}\rho^{(\alpha)}U_{\alpha}^{\dagger}-\rho^{(\alpha)})\right].\label{eq:103 delta_alpha<O>}
\end{align}
The second line in Eq. (\ref{eq:103 delta_alpha<O>}) follows from
the fact that $U_{\alpha}|\psi\rangle=|\psi\rangle$ for $|\psi\rangle\notin E_{\alpha}$,
and consequently these eigenstates do not contribute to $\Delta_{\alpha}\left\langle O\right\rangle $. 

\textbf{Remark 3}. Throughout these appendices we will often replace
$\{\alpha\}_{\alpha}$ by a set of indices that are suitable to characterize
a particular set of unitaries $\{U_{\alpha}\}_{\alpha}$. As an example,
$\alpha$ can be replaced by a pair $\psi,\varphi$ or a related expression,
if the set $E_{\alpha}$ only contains the eigenstates $|\psi\rangle$
and $|\varphi\rangle$, i.e. $E_{\alpha}=\{|\psi\rangle,|\varphi\rangle\}$. 

\subsection{Population variations and currents}

If $U_{\alpha}$ is a two-level unitary, it is equivalent to a partial
swap $U_{|\psi\rangle\leftrightarrow|\varphi\rangle}$ between the
eigenstates $|\psi\rangle$ and $|\varphi\rangle$ that form the set
$E_{\alpha}$. Let us express $\alpha$ as $\psi,\varphi$ and $U_{|\psi\rangle\leftrightarrow|\varphi\rangle}$
as $U_{\psi,\varphi}$. Moreover, $\rho^{(\alpha)}$ is equivalent
to $\rho^{(\psi,\varphi)}=p_{\psi}|\psi\rangle\langle\psi|+p_{\varphi}|\varphi\rangle\langle\varphi|$.
This allows us to characterize the transformation $\rho^{(\psi,\varphi)}\rightarrow U_{\psi,\varphi}\rho^{(\psi,\varphi)}U_{\psi,\varphi}^{\dagger}$
using Eqs. (\ref{eq:11 U|i>}) and (\ref{eq:12 U|j>}). Denoting the
final populations as $p_{\psi}^{\prime}$ and $p_{\varphi}^{\prime}$,
we have that 
\begin{align}
U_{\psi,\varphi}\rho^{(\psi,\varphi)}U_{\psi,\varphi}^{\dagger} & =p_{\psi}^{\prime}|\psi\rangle\langle\psi|+p_{\varphi}^{\prime}|\varphi\rangle\langle\varphi|\nonumber \\
 & \quad+\sqrt{(1-r)r}\left(|\psi\rangle\langle\varphi|+|\varphi\rangle\langle\psi|\right),\label{eq:104 transformation on rho(i,j) from a two-level unitary}
\end{align}
where $p_{\psi}^{\prime}=p_{\psi}+r(p_{\varphi}-p_{\psi})$ is given
in (\ref{eq:13 p'j}) and $p_{\varphi}^{\prime}=p_{\psi}+p_{\varphi}-p_{\psi}^{\prime}$
. 

The observable $O_{k^{\prime}}\equiv|k_{v}^{\prime}\rangle\langle k_{v}^{\prime}|$
characterizes the catalyst population $\left\langle |k_{v}^{\prime}\rangle\langle k_{v}^{\prime}|\right\rangle $.
From Eq. (\ref{eq:104 transformation on rho(i,j) from a two-level unitary})
we obtain: 
\begin{equation}
\textrm{Tr}\left[O_{k^{\prime}}U_{\psi,\varphi}\rho^{(\psi,\varphi)}U_{\psi,\varphi}^{\dagger}\right]=p_{\psi}^{\prime}|\langle k_{v}^{\prime}|\psi\rangle|^{2}+p_{\varphi}^{\prime}|\langle k_{v}^{\prime}|\varphi\rangle|^{2}.\label{eq:105 final catalyst population Tr=00005BOk'rho'(i,j)=00005D}
\end{equation}
Hence, 
\begin{align}
\Delta_{\psi,\varphi}\left\langle O_{k^{\prime}}\right\rangle  & =\textrm{Tr}\left[O_{k^{\prime}}\left(U_{\psi,\varphi}\rho^{(\psi,\varphi)}U_{\psi,\varphi}^{\dagger}-\rho^{(\psi,\varphi)}\right)\right]\nonumber \\
 & =\left(p_{\psi}^{\prime}-p_{\psi}\right)|\langle k_{v}^{\prime}|\psi\rangle|^{2}+\left(p_{\varphi}^{\prime}-p_{\varphi}\right)|\langle k_{v}^{\prime}|\varphi\rangle|^{2}\nonumber \\
 & =r(p_{\varphi}-p_{\psi})\left[|\langle k_{v}^{\prime}|\psi\rangle|^{2}-|\langle k_{v}^{\prime}|\varphi\rangle|^{2}\right]\nonumber \\
 & =J_{|\varphi\rangle\rightarrow|\psi\rangle}\left[|\langle k_{v}^{\prime}|\psi\rangle|^{2}-|\langle k_{v}^{\prime}|\varphi\rangle|^{2}\right],\label{eq:106 Delta_i,j<Ok'>}
\end{align}
where the third line follows from the definition of the current $J_{|\varphi\rangle\rightarrow|\psi\rangle}$
(cf. Eq. (\ref{eq:14 current})). 

\textbf{Remark 4}. In this work, the unitaries that lead to catalytic
transformations are composed of two-level unitaries $U_{|\psi\rangle\leftrightarrow|\varphi\rangle}$
such that $\textrm{Tr}_{ch}|\psi\rangle\langle\varphi|=\textrm{Tr}_{v}|\psi\rangle\langle\varphi|=0$.
Accordingly, $|\psi\rangle=|\psi_{ch}k'_{v}\rangle$ and $|\varphi\rangle=|\varphi_{ch}k_{v}^{\prime\prime}\rangle$,
where $|\psi_{ch}\rangle$, $|\varphi_{ch}\rangle$ are different
eigenstates of $\rho_{ch}$ and $|k'_{v}\rangle$, $|k_{v}^{\prime\prime}\rangle$
are different eigenstates of $\rho_{v}$.\textcolor{red}{{} }

A consequence of Remark 4 is that if $|\langle k_{v}^{\prime}|\psi\rangle|=1$
then $|\langle k_{v}^{\prime}|\varphi\rangle|=0$ and vice versa.
Accordingly, either $\Delta_{\psi,\varphi}\left\langle O_{k^{\prime}}\right\rangle =J_{|\varphi\rangle\rightarrow|\psi\rangle}$
(if $|\langle k_{v}^{\prime}|\psi\rangle|=1$), or $\Delta_{\psi,\varphi}\left\langle O_{k^{\prime}}\right\rangle =-J_{|\varphi\rangle\rightarrow|\psi\rangle}$
(if $|\langle k_{v}^{\prime}|\varphi\rangle|=1$). Since $J_{|\varphi\rangle\rightarrow|\psi\rangle}=-J_{|\psi\rangle\rightarrow|\varphi\rangle}$
(cf. Eq. (\ref{eq:14 current})), we can rewrite Eq. (\ref{eq:106 Delta_i,j<Ok'>})
as 
\begin{equation}
\Delta_{\psi,\varphi}\left\langle O_{k^{\prime}}\right\rangle =\begin{cases}
\begin{array}{c}
J_{|\varphi\rangle\rightarrow|\psi\rangle},\textrm{ if }|\langle k_{v}^{\prime}|\psi\rangle|=1,\\
J_{|\psi\rangle\rightarrow|\varphi\rangle},\textrm{ if }|\langle k_{v}^{\prime}|\varphi\rangle|=1,\\
0,\textrm{ otherwise.}
\end{array}\end{cases}\label{eq:107 Delta_i,j<Ok'> in terms of global current}
\end{equation}
Consider now a set of partial swaps $\{U_{\psi,\varphi}\}_{\psi,\varphi\textrm{ free}}$
acting on orthogonal subspaces $\mathcal{H}_{\psi,\varphi}=\textrm{span}\{|\psi\rangle,|\varphi\rangle\}$.
Taking into account Eq. (\ref{eq:102 delta<O>}), for $U=\oplus_{\psi,\varphi\textrm{ free}}U_{\psi,\varphi}$
we obtain:
\begin{align}
\Delta\left\langle O_{k^{\prime}}\right\rangle  & =\textrm{Tr}\left[O_{k^{\prime}}\left(U\rho U^{\dagger}-\rho\right)\right]\nonumber \\
 & =\sum_{\psi,\varphi\textrm{ free}}\Delta_{\psi,\varphi}\left\langle O_{k^{\prime}}\right\rangle ,\label{eq:108 total Delta_i,j<Ok'>}
\end{align}
where $\Delta_{\psi,\varphi}\left\langle O_{k^{\prime}}\right\rangle $
obeys Eq. (\ref{eq:107 Delta_i,j<Ok'> in terms of global current}). 

\subsection{Diagonal form of the final catalyst state }

Before moving forward, it is important to point out a fact that underpins
all our forthcoming analysis on catalytic transformations. Based on
Proposition 2, we can fully characterize catalytic transformations
in terms of the populations in the eigenbasis of the initial state
$\rho_{v}$. 

\textbf{Proposition 2}. \textit{For any unitary} $U=\oplus_{\psi,\varphi\textrm{ free}}U_{\psi,\varphi}$\textit{
where each pair of eigenstates $|\psi\rangle$ and $|\varphi\rangle$
adhere to Remark 4, the catalyst state} $\rho'_{v}=\textrm{Tr}_{ch}\left(U\rho U^{\dagger}\right)$\textit{
is diagonal in the eigenbasis of the initial state $\rho_{v}$. Moreover,
this is true for any state $\rho=\rho_{ch}\otimes\rho_{v}$, irrespective
of the form of $\rho_{ch}$.}

\textit{Proof}. The action of direct sum $U=\oplus_{\psi,\varphi\textrm{ free}}U_{\psi,\varphi}$
on $\rho$ is given in terms of the action of each $U_{\psi,\varphi}$
on the corresponding projection. That is, 
\begin{equation}
U\rho U^{\dagger}=\sum_{\psi,\varphi\textrm{ free}}U_{\psi,\varphi}\rho^{(\psi,\varphi)}U_{\psi,\varphi}^{\dagger},\label{eq:108.1 final catalyst diagonal state}
\end{equation}
with $U_{\psi,\varphi}\rho^{(\psi,\varphi)}U_{\psi,\varphi}^{\dagger}$
given in Eq. (\ref{eq:104 transformation on rho(i,j) from a two-level unitary}).
Keeping in mind Remark 4, we have that $\textrm{Tr}_{ch}|\psi\rangle\langle\varphi|=0$
for any pair $|\psi\rangle,|\varphi\rangle$. Therefore, (\ref{eq:104 transformation on rho(i,j) from a two-level unitary})
and (\ref{eq:108.1 final catalyst diagonal state}) yields the state
$\textrm{Tr}_{ch}\left(U\rho U^{\dagger}\right)=\sum_{\psi,\varphi\textrm{ free}}\left(p_{\psi}^{\prime}\textrm{Tr}_{ch}|\psi\rangle\langle\psi|+p_{\varphi}^{\prime}\textrm{Tr}_{ch}|\varphi\rangle\langle\varphi|\right)$,
which is diagonal in the eigenbasis of $\rho_{v}$. 

\section{LOOPS AND CATALYTIC TRANSFORMATIONS }

For global eigenstates $|\psi\rangle=|i_{c}^{\prime}j_{h}^{\prime}k_{v}^{\prime}\rangle$
and $|\varphi\rangle=|i_{c}^{\prime\prime}j_{h}^{\prime\prime}k_{v}^{\prime\prime}\rangle$,
a two-level unitary $U_{\psi,\varphi}$ generates local currents between
all the pairs of local eigenstates. This is a consequence of the fact
that $\textrm{span}\{|\psi\rangle,|\varphi\rangle\}$ is contained
into $\textrm{span}\{|i_{c}^{\prime}\rangle,|i_{c}^{\prime\prime}\rangle\}\otimes\textrm{span}\{|j_{h}^{\prime}\rangle,|j_{h}^{\prime\prime}\rangle\}\otimes\textrm{span}\{|k_{v}^{\prime}\rangle,|k_{v}^{\prime\prime}\rangle\}$,
and therefore $U_{\psi,\varphi}$ maps the later subspace into itself
(cf. Remark 2). More specifically, this implies that $\Delta_{\psi,\varphi}\left\langle |i_{c}^{\prime}\rangle\langle i_{c}^{\prime}|\right\rangle =-\Delta_{\psi,\varphi}\left\langle |i_{c}^{\prime\prime}\rangle\langle i_{c}^{\prime\prime}|\right\rangle $,
$\Delta_{\psi,\varphi}\left\langle |j_{h}^{\prime}\rangle\langle j_{h}^{\prime}|\right\rangle =-\Delta_{\psi,\varphi}\left\langle |j_{h}^{\prime\prime}\rangle\langle k_{h}^{\prime\prime}|\right\rangle $,
and $\Delta_{\psi,\varphi}\left\langle |k_{v}^{\prime}\rangle\langle k_{v}^{\prime}|\right\rangle =-\Delta_{\psi,\varphi}\left\langle |k_{v}^{\prime\prime}\rangle\langle k_{v}^{\prime\prime}|\right\rangle $.
Therefore, the variation $\Delta_{\psi,\varphi}\left\langle |k_{v}^{\prime}\rangle\langle k_{v}^{\prime}|\right\rangle $
is associated with a population transfer from $|k_{v}^{\prime\prime}\rangle$
to $|k_{v}^{\prime}\rangle$, described by a local current $J_{|k_{v}^{\prime\prime}\rangle\rightarrow|k_{v}^{\prime}\rangle}$.
Likewise, the population variations $\Delta_{\psi,\varphi}\left\langle |i_{c}^{\prime}\rangle\langle i_{c}^{\prime}|\right\rangle $
and $\Delta_{\psi,\varphi}\left\langle |j_{h}^{\prime}\rangle\langle j_{h}^{\prime}|\right\rangle $
are also associated with local currents $J_{|i_{c}^{\prime\prime}\rangle\rightarrow|i_{c}^{\prime}\rangle}$
and $J_{|j_{h}^{\prime\prime}\rangle\rightarrow|j_{h}^{\prime}\rangle}$,
respectively. 

The same calculations that allowed us to derive Eq. (\ref{eq:107 Delta_i,j<Ok'> in terms of global current})
can be straightforwardly applied to express $\Delta_{\psi,\varphi}\left\langle |i_{c}^{\prime}\rangle\langle i_{c}^{\prime}|\right\rangle $
and $\Delta_{\psi,\varphi}\left\langle |j_{h}^{\prime}\rangle\langle j_{h}^{\prime}|\right\rangle $
in terms of the global current $J_{|\varphi\rangle\rightarrow|\psi\rangle}$.
In this way, we have that:
\begin{align}
\Delta_{\psi,\varphi}\left\langle O_{k^{\prime}}\right\rangle  & =J_{|k_{v}^{\prime\prime}\rangle\rightarrow|k_{v}^{\prime}\rangle}=J_{|\varphi\rangle\rightarrow|\psi\rangle},\label{eq:109 Jk''-->k'}\\
\Delta_{\psi,\varphi}\left\langle |i_{c}^{\prime}\rangle\langle i_{c}^{\prime}|\right\rangle  & =J_{|i_{c}^{\prime\prime}\rangle\rightarrow|i_{c}^{\prime}\rangle}=J_{|\varphi\rangle\rightarrow|\psi\rangle},\label{eq: 110 Ji''-->i'}\\
\Delta_{\psi,\varphi}\left\langle |j_{h}^{\prime}\rangle\langle j_{h}^{\prime}|\right\rangle  & =J_{|j_{h}^{\prime\prime}\rangle\rightarrow|j_{h}^{\prime}\rangle}=J_{|\varphi\rangle\rightarrow|\psi\rangle},\label{eq: 111 Jj''-->j'}
\end{align}
where we have reserved the simplified notation $O_{k^{\prime}}$ for
the projector $|k_{v}^{\prime}\rangle\langle k_{v}^{\prime}|$, as
Eq. (\ref{eq:109 Jk''-->k'}) will be the most utilized among Eqs.
(\ref{eq:109 Jk''-->k'})-(\ref{eq: 111 Jj''-->j'}).

These equations provide the equalities in Eqs. (\ref{eq:19.1 local current 1})-(\ref{eq:19.3 local current 3})
of the main text. The upper bounds in (\ref{eq:19.1 local current 1})-(\ref{eq:19.3 local current 3})
follow by realizing that $J_{|\varphi\rangle\rightarrow|\psi\rangle}$
is maximum or minimum for $r=1$, in which case $J_{|\varphi\rangle\rightarrow|\psi\rangle}=\mathcal{J}_{|\varphi\rangle\rightarrow|\psi\rangle}$. 

\textbf{Remark 5}.\textit{ }A local current is essentially a population
transfer between a pair of local eigenstates, and it is not restricted
to two-level unitaries $U_{\psi,\varphi}$. For example, any unitary
$U$ that acts on a subspace $\mathcal{H}_{c}\otimes\mathcal{H}_{h}\otimes\textrm{span}\{|k_{v}^{\prime\prime}\rangle,|k_{v}^{\prime}\rangle\}$
generates a local current $J_{|k_{v}^{\prime\prime}\rangle\rightarrow|k_{v}^{\prime}\rangle}=\Delta\bigl\langle O_{k'}\bigr\rangle=-\Delta\bigl\langle O_{k_{v}^{\prime\prime}}\bigr\rangle$.
Likewise, any unitary $U$ that acts on a subspace $\textrm{span}\{|i_{c}^{\prime\prime}\rangle,|i_{c}^{\prime}\rangle\}\otimes\mathcal{H}_{h}\otimes\mathcal{H}_{v}$
generates a local current $J_{|i_{c}^{\prime\prime}\rangle\rightarrow|i_{c}^{\prime}\rangle}=\Delta\bigl\langle|i'_{c}\rangle\langle i'_{c}|\bigr\rangle-\Delta\bigl\langle|i_{c}^{\prime\prime}\rangle\langle i_{c}^{\prime\prime}|\bigr\rangle$. 

In particular, consider a set of local unitaries $\{U_{\psi,\varphi}\}_{\psi,\varphi\textrm{ free}}$,
such that $U_{\psi,\varphi}$ acts on the subspace $\textrm{span}\{|\psi_{ch}k_{v}^{\prime}\rangle,|\varphi_{ch}k_{v}^{\prime\prime}\rangle\}$
(where $\psi$ and $\varphi$ are fixed). Since $\oplus_{\psi,\varphi\textrm{ free}}U_{\psi,\varphi}$
acts on $\mathcal{H}_{c}\otimes\mathcal{H}_{h}\otimes\textrm{span}\{|k_{v}^{\prime\prime}\rangle,|k_{v}^{\prime}\rangle\}$,
as per Remark 2, it generates a current $J_{|k_{v}^{\prime\prime}\rangle\rightarrow|k_{v}^{\prime}\rangle}$
(Remark 5). Using Eqs. (\ref{eq:108 total Delta_i,j<Ok'>}) and (\ref{eq:109 Jk''-->k'}),
we have that 

\begin{equation}
J_{|k_{v}^{\prime\prime}\rangle\rightarrow|k_{v}^{\prime}\rangle}=\sum_{\psi,\varphi\textrm{ free}}J_{|\varphi\rangle\rightarrow|\psi\rangle}.\label{eq:111.1 local catalyst current as sum of global currents}
\end{equation}
Moreover, we can define the ``swap local current'' 
\begin{equation}
\mathcal{J}_{|k_{v}^{\prime\prime}\rangle\rightarrow|k_{v}^{\prime}\rangle}\equiv\sum_{\psi,\varphi\textrm{ free}}\mathcal{J}_{|\varphi\rangle\rightarrow|\psi\rangle},\label{eq:111.2 swap catalyst current as sum of swap global currents}
\end{equation}
which is produced by the direct sum of swaps $\oplus_{\psi,\varphi\textrm{ free}}\mathcal{U}_{\psi,\varphi}$.
The current $J_{|k_{v}^{\prime\prime}\rangle\rightarrow|k_{v}^{\prime}\rangle}$
and the corresponding swap current $\mathcal{J}_{|k_{v}^{\prime\prime}\rangle\rightarrow|k_{v}^{\prime}\rangle}$
in (\ref{eq:19.3 local current 3}) can be recovered from (\ref{eq:111.1 local catalyst current as sum of global currents})
and (\ref{eq:111.2 swap catalyst current as sum of swap global currents}),
when the direct sums $\oplus_{\psi,\varphi\textrm{ free}}U_{\psi,\varphi}$
and $\oplus_{\psi,\varphi\textrm{ free}}\mathcal{U}_{\psi,\varphi}$
reduce to a single two-level unitary. 

Equipped with the general concept of local current characterized in
Remark 5, we can now introduce the most important element for the
analysis of catalytic transformations. Namely, the ``loop''. To
define this element, we must first define the related concept of ``chain''. 

\textbf{Definition 5 (chain)}. Let $\{k^{(m)}\}_{m=1}^{M}$ be a set
of indices, where $k^{(m)}\in\{1,2,...,d_{v}\}$, and let $J_{|k_{v}^{(m+1)}\rangle\rightarrow|k_{v}^{(m)}\rangle}$
be a local current generated by a unitary that acts on $\mathcal{H}_{c}\otimes\mathcal{H}_{h}\otimes\textrm{span}\{|k_{v}^{(m+1)}\rangle,|k_{v}^{(m)}\rangle\}$.
A chain is a set $\{J_{|k_{v}^{(m+1)}\rangle\rightarrow|k_{v}^{(m)}\rangle}\}_{m=1}^{M-1}$,
where $J_{|k_{v}^{(m+1)}\rangle\rightarrow|k_{v}^{(m)}\rangle}>0$
for $1\leq m\leq M-1$. 

Intuitively, the ``direction'' pointed out by the currents in a
chain leads us from the eigenstate $|k_{v}^{(M)}\rangle$ towards
the eigenstate $|k_{v}^{(1)}\rangle$, as illustrated in Fig. 12 .
To capture this intuition, we employ the notation $\mathbf{ch}_{|k_{v}^{(M)}\rangle\rightarrow|k_{v}^{(1)}\rangle}$,
and refer to $|k_{v}^{(M)}\rangle$ and $|k_{v}^{(1)}\rangle$ (or
$k_{v}^{(M)}$ and $k_{v}^{(1)}$ in the $\textrm{ln}(p^{ch})\times\textrm{ln}(p^{v})$
diagram) as the ``origin'' and the ``end'' of the chain, respectively.

\textbf{Definition 6 (loop)}. A loop is a set of currents $\mathbf{ch}_{|k_{v}^{(M)}\rangle\rightarrow|k_{v}^{(1)}\rangle}\cup\{J_{|k_{v}^{(1)}\rangle\rightarrow|k_{v}^{(M)}\rangle}\}$,
where $J_{|k_{v}^{(1)}\rangle\rightarrow|k_{v}^{(M)}\rangle}>0$.
This set can be described using the equivalent notations $\mathbf{loop}_{|k_{v}^{(M)}\rangle\rightarrow|k_{v}^{(M)}\rangle}=\mathbf{loop}_{|k_{v}^{(1)}\rangle\rightarrow|k_{v}^{(1)}\rangle}$.
Moreover, it can be seen as a ``closed chain'', where $|k_{v}^{(M)}\rangle$
and $|k_{v}^{(1)}\rangle$ play both the roles of origin and end.
\begin{figure}
\centering{}\includegraphics[scale=0.74]{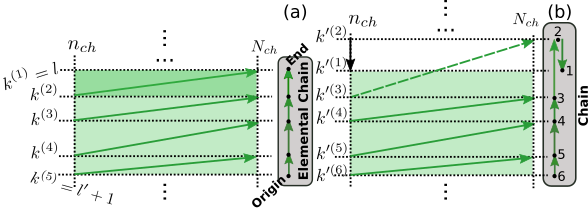}\caption{Two examples of chains $\mathbf{ch}_{|(l'+1)_{v}\rangle\rightarrow|l_{v}\rangle}$
and associated global currents in the $\textrm{ln}(p^{ch})\times\textrm{ln}(p^{v})$
diagram. By including a positive current $J_{|l_{v}\rangle\rightarrow|(l'+1)_{v}\rangle}$,
both chains can from a loop $\mathbf{loop}_{|l_{v}\rangle\rightarrow|l_{v}\rangle}$
(see e.g. Fig. 3(a)). (a) \textit{Elemental} chain (see Definition
7) generated by a unitary $\oplus_{k=k^{(1)}}^{k^{(5)}-1}V_{k}$ ,
where $V_{k}=V_{|n_{ch}(k+1)_{v}\rangle\leftrightarrow|N_{ch}k_{v}\rangle}$
and $\{k^{(1)},k^{(5)}\}=\{l,l'+1\}$. The local currents in this
kind of chain connect all the consecutive rows from $l'+1$ to $l$.
(b) Another chain with the same origin ($l'+1$) and end ($l$) of
the elemental chain. Here, the end is reached by replacing the global
current $J_{|n_{ch}k_{v}^{(2)}\rangle\rightarrow|N_{ch}k_{v}^{(1)}\rangle}$(arrow
inside the darker rectangle) in (a), by $J_{|n_{ch}k_{v}^{\prime(3)}\rangle\rightarrow|N_{ch}k_{v}^{\prime(2)}\rangle}$
(dashed green arrow) and $J_{|n_{ch}k_{v}^{\prime(2)}\rangle\rightarrow|n_{ch}k_{v}^{\prime(1)}\rangle}$
(black arrow). Since the local current $J_{|k_{v}^{\prime(3)}\rangle\rightarrow|k_{v}^{\prime(2)}\rangle}$
does not connect consecutive rows, this chain is not elemental. }
\end{figure}

\subsection{Loop generation via a direct sum of unitaries}

In Fig. 12 we apply the $\textrm{ln}(p^{ch})\times\textrm{ln}(p^{v})$
diagram to illustrate two sets of currents $\{J_{|k_{v}^{(m+1)}\rangle\rightarrow|k_{v}^{(m)}\rangle}\}_{m=1}^{M-1}$
that give rise to chains $\mathbf{ch}_{|(l'+1)_{v}\rangle\rightarrow|l_{v}\rangle}$
with a common origin $l'+1$ and end $l$. The chain in (a) was already
illustrated in Fig. 3, and is an example of the following set: 

\textbf{Definition 7 (elemental chain)}. An elemental chain is a chain
$\mathbf{ch}_{|k^{(M)}\rangle\rightarrow|k^{(1)}\rangle}=\{J_{|(k+1)_{v}\rangle\rightarrow|k_{v}\rangle}\}_{k=k^{(1)}}^{k^{(M)}-1}$,
where all the currents connect \textit{consecutive} eigenstates $\{|k_{v}\rangle\}_{k=k^{(1)}}^{k^{(M)}}$,
and no eigenstate $|k_{v}\rangle$ with $k\leq k^{(1)}-1$ or $k\geq k^{(M)}+1$
is connected. 

Figure 12(b) shows that a chain $\mathbf{ch}_{|(l'+1)_{v}\rangle\rightarrow|l_{v}\rangle}$
is not restricted to its elemental version. Taking this into account,
the purpose of the present section is twofold. On the one hand, we
will set necessary conditions for the generation of a loop through
a proper (direct) sum of unitaries. On the other hand, we will show
that any chain $\mathbf{ch}_{|k_{v}^{(M)}\rangle\rightarrow|k_{v}^{(1)}\rangle}$
in this loop can exist \textit{only if} the elemental chain exists.
These results are fundamental to characterize the existence of catalytic
transformations in our framework. 

\subsubsection{Equations for catalyst population variations }

Consider a set of subspaces $\{\mathcal{H}_{k^{(m)}}\}_{m=1}^{M}$,
such that such that
\begin{equation}
\mathcal{H}_{k^{(m)}}\equiv\textrm{span}\{|\psi_{ch}\rangle\}_{\psi=n}^{N}\otimes\textrm{span}\{|k_{v}^{(m+1)}\rangle,|k_{v}^{(m)}\rangle\}\label{eq:111.3 Definition of subspace H_(k^(m))}
\end{equation}
for $m\in\{1,2,...,M\}$, where we introduce a periodic boundary condition
$k^{(M+1)}=k^{(1)}$. This periodicity condition implies that $\mathcal{H}_{k^{(M)}}=\mathcal{H}_{c}\otimes\mathcal{H}_{h}\otimes\textrm{span}\{|k_{v}^{(1)}\rangle,|k_{v}^{(M)}\rangle\}.$
Moreover, let $\{V_{k^{(m)}}\}_{m=1}^{M-1}$ and $U_{k^{(M)}}$ be
unitaries whose domain subspaces are $\{\mathcal{H}_{k^{(m)}}\}_{m=1}^{M-1}$
and $\mathcal{H}_{k^{(M)}}$, respectively: 
\begin{align}
V_{k^{(m)}} & :\mathcal{H}_{k^{(m)}}\rightarrow\mathcal{H}_{k^{(m)}},\textrm{ for }m\in\{1,2,...,M-1\},\nonumber \\
U_{k^{(M)}} & :\mathcal{H}_{k^{(M)}}\rightarrow\mathcal{H}_{k^{(M)}}.\label{eq:111.4 V_(k^(m)) and U_k^(M)}
\end{align}
Importantly, both $V_{k^{(m)}}$ and $U_{k^{(M)}}$ \textit{do not}
have to be two-level unitaries. 

\textbf{Remark 6}. For the sake of brevity, in the remainder of these
appendices the indices $n$ and $N$ in the subspaces $\{\mathcal{H}_{k^{(m)}}\}_{m=1}^{M}$
will be left implicit. 

Our goal now is to set the stage for the generation of a loop via
a direct sum: 
\begin{equation}
U=U_{k^{(M)}}\oplus\left(\oplus_{m=1}^{M-1}V_{k^{(m)}}\right).\label{eq:112 U_k(M)+(+_mV_k(m))}
\end{equation}
Taking into account Remark 3, let us make the correspondence $\{\alpha\}_{\alpha}=\{k^{(m)}\}_{m=1}^{M}$.
In this way, the variation of the catalyst population $\left\langle O_{k^{(m')}}\right\rangle =\left\langle |k_{v}^{(m')}\rangle\langle k_{v}^{(m')}|\right\rangle $
reads (cf. (\ref{eq:102 delta<O>}))
\begin{equation}
\Delta\left\langle O_{k^{(m')}}\right\rangle =\sum_{m=1}^{M}\Delta_{k^{(m)}}\left\langle O_{k^{(m')}}\right\rangle ,\label{eq:113 Delta<O_k(m')>}
\end{equation}
where (cf. (\ref{eq:103 delta_alpha<O>}))
\begin{align}
\Delta_{k^{(M)}}\left\langle O_{k^{(m')}}\right\rangle  & =\textrm{Tr}\left[O_{k^{(m')}}\left(U_{k^{(M)}}\rho^{(k^{(M)})}U_{k^{(M)}}^{\dagger}\right.\right.\nonumber \\
 & \quad\left.\left.-\rho^{(k^{(M)})}\right)\right]\label{eq:114 Delta_k(M)<O_k(m')>}
\end{align}
and 
\begin{align}
\Delta_{k^{(m)}}\left\langle O_{k^{(m')}}\right\rangle  & =\textrm{Tr}\left[O_{k^{(m')}}\left(V_{k^{(m)}}\rho^{(k^{(m)})}V_{k^{(m)}}^{\dagger}\right.\right.\nonumber \\
 & \quad\left.\left.-\rho^{(k^{(m)})}\right)\right]\label{eq:115 Delta_k(m)<O_k(m')>}
\end{align}
for $1\leq m\leq M-1$. 

Since the supports of $V_{k^{(m)}}\rho^{(k^{(m)})}V_{k^{(m)}}^{\dagger}$
and $\rho^{(k^{(m)})}$ are both contained into $\mathcal{H}_{k^{(m)}}$,
$\Delta_{k^{(m)}}\left\langle O_{k^{(m')}}\right\rangle $ can be
different from zero only if $\mathcal{H}_{k^{(m)}}$ contains states
with a non-null probability to measure $|k_{v}^{(m')}\rangle$. Moreover,
the same argument applies to the variation $\Delta_{k^{(M)}}\left\langle O_{k^{(m')}}\right\rangle $.
Following the definition (\ref{eq:111.3 Definition of subspace H_(k^(m))}),
we thus have that only $\mathcal{H}_{k^{(m')}}$ and $\mathcal{H}_{k^{(m'-1)}}$
can provide a non-null contribution to $\Delta\left\langle O_{k^{(m')}}\right\rangle $
in Eq. (\ref{eq:113 Delta<O_k(m')>}). Using the periodic boundary
condition $k^{(M+1)}=k^{(1)}$ and setting also $k^{(0)}\equiv k^{(M)}$,
we can write $\Delta\left\langle O_{k^{(m')}}\right\rangle $ in the
compact manner: 

\begin{align}
\Delta\left\langle O_{k^{(m')}}\right\rangle  & =\Delta_{k^{(m'-1)}}\left\langle O_{k^{(m')}}\right\rangle +\Delta_{k^{(m')}}\left\langle O_{k^{(m')}}\right\rangle \nonumber \\
 & =J_{|k_{v}^{(m'-1)}\rangle\rightarrow|k_{v}^{(m')}\rangle}+J_{|k_{v}^{(m'+1)}\rangle\rightarrow|k_{v}^{(m')}\rangle}\nonumber \\
 & =J_{|k_{v}^{(m'-1)}\rangle\rightarrow|k_{v}^{(m')}\rangle}-{\color{red}{\normalcolor J_{|k_{v}^{(m')}\rangle\rightarrow|k_{v}^{(m'+1)}\rangle}}},\label{eq:116 Delta<O_k(m')> with local currents}
\end{align}
for $1\leq m'\leq M$. In the second line of Eq. (\ref{eq:116 Delta<O_k(m')> with local currents}),
we have resorted to Remark 5 to express the average variations in
the first line as local currents. 

The currents appearing in Eq. (\ref{eq:116 Delta<O_k(m')> with local currents})
form the set $\{J_{|k_{v}^{(m+1)}\rangle\rightarrow|k_{v}^{(m)}\rangle}\}_{m=1}^{M-1}\cup\{J_{|k_{v}^{(1)}\rangle\rightarrow|k_{v}^{(M)}\rangle}\}$.
If all these currents have the same magnitude, Eqs. (\ref{eq:116 Delta<O_k(m')> with local currents})
implies that $\Delta\left\langle O_{k^{(m')}}\right\rangle =0$ for
$m'\in\{1,2,...,M\}$. For $k^{(m')}$ such that $m'\notin\{1,2,...,M\}$,
the populations $\bigl\langle O_{k^{(m')}}\bigr\rangle$ remain trivially
unchanged under the action of $U$ (Eq. (\ref{eq:112 U_k(M)+(+_mV_k(m))})),
since it acts on a subspace that does not contain any state with a
non-null probability to measure $O_{k^{(m')}}$. This leads us to
the following observation:

\textbf{Remark 7}. If $U_{k^{(M)}}$ generates a positive current
$J_{|k_{v}^{(1)}\rangle\rightarrow|k_{v}^{(M)}\rangle}>0$, the transformation
$\rho\rightarrow\rho'=U\rho U^{\dagger}$ (where $U$ satisfies (\ref{eq:112 U_k(M)+(+_mV_k(m))}))
is catalytic \textit{only if} $U$ generates a uniform loop $\mathbf{ch}_{|k_{v}^{(M)}\rangle\rightarrow|k_{v}^{(1)}\rangle}\cup\{J_{|k_{v}^{(1)}\rangle\rightarrow|k_{v}^{(M)}\rangle}\}$,
defined as a loop where the currents in the chain $\mathbf{ch}_{|k_{v}^{(M)}\rangle\rightarrow|k_{v}^{(1)}\rangle}$
satisfy $J_{|k_{v}^{(m+1)}\rangle\rightarrow|k_{v}^{(m)}\rangle}=J_{|k_{v}^{(1)}\rangle\rightarrow|k_{v}^{(M)}\rangle}$.
If $\rho'_{v}=\textrm{Tr}_{ch}\rho'$ is diagonal in the eigenbasis
of $\rho_{v}$, a uniform loop also guarantees that $\rho\rightarrow\rho'=U\rho U^{\dagger}$
is catalytic. 

\subsubsection{Conditions for the generation of a uniform loop }

In Lemma 1 we provide necessary and sufficient conditions for the
generation of a uniform loop. To that end, we introduce the following
definition, which is yet another generalization of passivity. As can
be readily seen from Eq. (\ref{eq:119 maximally active definition}),
it recovers usual passivity with respect to an arbitrary observable
if $\mathcal{H}_{\alpha'}=\mathcal{H}$. 

\textbf{Definition 8 (passivity within a Hilbert subspace)}. Let $\{U_{\alpha'}\}_{U_{\alpha'}}=\{U:U\textrm{ acts on }\mathcal{\mathcal{H}}_{\alpha'}\}$
be the set of all the unitaries that map a \textit{fixed} subspace
$\mathcal{H}_{\alpha'}\subseteq\mathcal{H}$ into itself (not to confuse
with $\{U_{\alpha}\}_{\alpha}$, which denotes a set where each $U_{\alpha}$
acts on a different $\mathcal{H}_{\alpha}$). We say that a state
$\rho$ is passive with respect to the pair $\{O,\mathcal{H}_{\alpha'}\}$,
where $O$ is a general observable (global or local), iff $\textrm{\textrm{Tr}}\left(OU_{\alpha'}\rho U_{\alpha'}^{\dagger}\right)\geq\textrm{\textrm{Tr}}\left(O\rho\right)$
for any $U_{\alpha'}\in\{U_{\alpha'}\}_{U_{\alpha'}}$. 

The previous definition can be conveniently recast by decomposing
$\rho$ as $\rho=\rho^{(\alpha')}+\rho^{(\sim\alpha')}$ (cf. Eq.
(\ref{eq:101 decomposition of rho})), where $\rho^{(\sim\alpha')}=\sum_{|\psi\rangle\notin E_{\alpha'}}|\psi\rangle\langle\psi|\rho\sum_{|\psi\rangle\notin E_{\alpha'}}|\psi\rangle\langle\psi|$.
Since $U_{\alpha'}|\psi\rangle=|\psi\rangle$ for $|\psi\rangle\notin E_{\alpha'}$,
$\textrm{Tr}\left(OU_{\alpha'}\rho U_{\alpha'}^{\dagger}\right)=\textrm{Tr}\left(OU_{\alpha'}\rho^{(\alpha')}U_{\alpha'}^{\dagger}\right)+\textrm{Tr}\left(O\rho^{(\sim\alpha')}\right)$.
Accordingly, $\rho$ is passive with respect to $\{O,\mathcal{H}_{\alpha'}\}$
iff, for any $U_{\alpha'}$, 
\begin{equation}
\textrm{Tr}\left[O\left(U_{\alpha'}\varrho^{(\alpha')}U_{\alpha'}^{\dagger}\right)\right]\geq\textrm{Tr}\left[O\left(\varrho^{(\alpha')}\right)\right],\label{eq:119 maximally active definition}
\end{equation}
where $\varrho^{(\alpha')}=\rho^{(\alpha')}/\textrm{Tr}\left(\rho^{(\alpha')}\right)$
is the normalized version of $\varrho^{(\alpha')}$. 

Note also that a necessary condition for Eq. (\ref{eq:119 maximally active definition})
is that $[O,\varrho^{(\alpha')}]=0$. Therefore, $O$ must have an
eigendecomposition $O=\sum_{\psi}o_{\psi}|\psi\rangle\langle\psi|$.
Given that the supports of $\varrho^{(\alpha')}$ and $U_{\alpha'}\varrho^{(\alpha')}U_{\alpha'}^{\dagger}$
are both contained in $E_{\alpha'}$, we can replace $O$ by $\sum_{\psi:|\psi\rangle\in E_{\alpha'}}o_{\psi}|\psi\rangle\langle\psi|$
in Eq. (\ref{eq:119 maximally active definition}) (as the average
values over $\sum_{\psi:|\psi\rangle\notin E_{\alpha'}}o_{\psi}|\psi\rangle\langle\psi|$
are null). Hence, all the relevant operators in (\ref{eq:119 maximally active definition})
have eigenstates that belong to $\mathcal{H}_{\alpha'}$, and we can
apply standard passivity considering $\mathcal{H}_{\alpha'}$ as the
effective Hilbert space. This leads us to conclude that $\rho$ is
passive with respect to $\{O,\mathcal{H}_{\alpha'}\}$ iff, for $|\psi\rangle\in E_{\alpha'}$,
\begin{align}
o_{\psi}\leq o_{\varphi} & \Leftrightarrow\langle\psi|\varrho^{(\alpha')}|\psi\rangle\geq\langle\varphi|\varrho^{(\alpha')}|\varphi\rangle\nonumber \\
 & \Leftrightarrow\langle\psi|\rho^{(\alpha')}|\psi\rangle\geq\langle\varphi|\rho^{(\alpha')}|\varphi\rangle.\label{eq:120 m.a. in terms of eigenvalues}
\end{align}

\textbf{Lemma 1}. \textit{Let $\rho=\rho_{ch}\otimes\rho_{v}$ be
a state, where $\rho_{ch}=\sum_{\psi=1}^{d_{c}d_{h}}p_{\psi}^{ch}|\psi_{ch}\rangle\langle\psi_{ch}|$
and $\rho_{v}=\sum_{k=1}^{d_{v}}|k_{v}\rangle\langle k_{v}|$. Then
the following statements are equivalent:}
\begin{enumerate}
\item \textit{The unitary $\oplus_{m=1}^{M-1}V_{k^{(m)}}$ (cf. Eq. }(\ref{eq:112 U_k(M)+(+_mV_k(m))})\textit{)
generates a chain $\mathbf{ch}_{|k_{v}^{(M)}\rangle\rightarrow|k_{v}^{(1)}\rangle}$.}
\item \textit{$\rho$ is }\textbf{\textit{not}}\textit{ passive with respect
to $\{{\color{red}{\normalcolor O_{k^{(m+1)}}}},\mathcal{H}_{k^{(m)}}\}$,
for all $m\in\{1,2,...,M-1\}$. }
\item \textit{$p_{n}^{ch}p_{k^{(m+1)}}^{v}>p_{N}^{ch}p_{k^{(m)}}^{v}$,
for all $m\in\{1,2,...,M-1\}$. }
\end{enumerate}
\textit{Proof}. We prove the implications $1\Rightarrow2\Rightarrow3\Rightarrow1$.
For the implication $1\Rightarrow2$, we note that since $J_{|k_{v}^{(m+1)}\rangle\rightarrow|k_{v}^{(m)}\rangle}$
is the population transfer due to a unitary $V_{k^{(m)}}:\mathcal{H}_{k^{(m)}}\rightarrow\mathcal{H}_{k^{(m)}}$,
the positivity of $J_{|k_{v}^{(m+1)}\rangle\rightarrow|k_{v}^{(m)}\rangle}$
is tantamount to say that it is possible to reduce $\left\langle O_{k^{(m+1)}}\right\rangle $
through some unitary of this kind. That is, that $\rho$ is not passive
with respect to $\{O_{k^{(m+1)}},\mathcal{H}_{k^{(m)}}\}$. 

To prove the implication $2\Rightarrow3$ we apply Eq. (\ref{eq:120 m.a. in terms of eigenvalues})
to show that the negation of 3 means that $\rho$ is passive with
respect to $\{O_{k^{(m+1)}},\mathcal{H}_{k^{(m)}}\}$. To that end,
we use the eigendecomposition 
\begin{align}
O_{k^{(m+1)}} & ={\color{red}{\normalcolor \mathbb{I}_{ch}}}\otimes|k_{v}^{(m+1)}\rangle\langle k_{v}^{(m+1)}|\nonumber \\
 & =\sum{}_{\psi=1}^{d_{c}d_{h}}|\psi_{ch}k_{v}^{(m+1)}\rangle\langle\psi_{ch}k_{v}^{(m+1)}|,\label{eq:121 eigendecomp of O_k(m+1)}
\end{align}
which gives $O_{k^{(m+1)}}$ in terms of the global eigenstates $\{|\psi\rangle\}_{\psi}$.
Similarly, 

\begin{align}
\rho^{(k^{(m)})} & =\sum_{\psi=n}^{N}p_{\psi}^{ch}p_{k^{(m+1)}}^{v}|\psi_{ch}k^{(m+1)}\rangle\langle\psi_{ch}k^{(m+1)}|\nonumber \\
 & \quad+\sum_{\psi=n}^{N}p_{\psi}^{ch}p_{k^{(m)}}^{v}|\psi_{ch}k^{(m)}\rangle\langle\psi_{ch}k^{(m)}|.\label{eq:122 eigendecomp of rho^k(m)}
\end{align}

If $p_{n}^{ch}p_{k^{(m+1)}}^{v}\leq p_{N}^{ch}p_{k^{(m)}}^{v}$ for
some value of $m$ (negation of Statement 3 in Lemma 1), the non-increasing
order of $\{p_{\psi}^{ch}\}_{\psi=1}^{d_{c}d_{h}}$ implies $p_{\varphi}^{ch}p_{k^{(m+1)}}^{v}\leq p_{\psi}^{ch}p_{k^{(m)}}^{v}$,
for all $\psi,\varphi\in\{n,n+1,...,N\}$. Accordingly, all the eigenvalues
in the first line of (\ref{eq:122 eigendecomp of rho^k(m)}) are lower
or equal than the eigenvalues in the second line. In contrast, $O_{k^{(m+1)}}$
has eigenvalue 1 for the eigenstates $|\psi_{ch}k^{(m+1)}\rangle$,
and zero otherwise (cf. Eq. (\ref{eq:121 eigendecomp of O_k(m+1)})).
This shows that $O_{k^{(m+1)}}$ and $\rho^{(k^{(m)})}$ satisfy Eq.
(\ref{eq:120 m.a. in terms of eigenvalues}), and therefore $\rho$
is passive with respect to $\{O_{k^{(m+1)}},\mathcal{H}_{k^{(m)}}\}$. 

Finally, the inequality $p_{n}^{ch}p_{k^{(m+1)}}^{v}>p_{N}^{ch}p_{k^{(m)}}^{v}$
implies that the partial swap $V_{k^{(m)}}=V_{|n_{ch}k_{v}^{(m+1)}\rangle\leftrightarrow|N_{ch}k_{v}^{(m)}\rangle}$
generates a current $J_{|k_{v}^{(m+1)}\rangle\rightarrow|k_{v}^{(m)}\rangle}>0$.
If this is true for all $m\in\{1,2,...,M-1\}$, the set $\{J_{|k_{v}^{(m+1)}\rangle\rightarrow|k_{v}^{(m)}\rangle}\}_{m=1}^{M-1}$
is a chain $\mathbf{ch}_{|k_{v}^{(M)}\rangle\rightarrow|k_{v}^{(1)}\rangle}$.
This proves the implication $3\Rightarrow1$.

\subsection{Elemental chain and loops }

We have seen that a loop $\mathbf{loop}_{|k^{(M)}\rangle\rightarrow|k^{(M)}\rangle}$
is composed by a current $J_{|k^{(1)}\rangle\rightarrow|k^{(M)}\rangle}$
and a chain $\mathbf{ch}_{|k^{(M)}\rangle\rightarrow|k^{(1)}\rangle}$.
Therefore, given the condition that $J_{|k^{(1)}\rangle\rightarrow|k^{(M)}\rangle}>0$
exists, the existence of $\mathbf{loop}_{|k^{(M)}\rangle\rightarrow|k^{(M)}\rangle}$
reduces to the existence of $\mathbf{ch}_{|k^{(M)}\rangle\rightarrow|k^{(1)}\rangle}$.
While different sets of currents can give rise to the same kind of
chain $\mathbf{ch}_{|k^{(M)}\rangle\rightarrow|k^{(1)}\rangle}$,
as illustrated in Fig. 12, the elemental chain is particularly important.
Specifically, the following lemma states that a general chain $\mathbf{ch}_{|k^{(M)}\rangle\rightarrow|k^{(1)}\rangle}$
exists only if the elemental chain exists. Keeping in mind our previous
comments, under the assumption that $J_{|k^{(1)}\rangle\rightarrow|k^{(M)}\rangle}>0$
we have also that the elemental chain is necessary for the existence
of $\mathbf{loop}_{|k^{(M)}\rangle\rightarrow|k^{(M)}\rangle}$. 

\textbf{Lemma 2}. \textit{Suppose that we want to generate a chain
$\mathbf{ch}_{|k^{(M)}\rangle\rightarrow|k^{(1)}\rangle}=\{J_{|k_{v}^{(m+1)}\rangle\rightarrow|k_{v}^{(m)}\rangle}\}_{m=1}^{M-1}$
via a direct sum of unitaries $\oplus_{m=1}^{M-1}V_{k^{(m)}}$. This
is possible only if the }\textbf{\textit{elemental}}\textit{ chain
$\{J_{|(k+1)_{v}\rangle\rightarrow|k_{v}\rangle}\}_{k=k^{(1)}}^{k^{(M)}-1}$
can be generated through $\oplus_{k=k^{(1)}}^{k^{(M)}-1}V_{k}$}. 

\textit{Proof}. Note first that $\oplus_{k=k^{(1)}}^{k^{(M)}-1}V_{k}$
is a unitary of the form $\oplus_{m=1}^{M-1}V_{k^{(m)}}$, for indices
that satisfy $k^{(m+1)}=k^{(m)}+1$. Therefore, Lemma 1 also sets
necessary conditions for the generation of the elemental chain $\{J_{|(k+1)_{v}\rangle\rightarrow|k_{v}\rangle}\}_{k=k^{(1)}}^{k^{(M)}-1}$. 

Suppose that $\{J_{|(k+1)_{v}\rangle\rightarrow|k_{v}\rangle}\}_{k=k^{(1)}}^{k^{(M)}-1}$
cannot be generated through $\oplus_{m=1}^{M-1}V_{k^{(m)}}$. That
is, 
\begin{align}
J_{|(k'+1)_{v}\rangle\rightarrow|k'_{v}\rangle} & \leq0\textrm{ for some}\nonumber \\
k' & \in\{k^{(1)},k^{(1)}+1,...,k^{(M)}-1\}\label{eq:123 negative of current for k'}
\end{align}
and, according to Lemma 1 (equivalence 1$\Leftrightarrow$3), 
\begin{equation}
p_{n}^{ch}p_{k'+1}^{v}\leq p_{N}^{ch}p_{k'}^{v}.\label{eq:124 non-existence of chain with eigenvalues}
\end{equation}

Using the conventional sorting $p_{k+1}^{v}\leq p_{k}^{v}$, the inequality
(\ref{eq:124 non-existence of chain with eigenvalues}) also yields
\begin{equation}
p_{n}^{ch}p_{k^{(m+1)}}^{v}\leq p_{N}^{ch}p_{k^{(m)}}^{v},\label{eq:125 non-existence of chain with eigenvalues}
\end{equation}
for $k^{(m+1)}\geq k'+1$ and $k^{(m)}\leq k^{\prime}$. Hence, for
an \textit{arbitrary} current $J_{|k_{v}^{(m+1)}\rangle\rightarrow|k_{v}^{(m)}\rangle}$
to be positive the following conditions are necessary: 

\begin{align}
k^{(m+1)} & \geq k'+1\textrm{ and }k^{(m)}\geq k^{\prime}+1,\nonumber \\
\textrm{or }k^{(m+1)} & \leq k^{\prime}\textrm{ and }k^{(m)}\leq k^{\prime}.\label{eq:126 conditions for J|k(m+1)>-->|k(m)> positive}
\end{align}

Now, the currents in a general chain $\{J_{|k_{v}^{(m+1)}\rangle\rightarrow|k_{v}^{(m)}\rangle}\}_{m=1}^{M-1}$
are labeled by an arbitrary set of indices $\{k^{(m)}\}_{m=1}^{M-1}$.
Since $k^{(1)}\leq k'$ (cf. (\ref{eq:123 negative of current for k'})),
the second line of Eq. (\ref{eq:126 conditions for J|k(m+1)>-->|k(m)> positive})
implies that for the general chain $J_{|k_{v}^{(2)}\rangle\rightarrow|k_{v}^{(1)}\rangle}>0$
only if $k^{(2)}\leq k^{\prime}$, which in turn implies that $J_{|k_{v}^{(3)}\rangle\rightarrow|k_{v}^{(2)}\rangle}>0$
only if $k^{(3)}\leq k^{\prime}$ and so on. In other words, the recursive
application of (\ref{eq:126 conditions for J|k(m+1)>-->|k(m)> positive})
implies that, for any $m$, $J_{|k_{v}^{(m+1)}\rangle\rightarrow|k_{v}^{(m)}\rangle}>0$
only if $k^{(m+1)}\leq k'$. For $m=M-1$, we thus have that $J_{|k_{v}^{(M)}\rangle\rightarrow|k_{v}^{(M-1)}\rangle}>0$
only if $k^{(M)}\leq k'$. However, $k'\leq k^{(M)}-1$ according
to (\ref{eq:123 negative of current for k'}). This leads to the impossibility
of a positive current $J_{|k_{v}^{(M)}\rangle\rightarrow|k_{v}^{(M-1)}\rangle}$,
which is necessary for the existence of $\{J_{|k_{v}^{(m+1)}\rangle\rightarrow|k_{v}^{(m)}\rangle}\}_{m=1}^{M-1}$.

\section{THEOREM 1 }

\textbf{Theorem 1 (catalytic transformations)}. \textit{If $\rho=\rho_{c}\otimes\rho_{h}\otimes\rho_{v}$
is a quantum state that satisfies (}(\ref{eq:5 cooling with the catalyst})
\textit{in the main text) }

\textit{
\begin{equation}
\frac{p_{i'}^{c}}{p_{i'+1}^{c}}<\frac{p_{1}^{h}p_{1}^{v}}{p_{d_{h}}^{h}p_{d_{v}}^{v}},\label{eq:91.1 cooling with catalyst SM}
\end{equation}
the following statemens are equivalent: }
\begin{enumerate}
\item \textit{There exists a catalytic transformation $\rho\rightarrow\rho'=U\rho U^{\dagger}$,
where $U=U_{i^{\prime},l,l'}\oplus\left(\oplus_{\alpha}V_{\alpha}^{(2)}\right)$,
${\color{red}{\normalcolor U_{i^{\prime},l,l'}=U_{|i'_{c}d_{h}(l'+1)_{v}\rangle\leftrightarrow|(i'+1)_{c}d_{h}l_{v}\rangle}}}$
(}(\ref{eq:23 Ui',l,l'})\textit{ in the main text), and $\oplus_{\alpha}V_{\alpha}^{(2)}:\mathcal{H}_{n,N}\rightarrow\mathcal{H}_{n,N}$
is an arbitrary direct sum of two-level unitaries acting on }$\mathcal{H}_{n,N}=\textrm{span}\{|\psi_{ch}\rangle\}_{\psi=n}^{N}\otimes\mathcal{H}_{v}$\textit{.}
\item \textit{For $V_{|n_{ch}(k+1)_{v}\rangle\leftrightarrow|N_{ch}k_{v}\rangle}\neq\mathbb{I}$
(non-trivial partial swap), all the partial swaps $\{V_{|n_{ch}(k+1)_{v}\rangle\leftrightarrow|N_{ch}k_{v}\rangle}\}_{k=l}^{l'}$
generate positive currents $J_{|(k+1)_{v}\rangle\rightarrow|k_{v}\rangle}$
(i.e. $J_{|(k+1)_{v}\rangle\rightarrow|k_{v}\rangle}>0$). }
\end{enumerate}

\subsection{Proof}

\textbf{Sufficiency (implication $2\Rightarrow1$)}: By resorting
to Definition 7, we see that the set of positive currents $\{J_{|(k+1)_{v}\rangle\rightarrow|k_{v}\rangle}\}_{k=l}^{l'}$
constitutes an elemental chain $\mathbf{ch}_{|(l'+1)_{v}\rangle\rightarrow|l_{v}\rangle}$,
corresponding to $k^{(1)}=l$ and $k^{(M)}=l'+1$. We will show that
this chain can be used to generate a \textit{uniform} loop $\mathbf{loop}_{|l_{v}\rangle\rightarrow|l_{v}\rangle}=\mathbf{ch}_{|(l'+1)_{v}\rangle\rightarrow|l_{v}\rangle}\cup\{J_{|l_{v}\rangle\rightarrow|(l'+1)_{v}\rangle}\}$,
where $J_{|l_{v}\rangle\rightarrow|(l'+1)_{v}\rangle}>0$ is a local
current due to $U_{i^{\prime},l,l^{\prime}}$. 

The magnitude of each current depends on the swap intensities of $U_{i^{\prime},l,l^{\prime}}$
and $V_{|n_{ch}(k+1)_{v}\rangle\leftrightarrow|N_{ch}k_{v}\rangle}$.
Let $\mathcal{J}_{|l_{v}\rangle\rightarrow|(l'+1)_{v}\rangle}$ and
$\mathcal{J}_{|(k+1)_{v}\rangle\rightarrow|k_{v}\rangle}$ denote
the respective maximum currents, obtained by setting $r=1$ (cf. Eqs.
(\ref{eq:14 current}) and (\ref{eq:18 maxJ(i-->j)})). Moreover,
let us define the current 
\begin{align}
{\normalcolor \mathcal{J}_{\textrm{min}}} & {\normalcolor \equiv\textrm{min}\left\{ \mathcal{J}_{|l_{v}\rangle\rightarrow|(l'+1)_{v}\rangle},\right.}\nonumber \\
{\normalcolor } & {\normalcolor \quad\quad\quad\left.\textrm{min}_{l\leq k\leq l'}\mathcal{J}_{|(k+1)_{v}\rangle\rightarrow|k_{v}\rangle}\right\} .}\label{eq:127 J_min for loop}
\end{align}
A uniform loop $\mathbf{loop}_{|l_{v}\rangle\rightarrow|l_{v}\rangle}$
where all the currents satisfy $J_{|l_{v}\rangle\rightarrow|(l'+1)_{v}\rangle}=J_{|(k+1)_{v}\rangle\rightarrow|k_{v}\rangle}=\mathcal{J}_{\textrm{min}}$
for $l\leq k\leq l'$ can be obtained as follows. If a total swap
in $\{U_{i^{\prime},l,l^{\prime}},V_{|n_{ch}(k+1)_{v}\rangle\leftrightarrow|N_{ch}k_{v}\rangle}\}_{k=l}^{l'}$
generates a current of magnitude $\mathcal{J}_{\textrm{min}}$, then
$r=1$ is chosen. If it generates a current larger than $\mathcal{J}_{\textrm{min}}$,
then $r<1$ is tuned in such a way that the current of the corresponding
partial swap matches $\mathcal{J}_{\textrm{min}}$. With these swap
intensities, the unitary 
\begin{equation}
U=U_{i^{\prime},l,l^{\prime}}\oplus\left(\oplus_{k=l}^{l'}V_{|n_{ch}(k+1)_{v}\rangle\leftrightarrow|N_{ch}k_{v}\rangle}\right)\label{eq:128 explicit catalytic unitary}
\end{equation}
produces a uniform loop, and Remark 7 (see also Proposition 2) implies
that the transformation $\rho\rightarrow U\rho U^{\dagger}$ is catalytic. 

\textbf{Necessity (implication $1\Rightarrow2$)}: If $V_{|n_{ch}(k'+1)_{v}\rangle\leftrightarrow|N_{ch}k'_{v}\rangle}$
produces a negative current $J_{|(k'+1)_{v}\rangle\rightarrow|k'_{v}\rangle}\leq0$
for some $k'\in\{l,l+1,...,l',l'+1\}$, then $p_{n}^{ch}p_{k'+1}^{v}\leq p_{N}^{ch}p_{k'}^{v}$.
Thus, by applying the equivalence $1\Leftrightarrow3$ in Lemma 1
with the indices $\{k^{(1)},k^{(2)},...,k^{(M)}\}=\{l,l+1,...,l'+1\}$,
we have that the \textit{elemental} chain $\mathbf{ch}_{|(l'+1)_{v}\rangle\rightarrow|l_{v}\rangle}$
cannot be generated via $\oplus_{k=l}^{l'}V_{k}$. From Lemma 2, this
implies the impossibility to produce a \textit{general} chain $\mathbf{ch}_{|(l'+1)_{v}\rangle\rightarrow|l_{v}\rangle}$
through $\oplus_{m=1}^{M-1}V_{k^{(m)}}$. 

Our next step is to prove that any direct sum $\oplus_{\alpha\textrm{ free}}V_{\alpha}^{(2)}:{\color{red}{\normalcolor \mathcal{H}_{n,N}\rightarrow\mathcal{H}_{n,N}}}$
can be written as $\oplus_{m=1}^{M-1}V_{k^{(m)}}$, and therefore
the impossibility to generate the chain $\mathbf{ch}_{|(l'+1)_{v}\rangle\rightarrow|l_{v}\rangle}$
via $\oplus_{m=1}^{M-1}V_{k^{(m)}}$ also holds for $\oplus_{\alpha\textrm{ free}}V_{\alpha}^{(2)}:{\color{red}{\normalcolor \mathcal{H}_{n,N}\rightarrow\mathcal{H}_{n,N}}}$.
The key idea is that any subspace $\textrm{span}\{|\psi_{ch}\rangle\}_{\psi=n}^{N}\otimes\textrm{span}\{|k_{v}^{\prime}\rangle,|k_{v}^{\prime\prime}\rangle\}$
can be written as $\mathcal{H}_{k^{(m)}}=\textrm{span}\{|\psi_{ch}\rangle\}_{\psi=n}^{N}\otimes\textrm{span}\{|k_{v}^{(m)}\rangle,|k_{v}^{(m+1)}\rangle\}$,
if $k^{(m)}=k^{\prime}$ and $k^{(m+1)}=k^{\prime\prime}$. Hence,
a two-level unitary $V_{\alpha}^{(2)}:{\color{red}{\normalcolor \mathcal{H}_{n,N}\rightarrow\mathcal{H}_{n,N}}}$
is a particular case of the more general unitary $V_{k^{(m)}}:\mathcal{H}_{k^{(m)}}\rightarrow\mathcal{H}_{k^{(m)}}$.
This suffices to prove that any unitary $\oplus_{\alpha\textrm{ free}}V_{\alpha}^{(2)}$
acting on $\mathcal{H}_{n,N}$ can be expressed as $\oplus_{m=1}^{M-1}V_{k^{(m)}}$. 

\section{PROOF OF COROLLARY 2}

Corollary 2 is a consequence of Theorem 1, and provides sufficient
conditions for a catalytic and non-unital transformation. To prove
it we will resort to the concept of majorization between quantum states. 

\textbf{Definition 9 (majorization \citep{60Nielsen-majorization,59Majorization})}.\textit{
}Let $\varrho$ and $\sigma$ denote two generic quantum states defined
on some Hilbert space of dimension $d$, with respective eigenvalues
$\{r_{i}\}_{i=1}^{d}$ and $\{s_{i}\}_{i=1}^{d}$. Moreover, let $\{r_{i}^{\downarrow}\}_{i}$
and $\{s_{i}^{\downarrow}\}_{i}$ be the same eigenvalues arranged
in non-increasing order, i.e. $r_{i+1}^{\downarrow}\leq r_{i}^{\downarrow}$
and $s_{i+1}^{\downarrow}\leq s_{i}^{\downarrow}$. It is said that
``$\varrho$ majorizes $\sigma$'', formally written as $\varrho\succ\sigma$,
iff 
\begin{equation}
\sum_{i=1}^{j}r_{i}^{\downarrow}\geq\sum_{i=1}^{j}s_{i}^{\downarrow}\label{eq:9 majorization}
\end{equation}
for all $1\leq j\leq d$. 

From a physical viewpoint, majorization is useful to compare the degree
of purity between two quantum states. For example, Eq. (\ref{eq:9 majorization})
implies that a pure state majorizes any other state, while a fully
mixed state is majorized by any other state. 

\subsection{Non-unital transformations (cf. Definition 3 in the main text) and
majorization }

An important result of information theory \citep{60Nielsen-majorization,61Unital-maps}
states that $\varrho\succ\sigma$ iff there exist probabilities $\{\lambda_{i}\}_{i}$
and unitary matrices $\{U_{i}\}_{i}$, such that $\sigma=\sum_{i}\lambda_{i}U_{i}\varrho U_{i}^{\dagger}$.
Therefore, our definition of non-unital transformation is equivalent
to the condition that $\rho_{c}$ \textit{does not} majorize $\rho'_{c}$.
Taking into account the symbol for majorization, a transformation
of this kind can be described using the notation $\rho_{c}\nsucc\rho'_{c}$,
or $\rho'_{c}\nprec\rho{}_{c}$. 

The violation of majorization implies that at least one of the inequalities
(\ref{eq:9 majorization}) is not satisfied. In terms of the eigenvalues
of $\rho_{c}$ and $\rho'_{c}$, $\rho_{c}\nsucc\rho'_{c}$ is equivalent
to 
\begin{equation}
\sum_{i=1}^{i'}p_{i}^{c}<\sum_{l=1}^{i'}q_{i}^{c}\textrm{ for some}1\leq i'\leq d_{c}-1,\label{eq:9.1 violation of majorization}
\end{equation}
where we recall that $\{q_{i}^{c}\}_{i}$ denotes the eigenvalues
of $\rho'_{c}$. In the proof of Corollary 2 we will apply Lemma 3,
which allows us to characterize a class of non-unital transformations
using the final populations of the cold object. Moreover, this lemma
is proven using Eq. (\ref{eq:9.1 violation of majorization}). 

\textbf{Lemma 3}. \textit{Let $\rho_{c}\rightarrow\rho'_{c}$ be a
transformation such that $\sum_{i=1}^{i'}p_{i}^{\prime c}>\sum_{i=1}^{i'}p_{i}^{c}$
for some $i^{\prime}\in\{1,...,d_{c}-1\}$. Then $\rho_{c}\rightarrow\rho'_{c}$
is a non-unital transformation. }

\textit{Proof}. From the Schur-Horn theorem \citep{Horn}, the vector
of eigenvalues of $\rho'_{c}$ majorizes the vector formed by the
populations $\{p_{i}^{\prime c}\}_{i}$. That is, $\sum_{i=1}^{j}q_{i}^{c}\geq\sum_{i=1}^{j}\left(p_{i}^{\prime c}\right)^{\downarrow}$
for all $j\in\{1,2,...,d_{c}\}$, where $\left(p_{i}^{\prime c}\right)^{\downarrow}$
are the final populations in non-increasing order. Since $\sum_{i=1}^{j}\left(p_{i}^{\prime c}\right)^{\downarrow}\geq\sum_{i=1}^{j}p_{i}^{\prime c}$
for $1\leq j\leq d_{c}$, it follows that $\sum_{i=1}^{j}q_{i}^{c}\geq\sum_{i=1}^{j}p_{i}^{\prime c}$
for $1\leq j\leq d_{c}$. Therefore, if $\sum_{i=1}^{i'}p_{i}^{\prime c}>\sum_{i=1}^{i'}p_{i}^{c}$
for some $i'$, it follows that $\sum_{i=1}^{i'}q_{i}^{c}>\sum_{i=1}^{i'}p_{i}^{c}$,
as expressed in Eq. (\ref{eq:9.1 violation of majorization}). 

\subsection{Proof of Corollary 2 }

If (\ref{eq:91.1 cooling with catalyst SM}) holds, the unitary $U_{i',l,l'}$
increases the population $\bigl\langle|i_{c}^{\prime}\rangle\langle i_{c}^{\prime}|\bigr\rangle$
and reduces the population $\bigl\langle|(i^{\prime}+1)_{c}\rangle\langle(i^{\prime}+1)_{c}|\bigr\rangle$
by the amount

\begin{equation}
\Delta_{i',l,l'}\bigl\langle|i_{c}^{\prime}\rangle\langle i_{c}^{\prime}|\bigr\rangle=J_{|(i^{\prime}+1)_{c}\rangle\rightarrow|i_{c}^{\prime}\rangle}>0.\label{eq:28 Delta_cool<|i'c><i'c|> > 0 for non-unital transf}
\end{equation}
Furthermore, $\Delta_{i',l,l'}\bigl\langle|i_{c}\rangle\langle i_{c}|\bigr\rangle=0$
for $i\leq i^{\prime}-1$ and $i\geq i^{\prime}+1$. Accordingly,
\begin{equation}
\sum_{i=1}^{i'}p_{i}^{\prime c}-\sum_{i=1}^{i'}p_{i}^{c}=J_{|(i^{\prime}+1)_{c}\rangle\rightarrow|i_{c}^{\prime}\rangle}.\label{eq:28.1 increment of partial sum}
\end{equation}
This implies that the transformation $\rho\rightarrow\sigma=U_{i',l,l'}\rho U_{i',l,l'}^{\dagger}$
is non-unital, following Lemma 3. 

\textbf{Remark 8}. For a sufficiently large ratio $\frac{p_{l}^{v}}{p_{l'+1}^{v}}$,
and $\frac{p_{i'}^{c}}{p_{i'+1}^{c}}$ finite, the inequality $\frac{p_{i'}^{c}}{p_{i'+1}^{c}}<\frac{p_{1}^{h}p_{l}^{v}}{p_{d_{h}}^{h}p_{l'+1}^{v}}$
((\ref{eq:91.1 cooling with catalyst SM})) holds even if $p_{1}^{h}=p_{d_{h}}^{h}$.
That is, $\frac{p_{i'}^{c}}{p_{i'+1}^{c}}<\frac{p_{l}^{v}}{p_{l'+1}^{v}}$.
This implies that $U_{|i'_{c}(l'+1)_{v}\rangle\leftrightarrow|(i'+1)_{c}l_{v}\rangle}:\mathcal{H}_{c}\otimes\mathcal{H}_{v}\rightarrow\mathcal{H}_{c}\otimes\mathcal{H}_{v}$
also generates a current $J_{|(i^{\prime}+1)_{c}\rangle\rightarrow|i_{c}^{\prime}\rangle}=r(p_{i'+1}^{c}p_{l}^{v}-p_{i'}^{c}p_{l'+1}^{v})>0$.
In this case a non-unital transformation $\rho\rightarrow U_{|i'_{c}(l'+1)_{v}\rangle\leftrightarrow|(i'+1)_{c}l_{v}\rangle}\rho U_{|i'_{c}(l'+1)_{v}\rangle\leftrightarrow|(i'+1)_{c}l_{v}\rangle}^{\dagger}$
can be generated \textit{without using the hot object}, according
to Eq. (\ref{eq:28.1 increment of partial sum}). 

If $|n_{ch}\rangle=|1_{c}1_{h}\rangle$ and $|N_{ch}\rangle=|i'_{c}d_{h}\rangle$
(hypothesis of Corollary 2), we show now that the restoring unitary
$V_{\textrm{res},n,N}=\oplus_{k=l}^{l'}V_{|n_{ch}(k+1)_{v}\rangle\leftrightarrow|N_{ch}k_{v}\rangle}$
preserves the sum of populations $\sum_{i=1}^{i'}p_{i}^{c}$. Given
that all the partial swaps $V_{|1_{c}1_{h}(k+1)_{v}\rangle\leftrightarrow|i'_{c}d_{h}k_{v}\rangle}$
act on $\textrm{span}\{|1_{c}\rangle,|i'_{c}\rangle\}\otimes\mathcal{H}_{h}\otimes\mathcal{H}_{v}$,
$V_{\textrm{res},n,N}$ acts on the same subspace. Thus, $V_{\textrm{res},n,N}$
performs a population transfer $\Delta_{\textrm{res},n,N}\bigl\langle|i'_{c}\rangle\langle i'_{c}|\bigr\rangle=-\Delta_{\textrm{res},n,N}\bigl\langle|1_{c}\rangle\langle1_{c}|\bigr\rangle$.
Hence, 
\begin{align}
\Delta_{\textrm{res},n,N}\left(\sum_{i=1}^{i'}\bigl\langle|i_{c}\rangle\langle i_{c}|\bigr\rangle\right) & =\Delta_{\textrm{res},n,N}\bigl\langle|i'_{c}\rangle\langle i'_{c}|\bigr\rangle\nonumber \\
 & \quad+\Delta_{\textrm{res},n,N}\bigl\langle|1_{c}\rangle\langle1_{c}|=0.\label{eq:28.2 Delta_res,n,N=00003D0}
\end{align}
Using Eqs. (\ref{eq:102 delta<O>}) and (\ref{eq:103 delta_alpha<O>}),
with $O=\sum_{i=1}^{i'}|i_{c}\rangle\langle i_{c}|$, we have that
the total change in the sum $\sum_{i=1}^{i'}p_{i}^{c}$ reads 
\begin{align}
\Delta\left(\sum_{i=1}^{i'}\bigl\langle|i_{c}\rangle\langle i_{c}|\bigr\rangle\right) & =\Delta_{i',l,l'}\left(\sum_{i=1}^{i'}\bigl\langle|i_{c}\rangle\langle i_{c}|\bigr\rangle\right)\nonumber \\
 & \quad+\Delta_{\textrm{res},n,N}\left(\sum_{i=1}^{i'}\bigl\langle|i_{c}\rangle\langle i_{c}|\bigr\rangle\right)\nonumber \\
 & =\Delta_{i',l,l'}\bigl\langle|i_{c}^{\prime}\rangle\langle i_{c}^{\prime}|\bigr\rangle,\label{eq:28.4 Delta(partial sum)}
\end{align}
which is positive according to Eq. (\ref{eq:28 Delta_cool<|i'c><i'c|> > 0 for non-unital transf}).
Hence, the transformation $\rho\rightarrow\rho'=V_{\textrm{res},n,N}U_{i',l,l'}\rho U_{i',l,l'}^{\dagger}V_{\textrm{res},n,N}^{\dagger}$
is non-unital. 

On the other hand, if $|n_{ch}\rangle=|(i'+1)_{c}1_{h}\rangle$ and
$|N_{ch}\rangle=|d_{c}d_{h}\rangle$ (hypothesis of Corollary 2),
the restoring unitary $V_{\textrm{res},n,N}=\oplus_{k=l}^{l'}V_{|n_{ch}(k+1)_{v}\rangle\leftrightarrow|N_{ch}k_{v}\rangle}$
preserves the sum $\sum_{i=i'+1}^{d_{c}}p_{i}^{c}$. The argument
is completely analogous to the previous one. In this case, $V_{\textrm{res},n,N}$
acts on $\textrm{span}\{|(i'+1)_{c}\rangle,|d_{c}\rangle\}\otimes\mathcal{H}_{h}\otimes\mathcal{H}_{v}$,
and consequently it generates a population transfer $\Delta\bigl\langle|d_{c}\rangle\langle d_{c}|\bigr\rangle=-\Delta\bigl\langle|(i'+1)_{c}\rangle\langle(i'+1)_{c}|\bigr\rangle$.
This implies that $\Delta_{\textrm{res},n,N}\left(\sum_{i=i'+1}^{d_{c}}\bigl\langle|i_{c}\rangle\langle i_{c}|\bigr\rangle\right)=0$.
Therefore, the total change in $\sum_{i=1}^{i'}p_{i}^{c}$ is also
given by (\ref{eq:28.1 increment of partial sum}).

\section{PROOF OF THEOREM 2}

\subsection{Proof of Statement 1}

\subsubsection{Global unitary for the implementation of the transformation }

Consider a full-rank catalyst state $\rho_{v}$, whose eigenvalues
satisfy the condition 
\begin{equation}
\frac{p_{k}^{v}}{p_{k+1}^{v}}=\mu\textrm{ for }1\leq k\leq d_{v}-1,\label{eq:29.6 catalyst eigenvalues for Theorem 2-1}
\end{equation}
where $\mu$ is a parameter larger than 1 but noot to large, as discussed
later. From this equation it readily follows that 
\begin{equation}
\prod_{k=1}^{K-1}\left(\frac{p_{k}^{v}}{p_{k+1}^{v}}\right)=\frac{p_{1}^{v}}{p_{K}^{v}}=\mu^{K-1}.\label{eq:29.7 Catalyst eigenvalues in tems of mu}
\end{equation}

Since $\mu>1$, for large enough $d_{v}$ we have that 
\begin{equation}
\mu^{d_{v}-1}=\frac{p_{1}^{v}}{p_{d_{v}}^{v}}>\frac{p_{d_{h}}^{h}}{p_{1}^{h}}\frac{p_{i'}^{c}}{p_{i'+1}^{c}}.\label{eq:29.8 violation of passivity in terms of mu}
\end{equation}
A simple rearrangement of this inequality allows us to rewrite it
as Eq. (\ref{eq:91.1 cooling with catalyst SM}), with $l=1$ and
$l'=d_{v}-1$. This ensures that the partial swap $U_{i',1,d_{v}-1}$
(cf. Eq. (\ref{eq:23 Ui',l,l'})) is a cooling unitary. 

On the other hand, the condition $p_{1}^{h}>p_{d_{h}}^{h}$ also allows
us to choose $\mu$ sufficiently small to satisfy:
\begin{equation}
\frac{p_{1}^{h}}{p_{d_{h}}^{h}}>\frac{p_{k}^{v}}{p_{k+1}^{v}}=\mu.\label{eq:29. 9 positive J|k+1>-->|k> in terms of mu}
\end{equation}
This implies that all the partial swaps $\{V_{|1_{c}1_{h}(k+1)_{v}\rangle\leftrightarrow|1_{c}d_{h}k_{v}\rangle}\}_{k=1}^{d_{v}-1}$
generate positive currents $J_{|(k+1)_{v}\rangle\rightarrow|k_{v}\rangle}$.
Writing $V_{|1_{c}1_{h}(k+1)_{v}\rangle\leftrightarrow|1_{c}d_{h}k_{v}\rangle}$
as $V_{|n_{ch}(k+1)_{v}\rangle\leftrightarrow|N_{ch}k_{v}\rangle}$,
with $|n_{ch}\rangle=|1_{c}1_{h}\rangle$ and $|N_{ch}\rangle=|1_{c}d_{h}\rangle$,
Corollary 1 entails that 
\begin{equation}
U=U_{i',1,d_{v}-1}\oplus\left(\oplus_{k=1}^{d_{v}-1}V_{|1_{c}1_{h}(k+1)_{v}\rangle\leftrightarrow|1_{c}d_{h}k_{v}\rangle}\right)\label{eq:30 Explicit unitary}
\end{equation}
generates a catalytic and cooling transformation, if the swap intensities
are ajusted according to the prescription (\ref{eq:127 J_min for loop}). 

\subsubsection{Catalyst dimension and corresponding state }

Now we provide explicit expressions for $\mu$ and a catalyst dimension
$d_{v}^{\ast}$ that enables the transformation. To satisfy Eq. (\ref{eq:29.8 violation of passivity in terms of mu}),
we can choose $\mu=\frac{p_{1}^{h}}{p_{d_{h}}^{h}}-\epsilon$, with
$\epsilon$ sufficiently small to have $\mu>1$. Importantly, this
choice also guarantess that Eq. (\ref{eq:29. 9 positive J|k+1>-->|k> in terms of mu})
is fulfilled. With this expression for $\mu$ now we can derive a
catalyst dimension that is consistent with the aforementioned equations. 

By taking the natural logarithm of Eq. (\ref{eq:29.8 violation of passivity in terms of mu}),
we find that 

\begin{equation}
d_{v}>1+\frac{1}{\textrm{ln}(\mu)}\textrm{ln}\left(\frac{p_{d_{h}}^{h}}{p_{1}^{h}}\frac{p_{i'}^{c}}{p_{i'+1}^{c}}\right).\label{eq:30.1 Bound on catalyst dimension}
\end{equation}
This inequality is always satisfied by a dimension 
\begin{equation}
d_{v}^{\ast}\equiv\left\lceil 1+\frac{1}{\textrm{ln}(\mu)}\textrm{ln}\left(\frac{p_{d_{h}}^{h}}{p_{1}^{h}}\frac{p_{i'}^{c}}{p_{i'+1}^{c}}\right)\right\rceil +1,\label{eq:30.2 dv*}
\end{equation}
where $\bigl\lceil\cdot\bigr\rceil$ denotes the ceiling function.
It is worth noting that the addition of 1 is not necessary if the
ceiling function yields a number larger or equal than 2. However,
for $\left\lceil 1+\frac{1}{\textrm{ln}(\mu)}\textrm{ln}\left(\frac{p_{d_{h}}^{h}}{p_{1}^{h}}\frac{p_{i'}^{c}}{p_{i'+1}^{c}}\right)\right\rceil =1$
the addtion of 1 yields the smallest possible dimension $d_{v}^{\ast}=2$.
We also remark that $\frac{p_{d_{h}}^{h}}{p_{1}^{h}}\frac{p_{i'}^{c}}{p_{i'+1}^{c}}>1$,
due to the passivity of $\rho_{c}\otimes\rho_{h}$ (cf. Eq. (\ref{eq:passivity})). 

Finally, Eq. (\ref{eq:29.7 Catalyst eigenvalues in tems of mu}) provides
the eigenvalues of $\rho_{v}$. That is, $p_{K}^{v}=\frac{p_{1}^{v}}{\mu^{K-1}}$,
with $p_{1}^{v}$ derived from normalization $\sum_{K=1}^{d_{v}^{\ast}}p_{K}^{v}=1$. 

\subsection{Proof of Statement 2}

\subsubsection{Global unitary for the implementation of the transformation }

The core of the proof is very similar to that of the proof for Statement
1. In particular, we also assume catalyst eigenvalues that satisfy
Eqs. (\ref{eq:29.6 catalyst eigenvalues for Theorem 2-1}) and (\ref{eq:29.7 Catalyst eigenvalues in tems of mu}).
For $\mu>1$ and $d_{v}$ sufficiently large we have that 
\begin{equation}
\mu^{d_{v}-1}=\frac{p_{1}^{v}}{p_{d_{v}}^{v}}>\frac{p_{i'}^{c}}{p_{i'+1}^{c}}.\label{eq:31 violation of passivity in terms of mu, for non-unit trans}
\end{equation}
By setting $l=1$ and $l'=d_{v}-1$, it follows from Remark 8 that
$U_{|(i'+1)_{c}1_{v}\rangle\leftrightarrow|i'_{c}d_{v}\rangle}$ generates
a non-unital transformation. 

Our goal now is to show that a restoring unitary can be constructed
if $d_{c}\geq3$ and $p_{1}^{c}>p_{i'}^{c}$ or $p_{i'+1}^{c}>p_{d_{c}}^{c}$.
For $\mu$ sufficiently small (but larger than 1), 
\begin{equation}
\frac{p_{1}^{c}}{p_{i'}^{c}}>\frac{p_{k}^{v}}{p_{k+1}^{v}}=\mu\textrm{ if }p_{1}^{c}>p_{i'}^{c},\label{eq:31.2 positive J|k+1>-->|k> in terms of mu 1}
\end{equation}
or 
\begin{equation}
\frac{p_{i'+1}^{c}}{p_{d_{c}}^{c}}>\frac{p_{k}^{v}}{p_{k+1}^{v}}=\mu\textrm{ if }p_{i'+1}^{c}>p_{d_{c}}^{c}.\label{eq:31.3 positive J|k+1>-->|k> in terms of mu 2}
\end{equation}

Hence, 
\begin{equation}
V_{1}\equiv\oplus_{k=1}^{d_{v}-1}V_{|1_{c}(k+1)_{v}\rangle\leftrightarrow|i'_{c}k_{v}\rangle}\label{eq:31.4 V1}
\end{equation}
generates a chain $\mathbf{ch}_{|d_{v}\rangle\rightarrow|1_{v}\rangle}=\{J_{|(k+1)_{v}\rangle\rightarrow|k_{v}\rangle}\}_{k}$,
with $J_{|(k+1)_{v}\rangle\rightarrow|k_{v}\rangle}\propto p_{1}^{c}p_{k+1}^{v}-p_{i'}^{c}p_{k}^{v}>0$
if Eq. (\ref{eq:31.2 positive J|k+1>-->|k> in terms of mu 1}) holds,
and 

\begin{equation}
V_{2}\equiv\oplus_{k=1}^{d_{v}-1}V_{|(i'+1)_{c}(k+1)_{v}\rangle\leftrightarrow|d_{c}k_{v}\rangle}\label{eq:31.5 V2}
\end{equation}
generates a chain $\mathbf{ch}_{|d_{v}\rangle\rightarrow|1_{v}\rangle}=\{J_{|(k+1)_{v}\rangle\rightarrow|k_{v}\rangle}\}_{k}$,
with $J_{|(k+1)_{v}\rangle\rightarrow|k_{v}\rangle}\propto p_{i'+1}^{c}p_{k+1}^{v}-p_{d_{c}}^{c}p_{k}^{v}>0$
if Eq. (\ref{eq:31.3 positive J|k+1>-->|k> in terms of mu 2}) holds.
This chain forms a loop with the current $J_{|1_{v}\rangle\rightarrow|d_{v}\rangle}$
due to $U_{|(i'+1)_{c}1_{v}\rangle\leftrightarrow|i'_{c}d_{v}\rangle}$,
and thus $V_{1}$ ($V_{2}$) is a restoring unitary for the transformation
$\rho\rightarrow U_{|(i'+1)_{c}1_{v}\rangle\leftrightarrow|i'_{c}d_{v}\rangle}\rho U_{|(i'+1)_{c}1_{v}\rangle\leftrightarrow|i'_{c}d_{v}\rangle}^{\dagger}$,
if Eq. (\ref{eq:31.2 positive J|k+1>-->|k> in terms of mu 1}) ((\ref{eq:31.3 positive J|k+1>-->|k> in terms of mu 2}))
is satisfied. 

On the other hand, the partial swaps in $V_{1}$ transfer population
$\Delta_{k}\bigl\langle|i'_{c}\rangle\langle i'_{c}|\bigr\rangle=-\Delta_{k}\bigl\langle|1{}_{c}\rangle\langle1{}_{c}|\bigr\rangle$
between the eigenstates $|1_{c}\rangle$ and $|i'_{c}\rangle$, and
the partial swaps in $V_{2}$ transfer population $\Delta_{k}\bigl\langle|(i'+1)_{c}\rangle\langle(i'+1)_{c}|\bigr\rangle=-\Delta_{k}\bigl\langle|d_{c}\rangle\langle d_{c}|\bigr\rangle$
between the eigenstates $|(i'+1)_{c}\rangle$ and $|d_{c}\rangle$.
In this way, the sum $\sum_{i=1}^{i'}\left\langle |i_{c}\rangle\langle i_{c}|\right\rangle $
is conserved by both $V_{1}$ and $V_{2}$ (in particular, $V_{2}$
conserves $\sum_{i=1}^{i'}\left\langle |i_{c}\rangle\langle i_{c}|\right\rangle $
because it conserves the sum $\sum_{i=i'+1}^{d_{c}}\left\langle |i_{c}\rangle\langle i_{c}|\right\rangle $
and $\sum_{i=1}^{i'}\left\langle |i_{c}\rangle\langle i_{c}|\right\rangle +\sum_{i=i'+1}^{d_{c}}\left\langle |i_{c}\rangle\langle i_{c}|\right\rangle =1$
). Therefore, for 
\begin{equation}
U_{j}=U_{|(i'+1)_{c}1_{v}\rangle\leftrightarrow|i'_{c}d_{v}\rangle}\oplus V_{j},\label{eq:31.4 Uj, j=00003D1,2}
\end{equation}
$j=1,2$, only $U_{|(i'+1)_{c}1_{v}\rangle\leftrightarrow|i'_{c}d_{v}\rangle}$
contributes to $\Delta\left(\sum_{i=1}^{i'}\left\langle |i_{c}\rangle\langle i_{c}|\right\rangle \right)$
in the transformation $\rho\rightarrow U_{j}\rho U_{j}^{\dagger}$.
This allows us to conclude that this transformation is catalytic and
non-unital.

\subsubsection{Catalyst dimension and corresponding state}

The natural logarithm of Eq. (\ref{eq:31 violation of passivity in terms of mu, for non-unit trans})
yields the inequality: 

\begin{equation}
d_{v}>1+\frac{1}{\textrm{ln}(\mu)}\textrm{ln}\left(\frac{p_{i'}^{c}}{p_{i'+1}^{c}}\right),\label{eq:31.7 inequal for dv}
\end{equation}
which is satisfied by the dimension $d_{v}=d_{v}^{\ast}$ given by
\begin{equation}
d_{v}^{\ast}\equiv\left\lceil 1+\frac{1}{\textrm{ln}(\mu)}\textrm{ln}\left(\frac{p_{i'}^{c}}{p_{i'+1}^{c}}\right)\right\rceil +1.\label{eq:31.8 dv*}
\end{equation}
With this dimension, the catalyst eigenvalues satisfy Eq. (\ref{eq:29.7 Catalyst eigenvalues in tems of mu})
and the normalization condition $\sum_{k=1}^{d_{v}^{\ast}}p_{k}^{v}=1$. 

\subsubsection{Catalytic increment of the ground population of the cold object }

The following corollary states that a sufficiently large catalyst
allows to increase the population $\bigl\langle|1_{c}\rangle\langle1_{c}|\bigr\rangle$,
using a unitary $U$ that acts on $\mathcal{H}_{c}\otimes\mathcal{H}_{v}$.
This corollary is a consequence of Eqs. (\ref{eq:31 violation of passivity in terms of mu, for non-unit trans})
and (\ref{eq:31.3 positive J|k+1>-->|k> in terms of mu 2}). 

\textbf{Corollary 3}.\textit{ Given a state of the cold object $\rho_{c}=\sum_{i=1}^{d_{c}}p_{i}^{c}|i_{c}\rangle\langle i_{c}|$,
such that $p_{2}^{c}>p_{d_{c}}^{c}$, it is possible to catalytically
increase the ground population $p_{1}^{c}$.}

\textit{Proof}. The condition $p_{2}^{c}>p_{d_{c}}^{c}$ can be written
as $p_{i'+1}^{c}>p_{d_{c}}^{c}$, with $i'=1$. In this way, the unitary
$U_{|(i'+1)_{c}1_{v}\rangle\leftrightarrow|i'_{c}d_{v}\rangle}=U_{|2_{c}1_{v}\rangle\leftrightarrow|1_{c}d_{v}\rangle}$
transfers population $\Delta p_{1}^{c}=\Delta\bigl\langle|1_{c}\rangle\langle1_{c}|\bigr\rangle=J_{|2_{c}\rangle\rightarrow|1_{c}\rangle}>0$,
and Eq. (\ref{eq:31.3 positive J|k+1>-->|k> in terms of mu 2}) can
be applied to construct a restoring unitary $V_{2}$ (cf. (\ref{eq:31.5 V2}))
for the transformation $\rho=\rho_{c}\otimes\rho_{v}\rightarrow U_{|2_{c}1_{v}\rangle\leftrightarrow|1_{c}d_{v}\rangle}\rho_{c}\otimes\rho_{v}U_{|2_{c}1_{v}\rangle\leftrightarrow|1_{c}d_{v}\rangle}^{\dagger}$.
Since $V_{2}$ preserves the partial sum $\sum_{i=1}^{i'}\left\langle |i_{c}\rangle\langle i_{c}|\right\rangle =\left\langle |1_{c}\rangle\langle1_{c}|\right\rangle $,
the catalytic transformation $\rho\rightarrow U\rho U^{\dagger}$,
where $U=U_{|2_{c}1_{v}\rangle\leftrightarrow|1_{c}d_{v}\rangle}\oplus V_{2}$,
also transfers population $\Delta p_{1}^{c}=J_{|2_{c}\rangle\rightarrow|1_{c}\rangle}>0$. 

\section{GENERALIZATION OF RESTORING UNITARIES }

In this appendix we show how to ``merge'' several restoring unitaries
into a single restoring unitary that gives rise to catalytic transformations
studied later. To that end, let us first recall the basic mechanism
for catalyst restoration in our framework. Given a transformation
$\rho\rightarrow U_{k^{(M)}}\rho U_{k^{(M)}}^{\dagger}$ (cf. (\ref{eq:112 U_k(M)+(+_mV_k(m))})),
which generates a positive current $J_{|k_{v}^{(1)}\rangle\rightarrow|k_{v}^{(M)}\rangle}$,
we have previously constructed restoring unitaries that induce \textit{elemental}
chains $\mathbf{ch}_{|k_{v}^{(M)}\rangle\rightarrow|k_{v}^{(1)}\rangle}$.
In the proofs of Theorems 1 and 2 the indices $(k^{(1)},k^{(M)})$
are expressed as $(l,l'+1)$, with $l=1$ and $l'=d_{v}-1$ in the
case of (the proof of) Theorem 2. Moreover, the role of $U_{k^{(M)}}$
is taken by the two-level unitary $U_{i',l,l'}$. After characterizing
the associated elemental chains $\mathbf{ch}_{|(l'+1)_{v}\rangle\rightarrow|l_{v}\rangle}$,
we also provided a recipe to generate a \textit{uniform} loop $\mathbf{loop}_{|l_{v}\rangle\rightarrow|l_{v}\rangle}=\{J_{|l_{v}\rangle\rightarrow|(l'+1)_{v}\rangle}\}\cup\mathbf{ch}_{|(l'+1)_{v}\rangle\rightarrow|l_{v}\rangle}$
by properly tuning the swap intensities of $U_{i',l,l'}$ and the
partial swaps in 
\begin{equation}
V_{\textrm{res},n,N}=\oplus_{k=l}^{l'}V_{|n_{ch}(k+1)_{v}\rangle\leftrightarrow|N_{ch}k_{v}\rangle}.\label{eq:32 Vres,n,N}
\end{equation}

The generation of a uniform loop is the core of catalytic transformations
using arbitrary unitaries of the form (\ref{eq:112 U_k(M)+(+_mV_k(m))}),
as per Remark 7. In particular, the unitaries $U_{i',l,l'}\oplus V_{\textrm{res},n,N}$
that form the basis for Theorems 1 and 2 adhere to such a principle.
A unique feature of $U_{i',l,l'}\oplus V_{\textrm{res},n,N}$ is that
the local current produced by each of its partial swaps is associated
with a single \textit{global} current. For example, the current $J_{|(k+1)_{v}\rangle\rightarrow|k_{v}\rangle}$
due to $V_{|n_{ch}(k+1)_{v}\rangle\leftrightarrow|N_{ch}k_{v}\rangle}$
equals $J_{|n_{ch}(k+1)_{v}\rangle\rightarrow|N_{ch}k_{v}\rangle}$.
However, any direct sum of two-level unitaries that act on $\mathcal{H}_{c}\otimes\mathcal{H}_{h}\otimes\textrm{span}\{|(k+1)_{v}\rangle,|k_{v}\rangle\}$
also produces a local current $J_{|(k+1)_{v}\rangle\rightarrow|k_{v}\rangle}$,
according to Remark 5 . We take advantage of this possibility to extend
the restoring unitaries considered until now. As we will shortly see,
this generalization gives rise to loops where the local currents are
instead given by sums of global currents. 

\subsection{\textquotedblleft Addition\textquotedblright{} of restoring unitaries}

Suppose that there exist not only one but a set of pairs $(n,N)$
such that $V_{\textrm{res},n,N}$ are all valid restoring unitaries,
meaning that (for each pair $(n,N)$) for all $l\leq k\leq l'$ the
partial swaps $V_{|n_{ch}(k+1)_{v}\rangle\leftrightarrow|N_{ch}k_{v}\rangle}$
can generate currents $J_{|(k+1)_{v}\rangle\rightarrow|k_{v}\rangle}>0$.
Moreover, suppose also that for any two pairs $(n,N)$ and $(n',N')$
it holds that $n\neq n'$ and $N\neq N'$. Since this means that $V_{\textrm{res},n,N}$
and $V_{\textrm{res},n',N'}$ act on orthogonal subspaces, we can
construct a direct sum 
\begin{align}
\oplus_{n,N}V_{\textrm{res},n,N} & =\oplus_{n,N}\left(\oplus_{k=l}^{l'}V_{|n_{ch}(k+1)_{v}\rangle\leftrightarrow|N_{ch}k_{v}\rangle}\right)\nonumber \\
 & =\oplus_{k=l}^{l'}\left(\oplus_{n,N}V_{|n_{ch}(k+1)_{v}\rangle\leftrightarrow|N_{ch}k_{v}\rangle}\right)\nonumber \\
 & \equiv{\color{green}{\normalcolor \oplus_{k=l}^{l'}V_{k}}}.\label{eq:32.1 addition of Vres}
\end{align}

By definition, each $V_{k}=\oplus_{n,N}V_{|n_{ch}(k+1)_{v}\rangle\leftrightarrow|N_{ch}k_{v}\rangle}$
maps $\mathcal{H}_{c}\otimes\mathcal{H}_{h}\otimes\textrm{span}\{|(k+1)_{v}\rangle,|k_{v}\rangle\}$
into itself, and consequently it produces a current $J_{|(k+1)_{v}\rangle\rightarrow|k_{v}\rangle}$.
Following Eq. (\ref{eq:111.1 local catalyst current as sum of global currents}),
and associating each pair $\psi,\varphi$ ``free'' with a pair $(n,N)$,
we have that 
\begin{align}
\textrm{Tr}\left[O_{k}\left(V_{k}\rho V_{k}^{\dagger}-\rho\right)\right] & =J_{|(k+1)_{v}\rangle\rightarrow|k_{v}\rangle}\nonumber \\
 & =\sum_{n,N}J_{|n_{ch}(k+1)_{v}\rangle\rightarrow|N_{ch}k_{v}\rangle},\label{eq:32.2 local catalyst current J as sum of global currents}
\end{align}
and (cf. (\ref{eq:111.2 swap catalyst current as sum of swap global currents}))
\begin{equation}
\mathcal{J}_{|(k+1)_{v}\rangle\rightarrow|k_{v}\rangle}=\sum_{n,N}\mathcal{J}_{|n_{ch}(k+1)_{v}\rangle\rightarrow|N_{ch}k_{v}\rangle}.\label{eq:32.3 swap catalyst current}
\end{equation}
Since $J_{|n_{ch}(k+1)_{v}\rangle\rightarrow|N_{ch}k_{v}\rangle}$
is positive by assumption, for all $(n,N)$, it follows that $J_{|(k+1)_{v}\rangle\rightarrow|k_{v}\rangle}$in
Eq. (\ref{eq:32.2 local catalyst current J as sum of global currents})
is also positive. In this way, the addition (direct sum) of the unitaries
$V_{\textrm{res},n,N}$ results in a new restoring unitary $\oplus_{n,N}V_{\textrm{res},n,N}$,
with a corresponding chain $\mathbf{ch}_{|(l'+1)_{v}\rangle\rightarrow|l_{v}\rangle}=\{\sum_{n,N}J_{|n_{ch}(k+1)_{v}\rangle\rightarrow|N_{ch}k_{v}\rangle}\}_{k=l}^{l'}$. 

\subsection{Uniform loop from ${\normalcolor \oplus_{k=l}^{l'}V_{k}}=\oplus_{n,N}V_{\textrm{res},n,N}$}

Let us see now how a unitary of the form $U_{i',l,l'}\oplus\left(\oplus_{n,N}V_{\textrm{res},n,N}\right)$
can generate a uniform loop. The procedure is very similar to that
considered in the proofs of Theorems 1 and 2. Denoting as $\mathcal{J}_{|l_{v}\rangle\rightarrow|(l'+1)_{v}\rangle}$
the (local) swap current corresponding to $\mathcal{U}_{i',l,l'}$,
define 

\begin{equation}
\mathcal{J}_{\textrm{min}}\equiv\textrm{min}\left\{ \mathcal{J}_{|l_{v}\rangle\rightarrow|(l'+1)_{v}\rangle},\textrm{min}_{l\leq k\leq l'}\mathcal{J}_{|(k+1)_{v}\rangle\rightarrow|k_{v}\rangle}\right\} ,\label{eq:32.4 loop current Jmin}
\end{equation}
where $\mathcal{J}_{|(k+1)_{v}\rangle\rightarrow|k_{v}\rangle}$ satisfies
(\ref{eq:32.3 swap catalyst current}). 

Recalling that $\mathcal{J}_{|(k+1)_{v}\rangle\rightarrow|k_{v}\rangle}$
is the current obained from the direct sum of swaps $\oplus_{n,N}\mathcal{V}_{|n_{ch}(k+1)_{v}\rangle\leftrightarrow|N_{ch}k_{v}\rangle}$,
consider the following strategy. If $\oplus_{n,N}\mathcal{V}_{|n_{ch}(k+1)_{v}\rangle\leftrightarrow|N_{ch}k_{v}\rangle}$
produces a local current of magnitude $\mathcal{J}_{\textrm{min}}$,
then the swap intensities of all the $V_{|n_{ch}(k+1)_{v}\rangle\leftrightarrow|N_{ch}k_{v}\rangle}$
are set to $r=1$. Otherwise, some of the swap intensities are reduced
until $J_{|(k+1)_{v}\rangle\rightarrow|k_{v}\rangle}$ equals $\mathcal{J}_{\textrm{min}}$.
Note that this is always possible because $0\leq J_{|(k+1)_{v}\rangle\rightarrow|k_{v}\rangle}\leq\mathcal{J}_{|(k+1)_{v}\rangle\rightarrow|k_{v}\rangle}$,
with $J_{|(k+1)_{v}\rangle\rightarrow|k_{v}\rangle}=0$ obtained by
setting $r=0$ for all the $V_{|n_{ch}(k+1)_{v}\rangle\leftrightarrow|N_{ch}k_{v}\rangle}$.
The same recipe is applied to tune the swap intensity of $U_{i',l,l'}$. 

\textbf{Definition 10 (loop current)}. Since all the currents in a
uniform loop possess the same magnitude (cf. Remark 7), we denote
this magnitude as $J_{\textrm{loop}}$ and dub it ``loop current''.
Due to the result stated in the following lemma, we will often write
$J_{\textrm{loop}}$ as $\mathcal{J}_{\textrm{min}}$. 

\textbf{Proposition 3}. \textit{The maximum loop current coincides
with} $\mathcal{J}_{\textrm{min}}$\textit{. That is, }
\begin{align}
\mathcal{J}_{\textrm{min}} & =\textrm{max}_{\textrm{swap intensities of }U_{i',l,l'}\oplus\left(\oplus_{k=l}^{l'}V_{k}\right)}J_{\textrm{loop}}\nonumber \\
 & =\textrm{max}_{\textrm{swap intensities of }U_{i',l,l'}\oplus\left(\oplus_{n,N}V_{\textrm{res},n,N}\right)}J_{\textrm{loop}}\nonumber \\
 & \geq\textrm{max}_{\textrm{swap intensities of }U_{i',l,l'}\oplus V_{\textrm{res},n',N'}}J_{\textrm{loop}},\label{eq:32.5 lemma about Jmin}
\end{align}
\textit{where $V_{\textrm{res},n',N'}$ can be any single restoring
unitary included in the direct sum of the second line. }

\textit{Proof}. By construction, to achieve the current $\mathcal{J}_{\textrm{min}}$
it is necessary to perform total swaps for at least one of the currents
in the loop $\{J_{|l_{v}\rangle\rightarrow|(l'+1)_{v}\rangle}\}\cup\{J_{|(k+1)_{v}\rangle\rightarrow|k_{v}\rangle}\}_{k=l}^{l'}$.
Therefore, it is impossible to have a uniform loop such that $J_{\textrm{loop}}>\mathcal{J}_{\textrm{min}}$.
This proves the equality in the first line of (\ref{eq:32.5 lemma about Jmin})
(the second line is simply the definition (\ref{eq:32.1 addition of Vres})).
The inequality follows because, for any $(n',N')$, Eq. (\ref{eq:32.3 swap catalyst current})
implies that $\mathcal{J}_{|n'_{ch}(k+1)_{v}\rangle\rightarrow|N'_{ch}k_{v}\rangle}\leq\mathcal{J}_{|(k+1)_{v}\rangle\rightarrow|k_{v}\rangle}$,
and consequently $\textrm{min}\left\{ \mathcal{J}_{|l_{v}\rangle\rightarrow|(l'+1)_{v}\rangle},\textrm{min}_{l\leq k\leq l'}\mathcal{J}_{|n'_{ch}(k+1)_{v}\rangle\rightarrow|N'_{ch}k_{v}\rangle}\right\} \leq\mathcal{J}_{\textrm{min}}$. 

\textbf{Remark 9}. While the generalization here concerns restoring
unitaries, we can apply the same rationale to catalytic transformations
where $U_{i',l,l'}$ is also replaced by a direct sum of two-level
unitaries that acts on $\mathcal{H}_{c}\otimes\mathcal{H}_{h}\otimes\textrm{span}\{|(l'+1)_{v}\rangle,|l_{v}\rangle\}$.
In this case, $\mathcal{J}_{|l_{v}\rangle\rightarrow|(l'+1)_{v}\rangle}$
is given by a sum $\mathcal{J}_{|l_{v}\rangle\rightarrow|(l'+1)_{v}\rangle}=\sum_{\psi,\varphi\textrm{ free}}\mathcal{J}_{|\varphi_{ch}l_{v}\rangle\rightarrow|\psi_{ch}(l'+1)_{v}\rangle}$
(cf. (\ref{eq:111.2 swap catalyst current as sum of swap global currents})).
If all the currents $\mathcal{J}_{|\varphi_{ch}l_{v}\rangle\rightarrow|\psi_{ch}(l'+1)_{v}\rangle}$
are positive, $\mathcal{J}_{|l_{v}\rangle\rightarrow|(l'+1)_{v}\rangle}>0$
and Eqs. (\ref{eq:32.4 loop current Jmin}) and (\ref{eq:32.5 lemma about Jmin})
remain valid under this replacement. 

\section{MAXIMIZATION OF $\mathcal{J}_{\textrm{min}}$ AND OPTIMAL CATALYTIC COOLING OF A QUBIT USING ANOTHER QUBIT AS HOT OBJECT }

According to Theorem 1, given a transformation $\rho\rightarrow\sigma=U_{i',l,l'}\rho U_{i',l,l'}^{\dagger}$,
all the elements in a set $\{V_{\textrm{res},n,N}\}_{n,N}$ are restoring
unitaries as long as the partial swaps $\{V_{|n_{ch}(k+1)_{v}\rangle\leftrightarrow|N_{ch}k_{v}\rangle}\}_{k=l}^{l'}$
generate currents $J_{|(k+1)_{v}\rangle\rightarrow|k_{v}\rangle}>0$,
for \textit{any} pair $(n,N)$. Hence, the inequality $p_{n}^{ch}p_{k+1}^{v}>p_{N}^{ch}p_{k}^{v}$
must be valid for $l\leq k\leq l'$ and any pair $(n,N)$. For a \textit{fixed}
state $\rho=\rho_{c}\otimes\rho_{h}\otimes\rho_{v}^{\ast}$, it is
perfectly possible that its eigenvalues prevent to fulfill this condition,
even for a single pair $(n,N)$. 

However, the proof of Theorem 2 tells us that, under mild assumptions
regarding the state $\rho_{c}\otimes\rho_{h}$, we can always find
a full-rank catalyst state $\rho_{v}=\sum_{k=1}^{d_{v}^{\ast}}p_{k}^{v}|k_{v}\rangle\langle k_{v}|$
such that the transformation $\rho\rightarrow\sigma=U_{i',1,d_{v}^{\ast}-1}\rho U_{i',1,d_{v}^{\ast}-1}^{\dagger}$
possess at least one restoring unitary $V_{\textrm{res},n,N}$. This
fact is our starting point for the forthcoming analysis. For the sake
of clarity, in what follows we denote the aforementioned catalyst
state as $\rho_{v}^{\ast}$. 

The purpose of this appendix is threefold. First, we will identify
other unitaries $V_{\textrm{res},n,N}$ that recover the catalyst
from the transformation $\rho\rightarrow\sigma=U_{i',1,d_{v}^{\ast}-1}\rho U_{i',1,d_{v}^{\ast}-1}^{\dagger}$,
with $\rho=\rho_{c}\otimes\rho_{h}\otimes\rho_{v}^{\ast}$. Subsequently,
we will derive a new catalyst state $\bar{\rho}_{v}=\sum_{k=1}^{d_{v}^{\ast}}\bar{p}_{k}^{v}|k_{v}\rangle\langle k_{v}|$
whose eigenvalues maximize the loop current $\mathcal{J}_{\textrm{min}}$
in (\ref{eq:32.4 loop current Jmin}). The importance of this maximization
is encapsulated in the following remark:

\textbf{Remark 10}. Taking into account the correspondences $J_{|(i'+1)_{c}\rangle\rightarrow|i'_{c}\rangle}=J_{|1_{v}\rangle\rightarrow|d_{v}^{\ast}\rangle}$
and $J_{|1_{v}\rangle\rightarrow|d_{v}^{\ast}\rangle}=\mathcal{J}_{\textrm{min}}$
(once the swap intensities are tuned as indicated in the preceding
appendix), the maximization of $\mathcal{J}_{\textrm{min}}$ implies
the maximization of the current $J_{|(i'+1)_{c}\rangle\rightarrow|i'_{c}\rangle}$.
In particular, this optimizes the \textit{extracted heat} $-\Delta\bigl\langle H_{c}\bigr\rangle=J_{|(i'+1)_{c}\rangle\rightarrow|i'_{c}\rangle}(\varepsilon_{i'+1}^{c}-\varepsilon_{i'}^{c})$
in the case of cooling transformations, including the example considered
in Appendix I2. 

Finally, we will extend the maximization of $\mathcal{J}_{\textrm{min}}$
to catalytic unitaries of which $U_{i',1,d_{v}^{\ast}-1}\oplus\left(\oplus_{n,N}V_{\textrm{res},n,N}\right)$
is particular case, and initial states $\rho=\rho_{ch}\otimes\rho_{v}^{\ast}$,
where $\rho_{ch}$ is a general state of the hot and cold objects.

\subsection{Restoring unitaries for $\rho\rightarrow\sigma=U_{i',1,d_{v}^{\ast}-1}\rho U_{i',1,d_{v}^{\ast}-1}^{\dagger}$
(see also Fig. 13) }

\subsubsection{Cooling transformations }

In the case of cooling transformations we saw in Appendix G1 that
partial swaps $V_{|n_{ch}(k+1)_{v}\rangle\leftrightarrow|N_{ch}k_{v}\rangle}$
with $|n_{ch}\rangle=|1_{c}1_{h}\rangle$ and $|N_{ch}\rangle=|1_{c}d_{h}\rangle$
give rise to a restoring unitary $V_{\textrm{res},n,N}$. Since these
partial swaps generate positive currents $J_{|(k+1)_{v}\rangle\rightarrow|k_{v}\rangle}$
iff $p_{1}^{c}p_{1}^{h}p_{k+1}^{v}>p_{1}^{c}p_{d_{h}}^{h}p_{k+1}^{v}$,
we can easily identify other restoring unitaries. Specifically, the
inequality $p_{1}^{c}p_{1}^{h}p_{k+1}^{v}>p_{1}^{c}p_{d_{h}}^{h}p_{k+1}^{v}$
is equivalent to $p_{i}^{c}p_{1}^{h}p_{k+1}^{v}>p_{i}^{c}p_{d_{h}}^{h}p_{k+1}^{v}$,
for all $i\in\{1,2,...,d_{c}\}$. This implies that any partial swap
$V_{|n_{ch}(k+1)_{v}\rangle\leftrightarrow|N_{ch}k_{v}\rangle}$ with
$|n_{ch}\rangle=|i_{c}1_{h}\rangle$ and $|N_{ch}\rangle=|i_{c}d_{h}\rangle$
also induces a positive current $J_{|(k+1)_{v}\rangle\rightarrow|k_{v}\rangle}$,
and therefore any direct sum $\oplus_{k=1}^{d_{v}^{\ast}-1}V_{|i_{c}1_{h}(k+1)_{v}\rangle\leftrightarrow|i_{c}d_{h}k_{v}\rangle}$
provides a chain $\mathbf{ch}_{|d_{v}^{\ast}\rangle\rightarrow|1_{v}\rangle}$.
The addition of these restoring unitaries yields a restoring unitary
that acts on $\mathcal{H}_{h}\otimes\mathcal{H}_{v}$. Namely,
\begin{align}
V_{\textrm{res},hv} & =\oplus_{i=1}^{d_{c}}\left(\oplus_{k=1}^{d_{v}^{\ast}-1}V_{|i_{c}1_{h}(k+1)_{v}\rangle\leftrightarrow|i_{c}d_{h}k_{v}\rangle}\right)\nonumber \\
 & =\oplus_{k=1}^{d_{v}^{\ast}-1}V_{|1_{h}(k+1)_{v}\rangle\leftrightarrow|d_{h}k_{v}\rangle}.\label{eq:32.6 Vres,hv SM}
\end{align}

\begin{figure}
	
	\begin{centering}
		\includegraphics[scale=0.75]{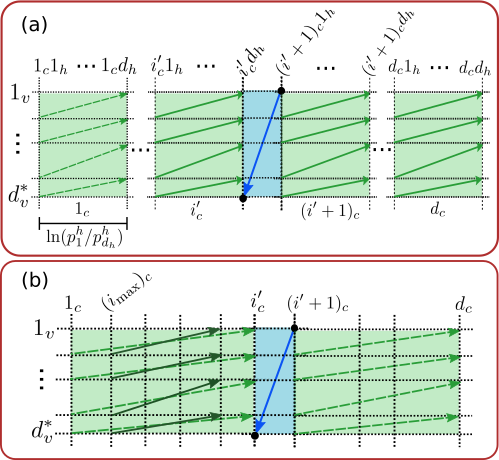}\caption{Currents induced by the addition of restoring unitaries, for cooling
			transformations (a), and (more general) non-unital transformations
			(b). The blue arrow depicts the global current generated by $U_{i',1,d_{v}^{\ast}-1}$,
			which takes the form $U_{i',1,d_{v}^{\ast}-1}=U_{|i'_{c}d_{v}^{\ast}\rangle\leftrightarrow|(i'+1)_{c}1_{v}\rangle}$
			in the case of (b). In both cases dashed green arrows correspond to
			the restoring unitaries derived in Appendix G. In (a) the vertical
			sequences of continuous arrows are replicas of the dashed arrows,
			and each corresponds to a different restoring unitary. Their addition
			yields $V_{\textrm{res},hv}$ in Eq. (\ref{eq:32.6 Vres,hv SM}).
			In (b) a cold object of dimension 10 (number of columns) is assumed,
			and the darker green arrows depict currents associated with an additional
			restoring unitary.\textcolor{green}{{} }}
		\par\end{centering}
\end{figure}

\subsubsection{Non-unital transformations }

For non-unital transformations, we already found two possible restoring
unitaries in Appendix G2 (cf. Eqs. (\ref{eq:31.4 V1}) and (\ref{eq:31.5 V2})).
If both $V_{1}$ and $V_{2}$ satisfy the required conditions (which
means that $p_{1}^{c}>p_{i'}^{c}$ \textit{and} $p_{i'+1}^{c}>p_{d_{c}}^{c}$),
we can immediately conclude that $V_{1}\oplus V_{2}$ is also a restoring
unitary. Note also that since $V_{1}=\oplus_{k=1}^{d_{v}^{\ast}-1}V_{|1_{c}(k+1)_{v}\rangle\leftrightarrow|i'_{c}k_{v}\rangle}$
acts on a subspace of $\mathcal{H}_{c}\otimes\mathcal{H}_{v}$, $\mathcal{H}_{n,N}$
is more properly described as $\mathcal{H}_{n,N}=\textrm{span}\{|i_{c}\rangle\}_{i=n}^{N}\otimes\mathcal{H}_{v}$,
where $|n_{c}\rangle=|1_{c}\rangle$ and $|N_{c}\rangle=|i'_{c}\rangle$.
Similarly, $V_{2}=\oplus_{k=1}^{d_{v}^{\ast}-1}V_{|(i'+1)_{c}(k+1)_{v}\rangle\leftrightarrow|d_{c}k_{v}\rangle}$
acts on a subspace $\mathcal{H}_{n,N}\subset\mathcal{H}_{c}\otimes\mathcal{H}_{v}$,
where $|n_{c}\rangle=|(i'+1)_{c}\rangle$ and $|N_{c}\rangle=|d_{c}\rangle$. 

Other restoring unitaries can be identified as follows: 
\begin{enumerate}
\item If $p_{1}^{c}>p_{i'}^{c}$, consider the set $\{(n,N)\}_{n,N}=\{(i,i'+1-i)\}_{i=1}^{i_{\textrm{max}}}$,
where $i_{\textrm{max}}$ is the maximum value of $i$ such that $p_{i}^{c}p_{k+1}^{v}-p_{i'+1-i}^{c}p_{k}^{v}>0$
for all $1\leq k\leq d_{v}^{\ast}-1$. In this way, all the partial
swaps $\{V_{|i_{c}(k+1)_{v}\rangle\leftrightarrow|(i'+1-i)_{c}k_{v}\rangle}\}_{i=1}^{i_{\textrm{max}}}$
generate positive currents $J_{|(k+1)_{v}\rangle\rightarrow|k_{v}\rangle}$,
and for $1\leq i\leq i_{\textrm{max}}$ any 
\begin{equation}
V_{i}\equiv\oplus_{k=1}^{d_{v}^{\ast}-1}V_{|i_{c}(k+1)_{v}\rangle\leftrightarrow|(i'+1-i)_{c}k_{v}\rangle}\label{eq:32.7 Vi for p_i(c)>p_i'(c)}
\end{equation}
 is a valid restoring unitary. Note that, for $i=1$, the previous
expression recovers the restoring unitary $V_{1}$.
\item If $p_{i'+1}^{c}>p_{d_{c}}^{c}$, consider the set $\{(n,N)\}_{n,N}=\{(i,d_{c}+i'+1-i)\}_{i=i'+1}^{i_{\textrm{max}}}$,
where $i_{\textrm{max}}$ is the maximum value of $i$ such that $p_{i}^{c}p_{k+1}^{v}-p_{d_{c}+i'+1-i}^{c}p_{k}^{v}>0$
for all $1\leq k\leq d_{v}^{\ast}-1$. In this way, all the partial
swaps $\{V_{|i_{c}(k+1)_{v}\rangle\leftrightarrow|(d_{c}+i'+1-i)_{c}k_{v}\rangle}\}_{i=i'+1}^{i_{\textrm{max}}}$
generate positive currents $J_{|(k+1)_{v}\rangle\rightarrow|k_{v}\rangle}$,
and for $i'+1\leq i\leq i{}_{\textrm{max}}$ any 
\begin{equation}
V_{i}\equiv\oplus_{k=1}^{d_{v}^{\ast}-1}V_{|i_{c}(k+1)_{v}\rangle\leftrightarrow|(d_{c}+i'+1-i)_{c}k_{v}\rangle}\label{eq:32.8 V_i(c) for p_(i'+1)>p_dc}
\end{equation}
 is a valid restoring unitary. Using this notation, $V_{i'+1}$ coincides
with $V_{2}$. 
\end{enumerate}

\subsection{Optimal cooling of a qubit using another qubit as hot object }

We resort now to results derived in the preceding section to address
the problem of maximizing the cooling of a qubit. Specifically, we
will consider a transformation $\rho=\rho_{c}\otimes\rho_{h}\otimes\rho_{v}\overset{\textrm{CC}}{\longrightarrow}U\rho U^{\dagger}$,
where $\rho_{c}$ and $\rho_{h}$ are both states of two-level systems,
and $U=U_{i',1,d_{v}^{\ast}-1}\oplus V_{\textrm{res},hv}$. For two-level
systems, the index $i'$ in the cooling unitary $U_{i',1,d_{v}^{\ast}-1}$
only admits the value $i'=1$. This yields $U_{1,1,d_{v}^{\ast}-1}=U_{|2_{c}1_{h}1_{v}\rangle\leftrightarrow|1_{c}2_{h}d_{v}^{\ast}\rangle}$,
and $V_{\textrm{res},hv}=\oplus_{k=1}^{d_{v}^{\ast}-1}V_{|1_{h}(k+1)_{v}\rangle\leftrightarrow|2_{h}k_{v}\rangle}$. 

Since $J_{|(i'+1)_{c}\rangle\rightarrow|i'_{c}\rangle}=\mathcal{J}_{\textrm{min}}$,
maximizing $J_{|(i'+1)_{c}\rangle\rightarrow|i'_{c}\rangle}$ is tantamount
to maximize the loop current $\mathcal{J}_{\textrm{min}}$ (cf. Remark
10). Specifically, we will optimize$\mathcal{J}_{\textrm{min}}$ with
respect to the eigenvalues of the state $\rho_{v}$. First, we will
show that $\mathcal{J}_{\textrm{min}}$ is maximized by eigenvalues
$\{\bar{p}_{k}^{v}\}_{k=1}^{d_{v}^{\ast}}$ such that $\mathcal{U}_{1,1,d_{v}^{\ast}-1}\oplus\left(\oplus_{k=1}^{d_{v}^{\ast}-1}\mathcal{V}_{|1_{h}(k+1)_{v}\rangle\leftrightarrow|2_{h}k_{v}\rangle}\right)$
generates a uniform loop. In other words, a maximum $\mathcal{J}_{\textrm{min}}$
is achieved when \textit{all the swap currents in (\ref{eq:32.4 loop current Jmin})
have the same magnitude}. For this to happen, the eigenvalues $\{\bar{p}_{k}^{v}\}_{k=1}^{d_{v}^{\ast}}$
must satisfy the set of linear equations 

\begin{equation}
\mathcal{J}_{|1_{v}\rangle\rightarrow|d_{v}^{\ast}\rangle}=\mathcal{J}_{|(k+1)_{v}\rangle\rightarrow|k_{v}\rangle},\textrm{ for }1\leq k\leq d_{v}^{\ast}-1.\label{eq:32.9 Eqs. for maximization of Jmin}
\end{equation}
Explicitly, 
\begin{equation}
a\bar{p}_{1}^{v}-b\bar{p}_{d_{v}^{\ast}}^{v}=c\bar{p}_{k+1}^{v}-d\bar{p}_{k}^{v},\label{eq:32.10 Eqs. for max. of Jmin explicit}
\end{equation}
for $1\leq k\leq d_{v}^{\ast}-1$, with coefficients 
\begin{equation}
\{a,b;c,d\}=\{p_{2}^{c}p_{1}^{h},p_{1}^{c}p_{2}^{h};p_{1}^{h},p_{2}^{h}\}.\label{eq:32.11 coefficients}
\end{equation}

In this way, the coefficients $a$ and $b$ reproduce the swap current
$\mathcal{J}_{|1_{v}\rangle\rightarrow|d_{v}^{\ast}\rangle}$ (generated
by $\mathcal{U}_{1,1,d_{v}^{\ast}-1}$), and the coefficients $c$
and $d$ reproduce the swap current $\mathcal{J}_{|(k+1)_{v}\rangle\rightarrow|k_{v}\rangle}$
(generated by $\mathcal{V}_{|1_{h}(k+1)_{v}\rangle\leftrightarrow|2_{h}k_{v}\rangle}$).
After proving that $\{\bar{p}_{k}^{v}\}_{k=1}^{d_{v}^{\ast}}$ are
optimal, we will proceed to their evaluation. 

\subsubsection{Proof of optimality of $\{\bar{p}_{k}^{v}\}_{k=1}^{d_{v}^{\ast}}$}

The proof that we will provide in the following is not restricted
to the example of optimal cooling considered here. More generally,
suppose that the swap currents at the r.h.s. of Eqs. (\ref{eq:32.9 Eqs. for maximization of Jmin})
are given by $\mathcal{J}_{|(k+1)_{v}\rangle\rightarrow|k_{v}\rangle}=c_{k+1}\bar{p}_{k+1}^{v}-d_{k}\bar{p}_{k}^{v}$,
where $c_{k+1}$ and $d_{k}$ are \textit{positive} coefficients that
may depend on $k$. Suppose also that the only condition on the coefficients
$a$ and $b$ in $\mathcal{J}_{|1_{v}\rangle\rightarrow|d_{v}^{\ast}\rangle}=a\bar{p}_{1}^{v}-b\bar{p}_{d_{v}^{\ast}}^{v}$
is that they are also positive. In this way, the generalized form
of (\ref{eq:32.9 Eqs. for maximization of Jmin}) and (\ref{eq:32.10 Eqs. for max. of Jmin explicit})
reads 

\begin{align}
\mathcal{J}_{|1_{v}\rangle\rightarrow|d_{v}^{\ast}\rangle} & =\mathcal{J}_{|(k+1)_{v}\rangle\rightarrow|k_{v}\rangle},\nonumber \\
\Leftrightarrow a\bar{p}_{1}^{v}-b\bar{p}_{d_{v}^{\ast}}^{v} & =c_{k+1}\bar{p}_{k+1}^{v}-d_{k}\bar{p}_{k}^{v},\label{eq:32.11.1 Generalized Eqs. for maximization of Jmin}
\end{align}
for $1\leq k\leq d_{v}^{\ast}-1$, with Eq. (\ref{eq:32.11 coefficients})
being a particular instance of the general coefficients $\{a,b;c_{k+1},d_{k}\}_{k=1}^{d_{v}^{\ast}-1}$
previously described. 

Now, let $\bar{\mathcal{J}}_{\textrm{min}}=\mathcal{J}_{|1_{v}\rangle\rightarrow|d_{v}^{\ast}\rangle}=\mathcal{J}_{|(k+1)_{v}\rangle\rightarrow|k_{v}\rangle}$
denote the loop current resulting from the solution of Eqs. (\ref{eq:32.11.1 Generalized Eqs. for maximization of Jmin}).
What we shall prove is that the current $\mathcal{J}_{\textrm{min}}$
obtained from \textit{any} other set of eigenvalues $\{p_{k}^{v}\}_{k=1}^{d_{v}^{\ast}}$
is such that $\mathcal{J}_{\textrm{min}}<\bar{\mathcal{J}}_{\textrm{min}}$.
To that end, we write $p_{k}^{v}$ as $p_{k}^{v}=\bar{p}_{k}^{v}+\delta_{k}$,
with the variations $\{\delta_{k}\}_{k=1}^{d_{v}^{\ast}}$ constrained
in such a way that $\{p_{k}^{v}\}_{k=1}^{d_{v}^{\ast}}$ constitutes
a proper probability distribution. This gives rise to the following
variations in the swap currents (cf. (\ref{eq:32.11.1 Generalized Eqs. for maximization of Jmin})):
\begin{align}
\delta\mathcal{J}_{|1_{v}\rangle\rightarrow|d_{v}^{\ast}\rangle} & =a\delta_{1}-b\delta_{d_{v}^{\ast}},\label{eq:32.12 deltaJcool}\\
\delta\mathcal{J}_{|(k+1)_{v}\rangle\rightarrow|k_{v}\rangle} & =c_{k+1}\delta_{k+1}-d_{k}\delta_{k}.\label{eq:32.13 delta Jcatal}
\end{align}
If we can show that at least one of these variations must be negative,
it follows that the loop current corresponding to $\{p_{k}^{v}\}_{k=1}^{d_{v}^{\ast}}$
satisfies: 
\begin{align}
\mathcal{J}_{\textrm{min}} & =\textrm{min}\left\{ \mathcal{J}_{|1_{v}\rangle\rightarrow|d_{v}^{\ast}\rangle}+\delta\mathcal{J}_{|1_{v}\rangle\rightarrow|d_{v}^{\ast}\rangle},\right.\nonumber \\
 & \quad\quad\quad\left.\textrm{min}_{l\leq k\leq l'}\left(\mathcal{J}_{|(k+1)_{v}\rangle\rightarrow|k_{v}\rangle}+\delta\mathcal{J}_{|(k+1)_{v}\rangle\rightarrow|k_{v}\rangle}\right)\right\} \nonumber \\
 & <\bar{\mathcal{J}}_{\textrm{min}},\label{eq:32.14 Jmin<Jmin optimal}
\end{align}
where $\mathcal{J}_{|1_{v}\rangle\rightarrow|d_{v}^{\ast}\rangle}$
and $\mathcal{J}_{|(k+1)_{v}\rangle\rightarrow|k_{v}\rangle}$ are
given in Eqs. (\ref{eq:32.11.1 Generalized Eqs. for maximization of Jmin}). 

Let us see that this is indeed the case. We start by pointing out
two (mutually exclusive) possibilities that the signs of the variations
$\{\delta_{k}\}_{k=1}^{d_{v}^{\ast}}$ must adhere to. Either $\{\delta_{k}\leq0\}_{k=1}^{k'-1}$
or $\{\delta_{k}>0\}_{k=1}^{k'-1}$, for $k'\in\{1,2,...,d_{v}^{\ast}\}$.
The first (second) option means that $k'-1$ is the maximum $k$ such
that $\delta_{k}\leq0$ ($\delta_{k}>0$), and thus $\delta_{k'}>0$
($\delta_{k'}\leq0$). Note also that $k'=d_{v}^{\ast}+1$ is not
included, because it is not possible to have $\{\delta_{k}\leq0\}_{k=1}^{d_{v}^{\ast}}$
or $\{\delta_{k}>0\}_{k=1}^{d_{v}^{\ast}}$ due to the normalization
constraint $\sum_{k}p_{k}^{v}=\sum_{k}\bar{p}_{k}^{v}+\sum_{k}\delta_{k}=1+\sum_{k}\delta_{k}=1$.
The examination of both options leads to: 
\begin{enumerate}
\item If $\{\delta_{k}\leq0\}_{k=1}^{k'-1}$, $\delta_{k'}>0$ and consequently
$\delta\mathcal{J}_{|(k'+1)_{v}\rangle\rightarrow|k'_{v}\rangle}>0$
only if $\delta_{k'+1}>0$ (cf. Eq. (\ref{eq:32.13 delta Jcatal})).
Therefore, $\delta\mathcal{J}_{|(k'+2)_{v}\rangle\rightarrow|(k'+1)_{v}\rangle}>0$
only if $\delta_{k'+2}>0$, and, by recursively applying this argument,
$\delta\mathcal{J}_{|d_{v}^{\ast}\rangle\rightarrow|(d_{v}^{\ast}-1)_{v}\rangle}>0$
only if $\delta_{d_{v}^{\ast}}>0$. In this way, we conclude that
$\delta_{1}\leq0$ and $\delta_{d_{v}^{\ast}}>0$. Using Eq. (\ref{eq:32.12 deltaJcool}),
this implies that $\delta\mathcal{J}_{|1_{v}\rangle\rightarrow|d_{v}^{\ast}\rangle}<0$. 
\item If $\{\delta_{k}>0\}_{k=1}^{k'-1}$, $\delta_{k'}\leq0$ and consequently
$\delta\mathcal{J}_{|k'_{v}\rangle\rightarrow|(k'-1)_{v}\rangle}<0$,
according to (\ref{eq:32.13 delta Jcatal}). 
\end{enumerate}
This completes the proof of Eq. (\ref{eq:32.14 Jmin<Jmin optimal}). 

\subsubsection{Evaluation of $\{\bar{p}_{k}^{v}\}_{k=1}^{d_{v}^{\ast}}$ and corresponding
cooling current}

The usefulness of the general proof given before will become apparent
in the next section and in Appendix L. Now, we return to example of
optimal cooling and focus on the solution of Eqs. (\ref{eq:32.10 Eqs. for max. of Jmin explicit}).
To that end, we will obtain a sum of the form $\sum_{k=1}^{k'}g_{k}\mathcal{J}_{|(k+1)_{v}\rangle\rightarrow|k_{v}\rangle}$,
with weights $g_{k}$ chosen in such a way that only terms proportional
to $\bar{p}_{k'+1}^{v}$ and $\bar{p}_{1}^{v}$ survive. For example,
by defining $\zeta\equiv c/d$, we have that: 
\begin{equation}
\mathcal{J}_{|2_{v}\rangle\rightarrow|1_{v}\rangle}+\zeta\mathcal{J}_{|3_{v}\rangle\rightarrow|2_{v}\rangle}=c\zeta\bar{p}_{3}^{v}-d\bar{p}_{1}^{v}.\label{eq:32.15 idetity with first two catalyst currents}
\end{equation}
As we will see, this method will lead us to a closed solution for
all $\bar{p}_{k'}^{v}\in\{\bar{p}_{k}^{v}\}_{k=1}^{d_{v}^{\ast}}$. 

Let us prove by induction that weights $g_{k}=\zeta^{k-1}$ yield
\begin{equation}
\sum_{k=1}^{k'}\zeta^{k-1}\mathcal{J}_{|(k+1)_{v}\rangle\rightarrow|k_{v}\rangle}=c\zeta^{k'-1}\bar{p}_{k'+1}^{v}-d\bar{p}_{1}^{v}.\label{eq:32.16 identity with sum of catalyst currents}
\end{equation}
The validity for $k'=2$ is manifested in (\ref{eq:32.15 idetity with first two catalyst currents}).
Assuming that Eq. (\ref{eq:32.16 identity with sum of catalyst currents})
holds for $k'$, for $k'+1$ it holds that 
\begin{align}
\sum_{k=1}^{k'+1}\zeta^{k-1}\mathcal{J}_{|(k+1)_{v}\rangle\rightarrow|k_{v}\rangle} & =\sum_{k=1}^{k'}\zeta^{k-1}\mathcal{J}_{|(k+1)_{v}\rangle\rightarrow|k_{v}\rangle}\nonumber \\
 & \quad+\zeta^{k'}\mathcal{J}_{|(k'+2)_{v}\rangle\rightarrow|(k'+1)_{v}\rangle}\nonumber \\
 & =c\zeta^{k'-1}\bar{p}_{k'+1}^{v}-d\bar{p}_{1}^{v}\nonumber \\
 & \quad+\zeta^{k'}\left(c\bar{p}_{k'+2}^{v}-d\bar{p}_{k'+1}^{v}\right)\nonumber \\
 & =c\zeta^{k'}\bar{p}_{k'+2}^{v}-d\bar{p}_{1}^{v},\label{eq:32.17 prove for general sum of catalyst currents}
\end{align}
where $\mathcal{J}_{|(k'+2)_{v}\rangle\rightarrow|(k'+1)_{v}\rangle}$
is written as $c\bar{p}_{k'+2}^{v}-d\bar{p}_{k'+1}^{v}$ (cf. (\ref{eq:32.10 Eqs. for max. of Jmin explicit}))
in the second equality. This shows that Eq. (\ref{eq:32.16 identity with sum of catalyst currents})
is also valid for any $k'\geq2$. For $k'=1$ it is straightforward
to check that the correct form of $\mathcal{J}_{|2_{v}\rangle\rightarrow|1_{v}\rangle}$
is also recovered. 

By isolating $\bar{p}_{k'+1}^{v}$ from (\ref{eq:32.16 identity with sum of catalyst currents})
we obtain: 
\begin{align}
\bar{p}_{k'+1}^{v} & =\frac{\zeta^{1-k'}}{c}\left[d\bar{p}_{1}^{v}+\sum_{k=1}^{k'}\zeta^{k-1}\mathcal{J}_{|(k+1)_{v}\rangle\rightarrow|k_{v}\rangle}\right]\nonumber \\
 & =\frac{\zeta^{1-k'}}{c}\left[d\bar{p}_{1}^{v}+\left(\sum_{k=1}^{k'}\zeta^{k-1}\right)\mathcal{J}_{|1_{v}\rangle\rightarrow|d_{v}^{\ast}\rangle}\right]\nonumber \\
 & =\frac{\zeta^{1-k'}}{c}\left[d\bar{p}_{1}^{v}+\frac{1-\zeta^{k'}}{1-\zeta}(a\bar{p}_{1}^{v}-b\bar{p}_{d_{v}^{\ast}}^{v})\right],\label{eq:32.18 p(k'+1) in terms of p1 and pdv}
\end{align}
where Eq. (\ref{eq:32.9 Eqs. for maximization of Jmin}) is applied
in the second line and in the third line we evaluate the geometric
series $\sum_{k=1}^{k'}\zeta^{k-1}$. The expression for $\bar{p}_{d_{v}^{\ast}}^{v}$
can be readily obtained by setting $k'=d_{v}^{\ast}-1$ in (\ref{eq:32.18 p(k'+1) in terms of p1 and pdv}),
and performing a few algebraic manipulations: 
\begin{align}
\bar{p}_{d_{v}^{\ast}}^{v} & =\left(d+a\frac{1-\zeta^{d_{v}^{\ast}-1}}{1-\zeta}\right)\frac{\bar{p}_{1}^{v}}{c\zeta^{d_{v}^{\ast}-2}+b\frac{1-\zeta^{d_{v}^{\ast}-1}}{1-\zeta}}\nonumber \\
 & =\frac{\left(d(\zeta-1)+a(\zeta^{d_{v}^{\ast}-1}-1)\right)\bar{p}_{1}^{v}}{c\zeta^{d_{v}^{\ast}-2}(\zeta-1)+b(\zeta^{d_{v}^{\ast}-1}-1)}.\label{eq:32.19 pdv in terms of p1}
\end{align}
If $\bar{p}_{d_{v}^{\ast}}^{v}$ in (\ref{eq:32.19 pdv in terms of p1})
is inserted back into (\ref{eq:32.18 p(k'+1) in terms of p1 and pdv}),
we obtain all the eigenvalues $\{\bar{p}_{k}^{v}\}_{k=2}^{d_{v}^{\ast}}$
in terms of $\bar{p}_{1}^{v}$ and the coefficients $\{a,b;c,d\}$.
Finally, the eigenvalue $\bar{p}_{1}^{v}$ can be deduced via probability
conservation $\bar{p}_{1}^{v}+\sum_{k'=1}^{d_{v}^{\ast}-1}\bar{p}_{k'+1}^{v}=1$. 

As implied by Remark 10, the cooling currrent $\mathcal{J}_{|(i'+1)_{c}\rangle\rightarrow|i'_{c}\rangle}$
can be obtained from $\mathcal{J}_{|(i'+1)_{c}\rangle\rightarrow|i'_{c}\rangle}=\mathcal{J}_{|1_{v}\rangle\rightarrow|d_{v}^{\ast}\rangle}$,
and thus $\mathcal{J}_{|(i'+1)_{c}\rangle\rightarrow|i'_{c}\rangle}=a\bar{p}_{1}^{v}-b\bar{p}_{d_{v}^{\ast}}^{v}$
(cf. (\ref{eq:32.9 Eqs. for maximization of Jmin}) and (\ref{eq:32.10 Eqs. for max. of Jmin explicit})).
After setting $a=p_{2}^{c}p_{1}^{h}$ and $b=p_{1}^{c}p_{2}^{h}$
and substituting $\bar{p}_{d_{v}^{\ast}}^{v}$ as given in (\ref{eq:32.19 pdv in terms of p1}),
we find: 
\begin{align}
\mathcal{J}_{|(i'+1)_{c}\rangle\rightarrow|i'_{c}\rangle} & =\frac{ac\zeta^{d_{v}^{\ast}-2}(\zeta-1)-bd(\zeta-1)}{c\zeta^{d_{v}^{\ast}-2}(\zeta-1)+b(\zeta^{d_{v}^{\ast}-1}-1)}\bar{p}_{1}^{v}\nonumber \\
 & =\frac{(\zeta-1)\left(p_{2}^{c}p_{1}^{h}p_{1}^{h}\zeta^{d_{v}^{\ast}-2}-p_{1}^{c}p_{2}^{h}p_{2}^{h}\right)}{p_{1}^{h}\zeta^{d_{v}^{\ast}-2}(\zeta-1)+p_{1}^{c}p_{2}^{h}(\zeta^{d_{v}^{\ast}-1}-1)}\bar{p}_{1}^{v}\nonumber \\
 & =\frac{(\zeta-1)p_{2}^{h}\left(p_{2}^{c}\zeta^{d_{v}^{\ast}}-p_{1}^{c}\right)}{\zeta^{d_{v}^{\ast}-1}(\zeta-1)+p_{1}^{c}(\zeta^{d_{v}^{\ast}-1}-1)}\bar{p}_{1}^{v}.\label{eq:32.20 Jcool(max)}
\end{align}
Since this is the maximum value of the cooling current, as per Remark
10, we express it as $J_{\textrm{cool}}^{\textrm{max}}$ in the main
text. 

\subsection{Maximization of the loop current for more general catalytic unitaries}

We conclude this appendix by illustrating a first scenario where the
optimality proof concerning Eqs. (\ref{eq:32.11.1 Generalized Eqs. for maximization of Jmin})
is useful. Although here we will obtain currents with coefficients
$c_{k+1}$ and $d_{k}$ that do not depend on $k$ (i.e. $c_{k+1}=c$
and $d_{k}=d$), the resulting set $\{a,b;c,d\}$ characterizes swap
currents $\{\mathcal{J}_{|1_{v}\rangle\rightarrow|d_{v}^{\ast}\rangle},\{\mathcal{J}_{|(k+1)_{v}\rangle\rightarrow|k_{v}\rangle}\}_{k=1}^{d_{v}^{\ast}-1}$
related to evolutions much more general than the unitary $\mathcal{U}_{1,1,d_{v}^{\ast}-1}\oplus\left(\oplus_{k=1}^{d_{v}^{\ast}-1}\mathcal{V}_{|1_{h}(k+1)_{v}\rangle\leftrightarrow|2_{h}k_{v}\rangle}\right)$
considered in Appendix I2. For these evolutions, the proof given in
I2 implies that the corresponding loop current is maximized by eigenvalues
$\{\bar{p}_{k}^{v}\}$ that solve Eqs. (\ref{eq:32.11.1 Generalized Eqs. for maximization of Jmin}).
While the motivation for this maximization may not always be evident,
it is of particular interest for catalytic and cooling transformations,
as highlighted in Remark 10. 

In what follows we consider catalytic unitaries where the restoring
unitary satisfies Eq. (\ref{eq:32.1 addition of Vres}) and the generalization
indicated by Remark 9 also takes place. By setting $l=1$ and $l'=d_{v}^{\ast}-1$,
we thus consider unitaries \textit{$U_{d_{v}^{\ast}}\oplus\left(\oplus_{k=1}^{d_{v}^{\ast}-1}V_{k}\right)$,
}where $U_{d_{v}^{\ast}}=\oplus_{\psi,\varphi\textrm{ free}}U_{|\psi\rangle\leftrightarrow|\varphi\rangle}$
acts on $\mathcal{H}_{c}\otimes\mathcal{H}_{h}\otimes\textrm{span}\{|1_{v}\rangle,|d_{v}^{\ast}\rangle\}$
and $V_{k}=\oplus_{n,N}V_{|n_{ch}(k+1)_{v}\rangle\leftrightarrow|N_{ch}k_{v}\rangle}$. 

\textbf{Proposition 4}. \textit{Let $U_{d_{v}^{\ast}}\oplus\left(\oplus_{k=1}^{d_{v}^{\ast}-1}V_{k}\right)$
be a unitary applied on }${\normalcolor \rho_{ch}\otimes\rho_{v}^{\ast}}$,\textit{
where $\rho_{ch}$ is a generic state of the compound $ch$ and $\rho_{v}^{\ast}=\sum_{k=1}^{d_{v}^{\ast}-1}p_{k}^{v}|k_{v}\rangle\langle k_{v}|$
is a full-rank catalyst state. Under the assumption that $U_{d_{v}^{\ast}}\oplus\left(\oplus_{k=1}^{d_{v}^{\ast}-1}V_{k}\right)$
generates a loop $\{J_{|1_{v}\rangle\rightarrow|d_{v}^{\ast}\rangle}\}\cup\{J_{|(k+1)_{v}\rangle\rightarrow|k_{v}\rangle}\}_{k=1}^{d_{v}^{\ast}-1}$,
the loop current} $\mathcal{J}_{\textrm{min}}=\textrm{min}\{\mathcal{J}_{|1_{v}\rangle\rightarrow|d_{v}^{\ast}\rangle},\mathcal{J}_{|(k+1)_{v}\rangle\rightarrow|k_{v}\rangle}\}_{k=1}^{d_{v}^{\ast}-1}$\textit{
is maximized by catalyst eigenvalues $\{\bar{p}_{k}^{v}\}_{k=1}^{d_{v}^{\ast}-1}$
that satisfy Eqs. (\ref{eq:32.11.1 Generalized Eqs. for maximization of Jmin}).
Accordingly, the optimal unitary is a permutation that we denote as
$\mathcal{U}_{d_{v}^{\ast}}\oplus\left(\oplus_{k=1}^{d_{v}^{\ast}-1}\mathcal{V}_{k}\right)$,
with }
\begin{align}
\mathcal{U}_{d_{v}^{\ast}} & =\oplus_{\psi,\varphi\textrm{ free}}\mathcal{U}_{|\varphi_{ch}1_{v}\rangle\leftrightarrow|\psi_{ch}d_{v}^{\ast}\rangle},\label{eq:32.21 permutation U_d*v}\\
\mathcal{V}_{k} & =\oplus_{n,N}\mathcal{V}_{|n_{ch}(k+1)_{v}\rangle\leftrightarrow|N_{ch}k_{v}\rangle},\label{eq:32.22 permutations V_k}
\end{align}
\textit{for $k\in\{1,2,...,d_{v}^{\ast}-1\}$. }

\textit{Proof}. Taking into account the proof given in Appendix I2,
we only need to express $\mathcal{J}_{|1_{v}\rangle\rightarrow|d_{v}^{\ast}\rangle}$
and $\mathcal{J}_{|(k+1)_{v}\rangle\rightarrow|k_{v}\rangle}$ as
in Eqs. (\ref{eq:32.11.1 Generalized Eqs. for maximization of Jmin}).
Specifically, we have that, if Eqs. (\ref{eq:32.21 permutation U_d*v})
and (\ref{eq:32.22 permutations V_k}) hold, 

\begin{align}
\mathcal{J}_{|1_{v}\rangle\rightarrow|d_{v}^{\ast}\rangle} & =\sum_{\psi,\varphi\textrm{ free}}\mathcal{J}_{|\varphi_{ch}1_{v}\rangle\rightarrow|\psi_{ch}d_{v}^{\ast}\rangle}\nonumber \\
 & =\sum_{\psi,\varphi\textrm{ free}}\left(p_{\varphi}^{ch}\bar{p}_{1}^{v}-p_{\psi}^{ch}\bar{p}_{d_{v}^{\ast}}^{v}\right)\nonumber \\
 & =\left(\sum_{\varphi\textrm{ free}}p_{\varphi}^{ch}\right)\bar{p}_{1}^{v}-\left(\sum_{\psi\textrm{ free}}p_{\psi}^{ch}\right)\bar{p}_{d_{v}^{\ast}}^{v}\nonumber \\
 & \equiv a\bar{p}_{1}^{v}-b\bar{p}_{d_{v}^{\ast}}^{v},\label{eq:32..23 J|1v>-->|d*v> for perm U_d*v}
\end{align}
where the first line obeys (\ref{eq:111.2 swap catalyst current as sum of swap global currents}),
and 
\begin{align}
\mathcal{J}_{|(k+1)_{v}\rangle\rightarrow|k_{v}\rangle} & =\sum_{n,N}\mathcal{J}_{|n_{ch}(k+1)_{v}\rangle\rightarrow|N_{ch}k_{v}\rangle}\nonumber \\
 & =\sum_{n,N}\left(p_{n}^{ch}\bar{p}_{k+1}^{v}-p_{N}^{ch}\bar{p}_{k}^{v}\right)\nonumber \\
 & =\left(\sum_{n}p_{n}^{ch}\right)\bar{p}_{k+1}^{v}-\left(\sum_{N}p_{N}^{ch}\right)\bar{p}_{k}^{v}\nonumber \\
 & \equiv c\bar{p}_{k+1}^{v}-d\bar{p}_{k}^{v},\label{eq:32.24  J|(k+1)v>-->|kv> for perm V_k}
\end{align}
where the first line obeys (\ref{eq:32.3 swap catalyst current}). 

Note that in this case the coefficients $c$ and $d$ are independent
of $k$ (i.e. $c_{k+1}=c$ and $d_{k}=d$). Furthermore, note also
that Eqs. (\ref{eq:32.21 permutation U_d*v}) and (\ref{eq:32.22 permutations V_k})
recover the unitary for optimal cooling $\mathcal{U}_{1,1,d_{v}^{\ast}-1}\oplus\left(\oplus_{k=1}^{d_{v}^{\ast}-1}\mathcal{V}_{|1_{h}(k+1)_{v}\rangle\leftrightarrow|2_{h}k_{v}\rangle}\right)$,
by setting $\mathcal{U}_{d_{v}^{\ast}}=\mathcal{U}_{|2_{c}1_{h}1_{v}\rangle\leftrightarrow|1_{c}d_{h}d_{v}^{\ast}\rangle}$
and $\mathcal{V}_{k}=\mathcal{V}_{|1_{h}(k+1)_{v}\rangle\leftrightarrow|2_{h}k_{v}\rangle}$. 

\textbf{Remark 11}. Since the coefficients $c$ and $d$ in (\ref{eq:32..23 J|1v>-->|d*v> for perm U_d*v})
and (\ref{eq:32.24  J|(k+1)v>-->|kv> for perm V_k}) are independent
of $k$, the optimal eigenvalues $\{\bar{p}_{k}^{v}\}_{k}$ are provided
by Eqs. (\ref{eq:32.18 p(k'+1) in terms of p1 and pdv}) and (\ref{eq:32.19 pdv in terms of p1}).
Moreover, if $\mathcal{J}_{|1_{v}\rangle\rightarrow|d_{v}^{\ast}\rangle}=\mathcal{J}_{|(i'+1)_{c}\rangle\rightarrow|i'_{c}\rangle}$
(which occurs e.g. if $\mathcal{U}_{d_{v}^{\ast}}=\mathcal{U}_{|i'_{c}d_{h}d_{v}^{\ast}\rangle\leftrightarrow|(i'+1)_{c}1_{h}1_{v}\rangle}$),
the resulting current $\mathcal{J}_{\textrm{min}}$ satisfies the
first line of (\ref{eq:32.20 Jcool(max)}). 

\section{PROOF OF THEOREM 3 AND EXAMPLE OF CATALYTIC COOLING ENHANCEMENT}

To prove Theorem 3 we shall employ the following lemma, which provides
an optimal unitary to cool the cold qubit using only the hot object.
This unitary maximizes the ground population of the cold qubit, keeping
in mind that reducing the energy of a two-level system is equivalent
to increase its ground population. In addition, we stress that in
this setting the state $\rho_{c}\otimes\rho_{h}$ is non-passive with
respect to $H_{c}$. 

\textbf{Lemma 4 (maximum cooling of a qubit using only the hot object)}.
\textit{If $\rho_{c}=\sum_{i=1}^{2}p_{i}^{c}|i_{c}\rangle\langle i_{c}|$
is a general state of a qubit and $\rho_{h}=\sum_{j=1}^{d_{h}}p_{j}^{h}|j_{h}\rangle\langle j_{h}|$
is a general state of a hot object, the ground population of the cold
qubit is maximized by a permutation }
\begin{equation}
\tilde{\mathcal{U}}\equiv\oplus_{j=1}^{j_{\textrm{max}}}\mathcal{U}{}_{|1_{c}(d_{h}-j_{\textrm{max}}+j)_{h}\rangle\leftrightarrow|2_{c}(j_{\textrm{max}}-j+1)_{h}\rangle},\label{eq:33 permutation for maximum cooling of a qubit}
\end{equation}
\textit{where} $j_{\textrm{max}}$\textit{ is the maximum value of
$j$ such that $p_{2}^{c}p_{j}^{h}>p_{1}^{c}p_{d_{h}-j+1}^{h}$. }

\textit{Proof}. We define the sets of indices $A=\{j\}_{1\leq j\leq j_{\textrm{max}}}$,
$A'=\{j\}_{j_{\textrm{max}}+1\leq j\leq d_{h}}$, $B=\{j\}_{d_{h}-j_{\textrm{max}}+1\leq j\leq d_{h}}$,
and $B'=\{j\}_{1\leq j\leq d_{h}-j_{\textrm{max}}}$. From these definitions,
we note immediately that $A'$ is the complement of $A$, and $B'$
is the complement of $B$. Accordingly, the state $\rho_{c}\otimes\rho_{h}$
can be expressed as 
\begin{align}
\rho_{c}\otimes\rho_{h} & =\sum_{j\in B}p_{1}^{c}p_{j}^{h}|1_{c}j_{h}\rangle\langle1_{c}j_{h}|+\sum_{j\in B'}p_{1}^{c}p_{j}^{h}|1_{c}j_{h}\rangle\langle1_{c}j_{h}|\nonumber \\
 & \quad+\sum_{j\in A}p_{2}^{c}p_{j}^{h}|2_{c}j_{h}\rangle\langle2_{c}j_{h}|+\sum_{j\in A'}p_{2}^{c}p_{j}^{h}|2_{c}j_{h}\rangle\langle2_{c}j_{h}|.\label{eq:33.1 rho_c x rho_h with A,A',B,B'}
\end{align}
According to (\ref{eq:33 permutation for maximum cooling of a qubit}),
the effect of $\tilde{\mathcal{U}}$ is to swap the eigenstates in
the set $\{|1_{c}j_{h}\rangle\}_{j\in B}$ with the eigenstates in
the set $\{|2_{c}j_{h}\rangle\}_{j\in A}$. For example, $|1_{c}(d_{h}-j_{\textrm{max}}+1)_{h}\rangle$
is swapped with $|2_{c}(j_{\textrm{max}})_{h}\rangle$, $|1_{c}(d_{h}-j_{\textrm{max}}+2)_{h}\rangle$
is swapped with $|2_{c}(j_{\textrm{max}}-1)_{h}\rangle$, and so forth,
up to $|1_{c}d_{h}\rangle$ is swapped with $|2_{c}1_{h}\rangle$.
Since this is equivalent to exchange the eigenvalues corresponding
to $\{|1_{c}j_{h}\rangle\}_{j\in B}$ with those corresponding to
$\{|2_{c}j_{h}\rangle\}_{j\in A}$, the transformed state $\tilde{\mathcal{U}}(\rho_{c}\otimes\rho_{h})\tilde{\mathcal{U}}^{\dagger}$
reads 
\begin{align}
\tilde{\mathcal{U}}(\rho_{c}\otimes\rho_{h})\tilde{\mathcal{U}}^{\dagger} & =\sum_{j\in A}p_{2}^{c}p_{j}^{h}|1_{c}j_{h}\rangle\langle1_{c}j_{h}|+\sum_{j\in B'}p_{1}^{c}p_{j}^{h}|1_{c}j_{h}\rangle\langle1_{c}j_{h}|\nonumber \\
 & \quad+\sum_{j\in B}p_{1}^{c}p_{j}^{h}|2_{c}j_{h}\rangle\langle2_{c}j_{h}|+\sum_{j\in A'}p_{2}^{c}p_{j}^{h}|2_{c}j_{h}\rangle\langle2_{c}j_{h}|.\label{eq:33.2 rho_c x rho_h evolved with optimal permutation}
\end{align}

Let us see now that $\tilde{\mathcal{U}}(\rho_{c}\otimes\rho_{h})\tilde{\mathcal{U}}^{\dagger}$
is passive with respect to $|2_{c}\rangle\langle2_{c}|$, which amounts
to the impossibility to further reduce $\bigl\langle|2_{c}\rangle\langle2_{c}|\bigr\rangle$
via unitary evolutions (i.e. $\Delta\bigl\langle|2_{c}\rangle\langle2_{c}|\bigr\rangle\geq0$
if $\tilde{\mathcal{U}}(\rho_{c}\otimes\rho_{h})\tilde{\mathcal{U}}^{\dagger}$
is evolved unitarily). Since $\Delta\bigl\langle|2_{c}\rangle\langle2_{c}|\bigr\rangle=-\Delta\bigl\langle|1_{c}\rangle\langle1_{c}|\bigr\rangle$
(by probability conservation), this also means that $\Delta\bigl\langle|1_{c}\rangle\langle1_{c}|\bigr\rangle\leq0$
for any unitary applied on $\tilde{\mathcal{U}}(\rho_{c}\otimes\rho_{h})\tilde{\mathcal{U}}^{\dagger}$,
and consequenly this state maximizes the ground population $\bigl\langle|1_{c}\rangle\langle1_{c}|\bigr\rangle.$

By definition of $j_{\textrm{max}}$, $p_{2}^{c}p_{j_{\textrm{max}}}^{h}>p_{1}^{c}p_{d_{h}-j_{\textrm{max}}+1}^{h}$.
Taking into account the ordering $p_{j}^{h}\geq p_{j+1}^{h}$, and
the definitions of $A$ and $B$, the inequality $p_{2}^{c}p_{j_{\textrm{max}}}^{h}>p_{1}^{c}p_{d_{h}-j_{\textrm{max}}+1}^{h}$
can be rewritten as 
\begin{equation}
\textrm{min}_{j\in A}p_{2}^{c}p_{j}^{h}>\textrm{max}_{j\in B}p_{1}^{c}p_{j}^{h}.\label{eq:33.3 comp bet A and B}
\end{equation}
Similarly, since $j=j_{\textrm{max}}$ is the maximum value such that
$p_{2}^{c}p_{j}^{h}>p_{1}^{c}p_{d_{h}-j+1}^{h}$, we also have that
$p_{2}^{c}p_{j_{\textrm{max}}+1}^{h}\leq p_{1}^{c}p_{d_{h}-j_{\textrm{max}}}^{h}$,
which is equivalent to 
\begin{equation}
\textrm{min}_{j\in B'}p_{1}^{c}p_{j}^{h}\geq\textrm{max}_{j\in A'}p_{2}^{c}p_{j}^{h}.\label{eq:33.4 comp bet B' and A'}
\end{equation}
Finally, the ordering $p_{j}^{h}\geq p_{j+1}^{h}$ and the definitions
of the sets $A$, $A'$, $B$, and $B'$, lead to: 
\begin{align}
\textrm{min}_{j\in A}p_{2}^{c}p_{j}^{h} & \geq\textrm{max}_{j\in A'}p_{2}^{c}p_{j}^{h},\label{eq:33.5 comp bet A and A'}\\
\textrm{min}_{j\in B'}p_{1}^{c}p_{j}^{h} & \geq\textrm{max}_{j\in B}p_{1}^{c}p_{j}^{h}.\label{eq:33.6 comp bet B' and B}
\end{align}
Equations (\ref{eq:33.3 comp bet A and B})-(\ref{eq:33.6 comp bet B' and B})
tell us that the smallest eigenvalue in the first line of (\ref{eq:33.2 rho_c x rho_h evolved with optimal permutation})
must be larger or equal than the largest eigenvalue in the second
line. Since $|2_{c}\rangle\langle2_{c}|=|2_{c}\rangle\langle2_{c}|\otimes\mathbb{I}_{h}$
has eigenvalue 0 for the eigenstates $\{|1_{c}j_{h}\rangle\}_{j}$,
and eigenvalue 1 for the eigenstates $\{|2_{c}j_{h}\rangle\}_{j}$
we conclude that $\tilde{\mathcal{U}}(\rho_{c}\otimes\rho_{h})\tilde{\mathcal{U}}^{\dagger}$
is passive with respect to $|2_{c}\rangle\langle2_{c}|$. 

\subsection{Proof of Theorem 3}

In the following we shall assume that the eigenvalues of the catalyst
satisfy Eq. (\ref{eq:29.6 catalyst eigenvalues for Theorem 2-1}).
That is, $\frac{p_{k}}{p_{k+1}}=\mu>1$, for $1\leq k\leq d_{v}^{\ast}-1$.
Moreover, the initial total state is $\rho=\rho_{c}\otimes\rho_{h}\otimes\rho_{v}$. 

Given the effect of $\tilde{\mathcal{U}}$, previously described,
$\tilde{\mathcal{U}}\otimes\mathbb{I}_{v}$ acts on the subspace 
\begin{equation}
\tilde{\mathcal{H}}=\textrm{span}\left(\{|2_{c}j_{h}\rangle\}_{j\in A}\cup\{|1_{c}j_{h}\rangle\}_{j\in B}\right)\otimes\mathcal{H}_{v}.\label{eq:33.7 subspace H tilda}
\end{equation}
Now we want to show that, if the conditions of Theorem 3 hold, we
can construct a catalytic and cooling unitary $U:\tilde{\mathcal{H}'}\rightarrow\tilde{\mathcal{H}'}$,
where 
\begin{equation}
\tilde{\mathcal{H}'}=\textrm{span}\left(\{|2_{c}j_{h}\rangle\}_{j\in A'}\cup\{|1_{c}j_{h}\rangle\}_{j\in B'}\right)\otimes\mathcal{H}_{v}.\label{eq:33.8 subspace H' tilda}
\end{equation}
Taking into account that $\tilde{\mathcal{H}}$ and $\tilde{\mathcal{H}'}$
are orthogonal, the direct sum $\tilde{\mathcal{U}}\oplus U$ yields
the total variation (cf. Eqs. (\ref{eq:102 delta<O>}) and (\ref{eq:103 delta_alpha<O>}))
$\Delta\left\langle H_{c}\right\rangle =\textrm{Tr}\left[H_{c}\left(\tilde{\mathcal{U}}\oplus U\rho\tilde{\mathcal{U}}^{\dagger}\oplus U^{\dagger}-\rho\right)\right]$:
\begin{equation}
\Delta\left\langle H_{c}\right\rangle =\Delta_{\tilde{\mathcal{U}}}\left\langle H_{c}\right\rangle +\Delta_{U}\left\langle H_{c}\right\rangle <\Delta_{\tilde{\mathcal{U}}}\left\langle H_{c}\right\rangle ,\label{eq:33.9 Delta<Hc> with U tilda and U}
\end{equation}
where $\Delta_{\tilde{\mathcal{U}}}\left\langle H_{c}\right\rangle =\textrm{Tr}\left[H_{c}\left(\tilde{\mathcal{U}}\rho_{c}\otimes\rho_{h}\tilde{\mathcal{U}}^{\dagger}-\rho_{c}\otimes\rho_{h}\right)\right]$
and $\Delta_{U}\left\langle H_{c}\right\rangle =\textrm{Tr}\left[H_{c}\left(U\rho U^{\dagger}-\rho\right)\right]$. 

As a preliminary result that will be used later, let us check that
\begin{equation}
j_{\textrm{max}}\leq d_{h}/2.\label{eq:40 jmax<=00003Ddh/2}
\end{equation}
If $j_{\textrm{max}}\geq d_{h}/2+1$, we have that $d_{h}-j_{\textrm{max}}+1\leq d_{h}/2<j_{\textrm{max}}$
and consequently $p_{j_{\textrm{max}}}^{h}\leq p_{d_{h}-j_{\textrm{max}}+1}^{h}$.
This implies that $p_{2}^{c}p_{j_{\textrm{max}}}^{h}\leq p_{1}^{c}p_{d_{h}-j_{\textrm{max}}+1}^{h}$,
in contradiction with the definition of $j_{\textrm{max}}$. 

\subsubsection{Proof for the case $p_{1}^{h}>p_{d_{h}/2}^{h}$}

Let us first provide the conditions for a cooling unitary $U_{\textrm{cool}}:\tilde{\mathcal{H}'}\rightarrow\tilde{\mathcal{H}'}$.
Since $\mu>1$, for sufficiently large $d_{v}$ we have that 
\begin{equation}
\mu^{d_{v}-1}=\frac{p_{1}^{v}}{p_{d_{v}}^{v}}>\frac{p_{1}^{c}p_{d_{h}-j_{\textrm{max}}}^{h}}{p_{2}^{c}p_{j_{\textrm{max}}+1}^{h}}.\label{eq:40.1 Condition for cooling unitary}
\end{equation}
In this way, the unitary $U_{\textrm{cool}}=U_{|2_{c}(j_{\textrm{max}}+1)_{h}1_{v}\rangle\leftrightarrow|1_{c}(d_{h}-j_{\textrm{max}})_{h}d_{v}\rangle}$
generates a cooling current $J_{|2_{c}\rangle\rightarrow|1_{c}\rangle}$,
and a local current $J_{|1_{v}\rangle\rightarrow|d_{v}\rangle}=J_{|2_{c}\rangle\rightarrow|1_{c}\rangle}$.
Moreover, both $|2_{c}(j_{\textrm{max}}+1)_{h}\rangle$ and $|1_{c}(d_{h}-j_{\textrm{max}})_{h}\rangle$
belong to $\textrm{span}\left(\{|2_{c}j_{h}\rangle\}_{j\in A'}\cup\{|1_{c}j_{h}\rangle\}_{j\in B'}\right)$,
which guarantees that $U_{\textrm{cool}}$ acts on $\tilde{\mathcal{H}'}$. 

Our next step is to obtain a chain $\mathbf{ch}_{|d_{v}\rangle\rightarrow|1_{v}\rangle}$
using a proper restoring unitary. From (\ref{eq:40 jmax<=00003Ddh/2})
and the ordering $p_{j}^{h}\geq p_{j+1}^{h}$ we have that $p_{d_{h}/2}^{h}\geq p_{d_{h}-j_{\textrm{max}}}^{h}$.
Hence, $p_{1}^{h}>p_{d_{h}-j_{\textrm{max}}}^{h}$. If $\mu>1$ is
sufficiently small, we can use the inequality $p_{1}^{h}>p_{d_{h}-j_{\textrm{max}}}^{h}$
to enforce the condition 
\begin{equation}
\frac{p_{1}^{c}p_{1}^{h}}{p_{1}^{c}p_{d_{h}-j_{\textrm{max}}}^{h}}>\frac{p_{k}^{v}}{p_{k+1}^{v}}=\mu.\label{eq:40.2 conditions for restoring unitary}
\end{equation}
This implies that the partial swaps in the direct sum
\begin{align}
V_{B'} & =\oplus_{k=1}^{d_{v}-1}V_{|1_{c}1_{h}(k+1)_{v}\rangle\leftrightarrow|1_{c}(d_{h}-j_{\textrm{max}})_{h}k_{v}\rangle}\nonumber \\
 & =|1_{c}\rangle\langle1_{c}|\otimes\left(\oplus_{k=1}^{d_{v}-1}V_{|1_{h}(k+1)_{v}\rangle\leftrightarrow|(d_{h}-j_{\textrm{max}})_{h}k_{v}\rangle}\right)\label{eq:40.3 VB'}
\end{align}
generates the desired chain $\mathbf{ch}_{|d_{v}\rangle\rightarrow|1_{v}\rangle}=\{J_{|(k+1)_{v}\rangle\rightarrow|k_{v}\rangle}\}_{k=1}^{d_{v}-1}$.
In addition, all the eigenstates ``connected'' by these partial
swaps belong to $\textrm{span}\{|1_{c}j_{h}\rangle\}_{j\in B'}\otimes\mathcal{H}_{v}$,
and therefore $V_{B'}$ also acts on $\tilde{\mathcal{H}'}$ (cf.
Remark 2). 

Noting that the eigenstates connected in Eq. (\ref{eq:40.3 VB'})
are orthogonal to those connected by $U_{\textrm{cool}}$, $U_{\textrm{cool}}$
and $V_{B'}$ act on orthogonal subspaces of $\tilde{\mathcal{H}'}$.
Hence, we can construct a direct sum $U=U_{\textrm{cool}}\oplus V_{B'}$,
with a corresponding loop $\mathbf{ch}_{|d_{v}\rangle\rightarrow|1_{v}\rangle}\cup\{J_{|1_{v}\rangle\rightarrow|d_{v}\rangle}\}$.
This loop can be made uniform through a tuning of swap intensities
analogous to that used in the sufficiency proof of Theorem 1. That
is, by choosing intensities such that $J_{|1_{v}\rangle\rightarrow|d_{v}\rangle}=J_{|(k+1)_{v}\rangle\rightarrow|k_{v}\rangle}=\mathcal{J}_{\textrm{min}}$,
where $\mathcal{J}_{\textrm{min}}$ is the minimum swap current in
the set $\{\mathcal{J}_{|1_{v}\rangle\rightarrow|d_{v}\rangle},\mathcal{J}_{|(k+1)_{v}\rangle\rightarrow|k_{v}\rangle}\}_{k}$.
This guarantees that the transformation $\rho\rightarrow U\rho U^{\dagger}$
is catalytic, as per Remark 7. 

Since $V_{B'}$ is a controlled unitary where the cold object is the
control, it does not change the average energy of the cold object.
Therefore, $\Delta_{U}\left\langle H_{c}\right\rangle =\Delta_{\textrm{cool}}\left\langle H_{c}\right\rangle +\Delta_{V_{B'}}\left\langle H_{c}\right\rangle =\Delta_{\textrm{cool}}\left\langle H_{c}\right\rangle $.
In this way, $\Delta_{U}\left\langle H_{c}\right\rangle $ in Eq.
(\ref{eq:33.9 Delta<Hc> with U tilda and U}) explicitly reads $\Delta_{U}\left\langle H_{c}\right\rangle =J_{|2_{c}\rangle\rightarrow|1_{c}\rangle}(\varepsilon_{1}^{c}-\varepsilon_{2}^{c})<0$. 

\subsubsection{Proof for the case $p_{d_{h}/2+1}^{h}>p_{d_{h}}^{h}$}

The proof is very similar to the previous one, with the only relevant
difference being related to the construction of the restoring unitary.
From Eq. (\ref{eq:40 jmax<=00003Ddh/2}), it follows that $j_{\textrm{max}}+1\leq d_{h}/2+1$
and therefore $p_{2}^{c}p_{j_{\textrm{max}}+1}^{h}\geq p_{2}^{c}p_{d_{h}/2+1}^{h}$.
In combination with the hypothesis $p_{d_{h}/2+1}^{h}>p_{d_{h}}^{h}$,
we obtain the inequality $p_{2}^{c}p_{j_{\textrm{max}}+1}^{h}>p_{d_{h}}^{h}$,
which can be applied to enforce the condition (for $\mu$ sufficiently
small): 
\begin{equation}
\frac{p_{2}^{c}p_{j_{\textrm{max}}+1}^{h}}{p_{2}^{c}p_{d_{h}}^{h}}>\frac{p_{k}^{v}}{p_{k+1}^{v}}=\mu.\label{eq:40.4 condition for restoring unitary VA'}
\end{equation}
Therefore, the partial swaps in the direct sum 

\begin{align}
V_{A'} & =\oplus_{k=1}^{d_{v}-1}V_{|2_{c}(j_{\textrm{max}}+1)_{h}(k+1)_{v}\rangle\leftrightarrow|2_{c}d_{h}k_{v}\rangle}\nonumber \\
 & =|2_{c}\rangle\langle2_{c}|\otimes\left(\oplus_{k=1}^{d_{v}-1}V_{|(j_{\textrm{max}}+1)_{h}(k+1)_{v}\rangle\leftrightarrow|d_{h}k_{v}\rangle}\right)\label{eq:40.5 VA'}
\end{align}
generate a chain $\mathbf{ch}_{|d_{v}\rangle\rightarrow|1_{v}\rangle}$.
Since these partial swaps connect eigenstates in $\textrm{span}\{|2_{c}j_{h}\rangle\}_{j\in A'}\otimes\mathcal{H}_{v}$,
$V_{A'}$ acts on $\tilde{\mathcal{H}'}$.

By choosing $d_{v}$ large enough, Eq. (\ref{eq:40.1 Condition for cooling unitary})
provides the same cooling unitary of the previous case, i.e. $U_{\textrm{cool}}=U_{|2_{c}(j_{\textrm{max}}+1)_{h}1_{v}\rangle\leftrightarrow|1_{c}(d_{h}-j_{\textrm{max}})_{h}d_{v}\rangle}$.
Once again, it can be readily checked that $U_{\textrm{cool}}$ and
$V_{A'}$ act on orthogonal subspaces of $\tilde{\mathcal{H}'}$.
Given that $V_{A'}$ does not change the average energy of the cold
qubit, $U=U_{\textrm{cool}}\oplus V_{A'}$ also gives rise to a catalytic
and cooling transformation that increases the cooling generated by
$\tilde{\mathcal{U}}$ (cf. (\ref{eq:33.9 Delta<Hc> with U tilda and U}))
in the amount $\Delta_{U}\left\langle H_{c}\right\rangle =\Delta_{\textrm{cool}}\left\langle H_{c}\right\rangle <0$. 

\subsection{Example of catalytic cooling enhancement with $d_{h}$ odd}

As explained in the main text, the optimal cooling of a qubit using
a three-level hot object leads to a passive state

\begin{align}
\sigma_{ch} & =|1_{c}\rangle\langle1_{c}|\otimes\left(\sum_{j=1}^{2}p_{1}^{c}p_{j}^{h}|j_{h}\rangle\langle j_{h}|+p_{2}^{c}p_{1}^{h}|3_{h}\rangle\langle3_{h}|\right)\nonumber \\
 & \quad+|2_{c}\rangle\langle2_{c}|\otimes\left(p_{1}^{c}p_{3}^{h}|1_{h}\rangle\langle1_{h}|+\sum_{j=2}^{3}p_{2}^{c}p_{j}^{h}|j_{h}\rangle\langle j_{h}|\right),\label{eq:40.6 passive state after optimal cooling}
\end{align}
where $\{p_{i}^{c}\}_{i=1}^{2}$ and $\{p_{j}^{h}\}_{j=1}^{3}$ are
respectively the eigenvalues of the initial states $\rho_{c}$ and
$\rho_{h}$. Hence, we can focus on the additional cooling provided
by a proper catalytic transformation $\sigma_{ch}\otimes\rho_{v}\overset{\textrm{CC}}{\longrightarrow}U\sigma_{ch}\otimes\rho_{v}U^{\dagger}$.
Since we consider a two-level catalyst, we will first identify a unitary
that generates a loop using a state $\rho_{v}^{\ast}=\sum_{k=1}^{d_{v}^{\ast}}p_{k}^{v}|k_{v}\rangle\langle k_{v}|=\sum_{k=1}^{2}p_{k}^{v}|k_{v}\rangle\langle k_{v}|$. 

\subsubsection{Unitary for the generation of the loop }

Under the degeneracy condition $\varepsilon_{1}^{h}=\varepsilon_{2}^{h}$,
we have that $p_{1}^{h}=p_{2}^{h}=\frac{e^{-\beta_{h}\varepsilon_{1}^{h}}}{2e^{-\beta_{h}\varepsilon_{1}^{h}}+e^{-\beta_{h}\varepsilon_{3}^{h}}}$.
Therefore, the eigenvalues of $\sigma_{ch}$ corresponding to the
eigenstates $|1_{c}3_{h}\rangle$ and $|2_{c}2_{h}\rangle$ (cf. Eq.
(\ref{eq:40.6 passive state after optimal cooling})) also satisfy
$p_{2}^{c}p_{1}^{h}=p_{2}^{c}p_{2}^{h}$. This condition allows us
to ``couple'' the eigenstates $|1_{c}3_{h}\rangle$ and $|2_{c}2_{h}\rangle$
to the catalyst eigenstates via a partial swap that generates a cooling
current for any $p_{1}^{v}$ and $p_{2}^{v}$ such that $p_{1}^{v}>p_{2}^{v}$.
Specifically, for $p_{1}^{v}>p_{2}^{v}$ the partial swap $U_{|2_{c}2_{h}1_{v}\rangle\leftrightarrow|1_{c}3_{h}2_{v}\rangle}$
produces a current $J_{|2_{c}\rangle\rightarrow|1_{c}\rangle}$ proportional
to $p_{2}^{c}p_{2}^{h}p_{1}^{v}-p_{2}^{c}p_{1}^{h}p_{2}^{v}=p_{2}^{c}p_{2}^{h}(p_{1}^{v}-p_{2}^{v})>0$.
The corresponding local current $J_{|1_{v}\rangle\rightarrow|2_{v}\rangle}$
is also positive. 

To complete the loop we only need to find a positive current $J_{|2_{v}\rangle\rightarrow|1_{v}\rangle}$.
As we show next, any of the partial swaps $V_{|1_{c}1_{h}2_{v}\rangle\leftrightarrow|1_{c}3_{h}1_{v}\rangle}$
or $V_{|2_{c}2_{h}2_{v}\rangle\leftrightarrow|2_{c}3_{h}1_{v}\rangle}$
can satisfy this condition. In particular, the partial swap $V_{|1_{c}1_{h}2_{v}\rangle\leftrightarrow|1_{c}3_{h}1_{v}\rangle}$
generates a positive current $J_{|1_{c}1_{h}2_{v}\rangle\rightarrow|1_{c}3_{h}1_{v}\rangle}$
if and only if the eigenvalue of $\sigma_{ch}\otimes\rho_{v}$ corresponding
to $|1_{c}1_{h}2_{v}\rangle$ is larger than the eigenvalue corresponding
to $|1_{c}3_{h}1_{v}\rangle$. That is, iff 
\begin{equation}
p_{1}^{c}p_{1}^{h}p_{2}^{v}>p_{2}^{c}p_{1}^{h}p_{1}^{v}\Leftrightarrow\frac{p_{1}^{c}}{p_{2}^{c}}>\frac{p_{1}^{v}}{p_{2}^{v}}.\label{eq:40.7 inequality for "left" rest current}
\end{equation}
Similarly, $V_{|2_{c}2_{h}2_{v}\rangle\leftrightarrow|2_{c}3_{h}1_{v}\rangle}$
generates a positive current $J_{|2_{c}2_{h}2_{v}\rangle\rightarrow|2_{c}3_{h}1_{v}\rangle}$
if and only if 
\begin{equation}
p_{2}^{c}p_{2}^{h}p_{2}^{v}>p_{2}^{c}p_{3}^{h}p_{1}^{v}\Leftrightarrow\frac{p_{2}^{h}}{p_{3}^{h}}>\frac{p_{1}^{v}}{p_{2}^{v}}.\label{eq:40.8 inequality for "right" rest current}
\end{equation}

If the eigenvalues $\{p_{k}^{v}\}_{k=1}^{2}$ are chosen in such a
way that the ratio $\frac{p_{1}^{v}}{p_{2}^{v}}$ satisfies $\frac{p_{1}^{v}}{p_{2}^{v}}=\textrm{min\ensuremath{\left\{  \frac{p_{1}^{c}}{p_{2}^{c}}-\epsilon,\frac{p_{2}^{h}}{p_{3}^{h}}-\epsilon\right\} } }$
(with $\epsilon$ sufficiently small to have $p_{1}^{v}>p_{2}^{v}$),
Eqs. (\ref{eq:40.7 inequality for "left" rest current}) and (\ref{eq:40.8 inequality for "right" rest current})
hold simultaneously. Hence, we have a positive current 
\begin{equation}
J_{|2_{v}\rangle\rightarrow|1_{v}\rangle}=J_{|1_{c}1_{h}2_{v}\rangle\rightarrow|1_{c}3_{h}1_{v}\rangle}+J_{|2_{c}2_{h}2_{v}\rangle\rightarrow|2_{c}3_{h}1_{v}\rangle},\label{eq:40.9 J|2v>-->|1v>}
\end{equation}
which guarantees that 
\begin{align}
U= & U_{|2_{c}2_{h}1_{v}\rangle\leftrightarrow|1_{c}3_{h}2_{v}\rangle}\oplus\left(V_{|1_{c}1_{h}2_{v}\rangle\leftrightarrow|1_{c}3_{h}1_{v}\rangle}\right.\nonumber \\
 & \oplus\left.V_{|2_{c}2_{h}2_{v}\rangle\leftrightarrow|2_{c}3_{h}1_{v}\rangle}\right)\label{eq:50 Unitary for loop}
\end{align}
produces a loop $\{J_{|2_{v}\rangle\rightarrow|1_{v}\rangle}\}\cup\{J_{|1_{v}\rangle\rightarrow|2_{v}\rangle}\}$. 

\subsubsection{Optimization of the cooling current }

Since (\ref{eq:50 Unitary for loop}) is an instance of the unitaries
considered in Proposition 4, the combination of this proposition with
Remark 10 implies that maximum cooling is obtained when all the two-level
unitaries in (\ref{eq:50 Unitary for loop}) are swaps (as indicated
in Eqs. (\ref{eq:37  U catalytic}) and (\ref{eq:38 Vres for U catalytic})
of the main text). Moreover, the corresponding cooling current $\mathcal{J}_{|2_{c}\rangle\rightarrow|1_{c}\rangle}$
can be obtained from (the first line of) Eq. (\ref{eq:32.20 Jcool(max)}),
according to Remark 11. For $d_{v}^{\ast}=2$, this equation takes
the simple form 
\begin{equation}
\mathcal{J}_{|2_{c}\rangle\rightarrow|1_{c}\rangle}=\left(\frac{ac-bd}{c+b}\right)\bar{p}_{1}^{v}.\label{eq:51 explicit cooling current}
\end{equation}
Remark 11 also allows us to apply (\ref{eq:32.19 pdv in terms of p1})
for the evaluation of $\bar{p}_{d_{v}^{\ast}}^{v}=\bar{p}_{2}^{v}$:
\begin{equation}
\bar{p}_{2}^{v}=\frac{\left(d+a\right)\bar{p}_{1}^{v}}{c+b}=1-\bar{p}_{1}^{v}\Rightarrow\bar{p}_{1}^{v}=\frac{c+b}{d+a+c+b}.\label{eq:52 explicit p1v (optimal)}
\end{equation}
This expression for $\bar{p}_{1}^{v}$ can be replaced into (\ref{eq:51 explicit cooling current})
to obtain 
\begin{equation}
\mathcal{J}_{|2_{c}\rangle\rightarrow|1_{c}\rangle}=\frac{ac-bd}{a+b+c+d}.\label{eq:53 explicit cooling current}
\end{equation}

Taking into account that $\mathcal{J}_{|1_{v}\rangle\rightarrow|2_{v}\rangle}$
and $\mathcal{J}_{|2_{v}\rangle\rightarrow|1_{v}\rangle}$ are respectively
generated by $\mathcal{U}_{|2_{c}2_{h}1_{v}\rangle\leftrightarrow|1_{c}3_{h}2_{v}\rangle}$
and $\mathcal{V}_{|1_{c}1_{h}2_{v}\rangle\leftrightarrow|1_{c}3_{h}1_{v}\rangle}\oplus\mathcal{V}_{|2_{c}2_{h}2_{v}\rangle\leftrightarrow|2_{c}3_{h}1_{v}\rangle}$,
the coefficients $\{a,b;c,d\}$ can be identified from 
\begin{align}
\mathcal{J}_{|1_{v}\rangle\rightarrow|2_{v}\rangle} & =p_{2}^{c}p_{2}^{h}\bar{p}_{1}^{v}-p_{2}^{c}p_{1}^{h}\bar{p}_{2}^{v},\nonumber \\
\mathcal{J}_{|2_{v}\rangle\rightarrow|1_{v}\rangle} & =\left(p_{1}^{c}p_{1}^{h}+p_{2}^{c}p_{2}^{h}\right)\bar{p}_{2}^{v}-\left(p_{2}^{c}p_{1}^{h}+p_{2}^{c}p_{3}^{h}\right)\bar{p}_{1}^{v}.\label{eq:54 swap catalyst currents}
\end{align}
Recalling also the condition $p_{1}^{h}=p_{2}^{h}$, a comparison
with Eqs. (\ref{eq:32..23 J|1v>-->|d*v> for perm U_d*v}) and (\ref{eq:32.24  J|(k+1)v>-->|kv> for perm V_k})
yields $a=p_{2}^{c}p_{1}^{h}$, $b=p_{2}^{c}p_{1}^{h}$, $c=p_{1}^{h}$,
and $d=p_{2}^{c}\left(p_{1}^{h}+p_{3}^{h}\right)$. Finally, a substitution
of these coefficients in (\ref{eq:53 explicit cooling current}) yields
\begin{align}
\mathcal{J}_{|2_{c}\rangle\rightarrow|1_{c}\rangle} & =\frac{p_{2}^{c}p_{1}^{h}(p_{1}^{c}p_{1}^{h}-p_{2}^{c}p_{3}^{h})}{2p_{2}^{c}p_{1}^{h}+p_{1}^{h}+p_{2}^{c}\left(p_{1}^{h}+p_{3}^{h}\right)}\nonumber \\
 & =\frac{p_{2}^{c}p_{1}^{h}(p_{1}^{c}p_{1}^{h}-p_{2}^{c}p_{3}^{h})}{p_{2}^{c}\left(2p_{1}^{h}+p_{3}^{h}\right)+p_{1}^{h}+p_{2}^{c}p_{1}^{h}}\nonumber \\
 & =\frac{p_{2}^{c}p_{1}^{h}(p_{1}^{c}p_{1}^{h}-p_{2}^{c}p_{3}^{h})}{p_{2}^{c}+p_{1}^{h}(1+p_{2}^{c})}.\label{eq:55 explicit cooling current}
\end{align}
Since $\Delta_{U}\bigl\langle H_{c}\bigr\rangle=\textrm{Tr}\left[H_{c}\left(U\sigma_{ch}\otimes\rho_{v}U^{\dagger}-\sigma_{ch}\otimes\rho_{v}\right)\right]=\mathcal{J}_{|2_{c}\rangle\rightarrow|1_{c}\rangle}(\varepsilon_{1}^{c}-\varepsilon_{2}^{c})$,
Eq. (\ref{eq:55 explicit cooling current}) leads to Eq. (\ref{eq:40 Jcool with total restoring current})
in the main text. 

\section{MAXIMUM COOLING OF A QUBIT USING $k$ HOT QUBITS }

The optimal cooling of a single qubit using $k$ hot qubits can be
obtained by applying Lemma 4. In this case, the $k$ hot qubits constitute
a hot object of dimension $d_{H}=2^{k}$ even. Since we are going
to employ the capital letter $H$ to label the hot object, when applying
Lemma 4 or any equation related to it $h$ must be replaced by $H$.
The corresponding state is given by $\rho_{H}=\rho_{h}^{\otimes k}$,
where $\rho_{h}=p_{1}^{h}|1_{h}\rangle\langle1_{h}|+p_{2}^{h}|2_{h}\rangle\langle2_{h}|$
is the state of a single hot qubit. The eigenstates of $\rho_{H}$
that describe $l\leq k$ excited qubits and $k-l$ qubits in the ground
state possess eigenvalue $(p_{1}^{h})^{k-l}(p_{2}^{h})^{l}$, and
degeneracy $\frac{k!}{l!(k-l)!}$. Moreover, from the inequality $p_{1}^{h}>p_{2}^{h}$
it readily follows that 
\begin{align}
(p_{1}^{h})^{k-l}(p_{2}^{h})^{l} & >\frac{p_{2}^{h}}{p_{1}^{h}}[(p_{1}^{h})^{k-l}(p_{2}^{h})^{l}]\nonumber \\
 & =(p_{1}^{h})^{k-(l+1)}(p_{2}^{h})^{l+1}.\label{eq:56 decreasing of eigenvalues w.r.t. l}
\end{align}

Let $\left\{ \left(p_{j}^{H}\right)^{\downarrow}\right\} _{j}$ and
$\left\{ \left(p_{j}^{H}\right)^{\uparrow}\right\} _{j}$ denote the
eigenvalues of $\rho_{H}$ in non-increasing order and in non-decreasing
order, respectively. In this way, the inequality $p_{2}^{c}p_{j}^{H}>p_{1}^{c}p_{d_{H}-j+1}^{H}$
in Lemma 4 can be written as 
\begin{equation}
p_{2}^{c}\left(p_{j}^{H}\right)^{\downarrow}>p_{1}^{c}\left(p_{j}^{H}\right)^{\uparrow},\label{eq:57 recast of inequality in Lemma 4}
\end{equation}
and we aim to find the maximum $j$ ($j_{\textrm{max}}$) that satisfies
this inequality to characterize the maximum extracted heat. 

By resorting to Eq. (\ref{eq:56 decreasing of eigenvalues w.r.t. l}),
we can provide explicit expressions for $\left\{ \left(p_{j}^{H}\right)^{\downarrow}\right\} _{j}$
and $\left\{ \left(p_{j}^{H}\right)^{\uparrow}\right\} _{j}$. Since
this equation tells us that the eigenvalues of $\rho_{H}$ are decreasing
with respect to $l$, the index $l$ naturally sets the non-increasing
sorting. Taking into account the degeneracy for $l$ fixed, we have
that 
\begin{equation}
\left\{ \left(p_{j}^{H}\right)^{\downarrow}\right\} _{j}\Leftrightarrow\bigcup_{l=0}^{k}\left\{ (p_{1}^{h})^{k-l}(p_{2}^{h})^{l}\textrm{ repeated }\frac{k!}{l!(k-l)!}\textrm{ times}\right\} ,\label{eq:58.1 non-increasing eigenvalues}
\end{equation}
where each set included in the union operation contains $\frac{k!}{l!(k-l)!}$
identical elements $(p_{1}^{h})^{k-l}(p_{2}^{h})^{l}$. If we substitute
$l$ by $k-l$, we obtain eigenvalues $(p_{1}^{h})^{l}(p_{2}^{h})^{k-l}$
that are increasing with respect to $l$. Moreover, under this substitution
the degeneracy is invariant, i.e. $\frac{k!}{l!(k-l)!}\rightarrow\frac{k!}{(k-l)!(k-(k-l))!}=\frac{k!}{(k-l)!l!}$.
This implies that 
\begin{equation}
\left\{ \left(p_{j}^{H}\right)^{\uparrow}\right\} _{j}\Leftrightarrow\bigcup_{l=0}^{k}\left\{ (p_{1}^{h})^{l}(p_{2}^{h})^{k-l}\textrm{ repeated }\frac{k!}{l!(k-l)!}\textrm{ times}\right\} .\label{eq:58.2 non-decreasing eigenvalues}
\end{equation}

Equations (\ref{eq:58.1 non-increasing eigenvalues}) and (\ref{eq:58.2 non-decreasing eigenvalues})
imply that, for any $j\in\{1,2,...,d_{h}\}$, the $j$th eigenvalue
counted in non-increasing order has the form $(p_{1}^{h})^{k-l}(p_{2}^{h})^{l}$,
and the $j$th eigenvalue counted in non-decreasing order has the
form $(p_{1}^{h})^{l}(p_{2}^{h})^{k-l}$, with a \textit{common} $l$
that depends on $j$. Under the assumption $p_{i}^{c}=p_{i}^{h}$
(state of the cold qubit identical to the state of each hot qubit),
Eq. (\ref{eq:57 recast of inequality in Lemma 4}) is thus equivalent
to 
\begin{align}
p_{1}^{c})^{k-l}(p_{2}^{c})^{l+1} & >(p_{1}^{c})^{l+1}(p_{2}^{c})^{k-l}\nonumber \\
\Leftrightarrow\left(\frac{p_{1}^{c}}{p_{2}^{c}}\right)^{k-2l-1} & >1.\label{eq:59 inequality in Lemma 4 in terms of l}
\end{align}
The maximum $j_{\textrm{max}}$ thus corresponds to the maximum $l=l_{\textrm{max}}$
that obeys this inequality. Let us check this separately for $k$
even and $k$ odd, taking into account that the l.h.s. of (the second
line of) (\ref{eq:59 inequality in Lemma 4 in terms of l}) is decreasing
w.r.t. $l$. 
\begin{itemize}
\item For $k$ even, $l=k/2-1$ yields $p_{1}^{c}/p_{2}^{c}>1$, whereas
$l=k/2$ would imply $p_{2}^{c}/p_{1}^{c}>1$. Hence, in this case
\begin{equation}
l_{\textrm{max}}=k/2-1.\label{eq:60.1 lmal for k even}
\end{equation}
 
\item For $k$ odd ($k\geq3$), $l=(k-3)/2$ yields $(p_{1}^{c}/p_{2}^{c})^{2}>1$,
whereas $l=(k-1)/2$ would imply $1>1$. Hence, in this case 
\begin{equation}
l_{\textrm{max}}=(k-3)/2.\label{eq:60.2 lmal for k odd}
\end{equation}
 
\end{itemize}
Including all the degeneracies from $l=0$ to $l=l_{\textrm{max}}$,
we also have that 
\begin{equation}
j_{\textrm{max}}=\sum_{l=0}^{l_{\textrm{max}}}\frac{k!}{l!(k-l)!}.\label{eq:61 jmax}
\end{equation}

The total extracted heat can be evaluated as 
\begin{equation}
-\Delta\bigl\langle H_{c}\bigr\rangle=\textrm{Tr}\left[H_{c}\left(\rho_{c}\otimes\rho_{H}-\tilde{\mathcal{U}}(\rho_{c}\otimes\rho_{H})\tilde{\mathcal{U}}^{\dagger}\right)\right],\label{eq:62 extracted heat}
\end{equation}
where $\rho_{c}=p_{1}^{c}|1_{c}\rangle\langle1_{c}|+p_{2}^{c}|2_{c}\rangle\langle2_{c}|$
is the state of the cold qubit and $\tilde{\mathcal{U}}(\rho_{c}\otimes\rho_{H})\tilde{\mathcal{U}}^{\dagger}$
is the passive state (\ref{eq:33.2 rho_c x rho_h evolved with optimal permutation}).
Using Eqs. (\ref{eq:33.1 rho_c x rho_h with A,A',B,B'}) and (\ref{eq:33.2 rho_c x rho_h evolved with optimal permutation}),
and assuming $H_{c}=|2_{c}\rangle\langle2_{c}|$, we have that 
\begin{equation}
-\Delta\bigl\langle H_{c}\bigr\rangle=\Delta p_{1}^{c}=p_{2}^{c}\sum_{j\in A}p_{j}^{H}-p_{1}^{c}\sum_{j\in B}p_{j}^{H},\label{eq:63 extracted heat explicit}
\end{equation}
with the sets $A=\{j\}_{1\leq j\leq j_{\textrm{max}}}$and $B=\{j\}_{d_{H}-j_{\textrm{max}}+1\leq j\leq d_{H}}$
defined at the beginning of the proof of Lemma 4. 

Since the indices in $A$ label the largest $j_{\textrm{max}}$ eigenvalues
of $\rho_{H}$, and the indices in $B$ label the smallest $j_{\textrm{max}}$
eigenstates of $\rho_{H}$, the combination of Eqs. (\ref{eq:58.1 non-increasing eigenvalues}),
(\ref{eq:58.2 non-decreasing eigenvalues}) and (\ref{eq:61 jmax})
leads to 
\begin{align}
\Delta p_{1}^{c} & =p_{2}^{c}\sum_{l=0}^{l_{\textrm{max}}}\frac{k!}{l!(k-l)!}(p_{1}^{h})^{k-l}(p_{2}^{h})^{l}\nonumber \\
 & \quad-p_{1}^{c}\sum_{l=0}^{l_{\textrm{max}}}\frac{k!}{l!(k-l)!}(p_{1}^{h})^{l}(p_{2}^{h})^{k-l}\nonumber \\
 & =\sum_{l=0}^{l_{\textrm{max}}}\frac{k!}{l!(k-l)!}\left[(p_{1}^{c})^{k-l}(p_{2}^{c})^{l+1}-(p_{1}^{c})^{l+1}(p_{2}^{c})^{k-l}\right],\label{eq:64 Delta_p1c}
\end{align}
where we haved used $p_{i}^{h}=p_{i}^{c}$ in the third line, and
$l_{\textrm{max}}$ is given by (\ref{eq:60.1 lmal for k even}) or
(\ref{eq:60.2 lmal for k odd}). Equation (\ref{eq:64 Delta_p1c})
is employed to plot the cooling coefficient $\xi_{\textrm{cool}}^{(k)}=\frac{\Delta p_{1}^{c}}{k}$
in Fig. 8 of the main text. For $k=2$, it is straightforward to check
that 
\begin{align}
\xi_{\textrm{cool}}^{(2)} & =\frac{\Delta p_{1}^{c}}{2}=\frac{(p_{1}^{c})^{2}p_{2}^{c}-p_{1}^{c}(p_{2}^{c})^{2}}{2}\nonumber \\
 & =\frac{(1-2p_{2}^{c})}{2}p_{1}^{c}p_{2}^{c},\label{eq:65 cooling coefficient for k=00003D2}
\end{align}
which coincides with Eq. (\ref{eq:44.1 conjecture}) of the main text. 

\section{PROTOCOL FOR THE SYSTEMATIC INCREMENT OF THE LOOP CURRENT }

In this appendix we introduce another ``layer'' of optimization
for the loop current in catalytic transformations. Furthermore, we
will illustrate the utility of the method with an example of catalytic
cooling. To begin with, it is convenient to recall the two optimization
procedures considered until now, and depicted in Fig. 14(a). In this
figure, the array of black and gray arrows at the left represents
a loop of swap currents $\{\mathcal{J}_{|(k+1)_{v}\rangle\rightarrow|k_{v}\rangle}\}_{k=1}^{d_{v}^{\ast}-1}\cup\{\mathcal{J}_{|1_{v}\rangle\rightarrow|d_{v}^{\ast}\rangle}\}$.
Such a loop is generated by a permutation $\mathcal{U}_{d_{v}^{\ast}}\oplus\left(\oplus_{k=1}^{d_{v}^{\ast}-1}\mathcal{V}_{k}\right)$
(cf. (\ref{eq:32.21 permutation U_d*v}) and (\ref{eq:32.22 permutations V_k})),
applied on an initial state $\rho_{ch}\otimes\rho_{v}^{\ast}$ where
$\rho_{v}^{\ast}=\sum_{k=1}^{d_{v}^{\ast}}p_{k}^{v}|k_{v}\rangle\langle k_{v}|$.
The magnitude of each current is larger the darker the shade of the
corresponding arrow.

A first maximization of the loop current was characterized through
Eq. (\ref{eq:32.5 lemma about Jmin}), for the case of \textit{fixed}
eigenvalues $\{p_{k}^{v}\}_{k}$. According to Proposition 3, under
this constraint the maximum value of ${\color{red}{\normalcolor J_{\textrm{loop}}}}$
is achieved by tuning the swap intensities in such a way that all
the currents in the loop match $\mathcal{J}_{\textrm{min}}$, which
yields ${\color{red}{\normalcolor J_{\textrm{loop}}}}=\mathcal{J}_{\textrm{min}}$.
In Fig. 14(a) this tuning leads to the (no permutation!) unitary $U_{d_{v}^{\ast}}\oplus\left(\oplus_{k=1}^{d_{v}^{\ast}-1}V_{k}\right)$.
Although the optimization in (\ref{eq:32.5 lemma about Jmin}) does
not refer to $U_{d_{v}^{\ast}}$ (but instead to $U_{i',l,l'}$),
we note that the extension posed in Remark 9 includes direct sums
(\ref{eq:32.21 permutation U_d*v}). 

Subsequently, through Proposition 4 we stated that if $\mathcal{J}_{\textrm{min}}$
is maximized with respect to the catalyst eigenvalues then the optimal
unitary is in fact the permutation $\mathcal{U}_{d_{v}^{\ast}}\oplus\left(\oplus_{k=1}^{d_{v}^{\ast}-1}\mathcal{V}_{k}\right)$.
Therefore, the maximum of $\mathcal{J}_{\textrm{min}}$ is associated
with a pair $(\mathcal{U},\{\bar{p}_{k}^{v}\}_{k=1}^{d_{v}^{\ast}})$,
where $\{\bar{p}_{k}^{v}\}_{k=1}^{d_{v}^{\ast}}$ are the optimal
eigenvalues (cf. Eq. (\ref{eq:32.11.1 Generalized Eqs. for maximization of Jmin}))
and we introduce the simplified notation $\mathcal{U}\equiv\mathcal{U}_{d_{v}^{\ast}}\oplus\left(\oplus_{k=1}^{d_{v}^{\ast}-1}\mathcal{V}_{k}\right)$. 

\subsection{The protocol }

The protocol works on a basis similar to the addition of restoring
unitaries considered in Appendix H. However, in this case even \textit{single}
two-level unitaries can be added (with ``addition'' meaning direct
sum), and we also include a operation of ``subtraction'' defined
as follows. If $U=U_{\alpha}\oplus U_{\alpha'}$, the subtraction
of $U_{\alpha'}$ from $U$ is denoted as $U\ominus U_{\alpha'}$
and yields $U\ominus U_{\alpha'}=U_{\alpha}$. Hence, the effect of
this operation is to replace the block $U_{\alpha'}$ by the corresponding
identity matrix. 

The starting point for the protocol is the pair $(\mathcal{U},\{\bar{p}_{k}^{v}\}_{k=1}^{d_{v}^{\ast}})$,
characterized before, and it proceeds by adding or subtracting swaps
from $\mathcal{U}$. The criteria to decide how to perform these operations
will be explained in Step 1. Furthermore, each modification of $\mathcal{U}$
is accompanied by the derivation of a new set of catalyst eigenvalues
adapted to the new permutation (Step 2). After $m$ repetitions of
these steps, the output of the protocol is a new pair $(\mathcal{U}(m),\{\bar{p}_{k}^{v}(m)\}_{k=1}^{d_{v}^{\ast}})$
whose loop current $\mathcal{J}_{\textrm{min}}(m)$ satisfies $\mathcal{J}_{\textrm{min}}(m)\geq\mathcal{J}_{\textrm{min}}$.
More precisely, this means that the action of $\mathcal{U}(m)$ on
the state $\rho_{ch}\otimes\bar{\rho}_{v}(m)$, where $\bar{\rho}_{v}(m)=\sum_{k=1}^{d_{v}^{\ast}}\bar{p}_{k}^{v}(m)|k_{v}\rangle\langle k_{v}|$,
produces the loop current $\mathcal{J}_{\textrm{min}}(m)\geq\mathcal{J}_{\textrm{min}}$. 

To present the two steps that compose each round it is convenient
to rewrite Eqs. (\ref{eq:32..23 J|1v>-->|d*v> for perm U_d*v}) and
(\ref{eq:32.24  J|(k+1)v>-->|kv> for perm V_k}) in a more compact
manner. In particular, we adopt a ``periodic boundary condition''
$d_{v}^{\ast}+1\Leftrightarrow1$ that allows us to express $\{\mathcal{J}_{|(k+1)_{v}\rangle\rightarrow|k_{v}\rangle}\}_{k=1}^{d_{v}^{\ast}-1}\cup\{\mathcal{J}_{|1_{v}\rangle\rightarrow|d_{v}^{\ast}\rangle}\}$
as $\{\mathcal{J}_{|(k+1)_{v}\rangle\rightarrow|k_{v}\rangle}\}_{k=1}^{d_{v}^{\ast}}$.
Instead of the constant coefficients in the aforementioned equations,
we will have coefficients $\{a(m),b(m);c_{k+1}(m),d_{k}(m)\}_{k=1}^{d_{v}^{\ast}-1}$
which depend not only on $k$ (cf. (\ref{eq:32.11.1 Generalized Eqs. for maximization of Jmin}))
but also on $m$. Since we aim to recombine (\ref{eq:32..23 J|1v>-->|d*v> for perm U_d*v})
and (\ref{eq:32.24  J|(k+1)v>-->|kv> for perm V_k}) into a single
expression, we also set $c_{d_{v}^{\ast}+1}(m)\equiv a(m)$ and $d_{d_{v}^{\ast}}(m)\equiv b(m)$,
so that $\{a(m),b(m);c_{k+1}(m),d_{k}(m)\}_{k=1}^{d_{v}^{\ast}-1}$
is equivalent to $\{c_{k+1}(m),d_{k}(m)\}_{k=1}^{d_{v}^{\ast}}$. 

Using the notations indicated above, and associating $m=0$ (zero
rounds) to the pair $(\mathcal{U},\{\bar{p}_{k}^{v}\}_{k=1}^{d_{v}^{\ast}})$
(i.e. $(\mathcal{U}(0),\{\bar{p}_{k}^{v}(0)\}_{k=1}^{d_{v}^{\ast}})\equiv(\mathcal{U},\{\bar{p}_{k}^{v}\}_{k=1}^{d_{v}^{\ast}})$),
the action of $\mathcal{U}(m)$ on $\rho_{ch}\otimes\bar{\rho}_{v}(m)$
produces currents 
\begin{align}
\mathcal{J}_{|(k+1)_{v}\rangle\rightarrow|k_{v}\rangle} & =\sum_{\psi,\varphi\textrm{ free}}\mathcal{J}_{|\varphi_{ch}(k+1)_{v}\rangle\rightarrow|\psi_{ch}k_{v}\rangle}\nonumber \\
 & =\sum_{\psi,\varphi\textrm{ free}}\left(p_{\varphi}^{ch}\bar{p}_{k+1}^{v}(m)-p_{\psi}^{ch}\bar{p}_{k}^{v}(m)\right)\nonumber \\
 & =\left(\sum_{\varphi\textrm{ free}}p_{\varphi}^{ch}\right)\bar{p}_{k+1}^{v}(m)\nonumber \\
 &\quad-\left(\sum_{\psi\textrm{ free}}p_{\psi}^{ch}\right)\bar{p}_{k}^{v}(m)\nonumber \\
 & =c_{k+1}(m)\bar{p}_{k+1}^{v}(m)-d_{k}(m)\bar{p}_{k}^{v}(m),\label{eq:66 recast of loop currents for protocol}
\end{align}
for all $k\in\{1,2,...,d_{v}^{\ast}\}$. If $m=0$ (initial pair),
Eqs. (\ref{eq:32..23 J|1v>-->|d*v> for perm U_d*v}) and (\ref{eq:32.24  J|(k+1)v>-->|kv> for perm V_k})
can be recovered if we let the indices $(\varphi,\psi)$ to coincide
with $(n,N)$ for $1\leq k\leq d_{v}^{\ast}-1$ and set $\{c_{k+1}(0),d_{k}(0)\}_{k=1}^{d_{v}^{\ast}-1}=\{c,d\}_{k=1}^{d_{v}^{\ast}-1}$. 

In the following, all the swaps will be denoted as $\mathcal{U}_{|\varphi_{ch}(k+1)_{v}\rangle\leftrightarrow|\psi_{ch}k_{v}\rangle}$
(including those that previously would be written using $\mathcal{V}$
instead of $\mathcal{U}$). Each round of the protocol is an iteration
of the following steps: 

\begin{figure}
\begin{centering}
\includegraphics[scale=0.75]{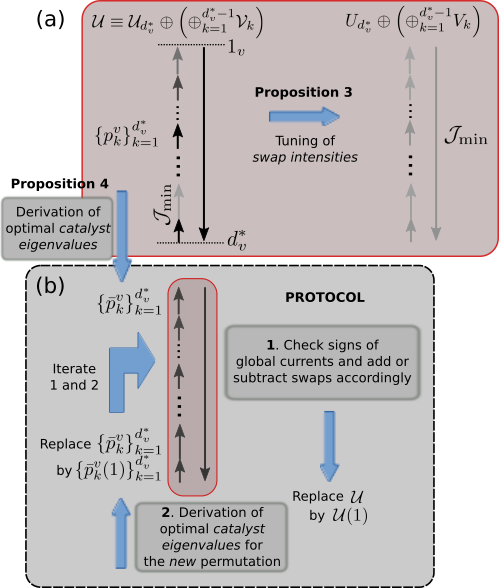}\caption{Different methods to optimize the loop current. The arrays of arrows
represent loops $\{\mathcal{J}_{|(k+1)_{v}\rangle\rightarrow|k_{v}\rangle}\}_{k=1}^{d_{v}^{\ast}-1}\cup\{\mathcal{J}_{|1_{v}\rangle\rightarrow|d_{v}^{\ast}\rangle}\}$.
(a) The two optimization methods described by Propositions 3 and 4.
(b) First round of the protocol presented in this appendix. }
 
\par\end{centering}
\end{figure}

\begin{enumerate}
\item \textbf{Check signs of global currents and add or subtract swaps accordingly}.
The optimization that leads from the eigenvalues $\{p_{k}^{v}\}$
to $\{\bar{p}_{k}^{v}\}$ (step connecting Fig. 14(a) with 14(b))
may have an undesired effect. Specifically, it is possible that some
global currents in Eq. (\ref{eq:66 recast of loop currents for protocol})
become\textit{ negative}, after the substitution of $\{p_{k}^{v}\}$
by $\{\bar{p}_{k}^{v}\}=\{\bar{p}_{k}^{v}(0)\}$. That is, 
\begin{align}
\mathcal{J}_{|\varphi'_{ch}(k+1)_{v}\rangle\rightarrow|\psi'_{ch}k_{v}\rangle} & \leq0\textrm{ for }\{(\psi',\varphi')\}_{\psi',\varphi'}\nonumber \\
\Leftrightarrow p_{\varphi'}^{ch}\bar{p}_{k+1}^{v}(0)-p_{\psi'}^{ch}\bar{p}_{k}^{v}(0) & \leq0\textrm{ for }\{(\psi',\varphi')\}_{\psi',\varphi'},\label{eq:67 new negative currents}
\end{align}
where $\{(\psi',\varphi')\}_{\psi',\varphi'}$ is a subset of indices
included in the sum $\sum_{\psi,\varphi\textrm{ free}}$. These currents
can be removed by subtracting the corresponding swaps $\{\mathcal{U}_{|\varphi'_{ch}(k+1)_{v}\rangle\leftrightarrow|\psi'_{ch}k_{v}\rangle}\}_{\psi',\varphi'}$
from $\mathcal{U}=\mathcal{U}(0)$, thereby increasing $\mathcal{J}_{|(k+1)_{v}\rangle\rightarrow|k_{v}\rangle}$.
\\
Conversely, it is also possible that the transition from $\{p_{k}^{v}\}$
to $\{\bar{p}_{k}^{v}(0)\}$ gives rise to new \textit{positive} global
currents. That is, 
\begin{align}
\mathcal{J}_{|\varphi_{ch}^{\prime\prime}(k+1)_{v}\rangle\rightarrow|\psi_{ch}^{\prime\prime}k_{v}\rangle} & >0\textrm{ for }\{(\psi^{\prime\prime},\varphi^{\prime\prime})\}_{\psi^{\prime\prime},\varphi^{\prime\prime}}\nonumber \\
\Leftrightarrow p_{\varphi^{\prime\prime}}^{ch}\bar{p}_{k+1}^{v}(0)-p_{\psi^{\prime\prime}}^{ch}\bar{p}_{k}^{v}(0) & >0\textrm{ for }\{(\psi^{\prime\prime},\varphi^{\prime\prime})\}_{\psi^{\prime\prime},\varphi^{\prime\prime}},\label{eq:68 new positive currents}
\end{align}
where $\{(\psi^{\prime\prime},\varphi^{\prime\prime})\}_{\psi^{\prime\prime},\varphi^{\prime\prime}}$
is a subset of indices \textit{not} included in the sum $\sum_{\psi,\varphi\textrm{ free}}$.
These currents can be added to $\mathcal{J}_{|(k+1)_{v}\rangle\rightarrow|k_{v}\rangle}$
by adding the corresponding swaps $\{\mathcal{U}_{|\varphi_{ch}^{\prime\prime}(k+1)_{v}\rangle\leftrightarrow|\boldsymbol{i}_{ch}^{\prime\prime}k_{v}\rangle}\}_{\psi^{\prime\prime},\varphi^{\prime\prime}}$
to $\mathcal{U}=\mathcal{U}(0)$, thereby increasing $\mathcal{J}_{|(k+1)_{v}\rangle\rightarrow|k_{v}\rangle}$. 
\item \textbf{Derivation of optimal catalyst eigenvalues}. The addition
and/or subtraction of swaps yield a new permutation $\mathcal{U}(1)$.
This also modifies the currents in (\ref{eq:66 recast of loop currents for protocol}),
since the addition and/or subtraction of global currents leads to
new coefficients $c_{k+1}(1)$ and $d_{k}(1)$ given by 
\begin{align}
c_{k+1}(1) & =c_{k+1}-\sum_{\varphi'}p_{\varphi'}^{ch}+\sum_{\varphi^{\prime\prime}}p_{\varphi^{\prime\prime}}^{ch},\nonumber \\
d_{k}(1) & =d_{k}-\sum_{\psi'}p_{\psi'}^{ch}+\sum_{\psi^{\prime\prime}}p_{\psi^{\prime\prime}}^{ch}.\label{eq:69 new coefficients}
\end{align}
With these coefficients, Proposition 4 allows us to obtain the new
loop current $\mathcal{J}_{\textrm{min}}(1)$ (associated with $\mathcal{U}(1)$),
by solving the equations 

\begin{align}
c_{d_{v}^{\ast}+1}(1)\bar{p}_{k+1}^{v}(1)-d_{d_{v}^{\ast}}(1)\bar{p}_{k}^{v}(1)=&c_{k+1}(1)
\bar{p}_{k+1}^{v}(1)\nonumber\\ &
-d_{k}(1)\bar{p}_{k}^{v}(1),\label{eq:70 optimal solution for new coefficients}
\end{align}
for all $k\in\{1,2,...,d_{v}^{\ast}\}$. 
\end{enumerate}
Let us see now why $\mathcal{J}_{\textrm{min}}(1)\geq\mathcal{J}_{\textrm{min}}$.
Before the application of Step 1, $\mathcal{J}_{\textrm{min}}=\mathcal{J}_{|(k+1)_{v}\rangle\rightarrow|k_{v}\rangle}$
for all $k\in\{1,2,...,d_{v}^{\ast}\}$. Since the addition and/or
subtraction of swaps in this step increases some currents $\mathcal{J}_{|(k+1)_{v}\rangle\rightarrow|k_{v}\rangle}$
in the set $\{\mathcal{J}_{|(k+1)_{v}\rangle\rightarrow|k_{v}\rangle}\}_{k=1}^{d_{v}^{\ast}}$,
the magnitude of these currents is larger than $\mathcal{J}_{\textrm{min}}$.
However, such an increment cannot be harnessed using the eigenvalues
$\{\bar{p}_{k}^{v}(0)\}_{k}$, because they are optimal in combination
with $\mathcal{U}(0)$ and not with $\mathcal{U}(1)$. In fact, due
to the non-uniform magnitude of the new currents Proposition 3 tells
us that, if the eigenvalues $\{\bar{p}_{k}^{v}(0)\}_{k}$ are not
modified, the maximum loop current is still $\mathcal{J}_{\textrm{min}}$,
and achieving it requires to change some swaps in $\mathcal{U}(1)$
by partial swaps so as to implement the suitable tuning (cf. Eq. (\ref{eq:32.5 lemma about Jmin})). 

On the other hand, Proposition 4 implies that the resulting unitary
is suboptimal, and the optimal solution is instead provided by the
pair $(\mathcal{U}(1),\{\bar{p}_{k}^{v}(1)\}_{k=1}^{d_{v}^{\ast}})$.
This justifies the implementation of Step 2 and guarantees that $\mathcal{J}_{\textrm{min}}(1)\geq\mathcal{J}_{\textrm{min}}$.
After the first round, it is possible that the optimization leading
from $\{\bar{p}_{k}^{v}(0)\}_{k}$ to $\{\bar{p}_{k}^{v}(1)\}_{k}$
generates again (undesired) negative currents or (beneficial) positive
currents. This prepares the stage for a next round, with an output
$(\mathcal{U}(2),\{\bar{p}_{k}^{v}(2)\}_{k=1}^{d_{v}^{\ast}})$ such
that $\mathcal{J}_{\textrm{min}}(2)\geq\mathcal{J}_{\textrm{min}}(1)$.
Since the loop current can never decrease after each round, $\mathcal{J}_{\textrm{min}}(m)\geq\mathcal{J}_{\textrm{min}}$
for the $m$th round. The iteration finishes when no new negative
neither positive currents appear. 

\subsection{Illustrative example: Application of the protocol for the catalytic
cooling of a qubit, using a two-level catalyst and a four-level hot
object}

Consider a cold qubit and a four-level hot object in the initial states
\begin{align}
\rho_{c} & =\sum_{i=1}^{2}p_{i}^{c}|i_{c}\rangle\langle i_{c}|\nonumber \\
 & =0.897|1_{c}\rangle\langle1_{c}|+0.103|2_{c}\rangle\langle2_{c}|,\label{eq:71 rhoc}\\
\rho_{h} & =\sum_{j=1}^{4}p_{j}^{h}|j_{c}\rangle\langle j_{c}|\nonumber \\
 & =0.538|1_{h}\rangle\langle1_{h}|+0.3|2_{h}\rangle\langle2_{h}|\nonumber \\
 & \quad+0.1|3_{h}\rangle\langle3_{h}|+0.062|4_{h}\rangle\langle4_{h}|.\label{eq:72 rhoh}
\end{align}
Since $\frac{p_{1}^{c}}{p_{2}^{c}}=8.708>\frac{p_{1}^{h}}{p_{4}^{h}}=8.677$,
the state $\rho_{c}\otimes\rho_{h}$ is passive with respect to $H_{c}$
(cf. Eq. (\ref{eq:4 passiv cond with populations})). After identifying
a permutation $\mathcal{U}(0)$ that cools down the cold qubit using
a catalyst, we will apply the protocol previously introduced to derive
a new permutation $\mathcal{U}(1)$ (and new catalyst eigenvalues)
that outperforms $\mathcal{U}(0)$. This is possible because, as we
will see, the loop currents $\mathcal{J}_{\textrm{min}}$ and $\mathcal{J}_{\textrm{min}}(1)>\mathcal{J}_{\textrm{min}}$
give the magnitude of the corresponding cooling currents. 

\subsubsection{Initial pair $(\mathcal{U}(0),\{\bar{p}_{k}^{v}(0)\}_{k=1}^{d_{v}^{\ast}})$ }

First, let us see that $d_{v}^{\ast}=2$ suffices to have a unitary
\textit{$U_{d_{v}^{\ast}}\oplus\left(\oplus_{k=1}^{d_{v}^{\ast}-1}V_{k}\right)=U_{2}\oplus V_{1}$
}that\textit{ }generates a loop $\{J_{|1_{v}\rangle\rightarrow|2_{v}\rangle}\}\cup\{J_{|2_{v}\rangle\rightarrow|1_{v}\rangle}\}$,
which will allow us to apply Proposition 4 to derive $\{\bar{p}_{k}^{v}(0)\}_{k=1}^{d_{v}^{\ast}}=\{\bar{p}_{k}^{v}(0)\}_{k=1}^{2}$.
For the sake of brevity, we show directly that the direct sum of 
\begin{align}
U_{2} & =U_{|2_{c}1_{h}1_{v}\rangle\leftrightarrow|1_{c}4_{h}2_{v}\rangle},\label{eq:73.1 U2}\\
V_{1} & =\left(\oplus_{i=1}^{2}V_{|i_{c}1_{h}2_{v}\rangle\leftrightarrow|i_{c}4_{h}1_{v}\rangle}\right)\oplus\left(\oplus_{i=1}^{2}V_{|i_{c}2_{h}2_{v}\rangle\leftrightarrow|i_{c}3_{h}1_{v}\rangle}\right)\nonumber \\
 & =V_{|1_{h}2_{v}\rangle\leftrightarrow|4_{h}1_{v}\rangle}\oplus V_{|2_{h}2_{v}\rangle\leftrightarrow|3_{h}1_{v}\rangle},\label{eq:73.2 V1}
\end{align}
satisfies this requirement, when applied on an initial state $\rho_{c}\otimes\rho_{h}\otimes\rho_{v}^{\ast}$,
with $\rho_{v}^{\ast}=\sum_{k=1}^{2}p_{k}^{v}|k_{v}\rangle\langle k_{v}|$
(and $\rho_{x=c,h}$ given in Eqs. (\ref{eq:71 rhoc}) and (\ref{eq:72 rhoh})). 

Specifically, we can check that for eigenvalues $\{p_{1}^{v},p_{2}^{v}\}=\{2/3,1/3\}$
all the (global) currents produced by the partial swaps in (\ref{eq:73.1 U2})
and (\ref{eq:73.2 V1}) are positive. For $i=1,2$, the currents due
to $V_{1}$ read 
\begin{align}
J_{|i_{c}1_{h}2_{v}\rangle\rightarrow|i_{c}4_{h}1_{v}\rangle} & =r_{1,i}\mathcal{J}_{|i_{c}1_{h}2_{v}\rangle\rightarrow|i_{c}4_{h}1_{v}\rangle}\nonumber \\
 & =r_{1,i}p_{i}^{c}\left(p_{1}^{h}p_{2}^{v}-p_{4}^{h}p_{1}^{v}\right)=r_{1,i}(0.123)>0,\label{eq:74.1 global currents for V1}\\
J_{|i_{c}2_{h}2_{v}\rangle\rightarrow|i_{c}3_{h}1_{v}\rangle} & =\tilde{r}_{1,i}\mathcal{J}_{|i_{c}2_{h}2_{v}\rangle\rightarrow|i_{c}3_{h}1_{v}\rangle}\nonumber \\
 & =\tilde{r}_{1,i}p_{i}^{c}\left(p_{2}^{h}p_{2}^{v}-p_{3}^{h}p_{1}^{v}\right)=\tilde{r}_{1,i}(0.029)>0,\label{eq:74.2 global currents for V1}
\end{align}
where $r_{1,i}$ and $\tilde{r}_{1,i}$ are the corresponding swap
intensities. Moreover, a partial swap $U_{2}$ with intensity $r_{2}$
generates a current 
\begin{align}
J_{|2_{c}1_{h}1_{v}\rangle\rightarrow|1_{c}4_{h}2_{v}\rangle} & =r_{2}\mathcal{J}_{|2_{c}1_{h}1_{v}\rangle\rightarrow|1_{c}4_{h}2_{v}\rangle}\nonumber \\
 & =r_{2}\left(p_{2}^{c}p_{1}^{h}p_{1}^{v}-p_{1}^{c}p_{4}^{h}p_{2}^{v}\right)\nonumber \\
 & =r_{2}(0.018)>0.\label{eq:75 global current for U2}
\end{align}
In this way, 
\begin{align}
J_{|1_{v}\rangle\rightarrow|2_{v}\rangle} & =J_{|2_{c}1_{h}1_{v}\rangle\rightarrow|1_{c}4_{h}2_{v}\rangle},\label{eq:76.1 J|1v>-->|2v>}\\
J_{|2_{v}\rangle\rightarrow|1_{v}\rangle} & =\sum_{i=1}^{2}\left(J_{|i_{c}1_{h}2_{v}\rangle\rightarrow|i_{c}4_{h}1_{v}\rangle}+J_{|i_{c}2_{h}2_{v}\rangle\rightarrow|i_{c}3_{h}1_{v}\rangle}\right),\label{eq:76.2 J|2v>-->|1v>}
\end{align}
are also positive. 

The optimal $\{\bar{p}_{k}^{v}(0)\}_{k=1}^{2}$ are obtained from
the condition $\mathcal{J}_{|1_{v}\rangle\rightarrow|2_{v}\rangle}=\mathcal{J}_{|2_{v}\rangle\rightarrow|1_{v}\rangle}$
(Eqs. (\ref{eq:32.11.1 Generalized Eqs. for maximization of Jmin})).
By setting $r_{1,i}=\tilde{r}_{1,i}=r_{2}=1$ in Eqs. (\ref{eq:74.1 global currents for V1}),
(\ref{eq:74.2 global currents for V1}) and substuting the resulting
expressions into (\ref{eq:76.1 J|1v>-->|2v>}) and (\ref{eq:76.2 J|2v>-->|1v>}),
$\mathcal{J}_{|1_{v}\rangle\rightarrow|2_{v}\rangle}=\mathcal{J}_{|2_{v}\rangle\rightarrow|1_{v}\rangle}$
is equivalent to 
\begin{equation}
\left(p_{2}^{c}p_{1}^{h}\right)\bar{p}_{1}^{v}(0)-\left(p_{1}^{c}p_{4}^{h}\right)\bar{p}_{2}^{v}(0)=\left(p_{1}^{h}+p_{2}^{h}\right)\bar{p}_{2}^{v}(0)-\left(p_{3}^{h}+p_{4}^{h}\right)\bar{p}_{1}^{v}(0).\label{eq:77 Eq. for first optimization of eigenvalues}
\end{equation}
Hence, 
\begin{equation}
\bar{p}_{2}^{v}(0)=\frac{(p_{4}^{h}+p_{3}^{h})+p_{2}^{c}p_{1}^{h}}{(p_{1}^{h}+p_{2}^{h})+p_{1}^{c}p_{4}^{h}}\bar{p}_{1}^{v}(0)\Rightarrow\bar{p}_{1}^{v}(0)=0.803.\label{eq:78 first optimal eigenvalues}
\end{equation}
The optimal unitary is $\mathcal{U}(0)=\mathcal{U}_{2}\oplus\mathcal{V}_{1}$,
according to Proposition 4, and the resulting loop current is 
\begin{align}
\mathcal{J}_{\textrm{min}} & =\mathcal{J}_{|1_{v}\rangle\rightarrow|2_{v}\rangle}\nonumber \\
 & =\left(p_{2}^{c}p_{1}^{h}\right)\bar{p}_{1}^{v}(0)-\left(p_{1}^{c}p_{4}^{h}\right)\bar{p}_{2}^{v}(0)\nonumber \\
 & =0.033.\label{eq:79 first loop current}
\end{align}

\subsubsection{Derivation of $(\mathcal{U}(1),\{\bar{p}_{k}^{v}(1)\}_{k=1}^{d_{v}^{\ast}})$
through the protocol (see also Fig. 15) }

First round:  
\begin{enumerate}
\item To check the signs of the global currents under the action of $\mathcal{U}(0)$,
we must replace $\{p_{k}^{v}\}_{k}$ in Eqs. (\ref{eq:74.1 global currents for V1})-(\ref{eq:75 global current for U2})
by the new eigenvalues $\{\bar{p}_{k}^{v}(0)\}_{k}$, and then check
whether the signs of the inequalities is maintained or inverted. For
the currents $\mathcal{J}_{|i_{c}1_{h}2_{v}\rangle\rightarrow|i_{c}4_{h}1_{v}\rangle}$
and $\mathcal{J}_{|i_{c}2_{h}2_{v}\rangle\rightarrow|i_{c}3_{h}1_{v}\rangle}$
we have that 
\begin{align}
(p_{1}^{h})\bar{p}_{2}^{v}(0)-(p_{4}^{h})\bar{p}_{1}^{v}(0) & =0.055>0,\nonumber \\
(p_{2}^{h})\bar{p}_{2}^{v}(0)-(p_{3}^{h})\bar{p}_{1}^{v}(0) & =-0.021<0.\label{eq:80 check of signs for currents due to V1}
\end{align}
On the other hand, 
\begin{equation}
(p_{2}^{c}p_{1}^{h})\bar{p}_{1}^{v}(0)-(p_{1}^{c}p_{4}^{h})\bar{p}_{2}^{v}(0)=0.033>0.\label{eq:81 check of signs for current due to U2}
\end{equation}
Therefore, only the currents corresponding to $\mathcal{V}_{|i_{c}2_{h}2_{v}\rangle\leftrightarrow|i_{c}3_{h}1_{v}\rangle}$
become negative, and these swaps must be \textit{subtracted} from
$\mathcal{U}(0)$. \\
Next, we can see that the \textit{new} swap $\mathcal{U}_{|2_{c}2_{h}1_{v}\rangle\leftrightarrow|1_{c}3_{h}2_{v}\rangle}$
generates a positive current 
\begin{equation}
\mathcal{J}_{|2_{c}2_{h}1_{v}\rangle\rightarrow|1_{c}3_{h}2_{v}\rangle}=(p_{2}^{c}p_{2}^{h})\bar{p}_{1}^{v}(0)-(p_{1}^{c}p_{3}^{h})\bar{p}_{2}^{v}(0)=0.007.\label{eq:82 New positive current}
\end{equation}
By \textit{adding} this swap to $\mathcal{U}(0)\ominus\left(\oplus_{i=1}^{2}\mathcal{V}_{|i_{c}2_{h}2_{v}\rangle\leftrightarrow|i_{c}3_{h}1_{v}\rangle}\right)$,
the new permutation reads 
\begin{align}
\mathcal{U}(1) & =\left(\mathcal{U}_{|2_{c}1_{h}1_{v}\rangle\leftrightarrow|1_{c}4_{h}2_{v}\rangle}\oplus\mathcal{U}_{|2_{c}2_{h}1_{v}\rangle\leftrightarrow|1_{c}3_{h}2_{v}\rangle}\right)\nonumber \\
 & \quad\oplus\left(\oplus_{i=1}^{2}\mathcal{V}_{|i_{c}1_{h}2_{v}\rangle\leftrightarrow|i_{c}4_{h}1_{v}\rangle}\right).\label{eq:83 New optimal permutation}
\end{align}
\item Under the action of $\mathcal{U}(1)$, $\mathcal{J}_{|1_{v}\rangle\rightarrow|2_{v}\rangle}$
and $\mathcal{J}_{|2_{v}\rangle\rightarrow|1_{v}\rangle}$ are given
by $\mathcal{J}_{|1_{v}\rangle\rightarrow|2_{v}\rangle}=\mathcal{J}_{|2_{c}1_{h}1_{v}\rangle\rightarrow|1_{c}4_{h}2_{v}\rangle}+\mathcal{J}_{|2_{c}2_{h}1_{v}\rangle\rightarrow|1_{c}3_{h}2_{v}\rangle}$
and $\mathcal{J}_{|2_{v}\rangle\rightarrow|1_{v}\rangle}=\sum_{i=1}^{2}\mathcal{J}_{|i_{c}1_{h}2_{v}\rangle\rightarrow|i_{c}4_{h}1_{v}\rangle}$.
Hence, the equation $\mathcal{J}_{|1_{v}\rangle\rightarrow|2_{v}\rangle}=\mathcal{J}_{|2_{v}\rangle\rightarrow|1_{v}\rangle}$
to find the new eigenvalues $\{\bar{p}_{k}^{v}(1)\}_{k}$ takes the
form 
\begin{align}
\left[p_{2}^{c}(p_{1}^{h}+p_{2}^{h})\right]\bar{p}_{1}^{v}(1)-\left[p_{1}^{c}(p_{3}^{h}+p_{4}^{h})\right]\bar{p}_{2}^{v}(1) & =\left(p_{1}^{h}\right)\bar{p}_{2}^{v}(1)\nonumber \\
 & \quad-\left(p_{4}^{h}\right)\bar{p}_{1}^{v}(1).\label{eq:84 Eq. for new  optimization of eigenvalues}
\end{align}
The corresponding solution is 
\begin{equation}
\bar{p}_{2}^{v}(1)=\frac{p_{2}^{c}(p_{1}^{h}+p_{2}^{h})+p_{4}^{h}}{p_{1}^{c}(p_{3}^{h}+p_{4}^{h})+p_{1}^{h}}\bar{p}_{1}^{v}(1)\Rightarrow\bar{p}_{1}^{v}(1)=0.821.\label{eq:85 New optimal eigenvalues}
\end{equation}
This provides the loop current 
\begin{align}
\mathcal{J}_{\textrm{min}}(1) & =\mathcal{J}_{|1_{v}\rangle\rightarrow|2_{v}\rangle}\nonumber \\
 & =\left[p_{2}^{c}(p_{1}^{h}+p_{2}^{h})\right]\bar{p}_{1}^{v}(1)-\left[p_{1}^{c}(p_{3}^{h}+p_{4}^{h})\right]\bar{p}_{2}^{v}(1)\nonumber \\
 & =0.045.\label{eq:86 New loop current}
\end{align}
\end{enumerate}
Although we do not write it explicitly, it is not difficult to verify
that with the new eigenvalues $\{\bar{p}_{k}^{v}(1)\}_{k}$ all the
global currents generated by $\mathcal{U}(1)$ remain positive, as
seen in Fig. 15(c). Since no new positive current is recognized, the
protocol finishes after its first round. 

\subsubsection{Improvement of cooling }

By comparing Eqs. (\ref{eq:79 first loop current}) and (\ref{eq:86 New loop current}),
we see that $\mathcal{J}_{\textrm{min}}(1)>\mathcal{J}_{\textrm{min}}$.
Since both swaps in the first line of (\ref{eq:83 New optimal permutation})
transfer population from $|2_{c}\rangle$ to $|1_{c}\rangle$, we
have that $\mathcal{J}_{|2_{c}\rangle\rightarrow|1_{c}\rangle}=\mathcal{J}_{|1_{v}\rangle\rightarrow|2_{v}\rangle}$,
and therefore the cooling current associated with $\mathcal{U}(1)$
reads $\mathcal{J}_{|2_{c}\rangle\rightarrow|1_{c}\rangle}=\mathcal{J}_{\textrm{min}}(1)$.
Similarly, $\mathcal{J}_{|2_{c}\rangle\rightarrow|1_{c}\rangle}=\mathcal{J}_{|1_{v}\rangle\rightarrow|2_{v}\rangle}=\mathcal{J}_{\textrm{min}}$
for $\mathcal{U}(0)$. In addition, the restoring unitary $\oplus_{i=1}^{2}\mathcal{V}_{|i_{c}1_{h}2_{v}\rangle\leftrightarrow|i_{c}4_{h}1_{v}\rangle}=\mathcal{V}_{|1_{h}2_{v}\rangle\leftrightarrow|4_{h}1_{v}\rangle}$
acts on $\mathcal{H}_{h}\otimes\mathcal{H}_{v}$, and consequently
the transformation $\rho_{c}\otimes\rho_{h}\otimes\bar{\rho}_{v}(1)\overset{\textrm{CC}}{\longrightarrow}\mathcal{U}(1)\rho_{c}\otimes\rho_{h}\otimes\bar{\rho}_{v}(1)\mathcal{U}^{\dagger}(1)$
extracts heat $-\Delta\bigl\langle H_{c}\bigr\rangle=\mathcal{J}_{\textrm{min}}(1)(\varepsilon_{2}^{c}-\varepsilon_{1}^{c})$.
Since $\mathcal{J}_{\textrm{min}}(1)>\mathcal{J}_{\textrm{min}}$,
the transformation $\rho_{c}\otimes\rho_{h}\otimes\bar{\rho}_{v}(0)\overset{\textrm{CC}}{\longrightarrow}\mathcal{U}(0)\rho_{c}\otimes\rho_{h}\otimes\bar{\rho}_{v}(0)\mathcal{U}^{\dagger}(0)$
extracts \textit{less} heat $-\Delta\bigl\langle H_{c}\bigr\rangle=\mathcal{J}_{\textrm{min}}(\varepsilon_{2}^{c}-\varepsilon_{1}^{c})$. 

\begin{figure}
\centering{}\includegraphics[scale=0.8]{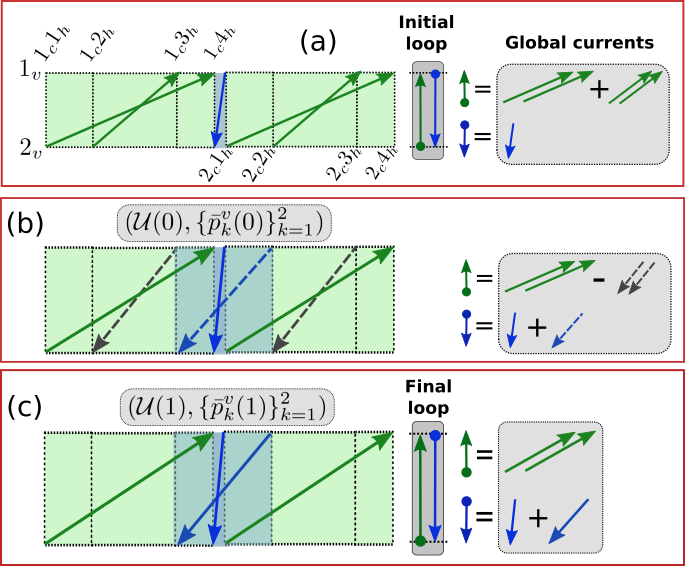}\caption{Application of the protocol in this appendix, to increase the heat
extracted in a catalytic and cooling transformation. The cold qubit
and four-level hot object start in the passive state $\rho_{c}\otimes\rho_{h}$
described by Eqs. (\ref{eq:71 rhoc}) and (\ref{eq:72 rhoh}). (a)
Initially, a two-level catalyst with eigenvalues $\{p_{1}^{v},p_{2}^{v}\}=\{2/3,1/3\}$
generates of a loop $\{J_{|1_{v}\rangle\rightarrow|2_{v}\rangle}\}\cup\{J_{|2_{v}\rangle\rightarrow|1_{v}\rangle}\}$
via the unitary $U_{2}\oplus V_{1}$ (cf. (\ref{eq:73.1 U2}) and
(\ref{eq:73.2 V1})). All the green arrows in the $\textrm{ln}(p^{ch})\times\textrm{ln}(p^{v})$
diagram contribute to $J_{|2_{v}\rangle\rightarrow|1_{v}\rangle}$.
(b) Optimizing the loop current with respect to the catalyst eigenvalues
yields $\{\bar{p}_{1}^{v}(0),\bar{p}_{2}^{v}(0)\}=\{0.803,0.197\}$
and the corresponding permutation $\mathcal{U}(0)$. This optimization
also produces negative currents $\mathcal{J}_{|i_{c}2_{h}2_{v}\rangle\rightarrow|i_{c}3_{h}1_{v}\rangle}$
for $i=0,1$ (gray dashed arrows), and a new positive current $\mathcal{J}_{|2_{c}2_{h}1_{v}\rangle\rightarrow|1_{c}3_{h}2_{v}\rangle}$
(blue dashed arrow). (c) After subtracting the swaps $\mathcal{V}_{|i_{c}2_{h}2_{v}\rangle\leftrightarrow|i_{c}3_{h}1_{v}\rangle}$
and adding $\mathcal{U}_{|2_{c}2_{h}1_{v}\rangle\leftrightarrow|1_{c}3_{h}2_{v}\rangle}$,
a new optimization yields $\{\bar{p}_{1}^{v}(1),\bar{p}_{2}^{v}(1)\}=\{0.821,0.179\}$
and the corresponding permutation $\mathcal{U}(1)$. The final loop
current coincides with the cooling current (blue arrow in the final
loop) and increases the extracted heat. }
 
\end{figure}

\bibliographystyle{plainnat}

\begin{thebibliography}{0}
\providecommand{\natexlab}[1]{#1}
\providecommand{\url}[1]{\texttt{#1}}
\expandafter\ifx\csname urlstyle\endcsname\relax
  \providecommand{\doi}[1]{doi: #1}\else
  \providecommand{\doi}{doi: \begingroup \urlstyle{rm}\Url}\fi

\end{thebibliography}


\begin{thebibliography}{99}
\bibitem{1catalytic-entanglement}D. Jonathan and M. B. Plenio, Entanglement-Assisted
Local Manipulation of Pure Quantum States, \href{https://doi.org/10.1103/PhysRevLett.83.3566}{Phys. Rev. Lett. \textbf{83}, 3566 (1999).} DOI: \href{https://doi.org/10.1103/PhysRevLett.83.3566}{https://doi.org/10.1103/PhysRevLett.83.3566}


\bibitem{key-2catal-majorization}M. Klimesh, Inequalities that Collectively
Completely Characterize the Catalytic Majorization Relation,
\href{https://arxiv.org/abs/0709.3680}{arXiv:0709.3680v1 (2007).}

\bibitem{key-3Mathematical-structure-of-entanglement-catalysis}S.
Daftuar and M. Klimesh, Mathematical structure of entanglement catalysis,
\href{https://doi.org/10.1103/PhysRevA.64.042314}{
Phys. Rev. A \textbf{64}, 042314 (2001).}
DOI: \href{https://doi.org/10.1103/PhysRevA.64.042314}{https://doi.org/10.1103/PhysRevA.64.042314}

\bibitem{key-4Catalytic-transformations-for-bipartite-pure-states}S.
Turgut, Catalytic transformations for bipartite pure states, 
\href{https://doi.org/10.1088/1751-8113/40/40/012}{J. Phys.
A: Math. Theor. \textbf{40} 12185 (2007).}
DOI: \href{https://doi.org/10.1088/1751-8113/40/40/012}{https://doi.org/10.1088/1751-8113/40/40/012}  

\bibitem{key-5Necessary-conditions-for-entanglement-catalysts}Y.
R. Sanders and G. Gour, Necessary conditions for entanglement catalysts,
\href{https://doi.org/10.1103/PhysRevA.79.054302}
{ Phys. Rev. A \textbf{79}, 054302 (2009).}
DOI: \href{https://doi.org/10.1103/PhysRevA.79.054302}{https://doi.org/10.1103/PhysRevA.79.054302} 

\bibitem{key-6Catalytic-Coherence}J. Aberg, Catalytic Coherence,
\href{https://doi.org/10.1103/PhysRevLett.113.150402}
{Phys. Rev. Lett. \textbf{113}, 150402 (2014).}
DOI: \href{https://doi.org/10.1103/PhysRevLett.113.150402}{https://doi.org/10.1103/PhysRevLett.113.150402}

\bibitem{key-7Catalytic-coherence-transformations}K. Bu, U. Singh,
and J. Wu, Catalytic coherence transformations,
\href{https://doi.org/10.1103/PhysRevA.93.042326}
{Phys. Rev. A \textbf{93},
042326 (2016).} 
DOI: \href{https://doi.org/10.1103/PhysRevA.93.042326}{https://doi.org/10.1103/PhysRevA.93.042326}

\bibitem{key-8RTs-with-catalysts}A. Anshu, M.-H. Hsieh, and R. Jain,
Quantifying Resources in General Resource Theory with Catalysts, 
\href{https://doi.org/10.1103/PhysRevLett.121.190504}
{Phys.
Rev. Lett. \textbf{121}, 190504 (2018).}
DOI: \href{https://doi.org/10.1103/PhysRevLett.121.190504}
{https://doi.org/10.1103/PhysRevLett.121.190504}

\bibitem{key-8.1Muller-unitarity}P. Boes, J. Eisert, R. Gallego,
M. P. Müller, and H. Wilming, Von Neumann Entropy from Unitarity,
\href{https://doi.org/10.1103/PhysRevLett.122.210402}
{Phys. Rev. Lett. \textbf{122}, 210402 (2019).}
DOI: \href{https://doi.org/10.1103/PhysRevLett.122.210402}{https://doi.org/10.1103/PhysRevLett.122.210402}

\bibitem{key-8.2Rethinasamy}S. Rethinasamy and M. M. Wilde, Relative
entropy and catalytic relative majorization,
\href{https://doi.org/10.1103/PhysRevResearch.2.033455}
{Phys. Rev. Research \textbf{2},
033455 (2020).}
DOI: \href{https://doi.org/10.1103/PhysRevResearch.2.033455}{https://doi.org/10.1103/PhysRevResearch.2.033455} 

\bibitem{key-8.3Boes-catal-randomness}P. Boes, H. Wilming, R. Gallego,
and J. Eisert, Catalytic Quantum Randomness, 
\href{https://doi.org/10.1103/PhysRevX.8.041016}
{Phys. Rev. X \textbf{8},
041016 (2018).}
DOI: \href{https://doi.org/10.1103/PhysRevX.8.041016}{https://doi.org/10.1103/PhysRevX.8.041016}

\bibitem{key-8.4Renner-catal-decoupling}C. Majenz, M. Berta, F. Dupuis,
R. Renner, and M. Christandl, Catalytic Decoupling of Quantum Information,
\href{https://doi.org/10.1103/PhysRevLett.118.080503}
{Phys. Rev. Lett. \textbf{118}, 080503 (2017).}
DOI: \href{https://doi.org/10.1103/PhysRevLett.118.080503}{https://doi.org/10.1103/PhysRevLett.118.080503}

\bibitem{key-9asymmetry-with-correlated-catalysts}F. Ding, X. Hu,
and H. Fan, Amplifying asymmetry with correlated catalysts,
\href{https://doi.org/10.1103/PhysRevA.103.022403} 
{Phys. Rev. A \textbf{103}, 022403
(2020).}
DOI: \href{https://doi.org/10.1103/PhysRevA.103.022403}{https://doi.org/10.1103/PhysRevA.103.022403} 

\bibitem{key-9.1Wilming-entropy-and-rev-catalysis}H. Wilming, Entropy
and reversible catalysis,
\href{https://arxiv.org/abs/2012.05573} 
{arXiv:2012.05573 (2020).}

\bibitem{key-10second-laws}F. Brandao, M. Horodecki, N. Ng, J. Oppenheim,
and S. Wehner, The second laws of quantum thermodynamics,
\href{https://doi.org/10.1073/pnas.1411728112}
{PNAS \textbf{112},
3275 (2015).}
DOI: \href{https://doi.org/10.1073/pnas.1411728112}{https://doi.org/10.1073/pnas.1411728112}

\bibitem{11Limits-to-catalysis-in-quantum-thermodynamics}N. Ng, L.
Mancinska, C. Cirstoiu, J. Eisert, and S. Wehner, Limits to catalysis
in quantum thermodynamics, 
\href{https://doi.org/10.1088/1367-2630/17/8/085004}
{New J. Phys. \textbf{17}, 085004 (2015).}
DOI: \href{https://doi.org/10.1088/1367-2630/17/8/085004}{https://doi.org/10.1088/1367-2630/17/8/085004} 

\bibitem{12Oppenheim-passivity}C. Sparaciari, D. Jennings, and J.
Oppenheim, Energetic instability of passive states in thermodynamics,
\href{https://doi.org/10.1038/s41467-017-01505-4}
{Nat. Commun. \textbf{8}, 1895 (2017).}
DOI: \href{https://doi.org/10.1038/s41467-017-01505-4}{https://doi.org/10.1038/s41467-017-01505-4}

\bibitem{13Third-Law-of-Thermodynamics-as-a-Single-Inequality}H.
Wilming and R. Gallego, Third Law of Thermodynamics as a Single Inequality,
\href{https://doi.org/10.1103/PhysRevX.7.041033}
{Phys. Rev. X \textbf{7}, 041033 (2017).}
DOI: \href{https://doi.org/10.1103/PhysRevX.7.041033}{https://doi.org/10.1103/PhysRevX.7.041033}

\bibitem{14Muller-catalysts}M. P. Muller, Correlating Thermal Machines
and the Second Law at the Nanoscale,
\href{https://doi.org/10.1103/PhysRevX.8.041051}
{Phys. Rev. X \textbf{8}, 041051
(2018).} 
DOI: \href{https://doi.org/10.1103/PhysRevX.8.041051}{https://doi.org/10.1103/PhysRevX.8.041051}

\bibitem{15universal-cataysts}P. Lipka-Bartosik and P. Skrzypczyk,
All states are universal catalysts in quantum thermodynamics,
\href{https://doi.org/10.1103/PhysRevX.11.011061}
{Phys. Rev. X \textbf{11}, 011061 (2021).}
DOI: \href{https://doi.org/10.1103/PhysRevX.11.011061}{https://doi.org/10.1103/PhysRevX.11.011061} 

\bibitem{15.1Lost-stochastic-independence}M. Lostaglio, M. P. Muller,
and M. Pastena, Stochastic Independence as a Resource in Small-Scale
Thermodynamics,
\href{https://doi.org/10.1103/PhysRevLett.115.150402}
{Phys. Rev. Lett. \textbf{115}, 150402 (2015).} 
DOI: \href{https://doi.org/10.1103/PhysRevLett.115.150402}{https://doi.org/10.1103/PhysRevLett.115.150402}

\bibitem{15.2Bypassing-FTs}P. Boes, R. Gallego, N. H. Y. Ng, J. Eisert,
and H. Wilming, By-passing fluctuation theorems, 
\href{https://doi.org/10.22331/q-2020-02-20-231}
{Quantum \textbf{4},
231 (2020).} 
DOI: \href{https://doi.org/10.22331/q-2020-02-20-231}{https://doi.org/10.22331/q-2020-02-20-231}

\bibitem{15.3Karen}A. E. Allahverdyan and K. V. Hovhannisyan, Work
extraction from microcanonical bath,
\href{https://doi.org/10.1209/0295-5075/95/60004}
{EPL \textbf{95}, 60004 (2011).} 
DOI: \href{https://doi.org/10.1209/0295-5075/95/60004}{https://doi.org/10.1209/0295-5075/95/60004}

\bibitem{15.4Sagawa}N. Shiraishi and T. Sagawa, Quantum thermodynamics
of correlated-catalytic state conversion at small-scale,
\href{https://doi.org/10.1103/PhysRevLett.126.150502} 
{Phys. Rev. Lett. 126, 150502 (2021).}
DOI: \href{https://doi.org/10.1103/PhysRevLett.126.150502}{https://doi.org/10.1103/PhysRevLett.126.150502} 

\bibitem{15.5Muller-and-Pastena-Major}M. P. Muller and M. Pastena,
A Generalization of Majorization that Characterizes Shannon Entropy,
\href{https://doi.org/10.1109/TIT.2016.2528285}
{IEEE Transactions on Information Theory \textbf{62}, 1711 (2016).}
DOI: \href{https://doi.org/10.1109/TIT.2016.2528285}{https://doi.org/10.1109/TIT.2016.2528285}

\bibitem{16RT-Horodecki-and-Oppenheim}M. Horodecki and J. Oppenheim,
Fundamental limitations for quantum and nanoscale thermodynamics,
\href{https://doi.org/10.1038/ncomms3059}
{Nat. Commun. \textbf{4}, 2059 (2013).}
DOI: \href{https://doi.org/10.1038/ncomms3059}
{https://doi.org/10.1038/ncomms3059}

\bibitem{17RT-Brandao}F. Brandao, M. Horodecki, J. Oppenheim, J.
M. Renes, and R. W. Spekkens, Resource Theory of Quantum States Out
of Thermal Equilibrium, 
\href{https://doi.org/10.1103/PhysRevLett.111.250404}
{Phys. Rev. Lett. \textbf{111}, 250404 (2013).}
DOI: \href{https://doi.org/10.1103/PhysRevLett.111.250404}{https://doi.org/10.1103/PhysRevLett.111.250404}

\bibitem{18Lostaglio-coherence-RT}M. Lostaglio, D. Jennings, and
T. Rudolph, Description of quantum coherence in thermodynamic processes
requires constraints beyond free energy, 
\href{https://doi.org/10.1038/ncomms7383}
{Nat. Commun. \textbf{6},
6383 (2015).}
DOI: \href{https://doi.org/10.1038/ncomms7383}
{https://doi.org/10.1038/ncomms7383}


\bibitem{20Korzekwa-work-and-coherence}K. Korzekwa, M. Lostaglio,
J. Oppenheim, and D. Jennings, The extraction of work from quantum
coherence, 
\href{https://doi.org/10.1088/1367-2630/18/2/023045}
{New J. Phys. \textbf{18}, 023045 (2016).}
DOI: \href{https://doi.org/10.1088/1367-2630/18/2/023045}{https://doi.org/10.1088/1367-2630/18/2/023045}

\bibitem{21Lostaglio-element-TOs}M. Lostaglio, A. M. Alhambra, and
C. Perry, Elementary Thermal Operations, 
\href{https://doi.org/10.22331/q-2018-02-08-52}
{Quantum \textbf{2}, 52 (2018).}
DOI: \href{https://doi.org/10.22331/q-2018-02-08-52}{https://doi.org/10.22331/q-2018-02-08-52}  

\bibitem{22Lostaglio-review}M. Lostaglio, An introductory review
of the resource theory approach to thermodynamics,
\href{https://doi.org/10.1088/1361-6633/ab46e5} 
{Rep. Prog. Phys.
\textbf{82}, 114001 (2019).}
DOI: \href{https://doi.org/10.1088/1361-6633/ab46e5}{https://doi.org/10.1088/1361-6633/ab46e5} 

\bibitem{23quantum-thermo-review}J. Goold, M. Huber, A. Riera, L.
del Rio, and P. Skrzypczyk, The role of quantum information in thermodynamics\textemdash a
topical review,
\href{https://doi.org/10.1088/1751-8113/49/14/143001} 
{J. Phys. A: Math. Theor. \textbf{49}, 143001 (2016).}
DOI: \href{https://doi.org/10.1088/1751-8113/49/14/143001}{https://doi.org/10.1088/1751-8113/49/14/143001} 

\bibitem{24Anders-Quant-Therm}S. Vinjanampathy and J. Anders, Quantum
thermodynamics, 
\href{https://doi.org/10.1080/00107514.2016.1201896}
{Contemporary Physics \textbf{57}, 545 (2016).}
DOI: \href{https://doi.org/10.1080/00107514.2016.1201896}{https://doi.org/10.1080/00107514.2016.1201896}

\bibitem{24.1Alhambra-HBAC}A. M. Alhambra, M. Lostaglio, and C. Perry,
Heat-Bath Algorithmic Cooling with optimal thermalization strategies,
\href{https://doi.org/10.22331/q-2019-09-23-188}
{Quantum \textbf{3}, 188 (2019).}
DOI: \href{https://doi.org/10.22331/q-2019-09-23-188}{https://doi.org/10.22331/q-2019-09-23-188}  

\bibitem{24.2Muller-finite-baths}J. Scharlau and M. P. Muller, Quantum
Horn's lemma, finite heat baths, and the third law of thermodynamics,
\href{https://doi.org/10.22331/q-2018-02-22-54}
{Quantum \textbf{2}, 54 (2018).}
DOI: \href{https://doi.org/10.22331/q-2018-02-22-54}{https://doi.org/10.22331/q-2018-02-22-54} 

\bibitem{34Paz-review-cooling}Freitas N., Gallego R., Masanes L., Paz J.P. (2018) Cooling to Absolute Zero: The Unattainability Principle. In: Binder F., Correa L., Gogolin C., Anders J., Adesso G. (eds) Thermodynamics in the Quantum Regime. Fundamental Theories of Physics, vol 195. Springer, Cham. DOI: \href{http://doi.org/10.1007/978-3-319-99046-0_25}{http://doi.org/10.1007/978-3-319-99046-0-25} 

\bibitem{34.1Challeng-unat-principle}M. Kolar, D. Gelbwaser-Klimovsky,
R. Alicki, and G. Kurizki, Quantum Bath Refrigeration towards Absolute
Zero: Challenging the Unattainability Principle,
\href{https://doi.org/10.1103/PhysRevLett.109.090601} 
{Phys. Rev. Lett.
\textbf{109}, 090601 (2012).}
DOI: \href{https://doi.org/10.1103/PhysRevLett.109.090601}{https://doi.org/10.1103/PhysRevLett.109.090601} 

\bibitem{35Masanes-Oppenh-third-law}L. Masanes and J. Oppenheim,
A general derivation and quantification of the third law of thermodynamics,
\href{https://doi.org/10.1038/ncomms14538}
{Nat. Commun. \textbf{8}, 14538 (2017).}
DOI: \href{https://doi.org/10.1038/ncomms14538}{https://doi.org/10.1038/ncomms14538} 

\bibitem{35.1Kosloff-third-law}A. Levy, R. Alicki, and R. Kosloff,
Quantum refrigerators and the third law of thermodynamics,
\href{https://doi.org/10.1103/PhysRevE.85.061126}
{Phys. Rev.
E \textbf{85}, 061126 (2012).}
DOI: \href{https://doi.org/10.1103/PhysRevE.85.061126}{https://doi.org/10.1103/PhysRevE.85.061126}

\bibitem{35.2polariz-and-HBAC-with-n-steps}N. A. Rodríguez-Briones,
and R. Laflamme, Achievable Polarization for Heat-Bath Algorithmic
Cooling,
\href{https://doi.org/10.1103/PhysRevLett.116.170501}
{Phys. Rev. Lett. \textbf{116}, 170501 (2016).}
DOI: \href{https://doi.org/10.1103/PhysRevLett.116.170501}{https://doi.org/10.1103/PhysRevLett.116.170501} 

\bibitem{35.3limit-HBAC}L. J. Schulman, T. Mor, and Y. Weinstein,
Physical Limits of Heat-Bath Algorithmic Cooling,
\href{https://doi.org/10.1103/PhysRevLett.94.120501}
{Phys. Rev. Lett.
\textbf{94}, 120501 (2005).}
DOI: \href{https://doi.org/10.1103/PhysRevLett.94.120501}{https://doi.org/10.1103/PhysRevLett.94.120501} 

\bibitem{36Paz-fund-limits-for-cooling}N. Freitas and J. P. Paz, Fundamental limits for cooling of linear
quantum refrigerators,
\href{https://doi.org/10.1103/PhysRevE.95.012146}
{Phys. Rev. E \textbf{95} 012146 (2017).}
DOI: \href{https://doi.org/10.1103/PhysRevE.95.012146}{https://doi.org/10.1103/PhysRevE.95.012146} 

\bibitem{37Huber-cooling-bound1}F. Clivaz, R. Silva, G. Haack, J.
Bohr Brask, N. Brunner, and M. Huber, Unifying Paradigms of Quantum
Refrigeration: A Universal and Attainable Bound on Cooling,
\href{https://doi.org/10.1103/PhysRevLett.123.170605}
{Phys. Rev. Lett. \textbf{123}, 170605 (2019).}
DOI: \href{https://doi.org/10.1103/PhysRevLett.123.170605}{https://doi.org/10.1103/PhysRevLett.123.170605}  

\bibitem{39.5satirability-of-asymp-limit-in-HBAC}S. Raeisi, and M.
Mosca, Asymptotic Bound for Heat-Bath Algorithmic Cooling,
\href{https://doi.org/10.1103/PhysRevLett.114.100404} 
{Phys. Rev.
Lett. \textbf{114}, 100404 (2015).}
DOI: \href{https://doi.org/10.1103/PhysRevLett.114.100404}{https://doi.org/10.1103/PhysRevLett.114.100404}  

\bibitem{39.6HBAC-and-correlated-environm}N. A. Rodríguez-Briones,
J. Li, X. Peng, T. Mor, Y. Weinstein, and R. Laflamme, Heat-bath algorithmic
cooling with correlated qubit-environment interactions,
\href{https://doi.org/10.1088/1367-2630/aa8fe0}
{New J. Phys.
\textbf{19}, 113047 (2017).} 
DOI: \href{https://doi.org/10.1088/1367-2630/aa8fe0}{https://doi.org/10.1088/1367-2630/aa8fe0} 

\bibitem{38Huber-cooling-bound2}F. Clivaz, R. Silva, G. Haack, J.
Bohr Brask, N. Brunner, and M. Huber, Unifying paradigms of quantum
refrigeration: Fundamental limits of cooling and associated work costs,
\href{https://doi.org/10.1103/PhysRevE.100.042130}
{Phys. Rev. E \textbf{100}, 042130 (2019).}
DOI: \href{https://doi.org/10.1103/PhysRevE.100.042130}{https://doi.org/10.1103/PhysRevE.100.042130}

\bibitem{38.1Gaussian-TOs}A. Serafini, M. Lostaglio, S. Longden,
U. Shackerley-Bennett, C.-Y. Hsieh, and G. Adesso, Gaussian Thermal
Operations and The Limits of Algorithmic Cooling, 
\href{https://doi.org/10.1103/PhysRevLett.124.010602}
{Phys. Rev. Lett.
\textbf{124}, 010602 (2020).}
DOI: \href{https://doi.org/10.1103/PhysRevLett.124.010602}{https://doi.org/10.1103/PhysRevLett.124.010602}

\bibitem{39.7cooling-improv-with-memory-effects}P. Taranto, F. Bakhshinezhad,
P. Schuttelkopf, F. Clivaz, and M. Huber, Exponential improvement
for quantum cooling through finite memory effects, 
\href{https://doi.org/10.1103/PhysRevApplied.14.054005}
{Phys. Rev. Applied
\textbf{14}, 054005 (2020).}
DOI: \href{https://doi.org/10.1103/PhysRevApplied.14.054005}{https://doi.org/10.1103/PhysRevApplied.14.054005} 

\bibitem{38.2Refrigerator-dimension}R. Silva, G. Manzano, P. Skrzypczyk,
and N. Brunner, Performance of autonomous quantum thermal machines:
Hilbert space dimension as a thermodynamical resource, 
\href{https://doi.org/10.1103/PhysRevE.94.032120}
{Phys. Rev.
E \textbf{94}, 032120 (2020).}
DOI: \href{https://doi.org/10.1103/PhysRevE.94.032120}{https://doi.org/10.1103/PhysRevE.94.032120} 

\bibitem{39.2Smallest-refrigerators}N. Linden, S. Popescu, and P.
Skrzypczyk, How Small Can Thermal Machines Be? The Smallest Possible
Refrigerator, 
\href{https://doi.org/10.1103/PhysRevLett.105.130401}
{Phys. Rev. Lett. \textbf{105}, 130401 (2010).}
DOI: \href{https://doi.org/10.1103/PhysRevLett.105.130401}{https://doi.org/10.1103/PhysRevLett.105.130401}

\bibitem{39.3Mitchison-cooling-and-coherence}M. T. Mitchison, M.
P. Woods, J. Prior, and Marcus Huber, Coherence-assisted single-shot
cooling by quantum absorption refrigerators,
\href{https://doi.org/10.1088/1367-2630/17/11/115013} 
{New J. Phys. \textbf{17},
115013 (2015).} 
DOI: \href{https://doi.org/10.1088/1367-2630/17/11/115013}{https://doi.org/10.1088/1367-2630/17/11/115013} 

\bibitem{39.4entang-and-cooling}N. Brunner, M. Huber, N. Linden,
S. Popescu, R. Silva, and P. Skrzypczyk, Entanglement enhances cooling
in microscopic quantum refrigerators, 
\href{https://doi.org/10.1103/PhysRevE.89.032115}
{Phys. Rev. E \textbf{89}, 032115
(2014).}
DOI: \href{https://doi.org/10.1103/PhysRevE.89.032115}{https://doi.org/10.1103/PhysRevE.89.032115}

\bibitem{36.1Karen-limits-cooling}A. E. Allahverdyan, K. V. Hovhannisyan,
D. Janzing, and G. Mahler, Thermodynamic limits of dynamic cooling,
\href{https://doi.org/10.1103/PhysRevE.84.041109}
{Phys. Rev. E \textbf{84}, 041109 (2011).}
DOI: \href{https://doi.org/10.1103/PhysRevE.84.041109}{https://doi.org/10.1103/PhysRevE.84.041109}

\bibitem{39.8No-go-theorem-for-purity}Lian-Ao Wu, Dvira Segal, and
Paul Brumer, No-go theorem for ground state cooling given initial
system-thermal bath factorization,
\href{https://doi.org/10.1038/srep01824}
{Scientific Reports \textbf{3},
1824 (2013).}
DOI: \href{https://doi.org/10.1038/srep01824}{https://doi.org/10.1038/srep01824}

\bibitem{39.9cooling-with-virtual-environment}F. Ticozzi and L. Viola,
Quantum resources for purification and cooling: fundamental limits
and opportunities, 
\href{https://doi.org/10.1038/srep05192}
{Scientific Reports \textbf{4}, 5192 (2014).}
DOI: \href{https://doi.org/10.1038/srep05192}{https://doi.org/10.1038/srep05192} 

\bibitem{40Wolf-Improved-Landauer}D. Reeb and M. M. Wolf, An improved
Landauer principle with finite-size corrections, 
\href{https://doi.org/10.1088/1367-2630/16/10/103011}
{New J. Phys. \textbf{16},
103011 (2014).}
\href{https://doi.org/10.1088/1367-2630/16/10/103011}{https://doi.org/10.1088/1367-2630/16/10/103011}

\bibitem{40.1Raam-PD}R. Uzdin and S. Rahav, The Passivity Deformation
Approach for the Thermodynamics of Isolated Quantum Setups,
\href{https://doi.org/10.1103/PRXQuantum.2.010336} 
{PRX Quantum \textbf{2}, 010336
(2020).}
DOI: \href{https://doi.org/10.1103/PRXQuantum.2.010336}{https://doi.org/10.1103/PRXQuantum.2.010336}  

\bibitem{40.2KMS-passive-states}W. Pusz and S. L. Woronowicz, Passive
states and KMS states for general quantum systems, 
\href{https://doi.org/10.1007/BF01614224}
{Commun. Math. Phys.
\textbf{58}, 273 (1978).}
DOI: \href{https://doi.org/10.1007/BF01614224}{https://doi.org/10.1007/BF01614224} 

\bibitem{41work-and-passive-states}A. E. Allahverdyan, R. Balian,
and Th. M. Nieuwenhuizen, Maximal work extraction from finite quantum
systems, 
\href{https://doi.org/10.1209/epl/i2004-10101-2}
{Europhys. Lett. \textbf{67}, 565 (2004).}
DOI: \href{https://doi.org/10.1209/epl/i2004-10101-2}{https://doi.org/10.1209/epl/i2004-10101-2}  

\bibitem{42Paul-Sk-passivity-and-virt-temp}P. Skrzypczyk, R. Silva,
and N. Brunner, Passivity, complete passivity, and virtual temperatures,
\href{https://doi.org/10.1103/PhysRevE.91.052133}
{Phys. Rev. E \textbf{91}, 052133 (2015).}
DOI: \href{https://doi.org/10.1103/PhysRevE.91.052133}{https://doi.org/10.1103/PhysRevE.91.052133} 

\bibitem{43Raam}R. Uzdin and S. Rahav, Global Passivity in Microscopic
Thermodynamics, 
\href{https://doi.org/10.1103/PhysRevX.8.021064}
{Phys. Rev. X \textbf{8}, 021064 (2018).}
DOI: \href{https://doi.org/10.1103/PhysRevX.8.021064}{https://doi.org/10.1103/PhysRevX.8.021064} 

\bibitem{44Thermometry-review}M. Mehboudi, A. Sanpera, and L. A.
Correa, 
Thermometry in the quantum regime: recent theoretical progress,
\href{https://doi.org/10.1088/1751-8121/ab2828}
{J. Phys. A: Math. Theor. \textbf{52}, 30 (2019).}
DOI: \href{https://doi.org/10.1088/1751-8121/ab2828}{https://doi.org/10.1088/1751-8121/ab2828} 

\bibitem{45Giovanetti-metrology}V. Giovannetti, S. Lloyd, and L.
Maccone, Quantum Metrology, 
\href{https://doi.org/10.1103/PhysRevLett.96.010401}
{Phys. Rev. Lett. \textbf{96}, 010401 (2006).} 
DOI: \href{https://doi.org/10.1103/PhysRevLett.96.010401}{https://doi.org/10.1103/PhysRevLett.96.010401} 

\bibitem{46Advances-metrology}V. Giovannetti, S. Lloyd, and L. Maccone,
Advances in quantum metrology, 
\href{https://doi.org/10.1038/nphoton.2011.35}
{Nat. Phot. \textbf{5}, 222 (2011).}
DOI: \href{https://doi.org/10.1038/nphoton.2011.35}{https://doi.org/10.1038/nphoton.2011.35} 

\bibitem{47Paris-metrology}M. G. A. Paris, Quantum Estimation For
Quantum Technology, 
\href{https://doi.org/10.1142/S0219749909004839}
{Int. J. Quantum. Inform. \textbf{7}, 125 (2009).}
DOI: \href{https://doi.org/10.1142/S0219749909004839}{https://doi.org/10.1142/S0219749909004839}  

\bibitem{48quantum-sensing}C.\LyXThinSpace L. Degen, F. Reinhard,
and P. Cappellaro, Quantum sensing,
\href{https://doi.org/10.1103/RevModPhys.89.035002}
{Rev. Mod. Phys. \textbf{89}, 035002
(2017).}
DOI: \href{https://doi.org/10.1103/RevModPhys.89.035002}{https://doi.org/10.1103/RevModPhys.89.035002} 

\bibitem{48.1Brunelli-Qubit-therm-for-microm-resonators}M. Brunelli,
S. Olivares, and M. G. A. Paris, Qubit thermometry for micromechanical
resonators, 
\href{https://doi.org/10.1103/PhysRevA.84.032105}
{Phys. Rev. A \textbf{84}, 032105 (2011).}
DOI: \href{https://doi.org/10.1103/PhysRevA.84.032105}{https://doi.org/10.1103/PhysRevA.84.032105} 

\bibitem{48.2Brunelli-Qubit-therm-of-a-harm-oscillator}M. Brunelli,
S. Olivares, M. Paternostro, and M. G. A. Paris, Qubit-assisted thermometry
of a quantum harmonic oscillator,
\href{https://doi.org/10.1103/PhysRevA.86.012125} 
{Phys. Rev. A \textbf{86}, 012125 (2012).}
DOI: \href{https://doi.org/10.1103/PhysRevA.86.012125}{https://doi.org/10.1103/PhysRevA.86.012125} 

\bibitem{48.3Single-qubit-therm}S. Jevtic, D. Newman, T. Rudolph,
and T. M. Stace, Single-qubit thermometry, 
\href{https://doi.org/10.1103/PhysRevA.91.012331}
{Phys. Rev. A \textbf{91},
012331 (2015).}
DOI: \href{https://doi.org/10.1103/PhysRevA.91.012331}{https://doi.org/10.1103/PhysRevA.91.012331}  

\bibitem{48.4temp-estim-with-sequential-meas}A. De Pasquale, K. Yuasa,
and V. Giovannetti, Estimating temperature via sequential measurements,
\href{https://doi.org/10.1103/PhysRevA.96.012316}
{Phys. Rev. A \textbf{96}, 012316 (2017).}
DOI: \href{https://doi.org/10.1103/PhysRevA.96.012316}{https://doi.org/10.1103/PhysRevA.96.012316} 

\bibitem{48.5Non-eq-themometry-with-qubit-probes}V. Cavina, L. Mancino,
A. De Pasquale, I. Gianani, M. Sbroscia, R. I. Booth, E. Roccia, R.
Raimondi, V. Giovannetti, and M. Barbieri, Bridging thermodynamics
and metrology in nonequilibrium quantum thermometry,
\href{https://doi.org/10.1103/PhysRevA.98.050101}
{Phys. Rev. A
\textbf{98}, 050101(R) (2018).}
DOI: \href{https://doi.org/10.1103/PhysRevA.98.050101}{https://doi.org/10.1103/PhysRevA.98.050101} 

\bibitem{49Correa-thermometry}L. A. Correa, M. Mehboudi, G. Adesso,
and A. Sanpera, Individual Quantum Probes for Optimal Thermometry,
\href{https://doi.org/10.1103/PhysRevLett.114.220405}
{Phys. Rev. Lett. \textbf{114}, 220405 (2015).}
DOI: \href{https://doi.org/10.1103/PhysRevLett.114.220405}{https://doi.org/10.1103/PhysRevLett.114.220405} 

\bibitem{54fermometer}M. T. Mitchison, T. Fogarty, G. Guarnieri,
S. Campbell, T. Busch, and J. Goold, \textit{In Situ} Thermometry
of a Cold Fermi Gas via Dephasing Impurities, 
\href{https://doi.org/10.1103/PhysRevLett.125.080402}
{Phys. Rev. Lett. \textbf{125},
080402 (2020).}
DOI: \href{https://doi.org/10.1103/PhysRevLett.125.080402}{https://doi.org/10.1103/PhysRevLett.125.080402}

\bibitem{54.1therm-via-strong-coupling}L. A. Correa, M. Perarnau-Llobet,
K. V. Hovhannisyan, S. Hernandez-Santana, M. Mehboudi, and A. Sanpera,
Enhancement of low-temperature thermometry by strong coupling, 
\href{https://doi.org/10.1103/PhysRevA.96.062103}
{Phys. Rev. A \textbf{96}, 062103 (2017).}
DOI: \href{https://doi.org/10.1103/PhysRevA.96.062103}{https://doi.org/10.1103/PhysRevA.96.062103} 

\bibitem{55ancilla-assisted-thermometry}A. H. Kiilerich, A. De Pasquale,
and V. Giovannetti, Dynamical approach to ancilla-assisted quantum
thermometry, 
\href{https://doi.org/10.1103/PhysRevA.98.042124}
{Phys. Rev. A \textbf{98}, 042124 (2018).}
DOI: \href{https://doi.org/10.1103/PhysRevA.98.042124}{https://doi.org/10.1103/PhysRevA.98.042124} 

\bibitem{56collisional-thermom}S. Seah, S. Nimmrichter, D. Grimmer,
J. P. Santos, V. Scarani, and G. T. Landi, Collisional Quantum Thermometry,
\href{https://doi.org/10.1103/PhysRevLett.123.180602}
{Phys. Rev. Lett. \textbf{123}, 180602 (2019).} DOI: \href{https://doi.org/10.1103/PhysRevLett.123.180602}{https://doi.org/10.1103/PhysRevLett.123.180602}  

\bibitem{57Karen-CoarseGr-thermometry}K. V. Hovhannisyan, M. R. Jorgensen,
G. T. Landi, A. M. Alhambra, J. B. Brask, and Marti Perarnau-Llobet,
Optimal Quantum Thermometry with Coarse-grained Measurements,
\href{https://doi.org/10.1103/PRXQuantum.2.020322}
{PRX Quantum \textbf{2},
020322 (2021).}
DOI: \href{https://doi.org/10.1103/PRXQuantum.2.020322}{https://doi.org/10.1103/PRXQuantum.2.020322}  

\bibitem{60.1Aicki-batteries}R. Alicki and M. Fannes, Entanglement
boost for extractable work from ensembles of quantum batteries, 
\href{https://doi.org/10.1103/PhysRevE.87.042123}
{Phys. Rev. E \textbf{87}, 042123 (2013).}
DOI: \href{https://doi.org/10.1103/PhysRevE.87.042123}{https://doi.org/10.1103/PhysRevE.87.042123}

\bibitem{60.2Ext-work-from-correlations}M. Perarnau-Llobet, K. V.
Hovhannisyan, M. Huber, P. Skrzypczyk, N. Brunner, and A. Acin, Extractable
Work from Correlations, 
\href{https://doi.org/10.1103/PhysRevX.5.041011}
{Phys. Rev. X\textbf{ 5}, 041011 (2015).}
DOI: \href{https://doi.org/10.1103/PhysRevX.5.041011}{https://doi.org/10.1103/PhysRevX.5.041011} 

\bibitem{60.3Marti-most-energetic-passive-states}M. Perarnau-Llobet,
K. V. Hovhannisyan, M. Huber, P. Skrzypczyk, J. Tura, and A. Acin,
Most energetic passive states, 
\href{https://doi.org/10.1103/PhysRevE.92.042147}
{Phys. Rev. E \textbf{92}, 042147 (2015).}
DOI: \href{https://doi.org/10.1103/PhysRevE.92.042147}{https://doi.org/10.1103/PhysRevE.92.042147} 

\bibitem{60.4Gaussian-work-extr}E. G. Brown, N. Friis, and M. Huber,
Passivity and practical work extraction using Gaussian operations,
\href{https://doi.org/10.1088/1367-2630/18/11/113028}
{New J. Phys. \textbf{18}, 113028 (2016).}
DOI: \href{https://doi.org/10.1088/1367-2630/18/11/113028}{https://doi.org/10.1088/1367-2630/18/11/113028}

\bibitem{60.5Comment_on_PD}For a graphical characterization of this
condition see the diagrams developed in Ref. \citep{40.1Raam-PD}. 

\bibitem{43.1random-unitaries}K. M. R. Audenaert, and S. Scheel,
On random unitary channels, 
\href{https://doi.org/10.1088/1367-2630/10/2/023011}
{New J. Phys. \textbf{10}, 023011 (2008).}
DOI: \href{https://doi.org/10.1088/1367-2630/10/2/023011}{https://doi.org/10.1088/1367-2630/10/2/023011} 

\bibitem{60Nielsen-majorization}M. A. Nielsen, \textit{An introduction
to majorization and its applications to quantum mechanics}, Lecture
Notes, Department of Physics, Univesity of Queensland, Queensland
4072, Australia (2002). 

\bibitem{61Unital-maps}J. Watrous, \textit{The Theory of Quantum
Information} (Cambridge University Press, 2018). DOI: \href{https://doi.org/10.1017/9781316848142}{https://doi.org/10.1017/9781316848142}  

\bibitem{61.1.1a}J. Kolody\'{n}ski, \textit{Precision bounds in noisy
quantum metrology}, Ph.D. thesis, University of Warsaw (2015), \href{https://arxiv.org/abs/1409.0535}{arXiv:1409.0535v2 }. 

\bibitem{61.2a}We note that although $\textrm{max}_{U_{Pe}}|\partial_{\beta}q_{1}^{P}|=\textrm{max}\left\{ \bigl|\textrm{min}_{U_{Pe}}\partial_{\beta}q_{1}^{P}\bigr|,\textrm{max}_{U_{Pe}}\partial_{\beta}q_{1}^{P}\right\} $
we can restrict ourselves to the maximization of $\partial_{\beta}q_{1}^{P}$.
First, probability conservation $\partial_{\beta}q_{1}^{P}=-\partial_{\beta}q_{2}^{P}$
implies that $\bigl|\textrm{min}_{U_{Pe}}\partial_{\beta}q_{1}^{P}\bigr|=\textrm{max}_{U_{Pe}}\partial_{\beta}q_{2}^{P}$.
Since the maximum $\textrm{max}_{U_{Pe}}\partial_{\beta}q_{2}^{P}$
is taken over \textit{all} the unitaries $U_{Pe}$, it is equivalent
to first apply the local permutation $|1_{P}\rangle\leftrightarrow|2_{P}\rangle$
and then maximize over $U_{Pe}$. However, this permutation is also
equivalent to the label exchange $q_{1}^{P}\leftrightarrow q_{2}^{P}$,
which yields $\textrm{max}_{U_{Pe}}\partial_{\beta}q_{2}^{P}=\textrm{max}_{U_{Pe}}\partial_{\beta}q_{1}^{P}$.
Accordingly, $\textrm{max}_{U_{Pe}}|\partial_{\beta}q_{1}^{P}|=\textrm{max}_{U_{Pe}}\partial_{\beta}q_{1}^{P}$. 


\bibitem{59Majorization}A. W. Marshall, I. Olkin, and B. C. Arnold,
\textit{Inequalities: theory of majorization and its applications}
(Springer, 1979). 
DOI: \href{https://doi.org/10.1007/978-0-387-68276-1}{https://doi.org/10.1007/978-0-387-68276-1} 

\bibitem{Horn}A. Horn, Doubly stochastic matrices and the diagonal
of a rotation matrix, 
\href{https://doi.org/10.2307/2372705 }
{Am. J. Math.\textbf{76}, 620 (1954).} 
DOI: \href{https://doi.org/10.2307/2372705}{https://doi.org/10.2307/2372705} 
\end{thebibliography}

\end{document}